\documentclass[%
 reprint,
 superscriptaddress,
preprintnumbers,
nofootinbib,
 amsmath,amssymb,
 aps,
]{revtex4-1}

\usepackage{amsmath}
\usepackage{amssymb}
\usepackage{amsthm}
\usepackage{MnSymbol}
\usepackage{nicefrac}
\usepackage{amsfonts}
\usepackage{dsfont}
\usepackage{slashed}

\usepackage{placeins}

\usepackage[markup=nocolor]{changes}
\usepackage{natbib}

\usepackage{graphicx}

\usepackage{dcolumn}
\usepackage{bm}

\usepackage{color}

\usepackage[normalem]{ulem}

\newcommand{\bea}{\begin{eqnarray}}
\newcommand{\eea}{\end{eqnarray}}
\newcommand{\be}{\begin{equation}}
\newcommand{\ee}{\end{equation}}

\usepackage{url} 
\usepackage[colorlinks=true, linkcolor=blue]{hyperref}

\begin{document}

\title{Asymptotically safe quantum gravity and its phenomenology -- a review}
 
 \author{Astrid Eichhorn}
   \email{eichhorn@thphys.uni-heidelberg.de}
\affiliation{Institute for Theoretical Physics, Heidelberg University, Philosophenweg 16, 69120 Heidelberg, Germany}

\begin{abstract}
Asymptotically safe quantum gravity is an approach to quantum gravity. It is based on the premise that quantum field theory can describe the quantum nature of gravity in our universe. At its core lies quantum scale symmetry. 
This review provides an introduction to the key ideas of the approach and surveys the current status of the field. Over the last years, the field has taken large strides towards an increasingly realistic setting: First, compelling evidence for quantum scale symmetry exists in four-dimensional, Euclidean, pure gravity, establishing the Reuter fixed point robustly. Second, matter fields, including the Standard Model as well as beyond-Standard-Model-candidates, have been studied in depth, with increasingly conclusive evidence for quantum scale symmetry. Most recently, the final gap to a realistic description of quantum gravity is being closed, because Lorentzian spacetime signature can now be accounted for.\\
As a consequence of quantum scale symmetry in the ultraviolet, the approach is highly predictive at all scales. This review discusses the physics of asymptotic safety across all scales. Predictive power for particle physics, black holes and cosmology provides a clear pathway to confronting quantum gravity with current and near-future observations.\\
The review closes by discussing the connection to other approaches to quantum gravity. It advocates the perspective that such connections between approaches may lead us to an understanding of universal physical features of quantum gravity.
\end{abstract}

\maketitle

\tableofcontents

\newpage
\section{Motivation for and aim of this review}
Asymptotically safe gravity is an increasingly popular approach to quantum gravity, for which a number of reviews \cite{Niedermaier:2006wt,Reuter:2012id,Eichhorn:2018yfc,Pereira:2019dbn,Pawlowski:2020qer,Eichhorn:2022jqj,Eichhorn:2023xee} as well as books \cite{Percacci:2017fkn,Reuter:2019byg} already exist. However, the field is fast-evolving and while different aspects of the field have relatively recently been reviewed individually \cite{Pawlowski:2020qer,Eichhorn:2022gku,Eichhorn:2022bgu,Wetterich:2022ncl,Knorr:2022dsx,Morris:2022btf,Martini:2022sll,Pawlowski:2023gym,Saueressig:2023irs,Platania:2023srt,Bonanno:2024xne}, several new developments have since taken place, motivating an up-to-date review that covers the research field as a whole. Three among several examples of relatively recent important progress are the following: The inclusion of Lorentzian signature, long a stumbling block, is now possible. The understanding of ``running'' couplings and their physical implications, in terms of the predictive power of the theory on the one hand, and on the scaling properties of observables, has been clarified.  The derivation of the physical properties of asymptotic safety has given rise to new, promising results, in particular for particle physics.

Thus, the present review aims at providing an introduction to the main physics aspects, key conceptual ideas and phenomenological consequences, as well as open questions and current frontiers in the field. It is meant to complement previous reviews that provide a much more in-depth introduction to calculational and technical aspects \cite{Pawlowski:2020qer,Knorr:2022dsx,Morris:2022btf,Martini:2022sll,Pawlowski:2023gym,Saueressig:2023irs} and which are, together with the lecture notes \cite{Eichhorn:2020mte,Reichert:2020mja,Basile:2024oms}, recommended for readers whose aim is to perform concrete calculations in asymptotic safety. 

The present review is also meant as an update of the older introduction to the research field in \cite{Eichhorn:2018yfc}. We focus on recent developments and new results and also point out topics on which the understanding has changed significantly. Aspects of the phenomenology are discussed also in the more specialized reviews on the interplay with matter \cite{Eichhorn:2022gku}, and the status of black-hole research \cite{Eichhorn:2022bgu,Platania:2023srt} as well as cosmology \cite{Bonanno:2017pkg,Wetterich:2022ncl,Bonanno:2024xne}.
In addition, similarly to \cite{Bonanno:2020bil}, the present review addresses common questions about asymptotic safety as well as common misconceptions or confusions. 

This review is organized in eight main sections. Sec.~\ref{sec:Basics} provides an introduction to the main idea of asymptotic safety and its predictive power. The methods used to investigate it are surveyed in Sec.~\ref{sec:methods}. In Sec.~\ref{sec:FP}, we discuss the evidence for the existence of an asymptotically safe fixed point. 
Sec.~\ref{sec:technicalities_matter} focuses on a few selected, more technical aspects, that are particularly relevant.
Sec.~\ref{sec:FAQ} lists a number of common questions on asymptotic safety and provides (partial) answers. Sec.~\ref{sec:pheno} discusses the physical implications of asymptotic safety at all scales and for gravitational physics -- including black holes and cosmology -- as well as particle physics. Finally, Sec.~\ref{sec:relations} highlights a topic of increasing popularity, namely the relations to other approaches to quantum gravity and the motivation for searching for such relations. Throughout all sections, open questions and current frontiers are pointed out and are also summarized in the final Sec.~\ref{sec:frontiers}.

\section{Guide for the busy reader}\label{sec:guide}
A linear reading of this review from the first to the last page is one, but not the only way in which this review can be read.  
Instead a very short overview of the content of the section (\emph{...where we...}) provides support for readers interested only in particular topics. For aspects that are covered in more depth in another section, references to these sections are added, so the reader can follow the cross-references within the review.\\

Here, we propose four ``reading tracks'' in addition to a full reading of the entire review.\\

For readers interested in the \emph{phenomenology of asymptotic safety in particle physics}, we suggest a proper reading of Sec.~\ref{sec:Basics}, a quick perusal of Sec.~\ref{sec:FP} with a focus on subsec.~\ref{sec:mattermatters} and subsec.~\ref{sec:summing_up}. From Sec.~\ref{sec:technicalities_matter} this track takes in  \ref{sec:nearpert} and from Sec.~\ref{sec:FAQ},  \ref{sec:notionsofrunning}. Finally from Sec.~\ref{sec:pheno} the parts \ref{sec:particlephysics}, \ref{sec:particlescattering} and \ref{sec:principledparameterized}.\\

For readers interested in the \emph{phenomenology of asymptotic safety from the perspective of cosmology}, a suggested track includes Sec.~\ref{sec:Basics}, a quick perusal of Sec.~\ref{sec:FP} with a focus on subsec.~\ref{sec:mattermatters} and subsec.~\ref{sec:summing_up}. From Sec.~\ref{sec:technicalities_matter}, this track includes \ref{sec:nearpert} and from Sec.~\ref{sec:FAQ}, subsec.~\ref{sec:MG}. From Sec.~\ref{sec:pheno}, it includes the subsections \ref{sec:dm}, \ref{sec:cosmo} and \ref{sec:principledparameterized}.\\

For readers interested in understanding \emph{how asymptotic safety fits into our broader understanding of quantum gravity} within the different approaches, the reading track starts with Sec.~\ref{sec:Basics}, Sec.~\ref{sec:methods} and Sec.~\ref{sec:FP}. From Sec.~\ref{sec:technicalities_matter}, subsec.~\ref{sec:local_coarse_graining} and from Sec.~\ref{sec:FAQ}, subsec.~\ref{sec:MG} are recommended. Finally, all of Sec.~\ref{sec:relations} is included in this reading track.\\

For readers interested in the \emph{theoretical foundations of asymptotic safety as a theory of gravity}, the track includes Sec.~\ref{sec:Basics}, Sec.~\ref{sec:methods} and Sec.~\ref{sec:FP}, followed by Sec.~\ref{sec:technicalities_matter} and Sec.~\ref{sec:FAQ}. Finally, this track includes Sec.~\ref{sec:BH}.

\section{Theoretical basics of asymptotic safety in gravity}\label{sec:Basics}
\emph{...where we introduce asymptotic safety and how this property may solve the problems of a perturbative quantization of gravity within quantum field theory. }\\

Some of the most fascinating questions about nature can only be answered, once we understand quantum gravity. These include, for example: What is the origin of the universe? What truly happens inside the core of a black hole? What is the microscopic nature of elementary particles and what shapes their properties? 

To develop a quantum theory of gravity, we remember two established facts: First, our understanding of gravity in its classical regime is based on a field theory of the metric, namely General Relativity (GR). Currently, there is no observational evidence that is strictly incompatible with GR. Second, our understanding of the other fundamental forces (strong, electroweak) is based on a local quantum field theory, the Standard Model (SM) of particle physics. Currently, there is no observational evidence that is strictly incompatible with the local quantum field theory framework. Well-established physics beyond the SM, such as an explanation of dark matter, neutrino masses or the origin of matter-antimatter asymmetry, has various possible explanations that fit straightforwardly into the overarching framework for our understanding of nature, namely based on (quantum) field theories.\\
These two facts strongly suggest that our first attempt to understand quantum gravity should be as a quantum field theory. Only once this possibility is irrefutably ruled out, the necessity for an alternative framework in our understanding of nature  is established.\footnote{In many settings in physics, there are distinct frameworks that are physically equivalent. Thus, this argument for a quantum-field theoretic understanding of gravity does not automatically constitute an argument against the development of other frameworks -- we come back to the idea of physical universality between distinct quantum-gravity approaches in  Sec.~\ref{sec:relations}.}\\

Within quantum field theory, perturbative quantization is the simplest idea to try and thus, we start our review from perturbative quantum gravity and its problems and introduce asymptotic safety from there. Throughout, we focus on a covariant, path-integral approach, because this seems more naturally suited to quantum theory of spacetime which respects general covariance.\footnote{Just as in non-gravitational theories, there may well be an equivalent formulation within a canonical framework, we come back to this in Sec.~\ref{sec:LQG}.}

\subsection{Which two problems of perturbative quantum gravity does asymptotic safety set out to solve?}\label{sec:problems_of_perturbation_theory}
\emph{\dots where we review the two problems that arise in the perturbative quantization of gravity, namely perturbative non-renormalizability (aka loss of predictivity) and the growth of scattering cross-sections with energy. These problems have the same origin, namely the negative mass-dimension of the Newton coupling.}\\

The two problems discussed we will discuss below are of course not the only two problems that (asymptotically safe) quantum gravity sets out to solve, but, from the perspective of perturbative quantization, these are two critical problems that any theoretically viable theory must find some way of addressing. Therefore, they constitute a useful starting point for the construction of a quantum theory of gravity.

In a perturbative quantization of General Relativity (GR), we start from the $d$-dimensional Einstein-Hilbert action 
\begin{equation}
S_{\rm EH} = \frac{1}{16\pi \, G_N}\int d^dx\sqrt{-g}R,
\end{equation}
 and perturb the metric around a background $\bar{g}_{\mu\nu}$, typically chosen to be the flat spacetime metric $\eta_{\mu\nu}$, through a, typically linear, split 
 \begin{equation}
 g_{\mu\nu} = \eta_{\mu\nu}+ \sqrt{16 \pi\, G_N}h_{\mu\nu}.
 \end{equation}
This enables the formulation of a gauge-fixed path integral for $h_{\mu\nu}$ within the standard framework for gauge theories. The factor of $\sqrt{16 \pi G_N}$ ensures that $h_{\mu\nu}$ has mass-dimension $(d-2)/2$, just like the other bosonic fields in the Standard Model. After the split, Feynman rules for gravitons, encoded in $h_{\mu\nu}$, can be derived from $S_{\rm EH} $.\\

This theory is perturbatively non-renormalizable, i.e., requires an infinite number of free parameters to remove all ultraviolet divergences. This amounts to a loss of predictivity.
We can already see this problem from the superficial degree of divergence $\mathcal{D}$ of Feynman diagrams. It is given by 
\begin{equation}
\mathcal{D} = (d-2)L+2, 
\end{equation}
where $L$ denotes the loop-order of a diagram. In $d=4$, which is the dimension of most interest, the superficial degree of divergence increases with each successive loop order. We rely on diffeomorphism symmetry to identify the divergent diagrams with curvature invariants. 
In cutoff-regularization, where loop momenta are restricted to lie below $\Lambda_{\rm cutoff}$, the highest divergence of one-loop diagrams is $\sqrt{-g}\Lambda_{\rm cutoff}^4$.  The highest power of $\Lambda_{\rm cutoff}$ occurs, when all factors of momentum at each vertex in the Feynman diagram are given by loop momenta.  Thus, by diffeomorphism symmetry, this introduces the cosmological constant as a counterterm. A decrease of the power of $\Lambda_{\rm cutoff}$ must be compensated by an increase in the external momenta of a diagram. The curvature invariant corresponding to the next-lower divergence, $\Lambda_{\rm cutoff}^2$, occurring in one-loop diagrams with two external momenta, is $\sqrt{-g}R$. This leads to a renormalization of $G_N$. The problematic divergences are the lowest-order, logarithmic ones, because they require the highest power of curvature invariants. These logarithmic divergences are also what is visible of divergences when using dimensional regularization around $d=4$, because this scheme projects onto logarithmic divergences.\footnote{The power-law divergences are visible as poles around other dimensionalities than $d=4$, i.e., they can in principle be recovered with dimensional regularization.}
At one-loop order, these divergences were first found in \cite{tHooft:1974toh} and read $\sqrt{-g}R^2 \ln(\Lambda_{\rm cutoff})$,  $\sqrt{-g}R_{\mu\nu}R^{\mu\nu} \ln(\Lambda_{\rm cutoff})$ and $\sqrt{-g}R_{\mu\nu\kappa\lambda}R^{\mu\nu\kappa\lambda} \ln(\Lambda_{\rm cutoff})$. They can be reduced to two independent divergences using the Gauss-Bonnet theorem. The remaining terms require counterterms not of the form of the terms in the original Lagrangian.\footnote{They can be set to zero on shell in the absence of a cosmological constant and of matter. Thus, a pure theory of gravity on a flat background is perturbatively renormalizable in $d=4$, but a description of quantum gravity applicable to our universe is not.}

Similarly, logarithmic divergences are present at each loop order and are associated to terms with $2L+2$ derivatives, e.g.,  curvature-cubed term $\sqrt{-g}R_{\mu\nu\kappa\lambda}R^{\kappa\lambda\rho \sigma}R_{\rho\sigma}^{\,\,\,\,\,\,\,\mu\nu}$ at two loops \cite{Goroff:1985sz,vandeVen:1991gw} and so on. Renormalization requires us to introduce the corresponding higher-derivative terms as counterterms with free couplings, and for $L \rightarrow \infty$ we end up with a theory with infinitely many terms in the dynamics, each of which has a coupling that is a free parameter. As a consequence, the theory looses predictivity, because there are infinitely many free parameters.\footnote{An effective field theory treatment of this theory is still possible, making predictions with finite precision for scales below the Planck scale, see, e.g., \cite{Donoghue:1993eb,Donoghue:1994dn,Donoghue:2022eay}.}\newline

The first challenge of  quantum gravity therefore is to discover a principle which restores predictivity. 
In Sec.~\ref{sec:predpower}, we explain how asymptotic safety  fixes the values of couplings not only in the UV, but also imposes relations between the couplings in the IR, thereby restoring predictivity at all scales.\newline

Perturbative non-renormalizability originates in the negative mass-dimension of the Newton coupling. It is therefore closely related to another problem, namely that the (naive) interaction strength of gravitons grows without bounds. This can be seen from the argument that a dimensionless measure of the interaction strength of gravitons requires a product of Newton coupling and an appropriate quantity with units of energy squared, e.g., the center-of-mass-energy for $s$-channel scattering. Thus, in an $s$-channel process, $G_N \cdot s$ is the dimensionless combination that enters the scattering cross-section. This quantity clearly grows quadratically with energy and becomes of order 1 at the Planck scale. One can see this in the example of $e^{+}e^-\rightarrow \mu^+\mu^-$ scattering, which only has two $s$-channel contributions at tree level. One contribution comes from the exchange of a photon and is proportional to $\alpha^2 =\left( \frac{e^2}{4\pi}\right)^2$, with the QED coupling $e$. The other comes from the exchange of a graviton and is proportional to $G_N^2\, s^2$. In total, the tree-level cross-section reads
\begin{equation}
\sigma(s)= \frac{4 \pi \alpha^2}{3 s} + \frac{\pi}{20}G_N^2 \, s.
\end{equation}
The QED-contribution decreases with $s$, and satisfies the Froissart bound \cite{Froissart:1961ux}, according to which cross-sections in unitary, Lorentz invariant quantum field theories with microcausality and a mass-gap in four dimensions cannot increase faster than $\ln(s)$. In contrast, the gravitational contribution, which is negligible at typical particle-accelerator energies, grows linearly with $s$ and therefore not only dominates at high $s$, but also violates the Froissart bound. It is strictly speaking not clear whether or not one should expect the Froissart bound to hold \cite{Basile:2024oms}, but one may view it as the most conservative assumption one can make about a quantum-field theoretic description of gravity.\\
Similarly, other graviton-mediated scattering cross-sections, because they scale with $G_N^n$, $n\geq2$, have a problematic growth with energy.\newline

The second challenge of quantum gravity is therefore to discover a dynamical principle which limits the growth of the gravitational interaction strength with energy. We note that such a dynamical principle may also be expected to resolve curvature singularities -- a classic sign of the incompleteness of GR and key motivation for quantum gravity at a more non-perturbative level.\newline

In view of these two problems of perturbative quantum gravity, one may consider it necessary to modify or totally abandon the QFT framework for quantum gravity, or postulate that the (trans-) planckian regime is necessarily highly non-perturbative.
There is, however, also a more conservative option and this is asymptotically safe gravity. Asymptotic safety takes the existence of higher-order terms in the dynamics seriously. These higher-order terms can completely modify the behavior of scattering cross-sections (and other observables) at (trans-) planckian scales. Thus, there is a possibility that the high-energy behavior of observables becomes tame. At the same time, asymptotic safety supplies a principle that restores predictivity, such that infinitely many couplings are present in the theory, but they are all calculable as functions of a handfull of free parameters.

\subsection{What is asymptotic safety?}\label{sec:WhatisAS}
\emph{\dots where we explain the key concepts underlying asymptotic safety. Asymptotic safety entails quantum scale symmetry, i.e., a symmetry under changes of the scale that arises at the quantum level, generated by quantum fluctuations. We point out that it is crucial to carefully distinguish between distinct notions of scale dependence and in particular distinguish dependence on physical scales from dependence on the Renormalization Group scale.}\\

Asymptotic safety must realize two mechanisms:  first, a mechanism to render a QFT with infinitely many interaction terms predictive, by reducing the number of free parameters to a finite one\footnote{More precisely, quantum fluctuations always generate all possible interaction terms compatible with the symmetries. In perturbatively renormalizable and asymptotically free theories, the prefactors of these terms are calculable in terms of the couplings of the perturbatively renormalizable interactions. We are looking for a generalization of this mechanism.}; second, a mechanism to limit the growth of observables (e.g., scattering cross-sections, but also tidal gravitational forces etc.) with energy (or other appropriate physical scales), such that the theory remains physically viable at all scales. While we have motivated this second aspect using the example of scattering cross-sections, physical viability also requires black-hole spacetimes as well as cosmological spacetimes to be free of divergences in observable quantities and other unphysical behavior.

In research on asymptotic safety, the first mechanism, and the resulting predictive power, is reasonably well understood and quantified for gravity, see Sec.~\ref{sec:FP}. In contrast, the second mechanism, which produces well-behaved observables as a function of physical scales, has come into focus more recently, e.g., \cite{Knorr:2019atm,Draper:2020bop,Draper:2020knh,Knorr:2022lzn,Pastor-Gutierrez:2024sbt,Chiesa:2026tlz,Knorr:2026vax}, see Sec.~\ref{sec:particlescattering}.\\

\begin{figure*}[!htbp]
\begin{center}
\includegraphics[width=0.8\linewidth, clip=true, trim=1.5cm 4cm 3cm 16cm]{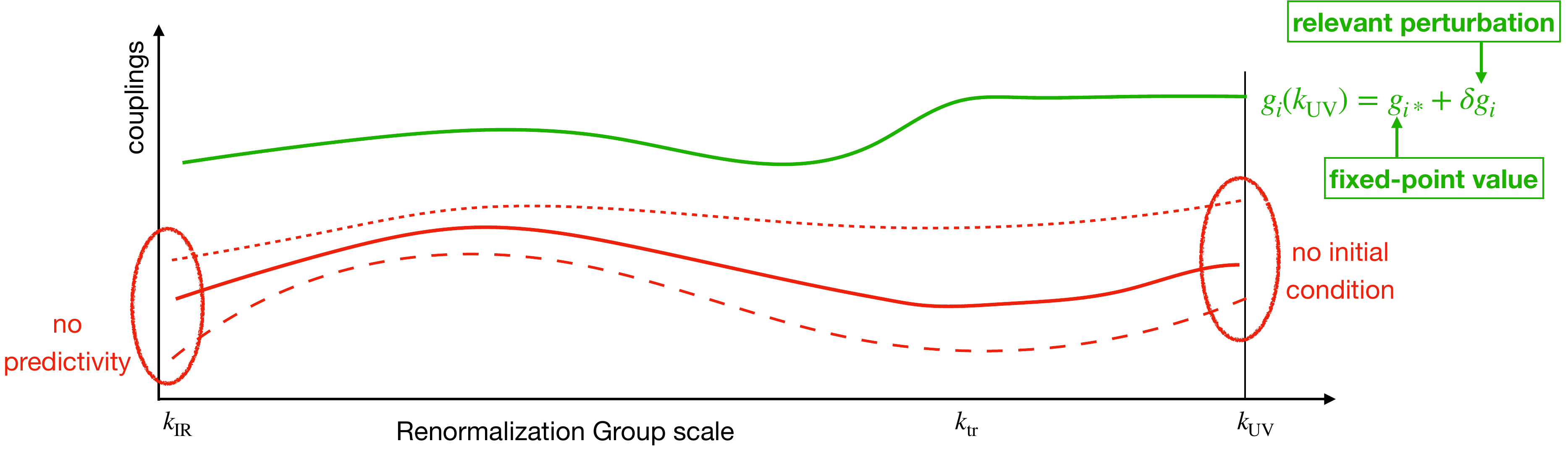}
\end{center}
\caption{\label{fig:prediction_illustration}We sketch how beta functions suffice to translate a given UV value of a coupling into its IR value. However, without the lack of a specific initial condition, there is no predictive power for the IR value (we assume the absence of IR fixed points). This changes if one starts from a UV fixed point, because then all initial conditions are determined by the set of relevant perturbations $\delta g_i$ of the fixed-point value. The choice of the amplitude of a relevant perturbation translates into the scale $k_{\rm tr}$, at which the couplings depart from a scale symmetric regime, in which they are constant.}
\end{figure*}

{\bf Effective action:}
Both mechanisms rely on scale symmetry being (approximately) realized at high energy/curvature scales. A useful quantity in QFT in terms of which these mechanisms can be discussed, is the effective action $\Gamma[\langle g_{\mu\nu}\rangle]$.
 Here $\langle g_{\mu\nu}\rangle$ is the expectation value of the metric. Because we are considering a quantum theory, a single classical field configuration of the metric, $g_{\mu\nu}$ is physically meaningless. Instead, all (on-shell and off-shell) field configurations are included in the path integral and only expectation values such as $\langle g_{\mu\nu}\rangle$ are physically meaningful.
The effective action is not a concept specific to gravity, but instead exists in any QFT and the discussion below holds analogously in non-gravitational QFTs, where the effective action is also used extensively.
The effective action $\Gamma[\langle g_{\mu\nu}\rangle]$ provides a description of a physical system, in which quantum effects are accounted for. Schematically, $\Gamma[\langle g_{\mu\nu}\rangle]$ is the result of performing the path integral
\begin{equation}
 \int \mathcal{D}g_{\mu\nu}e^{i\, S[g_{\mu\nu}]} \leadsto e^{i\Gamma[\langle g_{\mu\nu}\rangle]}, 
\end{equation}
where $S$ is the microscopic action and the arrow $\leadsto$ indicates that this is not an exact equality, but a schematic relation between $\Gamma$ and $S$. We provide the precise definition of $\Gamma$ below and here explain its role at a more heuristic level.
From this expression, it is plausible that the effective action plays a similar role to the classical action, just for the expectation value of the field, not a classical field configuration: it provides the equations of motion
\begin{equation}
\frac{\delta}{\delta \langle g_{\kappa\lambda}\rangle}\Gamma[\langle g_{\mu\nu}\rangle]=0.\label{eq:qmeom}
\end{equation}
These equations of motion for the expectation value $\langle g_{\mu\nu}\rangle$ arise from integrating over all (on-shell and off-shell) field configurations of the metric (modulo diffeomorphisms) in the path integral. In practice, given an effective action $\Gamma[\langle g_{\mu\nu}\rangle]$, the resulting equations of motion can be treated like classical equations of motion, i.e., $\langle g_{\mu\nu}\rangle$ is a particular field configuration that solves the differential equation given by Eq.~\eqref{eq:qmeom}, supplemented by the appropriate boundary/initial conditions for the system of interest.

The effective action also serves as the generator of all one-particle irreducible correlation functions, entering, e.g., scattering cross-sections, such that all physical information on the theory can be extracted from it.
To achieve this, formally, $\Gamma[\langle g_{\mu\nu}\rangle]$ is the Legendre transform of $\ln Z[J]$, where $Z[J]$ is the generating functional, 
\begin{equation}
Z[J] = \int \mathcal{D}g_{\mu\nu}\, e^{iS[g_{\mu\nu}]+ i\int d^dx \sqrt{-g}g_{\mu\nu}J^{\mu\nu}}.
\end{equation}
To render the generating functional well-defined, we must introduce ultraviolet (UV) and infrared (IR) regularizations.  From $Z[J]$, we define the effective action through a Legendre transform
\begin{equation}
\Gamma[\langle g_{\mu\nu}\rangle]=\underset{J}{\rm sup}\left(  -i \ln Z[J] -\int d^dx \sqrt{-{\rm det} \langle g_{\kappa\lambda}\rangle}\, J^{\mu\nu} \langle g_{\mu\nu}\rangle \right).\label{eq:effaction}
\end{equation}

The effective action in general contains all interactions compatible with the symmetries.\footnote{This is generically true, also when a theory is perturbatively renomalizable. In this case, all higher-order couplings are calculable functions of the couplings of perturbatively renormalizable interactions.} Because it is based on the path integral, all effective interactions are generated through quantum fluctuations, irrespective of which subsets of interactions with a given set of symmetries are included in the microscopic action. $\Gamma[\langle g_{\mu\nu}\rangle]$ thus contains infinitely many terms. For example, it contains higher-order curvature terms, which become dynamically important in settings with large curvature.
Each term in an expansion of the effective action is parameterized by a coupling.
\newline

{\bf Asymptotic-safety requirements:}
Asymptotic safety can be phrased as two requirements: First, $\Gamma[\langle g_{\mu\nu}\rangle]$ must be parameterized by only a finite number of free parameters that determine its couplings. Second, the correlators, which can be calculated from $\Gamma[\langle g_{\mu\nu}\rangle]$, must have the appropriate UV behavior, such that observables stay well-behaved in this limit. 

Both requirements are achieved through scale-symmetry and we explain below, how. We highlight that the two requirements pertain to different notions of scale. In older works, these two notions have sometimes been equated, which is not always correct \cite{Donoghue:2019clr,Bonanno:2020bil,Knorr:2026vax}.\\

{\bf Predictivity from scale symmetry:}
The first scale, commonly denoted $k$, tracks quantum fluctuations in the generating functional. It determines the scale of an IR cutoff, such that quantum fluctuations with momenta $p^2$ satisfying $k^2<p^2<\Lambda_{\rm cutoff}^2$ are integrated over in the path integral.\footnote{For $k^2$ to act as a cutoff on momenta, we must work in Euclidean signature. Recent progress on translating the Renormalization Group flow to Lorentzian signature is discussed in Sec.~\ref{sec:Lorentzian}. In addition, the momentum must be suitably generalized for the gravitational path integral, see Sec.~\ref{sec:local_coarse_graining}. The UV cutoff $\Lambda_{\rm cutoff}^2$ can be removed after renormalization.} This scale is the one relevant for predictivity: as $k$ changes, the couplings in the effective action, collectively denoted $\bar{g}_i$, evolve.
Their values in the IR limit $k_{\rm IR}\rightarrow 0$ follow from the Renormalization Group evolution, i.e., from integrating beta functions
\begin{equation}
\beta_{\bar{g}_i}=k\partial_k\, \bar{g}_i(k)\, \quad \Rightarrow \bar{g}_i(k_{\rm IR}) = \bar{g}_{i}(k_{\rm UV}) - \int_{k_{\rm IR}}^{k_{\rm UV}} \frac{dk}{k}\, \beta_{\bar{g}_i}.
\end{equation}
The IR value depends on the UV value $\bar{g}_i(k_{\rm UV})$ that is supplied. Thus, an infinite number of UV initial conditions $\bar{g}_i(k_{\rm UV})$ must be supplied, one for each coupling, cf.~Fig.~\ref{fig:prediction_illustration}. In a perturbative quantization of gravity, predictivity breaks down, because there is no relation between the initial data for the different couplings. 

\begin{figure}[!t]
\includegraphics[width=\linewidth,clip=true,trim=5cm 5cm 27cm 2cm]{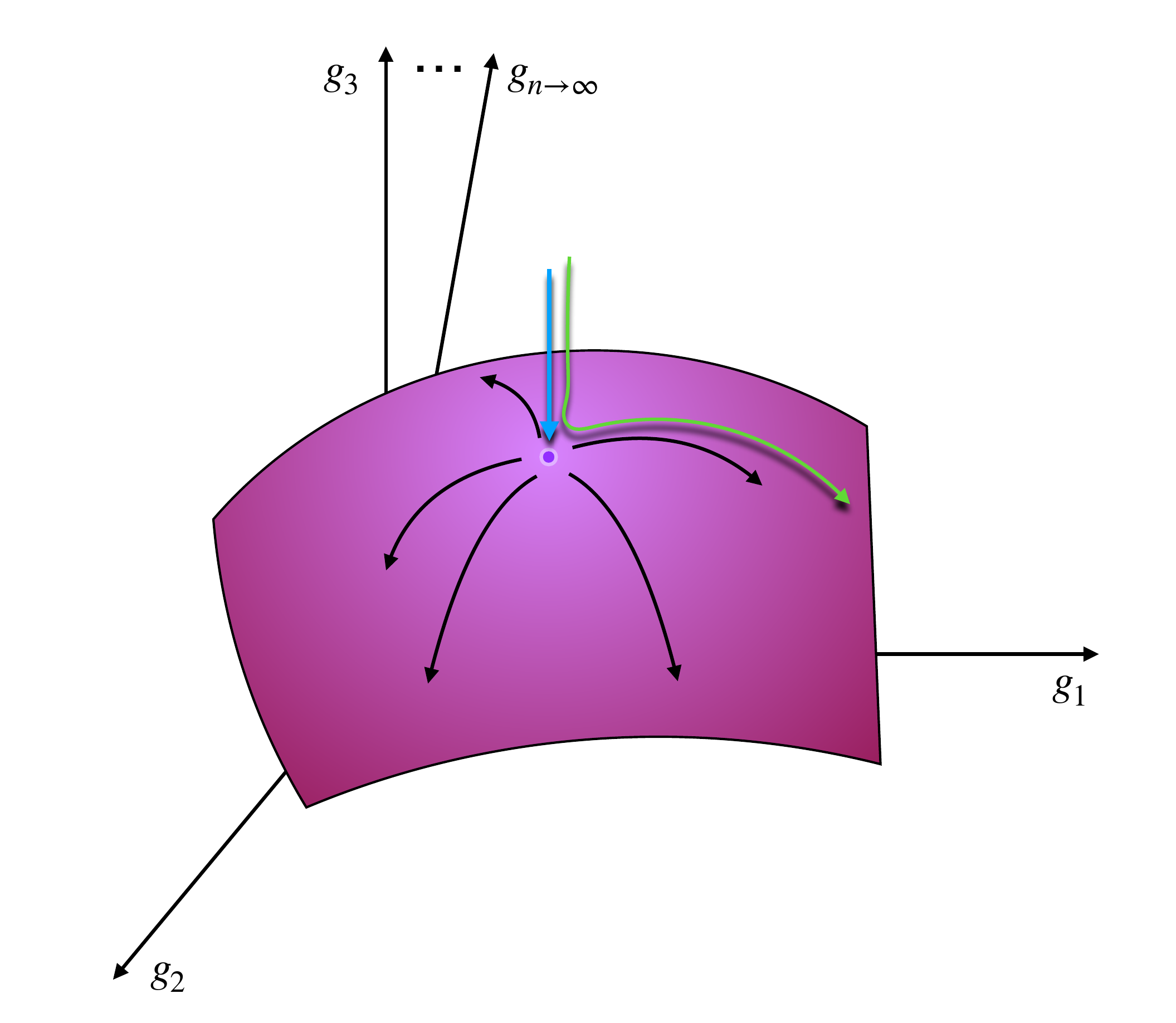}
\caption{\label{fig:critical_hypersurface}Illustration of the critical hypersurface (in purple) in the space of couplings. The fixed point (violet dot) has two relevant directions which generate the critical hypersurface. Black arrows indicate RG trajectories emanating from the fixed point. All other directions are irrelevant and thus constrained by the fixed-point requirement. The blue arrow indicates how the RG flow cannot leave the fixed point along such an irrelevant direction. The green trajectory is explained in Sec.~\ref{sec:eff_AS}.}
\end{figure}

To demand scale symmetry, we first have to remove all explicit scales from the system, i.e., for each coupling with non-vanishing mass dimension $d_{\bar{g}_i}= [\bar{g}_i]$, we define a dimensionless counterpart
\begin{equation}
g_i= \bar{g}_i \cdot k^{-d_{\bar{g}_i}}.
\end{equation}
Demanding scale symmetry solves the problem of predictivity as follows: at $k_{\rm UV}$, the couplings should be close to a fixed point of the beta functions, satisfying
\begin{equation}
\beta_{g_i}=0 \,\, \forall g_i\, \,{\rm at}\, \,g_{i}= g_{i\,\ast},
\end{equation}
where $g_{i\,\ast}$ denotes the fixed-point value. Thus, we obtain UV initial conditions that read
\begin{equation}
g_i(k_{\rm UV})= g_{i\, \ast} + \delta g_{i},
\end{equation}
where $\delta g_i$ is small and chosen to be a \emph{relevant perturbation of the fixed point}, cf.~Fig.~\ref{fig:prediction_illustration}. Intuitively, relevant perturbations of a fixed point are those along which the RG flow is driven away from the fixed point -- they act like ``source'' directions of the flow. In contrast, irrelevant perturbations are those for which the RG flow is driven back into the fixed point -- they act like ``sink'' directions of the flow, cf.~Fig.~\ref{fig:critical_hypersurface}. Because the irrelevant directions act like a ``sink'', the RG flow cannot depart from scale-symmetry along those directions. We discuss how to distinguish these two types of directions in Sec.~\ref{sec:predpower} and the evidence for a small, finite number of relevant perturbations in gravity in Sec.~\ref{sec:FP}.\\

{\bf Well-behaved observables from scale symmetry:}
The second scale with respect to which we require scale symmetry, is a physical scale. Whether this scale is, e.g.,  a curvature scale or momentum-scale scale depends on the physical system. The full effective action in gravity depends on curvatures and covariant derivatives, but, when considered in the flat space limit, the covariant derivatives give rise to momenta.
For the sake of simplicity, let us stay in the setup relevant for the calculation of scattering cross-sections, as discussed in Sec.~\ref{sec:problems_of_perturbation_theory}, i.e., we imagine taking the effective action and evaluating its flat-space limit, accounting for metric fluctuations about flat spacetime as our dynamical fields. When we derive Feynman rules not from a classical action with a finite number of terms, but from an effective action with infinitely many terms, there are many contributions to an $n$-point vertex. These have the same $n$ fields, but differ in their dependence on the momenta of the $n$ fields. We can summarize these in a momentum-dependent coupling
\begin{equation}
\bar{g}_i= \bar{g}_i(k,p_1,\dots,p_{n-1}),
\end{equation}
where we have explicitly indicated the RG scale dependence in the first argument and the dependence on physical momentum in the remaining arguments. Specifically for  the example we have given, a scattering vertex with $n$ fields, $g_i$ is a function of $n-1$ momenta due to momentum-conservation. The coupling can be made dimensionless by scaling with an appropriate power of any of the $n-1$ physical scales. 
Any physical observable depends on such dimensionless couplings. As long as all of these couplings remain finite at finite values of the physical scales, i.e., $p_i< \infty$, the observable stays bounded at all finite values of the scale. Scale symmetry achieves more: The dimensionless coupling is constant, and thus, the dimensionful coupling simply scales with physical scales according to its canonical dimension. Then, observables like scattering cross-sections scale according to their canonical dimension. Because a cross-section has units of area, in this regime it decreases quadratically with energy, such that consistency conditions like the Froissart bound can be satisfied.\\

{\bf In connection, the two requirements of scale symmetry ensure that i) the effective action is calculable in terms of a finite number of free parameters and ii) physical observables, which can all be derived from the effective action, remain bounded and well-behaved.}\footnote{More precisely, these requirements apply to the essential couplings, which are the ones that cannot be removed by a field redefinition. These are the only ones that enter observables, see \cite{Baldazzi:2021ydj,Baldazzi:2021orb}.}\\

In Weinberg's original discussion \cite{Weinberg:1980gg}, the two scale-dependences are equated. This is correct in many settings, in particular when the RG scale dependence is logarithmic and tracks the physical scale dependence. However, it is now understood that, in particular in a gravitational context, these scale dependences do not always track each other \cite{Donoghue:2019clr,Bonanno:2020bil,Buccio:2023lzo,Buccio:2024hys,Buccio:2025tci} and need to be distinguished from each other, as already done, e.g., in \cite{Christiansen:2014raa,Christiansen:2015rva,Denz:2016qks,Christiansen:2017bsy,Eichhorn:2018akn,Knorr:2021niv}, see Sec.~\ref{sec:notionsofrunning} for further details. Thus, the asymptotic-safety requirement, properly understood, actually consists out of two separate requirements.\\

In summary, asymptotically safe theories exhibit a UV regime determined by quantum scale symmetry. Scale symmetry means that dimensionless quantities are constant functions of the scale. To achieve classical scale symmetry, one sets dimensionful couplings to zero in the Lagrangian. Quantum scale symmetry provides a richer set of possibilities, because the dimensionless ratios of couplings have to be constant. Thus, dimensionful couplings scale towards zero or infinity, depending on their mass-dimensionality, but their dimensionless prefactor is constant. This can be achieved, if the contribution of quantum fluctuations to the scale dependence of couplings vanishes. \\

Because scale symmetry does not govern fundamental physics at scales that we have observational access to, such a scale-symmetric regime can only describe the UV regime of the theory. Distinct physical scales (e.g., the Higgs mass, the neutrino masses...) emerge in the IR regime of the theory, because we set initial conditions for the RG flow along relevant directions, slightly away from the fixed point at a finite UV scale. Thus, scale symmetry holds in the UV, but not in the IR. Predictive power remains for the IR, because of the scale-symmetric regime in the UV.

\subsection{Where does the predictive power of asymptotic safety come from?}\label{sec:predpower}
\emph{...where we review the notion of relevant and irrelevant directions and their respective connection to free parameters and predictable couplings. }\\

\begin{figure}[!t]
\includegraphics[width=\linewidth,clip=true,trim=4cm 24cm 47cm 3cm]{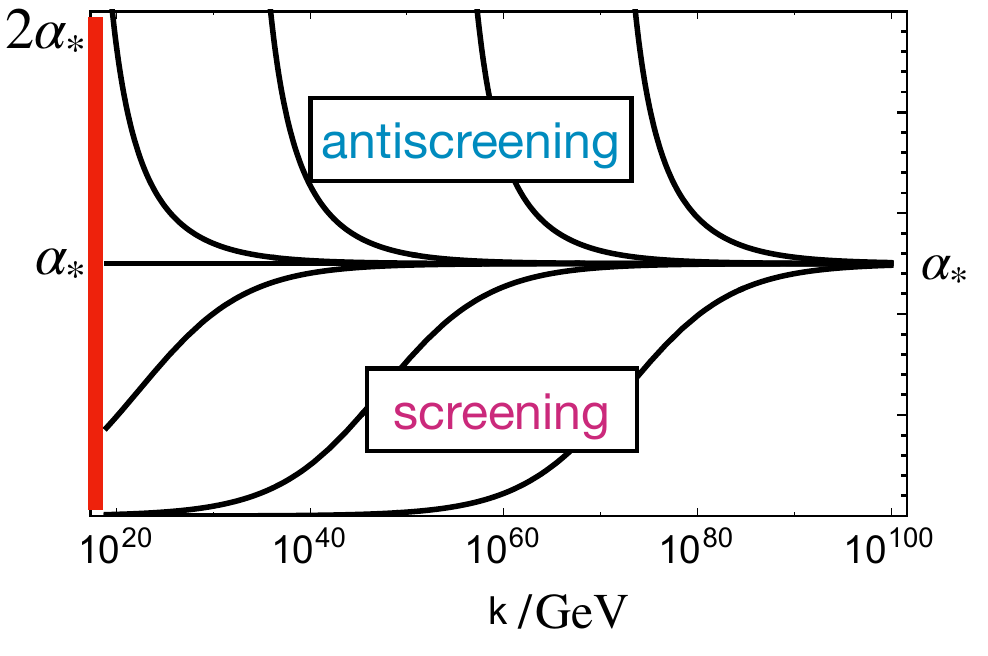}\\
\includegraphics[width=\linewidth,clip=true,trim=22.5cm 24cm 29cm 3cm]{predictive_power.pdf}\\
\caption{\label{fig:predictivepower} Using the simple beta functions $\beta_{\alpha}= \alpha(\alpha_{\ast}-\alpha)$ (upper panel) and $\beta_{\alpha}= -\alpha(\alpha_{\ast}-\alpha)$ (lower panel), we illustrate the physical mechanism behind the predictive power of asymptotic safety. A fixed point (at $\alpha=\alpha_{\ast}$) arises, because the different contributions in the beta functions come with different signs, and balance out at the fixed point. This balance can be achieved for either sign of the linear term. \\
Upper panel: If the linear term has a positive sign, i.e., encodes \emph{screening} of the coupling, the quadratic term must have a negative sign, i.e., encode \emph{antiscreening}. Thus, perturbations away from the fixed point grow under the RG flow and requiring a fixed point at very high scales is compatible with a range of couplings (indicated by the vertical red line) at the Planck scale. \\
Lower panel: If the linear term is negative and encodes antiscreening, the quadratic term must be positive and encode screening. Then, quantum fluctuations drive perturbations away from the fixed point to zero and restore scale symmetry. A single value of the coupling, the fixed-point value (green line) can be achieved at the Planck scale.\\
The mechanism generalizes to situations with more complex beta functions.}
\end{figure}

The predictive power of asymptotic safety is due to the irrelevant directions of the fixed point, which are those directions along which quantum fluctuations generate scale symmetry under the RG flow towards the IR, cf.~Fig.~\ref{fig:predictivepower}.

Given the set of beta functions $\beta_{g_i}$, we can linearize them about the fixed point at $\vec{g} = \vec{g}_{\ast}$ to obtain
\begin{eqnarray}
\beta_{g_i} &=& \beta_{g_i}\Big|_{\vec{g}= \vec{g}_{\ast}} + \sum_j \frac{\partial \beta_{g_i}}{\partial g_j}\Big|_{\vec{g}= \vec{g}_{\ast}} \left(g_j - g_{j\,\ast} \right)+ \mathcal{O}\left( g_j - g_{j\,\ast} \right)^2\nonumber\\
&=& 0 + \sum_{j}M_{ij} \left(g_j - g_{j\,\ast} \right)+ \mathcal{O}\left( g_j - g_{j\,\ast} \right)^2,
\end{eqnarray}
where we have used in the second line that the first term in the linearization vanishes at a fixed point and defined the stability matrix $M_{ij}$. This set of linearized differential equations has a solution in terms of the eigenvalues $-\theta_I$ and eigenvectors $V^I$ of the stability matrix:
\begin{equation}
g_{i}(t) =g_{i\, \ast} + \sum_I c_I\,V_i^{I} \, {\rm exp}\left(-\theta_I\, t\right), \quad \mbox{with } M_{ij}V_j^I =- \theta_I\, V_j^I.
\end{equation}
We have written the solution in terms of the RG-``time" $t= \ln(k/k_0)$, with $k_0$ an arbitrary reference scale. Changes in $k_0$ can be traded for changes in the $c_I$, which are constants of integration.
A subset of the $c_I$ corresponds to the free parameters that determine the physics of the theory, namely those which are associated to \emph{relevant} directions $V^I$ for which $\theta_I>0$. When $t$ decreases under the RG flow towards the IR, then $c_I$ multiplies an increasing term, i.e., all $g_i(t)$ receive a nonzero contribution determined by $c_I$, unless the $i$th component of the corresponding $V^I$ vanishes. For the simplest situation, when the eigenvectors $V^I$ are exactly aligned with the Cartesian basis in the theory space, each relevant direction is aligned with one coupling. Then, the IR value of that coupling follows by integrating the RG flow, starting from an initial condition determined in the linearized regime by $c_I$. The initial condition determines how far away from the fixed point the coupling is at a given UV reference scale, translating into a range of values a relevant coupling can achieve at lower energies, see upper panel in Fig.~\ref{fig:predictivepower}.s

In contrast, if $\theta_I<0$, then the term $c_I\, V_i^{I}{\rm exp}\left(-\theta_I\, t\right)$ is suppressed under the RG flow towards the IR, and $c_I$ never enters the low-energy value of a coupling. We call the corresponding eigendirection of the stability matrix an \emph{irrelevant} direction.
The associated $c_I$ is not a free parameter of the theory, in the sense that the physics does not depend on the choice of $c_I$. For the case when the stability matrix $M_{ij}$ is diagonal, the linearized RG flow results in the prediction $g_i(t) = g_{i\, \ast}$, see lower panel in Fig.~\ref{fig:predictivepower}. 
Beyond the linearized regime, this prediction is modified and $g_i(t)$ can in general depart from the fixed-point value, but is still predicted uniquely as a function of the values of the relevant couplings.

Finally, the case $\theta_I=0$ implies that higher-order terms in the beta function determine whether the coupling is marginally relevant or marginally irrelevant. The resulting scale-dependence of the coupling is logarithmic in $k$.

The relevant directions of a fixed point span its \emph{critical hypersurface}, cf.~Fig.~\ref{fig:critical_hypersurface}. In physical terms, the critical hypersurface parameterizes those effective field theories that provide effective, macroscopic descriptions of the asymptotically safe theory. Coupling values which lie outside the critical hypersurface parameterize effective field theories that cannot be UV completed in asymptotic safety, i.e., the critical hypersurface spans the \emph{landscape} of asymptotic safety\footnote{This holds within a fixed number of spacetime dimensions, fixed field content and local and global symmetries, all of which must be specified before the space of couplings can be determined. The full landscape is therefore larger, because one may, e.g., change the field content.
For more details, see our discussion of the swampland in asymptotic safety in Sec.~\ref{sec:AS_swamp}.} and coupling values off the critical hypersurface determine theories in the \emph{swampland}.

\begin{figure}[!t]
\begin{center}
\includegraphics[width=0.8\linewidth]{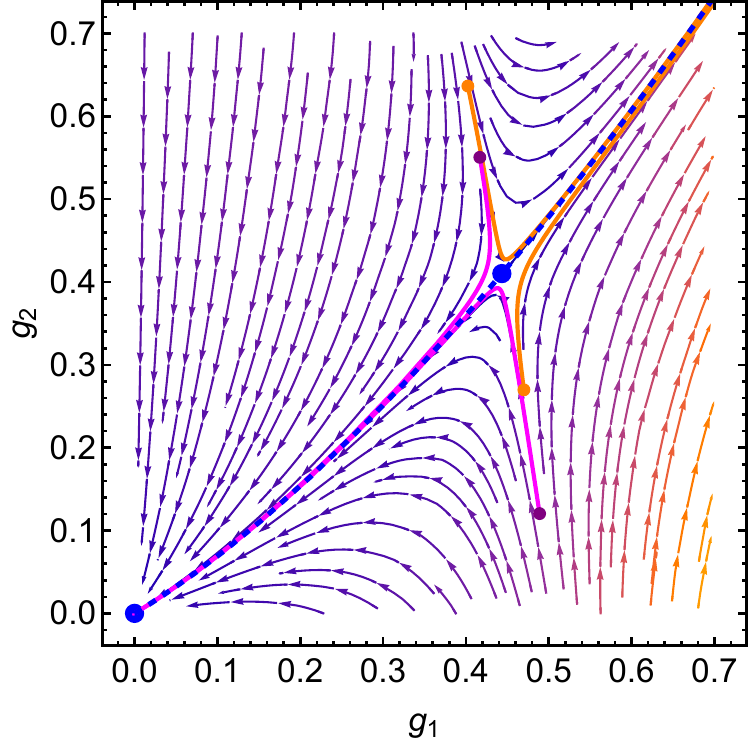}
\end{center}
\caption{\label{fig:fixedpointtrajectories} For purposes of illustration, we consider two beta functions $\beta_{g_1}=g_1-g_1\,g_2-3g_1^3$ and $\beta_{g_2}= g_2 +g_1\, g_2 - g_2^3-6 g_1^3$. The system has a Gaussian fixed point $(g_{1\,\ast}=0,g_{2\,\ast}=0)$ and an interacting fixed point at $(g_{1\,\ast}=0.443,g_{2\,\ast}=0.411)$, which are shown in blue. There is a positive critical exponent, $\theta_1= 1.70$, associated to the eigenvector $V^1=(-0.65,-0.76)$ and a negative critical exponent, $\theta_2=-1.46$, associated to the eigenvector $V^2=(0.16,-0.99)$. The arrows indicate the RG flow towards the IR. We show the critical hypersurface with blue dashed lines, which we obtain numerically by solving the set of differential equations $\beta_{g_1}= \partial_t g_1$ and $\beta_{g_2}= \partial_t g_2$ with initial condition $g_{1(2)}(t=10)= g_{1(2)\,\ast} \pm 0.2 \cdot V^1_{1(2)}\cdot {\rm exp}(-\theta_1 \cdot 10)$. The critical hypersurface consists of two parts, there is a separatrix connecting the interacting with the free fixed point (achieved by choosing the positive sign), and a trajectory which runs into a strongly-coupled regime in the IR (achieved by the negative sign in the expression for the couplings). \\
We also  show how initial conditions slightly off the critical hypersurface, given by  $g_{1(2)}(t=10)= g_{1(2)\,\ast} \pm 0.2 \cdot V^1_{1(2)}\cdot {\rm exp}(-\theta_1\cdot 10)+ a\, V^2_{1(2)}\cdot {\rm exp}(-\theta_2\cdot 10)$ result in virtually the same IR physics as trajectories that lie exactly in the critical hypersurface. We choose $a=10^{-2}$ and $a=-5\cdot 10^{-3}$ for the purple points and magenta trajectories, and $a=5\cdot 10^{-3}$ and $a=-8\cdot 10^{-3}$  for the orange points and trajectories.}
\end{figure}

The critical hypersurface is specified by relations between the different couplings in the theory. In practice, stability matrices are often not diagonal, and thus, relevant and irrelevant directions are not exactly aligned with couplings. Then, asymptotic safety determines values of a subset of couplings in terms of other couplings, see Fig.~\ref{fig:fixedpointtrajectories}.\newline

In parts of the literature, the explanation of free parameters uses the RG flow towards the UV. While, given a set of beta functions, the RG flow can mathematically be run in both directions, only one is physically meaningful: in nature, microphysics determines macrophysics, not vice-versa.\footnote{There are, of course, exceptions to this idea, related to UV-IR mixing, and/or systems for which physics at different scales does not decouple. Finally, in practice, the RG flow may also be a boundary-value problem, when the determination of different couplings occurs at different scales, see \cite{Eichhorn:2026djq}.} Therefore, we explain how predictivity can be seen from the RG flow towards the UV. 
Under the RG flow towards the UV, if $\theta_I>0$, the term $c_I\, V_i^I {\rm exp}(-\theta_I\cdot t)$ goes to zero, such that all choices of $c_I$ are viable. Conversely, for the case $\theta_I<0$, only $c_I=0$ results in a trajectory that ends in the fixed point, $g_i(t \rightarrow \infty) \rightarrow g_{i\,\ast}$. Thus, we again conclude that $c_I$ is a free parameter for a relevant direction. However, for $\theta_I<0$, we previously concluded that $c_I$ does not enter IR physics (and is therefore arbitrary, but not a free parameter of the physics), but now conclude that it should be zero. The difference lies in whether we demand an \emph{exact} fixed-point trajectory or merely demand that the RG flow originates in the linearized regime around the fixed point in the UV. An exact fixed-point trajectory is demanded in the argument based on the RG flow towards the UV. It should, however, be obvious that this can at best be a mathematical requirement, but not a requirement that is physically justified. To construct a theory of gravity which describes its quantum nature at and beyond the Planck scale, it is unnecessary (and not reasonable to expect) that we are constructing a theory that truly describes nature down to arbitrarily small distance scales. In practice, \emph{approximate} fixed-point trajectories, which spend some amount of RG ``time" very close to the fixed point in the UV, are sufficient for the theory to be asymptotically safe in practice.\footnote{The resulting notion of \emph{effective asymptotic safety} is discussed in more detail in Sec.~\ref{sec:eff_AS}.} For an approximate fixed-point trajectory, the values of $c_I$ associated to irrelevant directions specify the closest approach of the trajectory to the critical hypersurface. In Fig.~\ref{fig:fixedpointtrajectories} we illustrate an exact fixed-point trajectory, several approximate fixed-point trajectories (together with the associated $c_I$) and show how the resulting IR values of couplings all lie on (or very close to) the universal relation.

In addition, when we consider the RG flow towards the UV, we can be led to the incorrect conclusion that asymptotic safety requires a special choice of ``initial'' conditions for the value of the couplings in the IR: infinitely many $c_I$ have to be exactly zero, which one may interpret as ``special'' initial conditions. This is, of course, a misled conclusion, because nature works exactly the other way around. What looks like ``initial'' conditions for the RG flow is really the values \emph{automatically} reached by the RG flow towards the IR. In other words, microphysics determines the macrophysics, and highly predictive microphysics specifies macrophysics only up to very few free parameters. Thus, for an (approximate) fixed-point trajectory, the macrophysics is specified by very few relevant directions. Heuristically, this is not different from the insight that the Standard Model is built purely from renormalizable interactions, because it lies close enough to the Gaussian fixed point, such that higher-order interactions are all suppressed in the IR, \emph{irrespective of what their couplings at some high cutoff scale are}. In other words, there is no fine-tuning of the non-renormalizable, higher-order interactions in the SM; they are automatically small in the IR and the RG flow automatically ends up on the critical hypersurface of the Gaussian fixed point.
This Wilsonian explanation of the structure of the Standard Model carries over directly to interacting fixed points: the resulting IR physics is \emph{universal} in the sense that it lies on (or very close) the critical hypersurface of the fixed point, irrespective of the values of irrelevant couplings in the UV -- as long as a small handfull of parameters, namely those corresponding to relevant directions, are chosen close enough to the fixed point in the UV.

\subsection{What is effective asymptotic safety?}\label{sec:eff_AS}
\emph{...where we discuss the difference between effective and fundamental asymptotic safety. This is relevant from two perspectives. First, it makes the ambition of the asymptotic-safety paradigm more modest: Rather than building a fundamental theory that is assumed to literally hold at all scales, never to be superseded by an improved understanding of microphysics, one constructs an effective field theory with predictive power which has the ambition to elucidate the structure of gravity somewhat further into the UV, but does not come with the claim of being a fundamental theory.  Second, effective asymptotic safety provides a concrete proposal how other approaches to quantum gravity may be connected to asymptotic safety, which we come back to in Sec.~\ref{sec:relations}.}\\

\begin{figure}[!t]
\includegraphics[width=\linewidth,clip=true, trim=1cm 5cm 0cm 1cm]{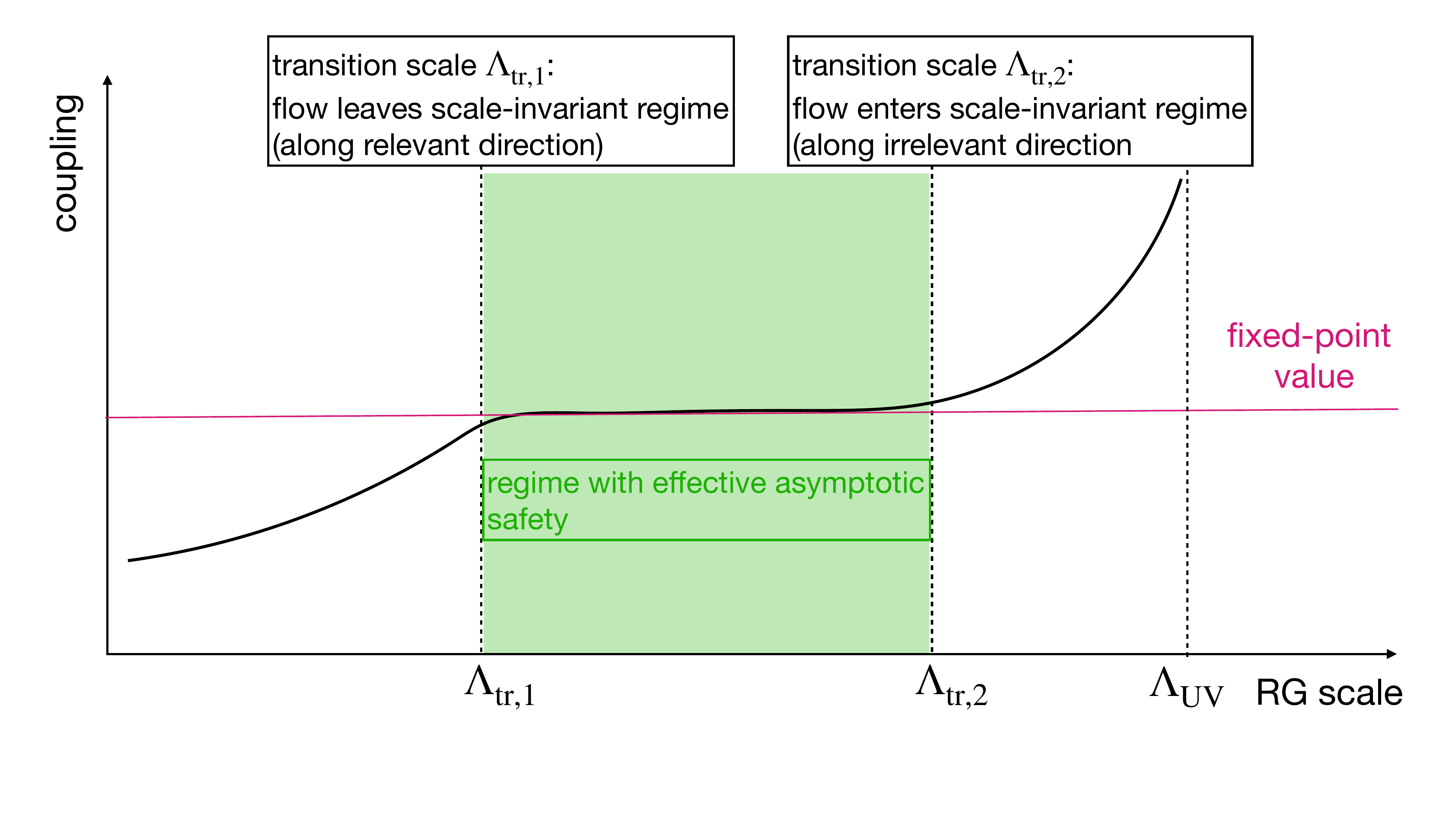}
\caption{\label{fig:effAS} We illustrate the RG flow of a coupling through an effectively asymptotically safe regime in which the coupling is very close to the fixed-point value.}
\end{figure}

Effective asymptotic safety is the scenario \cite{deAlwis:2019aud} that the fixed point governs the RG flow over a \emph{finite} range of scales, i.e., it is only realized approximately, but never exactly, see Fig.~\ref{fig:effAS}.

The name \emph{effective asymptotic safety} is intended to contrast with \emph{fundamental asymptotic safety}; the latter being the standard interpretation of the fixed point providing a UV \emph{completion} of the theory, i.e., the fixed point being reached exactly in the limit $k \rightarrow \infty$. In contrast, in effective asymptotic safety, the fixed point provides a UV \emph{extension}, but not a true UV completion of the system. Accordingly, the actual UV completion is not an asymptotically safe QFT, but might, e.g., be a string theory \cite{deAlwis:2019aud}. From the perspective of the UV completion, asymptotic safety is an (approximate) feature of the effective theory, hence the name \emph{effective asymptotic safety}. An example of an RG trajectory realizing effective, rather than fundamental, asymptotic safety is given by the green trajectory in Fig.~\ref{fig:critical_hypersurface}.\\
In the scenario of effective asymptotic safety, one assumes that the true UV completion has an effective description as a quantum field theory that is valid below a cutoff scale $\Lambda_{\rm UV}$. The UV completion determines the values of couplings in the effective theory at $\Lambda_{\rm UV}$. Effective asymptotic safety is realized if these values lie nearly within the basin of attraction of the fixed point, such that the RG flow is attracted towards the fixed point, as one flows towards the IR. Once the flow is close to the fixed point, all beta functions are close to zero and the flow slows down, exhibiting close-to-fixed-point scaling. If the initial condition was within the basin of attraction, the RG flow would approach the fixed point asymptotically and realize it for $k\rightarrow 0$. Because the initial condition is close to the basin of attraction, i.e., \emph{not within the IR critical surface of the fixed point}, this does not happen. Instead, a relevant direction starts to dominate and the RG flow leaves the neighborhood of the fixed point along this relevant direction. 

From the perspective of the deep IR, such a trajectory is essentially indistinguishable from a true fixed-point trajectory \cite{Percacci:2010af}. Hence, effective asymptotic safety generates universality: all UV completions which are compatible with effective asymptotic safety give rise to universal predictions in the IR, which are quantitatively close to the predictions associated to fundamental asymptotic safety. The degree of predictive power of effective asymptotic safety was quantified in \cite{Held:2020kze}.

\section{Methods for the study of asymptotic safety in gravity}\label{sec:methods}
\emph{...where we review the main methods used to study asymptotic safety, and also highlight the status of so far less used methods.}\\

Broadly speaking, there are techniques from three distinct categories that can be used to study asymptotic safety:
\begin{itemize}
\item[i)] continuum techniques, which include the functional Renormalization Group, perturbation theory and large $N$ techniques, 
 \item[ii)] discrete techniques, which include Causal and Euclidean Dynamical Triangulations and Regge calculus, 
 \item[iii)] combinatorial techniques, which include tensor models. 
 \end{itemize}
 The method used most extensively is the functional Renormalization Group, followed by Dynamical Triangulations. In contrast, perturbation theory has been mostly used close to two dimensions, large $N$ techniques have only been used very little, and the use of combinatorial techniques is at a very early, exploratory stage.

The use of such distinct methods is highly desirable due to the complementarity of the approximations and uncertainties.
For instance, in functional RG techniques, systematic uncertainties arise due to truncations of the terms in the effective action, but calculations proceed in the continuum and infinite-volume limit. In contrast, lattice simulations have no need for truncations\footnote{This assumes that all relevant interactions are known and included in the lattice action.}, but rely on a discretization and a finite volume. This introduces uncertainties. Clearly, the two methods are complementary; similarly, other techniques make yet other approximations.

\subsubsection{Functional Renormalization Group and other continuum methods}
{\bf Functional Renormalization Group:} This approach  \cite{Wetterich:1992yh} is aimed at calculating the effective action $\Gamma[\langle g_{\mu\nu}\rangle]$ and closely matches the discussion in Sec.~\ref{sec:WhatisAS}. Here, we review salient aspects of the method for the example of a scalar field $\phi$, and highlight some conceptual and technical challenges that arise in the application to gravity in Sec.~\ref{sec:local_coarse_graining}. More details and thorough discussions can be found in \cite{Gies:2006wv,Reuter:2012id,Reuter:2019byg,Reichert:2020mja,Pawlowski:2020qer,Pawlowski:2023gym,Saueressig:2023irs}.
Broadly speaking, the method enables us to compute the effective action $\Gamma$ from a functional differential equation in a two-step process which does not require knowledge of the microscopic action $S$. This is, of course, crucial for an application to asymptotic safety, because the microscopic action is determined by the values of couplings at the asymptotically safe fixed point. To derive the functional differential equation for $\Gamma$, one inserts an IR regulator into the path integral that cuts off the Euclidean path integral as a function of $k$. It is quadratic in the field, i.e., can be viewed like a $k$-dependent mass term
\begin{equation}
{\rm IR\!-\!cutoff}= \Delta S_k = \frac{1}{2}\int \frac{d^4p}{(2\pi)^4}\phi(p)R_k\left(\frac{p^2}{k^2} \right) \phi(-p),\label{eq:IRcutoff}
\end{equation}
which we have written in Fourier space and in four dimensions. The generalization to other dimensions is straightforward and in many settings, spacetime dimensionality $d$ can even be treated as a parameter.
$R_k(y)$ fulfills the properties $R_k(y)>0$ for $y<1$ and $R_k(y)=0$ for $y>1$. Hence, it suppresses quantum fluctuations with low momenta, $p^2<k^2$, whereas quantum fluctuations with momenta larger than the cutoff scale are unaffected and thus contribute to the path integral. A change in $k$ results in a change of the couplings, giving rise to beta functions. These can be derived explicitly from the Wetterich equation, which we introduce below.

\begin{figure}[!t]
\includegraphics[width=\linewidth,clip=true,trim=5cm 1cm 24cm 2cm]{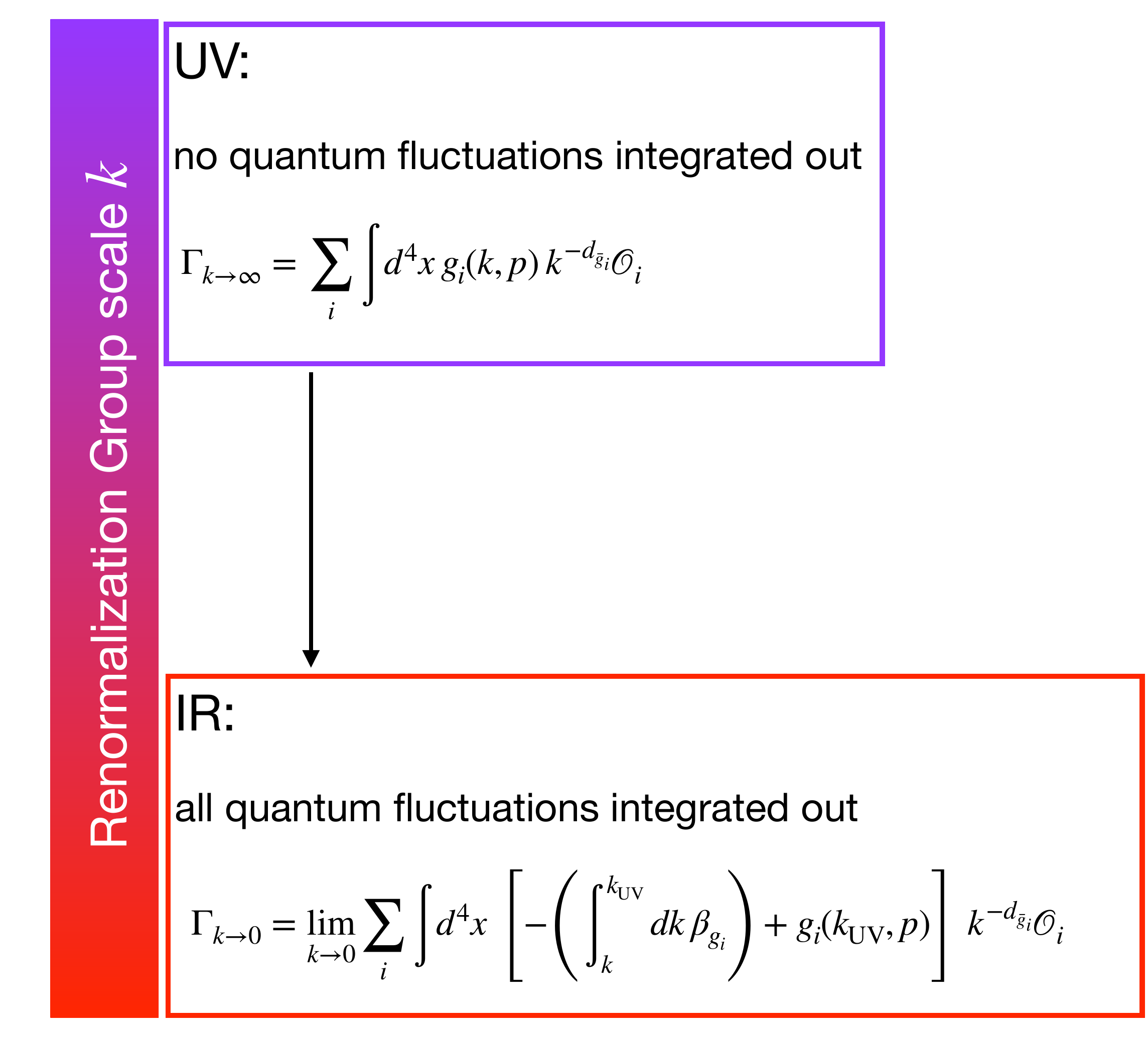}
\caption{\label{fig:FRG_schematic} We show a schematic illustration of how to obtain the effective action by integrating the corresponding couplings, using the beta functions and UV initial conditions determined by relevant perturbations $\delta g_i$ to be $g_i(k_{\rm UV})= g_{i\, \ast}+ \delta g_i$. To highlight that there is a separate dependence on physical scales, we schematically use the additional argument $p$ to a dependence on physical scales. In Sec.~\ref{sec:EuclideanFP} we come back to this momentum-dependence in more detail and explain how it is encoded in vertex functions or form factors. Here, we use $g_i(k,p)$ as a symbolic notation to indicate that the prefactors of the field-dependent terms in the effective action change, when we integrate over $k$, but also depend on the physical scales of the system.}
\end{figure}

In step one of using the functional RG, one searches for a fixed point in these beta functions. If a fixed point is found, it provides the input for step two, in which one integrates the RG flow towards the IR to obtain $\Gamma$, starting with initial conditions determined by relevant perturbations of the fixed point.  This gives rise to an effective action, in which the unphysical IR cutoff scale $k$ has been taken to zero, but which is expected to feature physical scale dependence. This scale dependence determines the behavior of physical observables at high scales. Through this two-step process,  the physical observables are ultimately determined by the finite number of free parameters that correspond to relevant directions of the fixed point, cf.~Fig.~\ref{fig:FRG_schematic}.

There is a potentially confusing aspect to this process, namely that it appears as if the effective action according to the definition Eq.~\eqref{eq:effaction} (and its generalization with the cutoff Eq.~\eqref{eq:IRcutoff}) can only be calculated if the classical action $S$ is already known. When one explores whether a theory may be asymptotically safe, the classical action is, however, not known a priori.
This point is resolved because the functional RG is based on a reformulation of the functional integral underlying the definition of $\Gamma_k$ into a functional differential equation.  This functional differential equation does not depend on $S$. It is called the Wetterich equation \cite{Wetterich:1992yh,Morris:1993qb} and was first formulated for gravity in a seminal paper by Martin Reuter \cite{Reuter:1996cp}. The Wetterich equation reads
\begin{equation}
k\,\partial_k\Gamma_k = \frac{1}{2} {\rm STr} \left(\Gamma_k^{(2)}+R_k \right)^{-1}k\, \partial_k R_k.\label{eq:floweq}
\end{equation}
Mathematically, this is a functional differential equation that determines $\Gamma_k$ in terms of its second functional derivative $\Gamma_k^{(2)}$, in which $\rm STr$ is a functional trace. A step-by-step-guide to a simple application of the Wetterich equation in gravity can be found, e.g., in \cite{Reichert:2020mja}.
Physically, the equation has a simple interpretation:  $\left(\Gamma_k^{(2)}+R_k \right)^{-1}$ is the propagator, derived from $\Gamma_k$, i.e., it is the propagator dressed by quantum fluctuations above $k^2$, and with a full dependence on $\langle \phi \rangle$. $\rm STr$ traces over the momenta of all quantum fluctuations, i.e., it can be understood as a loop integral. This loop integral is regularized in the UV through the regulator insertion $k\, \partial_k R_k$, because, due to our requirements for $R_k$, it holds that $k\, \partial_k R_k\rightarrow 0$ for $p^2/k^2>1$. There is also the IR cutoff $R_k$ in the regularized propagator, suppressing quantum fluctuations with $p^2/k^2\less 1$. Thus, the integration only has support for $p^2\approx k^2$.
Acting together, these various components therefore produce a loop integration, which accounts for quantum fluctuations only in the momentum shell $p^2/k^2 \approx 1$. Therefore, the change in $\Gamma_k$ is driven by quantum fluctuations with momenta close to $k^2$, which is in spirit a Wilsonian integration over subsequent momentum shells in the path integral.

From Eq.~\eqref{eq:floweq}, one can obtain the beta functions by selecting the appropriate $\mathcal{O}_i$ and then search for a fixed point and subsequently integrate the beta functions to obtain the effective action $\Gamma_{k\rightarrow 0}$. 

In practice, $\Gamma_k$ contains all terms compatible with the symmetries of the system -- in other words, infinitely many terms. The calculation of this infinite tower of beta functions is not possible in closed form; instead, one truncates the set of terms that enter $\Gamma_k$. Truncations do not need to be finite; infinite-dimensional truncations for gravity are found, e.g., in \cite{Codello:2008vh,Demmel:2015oqa,Falls:2017lst}, see \cite{Morris:2022btf,Pawlowski:2023gym} for reviews. Within a truncation, beta functions can be calculated, a fixed point can be searched for and, if found, characterized. Subsequent extensions of the truncation provide information on apparent convergence of results.

Given that asymptotic safety is based on a fixed point that is not Gaussian (i.e., free), one may wonder whether results obtained with the functional Renormalization Group are controllable and robust. The answer is, in fact, an emphatic yes! Examples for numerous characterizations of interacting fixed points in non-gravitational systems, reviewed in \cite{Dupuis:2020fhh}, provide blueprints for how this can be achieved, which we discuss further in Sec.~\ref{sec:truncations}. Although not clear in the initial years of the application of the Wetterich equation to gravity, a principle to set up consistent truncations has in the meantime been discovered, linked to the fact that the gravitational fixed point is near-perturbative, see Sec.~\ref{sec:nearpert}.

Finally, we point out that closely related to the Wetterich equation are proper-time flow equations \cite{Bonanno:2004sy,Bonanno:2025tfj,Bonanno:2025qsc,Bonanno:2025dry,Giacometti:2026zrs,Glaviano:2026lew}. In addition, there is also the Polchinski equation, which is related to the Wetterich equation through the Legendre transform between $\Gamma$ and $\ln Z$ and has been used in gravity in \cite{deAlwis:2017ysy}.\newline

{\bf Epsilon expansion:} This approach to determining fixed points is based on a continuous tracking of a fixed point as a function of spacetime dimension $d$. In this approach, one assumes that the fixed point a) is continuously deformed as a function of $d$ and b) becomes Gaussian when $d=d_c$, where $d_c$ is the critical dimension, i.e., the dimension when the coupling of interest is dimensionless. In $d=d_c$, its beta function therefore has no linear term. Above the critical dimension, the coupling $\bar{g}$ generically has negative mass dimension $d_{\bar{g}}$. In $d= d_c+ \epsilon$, i.e., above the critical spacetime dimension, the dimension of the coupling is $d_{\bar{g}}\sim \epsilon$. Thus the beta function of the dimensionless counterpart of the coupling, $g= \bar{g}k^{-d_{\bar{g}}}$
 is of the form
\begin{equation}
\beta_g=  + c_1\, \epsilon\, g + c_2\, g^2+ c_3 \, g^3+\dots,
\end{equation}
where $c_1>0$ is determined by the interaction in question and $c_2$ or $c_3$ may be zero or may have either sign. 
This has three fixed-point solutions, 
\begin{equation}
g_{\ast}=0,\quad g_{\ast}= \frac{-c_2}{2 c_3} \pm \sqrt{\frac{c_2^2}{4c_3^2}-\epsilon c_1}.
\end{equation}
Depending on the sign of the non-zero coefficient ($c_2$ or $c_3$), an interacting fixed point exists in $d>d_c$ or $d<d_c$ with a value that is parametrically controlled by $\epsilon$, such that it merges with the free fixed point for $d=d_c$. 

Well-known examples include $\lambda\, \phi^4$ theory, for which $c_2>0$ and thus an interacting fixed point exists for $d=4-\epsilon$ \cite{Wilson:1971dc}. This is the well-known Wilson-Fisher fixed point, which is the Ising model in $d=3$. Further examples are Yang-Mills theory, for which $c_2=0$ and $c_3<0$, such that an interacting fixed point exists for $d=4+\epsilon$ \cite{Gies:2003ic,Morris:2004mg} and the Gross-Neveu model (a theory with four-fermion interactions), for which $c_2<0$ and thus an interacting fixed point exists for $d=2+\epsilon$ \cite{Gawedzki:1985ed,Braun:2010tt}. Further references to these examples are given in \cite{Eichhorn:2018yfc}.
In all these examples, the first physically relevant value of $\epsilon$ is $\epsilon=1$. At this value, higher orders in the $\epsilon$ expansion are needed, such that a non-perturbative resummation, e.g., based on Padé approximants, can be performed. 

For the Newton coupling, the critical dimension is $d_c=2$. This is special, because in this dimensionality, the Einstein-Hilbert action is topological. Nevertheless, in $d=2+\epsilon$, a non-trivial beta function exists, for which $c_2<0$, see \cite{Jack:1990ey,Kawai:1989yh} for examples of early work, with more recent work in \cite{Martini:2021slj}, supporting the idea of asymptotic safety, see \cite{Martini:2022sll} for a review. Two-loop calculations have been done \cite{Aida:1994np}, however, higher-loop calculations would be needed to provide sufficient input to attempt a Padé resummation. In addition, there is gauge- and parameterization dependence off-shell\footnote{In perturbation theory, metric fluctuations around a background can be parameterized in different ways, e.g., linearly: $g_{\mu\nu} = \bar{g}_{\mu\nu}+ h_{\mu\nu}$ or exponentially: $g_{\mu\nu} = \bar{g}_{\mu\kappa}\, \exp(h_{..})^{\kappa}_{\nu}$.} in $c_2$. Thus,  one must work sufficiently generally in the corresponding parameters (e.g., gauge parameters) to ensure that indications for a fixed point are robust \cite{Falls:2017cze}, see Sec.~\ref{sec:gaugedep} and \ref{sec:parameterization}.\footnote{Beta functions are not observables, therefore, they may in principle be affected by choices of unphysical parameters (e.g., gauge parameters). Nevertheless, the existence of a fixed point is of course a physical statement about the theory and must therefore not depend on such choices.}\newline

{\bf Perturbation theory:} Perturbation theory can be applied to derive the beta function for the Newton coupling; however, not all regularization techniques work equally straightforwardly. Dimensional regularization, applied naively, does not work. Because it is a projection on logarithmic divergences, it produces a trivial beta function for the Newton coupling in four dimensions. This result should not be taken to imply that the Newton coupling does not depend on the RG scale. It simply means that the dependence is not logarithmic.

One can account for logarithmic divergences in other integer dimensions (which translate into power-law-divergences in four dimensions). Thereby, dimensional regularization can be adapted to produce a non-trivial beta function for the Newton coupling in four dimensions \cite{Falls:2024noj,Kluth:2024lar}. This beta function exhibits an interacting fixed point, at which the Newton coupling is relevant. The same result can be obtained in cutoff regularization \cite{Niedermaier:2009zz}. 

To understand the difference between cutoff regularization and dimensional regularization, a cutoff scheme has been developed in the functional RG, whereby results from dimensional regularization can be reproduced, because the regulator can be taken to zero after the calculation of the beta function \cite{Baldazzi:2020vxk}. In this limit, the gravitational fixed-point value diverges, but critical exponents stay finite \cite{deBrito:2022vbr}. This elucidates that, while the information on the gravitational fixed point and its properties are universal, i.e., independent of the regularization scheme, there are schemes in which results are more straightforward to interpret than in others.\newline

{\bf Large-$N$ techniques:} The idea behind large-$N$-techniques is that $N$ counts the number of fields (or more generally, degrees of freedom) in the system, and that the limit of $N \rightarrow \infty$ simplifies the calculations, because it removes contributions that have a subleading scaling with $N$. Ideally, this may even render calculations exact. For instance, for gravity-matter systems with $N$ matter fields, one may expect that 
the large-$N$ limit results in a dominance of matter fluctuations, such that gravity fluctuations can be neglected.\footnote{There are important caveats here: gravitational fluctuations come with a factor of the Newton coupling. The assumption that, at large $N$, this contribution is negligible, assumes that the Newton coupling (as well as other gravitational couplings) do not grow with $N$ such that they stay equally important at large $N$. In fact, a dynamical enhancement of gravitational fluctuations has been observed in some settings \cite{Meibohm:2015twa,Christiansen:2017cxa}\label{footnote1}.} In this spirit, \cite{Smolin:1981rm} found evidence for an asymptotically safe fixed point in four dimensions. Within the FRG, a large-$N$ limit can also in principle be taken, see, e.g., \cite{Dona:2013qba,Christiansen:2017cxa}. Because only vector fields antiscreen the Newton coupling, the large-$N$-limit for vectors should be under better control than for scalars or fermions; however, a comprehensive study of the large-$N$-limit has not yet been done and may be subtle due to the caveat mentioned in the footnote \ref{footnote1}.

\subsubsection{Discrete techniques}\label{sec:discretetechniques}
 Numerical simulations of the gravitational path integral begin by discretizing the configurations in the path integral. Thereby, the infinitely many degrees of freedom of the quantum field theory (which in the continuum are the values of the metric field (modulo diffeomorphisms) at each spacetime point) are reduced to finitely many degrees of freedom and numerical studies become possible. There is a choice of the discrete variables. For instance, in Regge calculus, one can loosely speaking think of the metric as the discrete variable, whereas dynamical triangulations discretize spacetime geometry more directly.

The main concept underlying numerical simulations of the gravitational path integral is the following: A continuum theory with an asymptotically safe fixed point has a higher-order phase transition, when the phase diagram of the discretized theory is spanned by the relevant couplings of the fixed point.\footnote{In the discretized setting, there is not necessarily a one-to-one correspondence of couplings of the discrete theory to specific couplings of the continuum theory, because of the differences in how the degrees of freedom of the theory can be encoded in the continuum and the discrete setting. However, if an interaction that is relevant and essential in the continuum is not in some way represented among the couplings of the discrete theory, one would not expect to be able to recover the asymptotically safe theory.} At a higher-order phase transition, the correlation length diverges, there is no longer a typical scale in the system and the physics becomes scale invariant.\footnote{This behavior is the correct one in theories on a fixed spacetime background; \cite{Ambjorn:2024qoe} highlights that the behavior of the correlation length may be different in a theory in which spacetime is dynamical.} This makes it possible to remove the UV cutoff without changing the physics.

To ensure that the resulting continuum theory is physically acceptable, the phase transition has to be approached from within a phase with acceptable physical properties. For instance, the phase should encode four-dimensional spacetimes, be compatible with General Relativity at low curvature scales and give rise to a viable cosmology, in particular admit a phase of accelerated expansion.

The most prominent approaches to numerical simulations of the gravitational path integral are causal and Euclidean dynamical triangulations, reviewed in \cite{Loll:2019rdj,Ambjorn:2022naa,Ambjorn:2024pyv,Loll:2025eks,Ambjorn:2026prt}, although early work on Regge calculus also exists \cite{Hamber:2009mt,Hamber:2015jja}. In dynamical triangulations, a $d$-dimensional spacetime is discretized into $d$-dimensional, flat simplices (e.g., triangles in two dimensions, tetrahedra in three dimensions and so on) and the curvature is encoded at the vertices.

Both causal and Euclidean dynamical triangulations are in practice based on a Euclidean weight $e^{-S}$ in the path integral. This makes it possible to use Monte-Carlo techniques to evaluate the path integral.
In Euclidean dynamical triangulations, the Euclidean weight is built into the definition of the theory, whereas, in Causal Dynamical Triangulations, one starts from the Lorentzian path integral with Lorentzian amplitude, $e^{iS}$, and includes all configurations that can be Wick-rotated.
Thus, a key difference between causal and Euclidean dynamical triangulations lies in the topology of the individual configurations and specifically in the behavior of spatial topology as a function of Euclidean ``time'': in the causal setting, each configuration can be Wick-rotated to a discrete Lorentzian spacetime and changes of spatial topology as a function of time are excluded in these configurations. In Euclidean dynamical triangulations, there is only a constraint on the overall topology (typically chosen to be that of $S^d$), which is also present in the causal setting.

In terms of the resulting phase diagram, while the EDT phase diagram contains two phases, there are more phases in CDTs.
In particular, a de Sitter like phase exists \cite{Ambjorn:2004qm}, on top of which fluctuations are relatively small \cite{Ambjorn:2008wc}, corroborating the idea that a physical background spacetime can emerge from a nonperturbative path integral \cite{Maas:2022lxv}. This phase is separated from other phases by a higher-order phase transition \cite{Ambjorn:2011cg,Ambjorn:2017tnl,Ambjorn:2018qbf}. To take a physical continuum limit, this transition must be approached from within the de Sitter like phase along a line along which a suitable physical quantity stays fixed, while the lattice scale is taken to zero. Such an approach corresponds to a Renormalization Group trajectory. Candidates for such trajectories have been investigated \cite{Ambjorn:2014gsa,Ambjorn:2019lrm,Ambjorn:2020rcn}. The most recent results recover the Gaussian fixed point as an IR fixed point and are compatible with the existence of an asymptotically safe fixed point \cite{Ambjorn:2024qoe}. 

In EDTs, so-called ``baby universes'' dominate and prevent a de-Sitter like phase. Interestingly, the so-called ``collapsed'' phase becomes more and more de-Sitter like as one moves along the first-order-transition line \cite{Bassler:2021pzt}. There is, therefore, some hope that the line features a higher-order endpoint, at which a physically acceptable continuum limit can be taken with de-Sitter-like behavior, see \cite{Laiho:2016nlp,Bassler:2021pzt,Dai:2021fqb,Dai:2023tud,Dai:2024vjc}.

\subsubsection{Tensor models and combinatorial techniques}\label{sec:tensormodels}
\begin{figure}[!t]
\begin{center}
\includegraphics[width=0.7\linewidth,clip=true,trim=4cm 16cm 45cm 2cm]{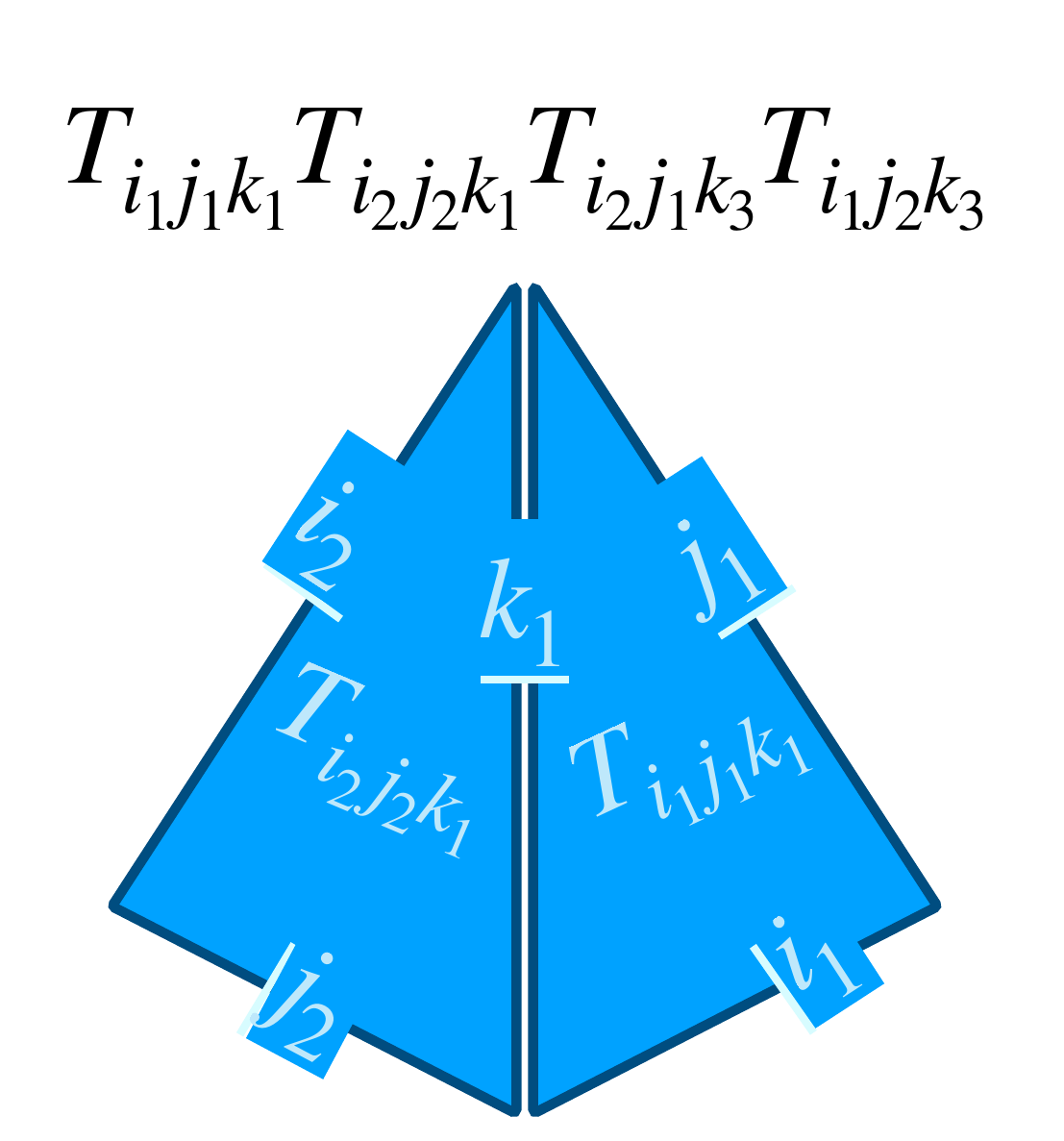}
\end{center}
\caption{\label{fig:tensormodel} In a rank-3 tensor model, a tensor invariant, such as the above quartic invariant, encodes a 3-dimensional building block of a piecewise flat random geometry. Each tensor represents a $(d-1)$ simplex, which is a triangle in the present case.}
\end{figure}

The combinatorics of gluing simplices into discrete, piecewise flat spacetimes can be encoded in tensor models \cite{Ambjorn:1990ge,Gross:1991hx,Sasakura:1990fs,Gurau:tensors}. A tensor model is a 0-dimensional field theory, because the tensors are not defined \emph{on} a spacetime, rather, they are ``building blocks" of spacetime: they encode the geometry of a spacetime at a combinatorial level. This is achieved by using a rank $d$ tensor for $d$-dimensional spacetimes. 
A tensor is associated to a vertex that sits in the center of a $d-1$-dimensional simplex. The \emph{dual} to the simplicial complex can be constructed by connecting the vertices through the $(d-2)$-dimensional faces of the simplicial complex, which is encoded in the contraction of two tensors through a common index, cf.~Fig.~\ref{fig:tensormodel} for the three-dimensional version of this construction. Interactions, represented by tensor invariants (with all indices contracted and the interaction invariant under the action of a group on a given index of the tensor), encode how $(d-1)$-dimensional simplices are glued to $d$-dimensional simplices. Finally, Feynman diagrams represent how $d$-simplices are glued along their common $(d-1)$-subsimplices, when two tensors are contracted by a propagator.

The Feynman diagrams of a tensor model correspond to the configurations in a path integral over triangulated, piecewise flat, simplicial manifolds. As an important advantage of tensor models, background independence is built in at the very foundations of the theory: there is no spacetime, the theory is a 0-dimensional field theory. Quantities like the metric are emergent, i.e., the metric is an effective degree of freedom, that one expects to arise in the continuum limit. It is even possible that suitable matter degrees of freedom emerge in this limit \cite{Abranches:2023tqe}, although it is far from clear that such matter content can be close to realistic matter, i.e., to the Standard Model.
A second important property of tensor models, which may be either a bug or a feature, is an automatic summation over topologies.

Taking inspiration from the two-dimensional case of matrix models, one may hope to recover the asymptotically safe fixed point in quantum gravity from a limit in which the tensor size $N$ is taken to infinity (such that the number of degrees of freedom becomes infinite, as in a non-zero-dimensional QFT), and the coupling is tuned to a critical value. In matrix models, such a limit exists and recovers the critical exponent that characterizes Liouville quantum gravity \cite{Ginsparg:1993is,DiFrancesco:1993cyw}. This limit can be recovered with a ``pre-geometric'' version of the functional RG, in which the matrix size $N$ plays the role of an ``RG scale'' \cite{Brezin:1992yc,Eichhorn:2013isa,Eichhorn:2014xaa,Castro:2020dzt}. This is an appropriate notion of ``scale'' in the absence of spacetime, because at the heart of Wilsonian renormalization lies the decimation of the degrees of freedom, in the sense that in the UV, there are many more degrees of freedom that are accessible than in the IR, where all the microscopic, UV degrees of freedom have already been integrated over. 

For tensor models, a systematic $1/N$ expansion, which is a prerequisite for the pregeometric RG, was discovered in \cite{Gurau:2010ba,Gurau:2011xq}, reviewed in \cite{Gurau:2011xp}. On this basis, the ``pregeometric RG'' for tensor models has been developed \cite{Eichhorn:2017xhy,Eichhorn:2018ylk}, see \cite{Eichhorn:2018phj} for a review.\footnote{There is also a pregeometric version of the Polchinski equation for these models \cite{Krajewski:2015clk,Krajewski:2016svb}.} This flow equation comes with a violation of certain symmetries in the tensor model, which have been studied, together with the corresponding Ward-identities, in \cite{Lahoche:2018ggd,Lahoche:2019vzy,Baloitcha:2020lha}.
Candidate fixed points have been identified, with critical exponents that are close enough to those found in continuum quantum gravity to warrant closer investigation \cite{Eichhorn:2019hsa,Castro:2026zch}. Robust evidence for the fixed point found with functional RG methods in the continuum, the Reuter fixed point, discussed below in Sec.~\ref{sec:FP}, has, as of yet, not been found in tensor models. Nevertheless, they remain a way of searching for asymptotic safety in gravity in a background-independent manner, such that their use in this context has also be advocated, e.g., in \cite{Pereira:2019dbn}.

\section{Existence of an asymptotically safe fixed point}\label{sec:FP}
\emph{...where we review the current status of asymptotic safety in gravity. In summary, there is compelling evidence for a fixed point with three relevant directions in Euclidean signature in pure gravity in four dimensions. It is considered as established that this fixed point exists, and thus it has been named ``Reuter fixed point" in honor of Martin Reuter who pioneered functional Renormalization Group techniques for quantum gravity \cite{Reuter:1996cp}.  Reuter also conducted early studies collecting evidence for the fixed point \cite{Lauscher:2001ya,Reuter:2001ag,Lauscher:2002sq}, complemented early on by other authors using Reuter's formulation of the Wetterich equation \cite{Dou:1997fg,Souma:1999at,Percacci:2003jz,Litim:2003vp}.\\
There is also strong evidence that the fixed point continues to exist, once matter fields are added. In particular, studies indicate that Standard Model matter can be added, leading to a deformation, but not a disappearance of the fixed point.\\
Finally, there is a recent effort to discover whether a Lorentzian-signature counterpart of the Reuter fixed point exists, with promising first results.}

\subsection{Fixed point in Euclidean signature and four spacetime dimensions for pure gravity}\label{sec:EuclideanFP}
For the roughly first two decades following the seminal paper \cite{Reuter:1996cp}, the search for an asymptotically safe fixed point  proceeded based on an expansion of the effective action in terms of (derivative) curvature invariants, sorted by the canonical dimension of the invariants,
\begin{eqnarray}
\Gamma&=& \int d^4x\,\sqrt{g}\Biggl(\frac{\Lambda}{8 \pi G_N} - \frac{1}{16\pi G_N}R + a_2\, R^2\nonumber\\
&+&b_2\, C_{\mu\nu\kappa\lambda}C^{\mu\nu\kappa\lambda} + a_3 R^3+ b_3 C_{\mu\nu\alpha \beta}C^{\alpha\beta\kappa\lambda}C_{\kappa\lambda}^{\,\,\,\,\,\,\,\,\mu\nu}+...  \Biggr)\nonumber\\
&+& \rm{top.~inv.}\,\,.\label{eq:effaction_expansion_dimension}
\end{eqnarray}
Based on this expansion, later supplemented by other expansion schemes, detailed below, compelling evidence for the existence of the fixed point in Euclidean gravity has been established.

\begin{figure}
\includegraphics[width=\linewidth]{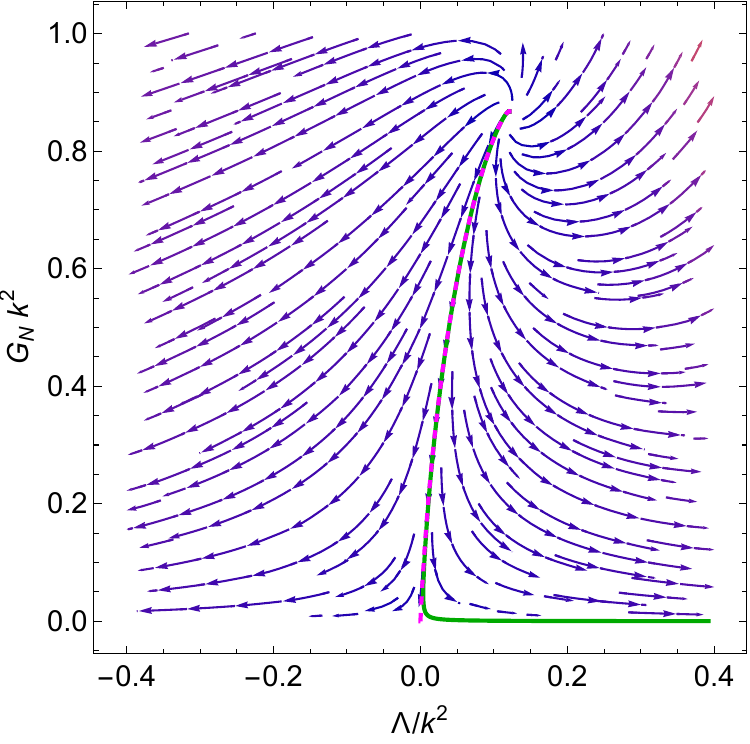}\\
\includegraphics[width=\linewidth]{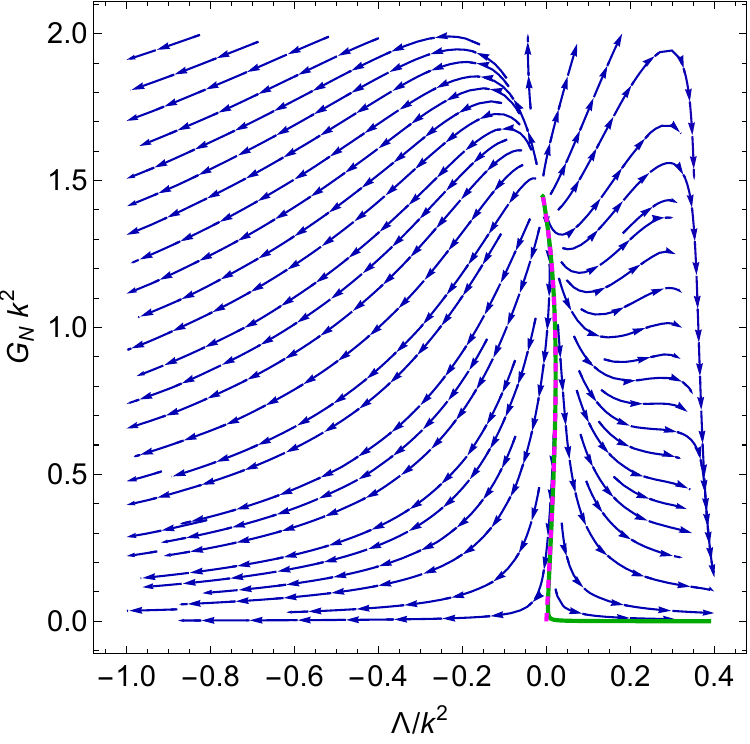}
\caption{\label{fig:EHplane} We show the RG flow in the $(G_N\, k^2)$-$(\Lambda k^{-2})$-plane, using the beta functions from \cite{Reuter:2012id} in the upper panel and \cite{Dona:2013qba} in the lower panel. In each panel, the dashed magenta line indicates the unique trajectory that connects the interacting fixed point in the UV to the free fixed point in the IR. The green line indicates a trajectory that produces the relation $G_N\cdot \Lambda = \rm const$ in the IR, with the constant a tiny value. The tininess of this value extracted from measurements gives rise to the cosmological-constant problem. Within asymptotic safety, this problem is part of the larger question whether the values of relevant couplings are determined by some more fundamental framework. Leaving such questions aside, the result that $\Lambda\cdot G_N \approx 10^{-120}$ can be \emph{accommodated} in asymptotic safety, by choosing a trajectory qualitatively similar to the green one, which in particular spends a long RG ``time" in the vicinity of the free fixed point.}
\end{figure}

Explicit calculations have been performed for numerous terms, including for all terms that are explicitly listed above; and consistently find a fixed point with three relevant directions, corresponding to $\Lambda$, $G_N$ (see Fig.~\ref{fig:EHplane}), as well as a superposition of $a_2$ and $b_2$ \cite{Benedetti:2009rx,Falls:2020qhj}. 
All interactions that have been included in this expansion of the effective action are shown in Tab.~\ref{tab:FP}, see also \cite{Saueressig:2023irs} for a review.

\begin{table*}[!t]
\begin{tabular}{|c|c|c|c|c|c|c|c|c|}
\hline
can.~dim. & can.~dim.  & & & & & &  \# of new & total \#\\
curv.~inv. & coup. & & & & & &   rel.~dir. & of rel.~dir.\\
\hline\hline
0 & 4 & $\mathds{1}$& -&- &-&-&  1&1 \\
\hline
2 & $2^{\ast}$ & $R$ &- &- &- & -&  1&2 \\
\hline
4 & 0 & $R^2$ & $C_{\mu\nu\kappa\lambda}C^{\mu\nu\kappa\lambda}$ &-&-&-&   1&3\\
& & \cite{Lauscher:2002sq,Falls:2020qhj} & \cite{Benedetti:2009rx,Falls:2020qhj}& & & &  & \\
\hline
6 & -2 & $R^3$ & $C_{\mu\nu\kappa\lambda}C^{\kappa\lambda\rho \sigma}C_{\rho\sigma}^{\,\,\,\,\mu\nu}$ &  $R\, R_{\mu\nu}R^{\mu\nu}$& $R R_{\mu\nu\kappa\lambda}R^{\mu\nu\kappa\lambda}$ &6 more & 0 & 3\\
& & \cite{Codello:2007bd,Falls:2018ylp}& \cite{Gies:2016con,Baldazzi:2023pep} & \cite{Falls:2017lst} & \cite{Kluth:2020bdv} $\clubsuit$ & & &\\
\hline
$\dots$& $\dots$& $\dots$& &$\dots$&$\dots$&$\dots$&0&3\\ 
\hline
42 & -38 &  $\dots$&  & $R (R_{\mu\nu}R^{\mu\nu})^{20}$ & $\dots$& &0&3\\ 
 & & & &   \cite{Falls:2017lst} &  & & & \\
\hline
$\dots$& $\dots$& $\dots$& &&$\dots$&$\dots$&0&3\\ 
\hline
140 & -136 & $R^{70}$ & & &$\dots$&$\dots$&0&3\\ 
& &  \cite{Falls:2018ylp} & & & & & & \\ 
\hline
$\dots$& $\dots$& & & &$\dots$&$\dots$&0&3\\ 
\hline
286 &-282 &  &  &  & $R (R_{\mu\nu\kappa\lambda}R^{\mu\nu\kappa\lambda})^{71}$ & & 0 & 3 (4)\\
& & & & &  \cite{Kluth:2020bdv} $\clubsuit$& & & \\
\hline\hline
\end{tabular}
\caption{\label{tab:FP}Curvature invariants and associated dimension of couplings (in $d=4$) and, starting at quadratic order, first as well as latest reference including the corresponding interaction. Entries with dots mean that the corresponding curvature invariant has been included to even higher order; empty entries mean that no calculation exists.
Relevant directions in general correspond to superpositions of couplings. We mark in each line, how many new relevant directions appear, when truncations are extended to the corresponding order in the canonical expansion; we also list the total number of relevant directions to this order. The asterisk on the canonical dimension of the coupling in the second line points to the choice of coupling: The square of the reduced Planck mass is related to the inverse Newton coupling; the former has dimension 2. The Euler invariant, a topological invariant, is not included in this table, but adds an exactly marginal direction. Note that we use interactions used in the corresponding references; a basis up to a given order is constructed, e.g., in \cite{Baldazzi:2023pep} and uses the Weyl tensor throughout, instead of the Riemann tensor, as we do in selected invariants.
The reference indicated with $\clubsuit$ reports a fixed point with three as well as one with four relevant directions.\\
We caution that some of the references project onto beta functions by choosing highly symmetric spacetime geometries (e.g., a 4-sphere), on which not all basis elements at a given order in the expansion can be disentangled.\\
Couplings computed in a vertex expansion are discussed below, but produce compatible results with those shown here.\\
According to the emerging pattern of \emph{near-perturbativity}, see Sec.~\ref{sec:nearpert}, all remaining couplings are expected to be irrelevant in $d=4$. We highlight that the quantitative values for critical exponents stabilize at higher orders in the expansion; e.g., at order $R^2$ in the expansion, there are three relevant directions, but the value of the third critical exponent is often significantly larger than when subsequent orders in the expansion are included.}
\end{table*}

For instance, Ref.~\cite{Falls:2020qhj} finds, in a truncation accounting for  $G$, $\Lambda$, $a_2$ and $b_2$ as well as the Euler invariant, the following set of critical exponents
\begin{equation}
\theta_{1,2}=2.35 \pm 1.68 i, \quad \theta_3=1.77,\quad \theta_4=0,\quad \theta_5=-3.20,\label{eq:critexpsRsquared}
\end{equation}
where the vanishing critical exponent is associated to the topological term.  All works listed in Tab.~\ref{tab:FP} find compatible results, with the caveat that in small truncations, critical exponents can vary significantly \cite{Knorr:2021slg}, and typically only stabilize at higher order in the truncation, as, e.g., in \cite{Falls:2013bv,Falls:2014tra,Falls:2017lst,Kluth:2020bdv}.

The two largest positive critical exponents are found to form a complex pair in many calculations, signaling the spiraling of RG trajectories about the fixed point, see upper panel Fig.~\ref{fig:EHplane}. There are, however, also calculations, which differ only by a somewhat different choice of regulator function or other auxiliary parameters, which produce real critical exponents, see, e.g., lower panel of Fig.~\ref{fig:EHplane}. Complex critical exponents are sometimes argued to be unphysical \cite{Bulycheva:2014twa,Jepsen:2020czw}, because a spiralling flow appears to violate the $a$- or $c$-theorem \cite{Zamolodchikov:1986gt,Cardy:1988cwa,Osborn:1989td,Komargodski:2011vj}, which state that, under a set of assumptions, there is a function of the couplings for four-dimensional theories that is monotonically decreasing under the RG flow.\footnote{In the context of the Wetterich equation \cite{Becker:2014pea} proposes a generalization that is based on a ``mode-counting'' property of the flowing effective action and applicable in the context of RG flows with complex critical exponents.} Explicit counter-examples to this argument are presented in \cite{Curtright:2011qg} and the question of whether imaginary parts of complex critical exponents are to be regarded as problematic in the context of gravity is not settled.  

Typically, the following parameterization of $G_N(k)k^2= G(k)$ captures the behavior along numerically integrated RG trajectories rather well \cite{Bonanno:2000ep}:
\begin{equation}
G_N(k) = \frac{1}{M_{\rm Planck}^2}\frac{1}{1+\frac{ k^2}{G_{\ast}\cdot M_{\rm Planck}^2}}.
\end{equation}
Herein, $M_{\rm Planck}$ denotes the low-energy value of the Planck mass and $G_{\ast}$ is the fixed-point value. The scale $G_{\ast}\cdot M_{\rm Planck}^2$ sets the transition to the fixed-point regime: a higher fixed-point value typically means a higher transition scale between fixed-point scaling $G_N \sim k^{-2}$ and classical scaling $G_N =\rm const$.

Higher-order terms, including the infamous two-loop counterterm $C_{\mu\nu\alpha \beta}C^{\alpha\beta\kappa\lambda}C_{\kappa\lambda}^{\,\,\,\,\,\,\,\,\mu\nu}$, are irrelevant \cite{Gies:2016con}, and, in fact
\begin{equation}
\theta_6=\theta_{C^3} = - 3.85
\end{equation}
was found in \cite{Baldazzi:2023pep}.\footnote{In Ref.~\cite{Baldazzi:2023pep} this critical exponent is $\theta_2$, because the minimal essential scheme \cite{Baldazzi:2021ydj,Baldazzi:2021orb} is used, in which the classical equations of motion are used to eliminate couplings, similarly to the elimination of one-loop counterterms through the equations of motion in the perturbative treatment of GR. Such couplings are called inessential. We label this critical exponent $\theta_6$ here to highlight that it is associated to another coupling than those listed in Eq.~\eqref{eq:critexpsRsquared}.} 

In \cite{Falls:2013bv, Falls:2014tra,Falls:2017lst,Falls:2018ylp,Kluth:2020bdv}, equations for $f(R)$, as well as $f(R_{\mu\nu}R^{\mu\nu})$ and $f(R_{\mu\nu\kappa\lambda})$ have been derived and the fixed point has been found and studied in Taylor expansions up to high orders, cf.~Tab.~\ref{tab:FP}. The critical exponents in these expansions approach canonical scaling, e.g.,
the critical exponents for $f(R)$ according to \cite{Falls:2018ylp} are
\begin{equation}
\theta_n = -a \, n+b \approx - 2.04\, n +2.91,
\end{equation}
such that the deviation from canonical scaling is small at high $n$,
\begin{equation}
\theta_n - d_{\bar{g}_n} = -2.04\, n+2.91-(-2n+4) \overset{n \rightarrow \infty}{\rightarrow} -0.04 n.
\end{equation}
An even faster approach to near-Gaussian behavior was observed in \cite{Falls:2017lst}, with $|\theta_n-d_{\bar{g}_n}| \sim 1.7 \cdot \exp(-n/6.5)$. We will come back to the implications of near-Gaussian scaling behavior in Sec.~\ref{sec:nearpert}. For now, we observe that it provides a bootstrap \cite{Falls:2013bv} to set up consistent truncations, by assuming that relevant interactions are those with the highest canonical scaling dimensions, of which there is only a small handful.\newline

Finally, one can go beyond truncations of a finite size, and keep a full function $f_k(R)$. This retains strictly more information than Taylor expansions of $f_k(R)$, because the large $R$-behavior of this function may not be captured by a Taylor expansion and thus has been studied extensively \cite{Benedetti:2013jk,Mitchell:2021qjr}, with a summary of the state-of-the-art provided in \cite{Morris:2022btf}. One can use a Sturm-Liouville analysis to establish properties of fixed-point solutions for $f_{\ast}(R)$ based on the differential equation for $f_k(R)$ \cite{Codello:2007bd,Codello:2008vh,Machado:2007ea,Benedetti:2012dx,Demmel:2012ub,Demmel:2014sga,Demmel:2015oqa,Ohta:2015efa,Ohta:2015fcu,Falls:2016msz,deBrito:2018hur} that the Wetterich equation produces. With this method, and the equations derived in \cite{Benedetti:2013jk,Mitchell:2021qjr}, one can establish that $f_k(R)$ has at most a discrete number of fixed points (rather than a continuous spectrum) and that the fixed point support a finite number of relevant operators \cite{Mitchell:2021qjr}.\newline

In passing, we note that in small truncations, there are various ways in which one can devise approximations that remove well-established fixed points through ad-hoc modifications, such as, e.g., an artificial field dependence of the regulator \cite{Bridle:2013sra}; thus, claims about the non-existence of the Reuter fixed point \cite{Branchina:2024lai} can, in view of the results in Tab.~\ref{tab:FP} as well as the results to be discussed below, not be taken seriously \cite{Bonanno:2025xdg,Held:2025vkd}.\\

There is one aspect to these calculations that we have not yet mentioned and that we will expand on in Sec.~\ref{sec:local_coarse_graining}, namely that an application of functional RG techniques to gravity requires us to introduce an auxiliary background metric. Therefore, the effective action at finite $k$ depends on two metrics, one of them an auxiliary background metric. The works we have discussed above all apply the background-approximation, in which derivatives of $\Gamma_k$ with respect to the background metric are not distinguished from those with respect to the physical metric. This approximation can be lifted in two ways: first, in so-called bimetric studies \cite{Manrique:2010mq,Manrique:2010am,Becker:2014qya}, which provide support that the approximation is not particularly severe and neither existence of the fixed point nor a large part of its properties depend on it.
Second, one may lift this approximation in combination with a vertex expansion of the effective action, \cite{Christiansen:2014raa,Christiansen:2015rva,Denz:2016qks,Eichhorn:2018akn,Eichhorn:2018nda,Knorr:2021niv,Bonanno:2021squ,Fehre:2021eob,Kher:2025rve,Pawlowski:2025etp,Chiesa:2026tlz}, see \cite{Pawlowski:2020qer,Pawlowski:2023gym} for reviews, which support the same conclusion, namely that the qualitative properties of the Reuter fixed point (existence and number of relevant directions) are captured correctly by the background-field approximation.\\

\begin{figure*}[!t]
\includegraphics[width=\linewidth,clip=true,trim=0cm 9cm 0cm 1cm]{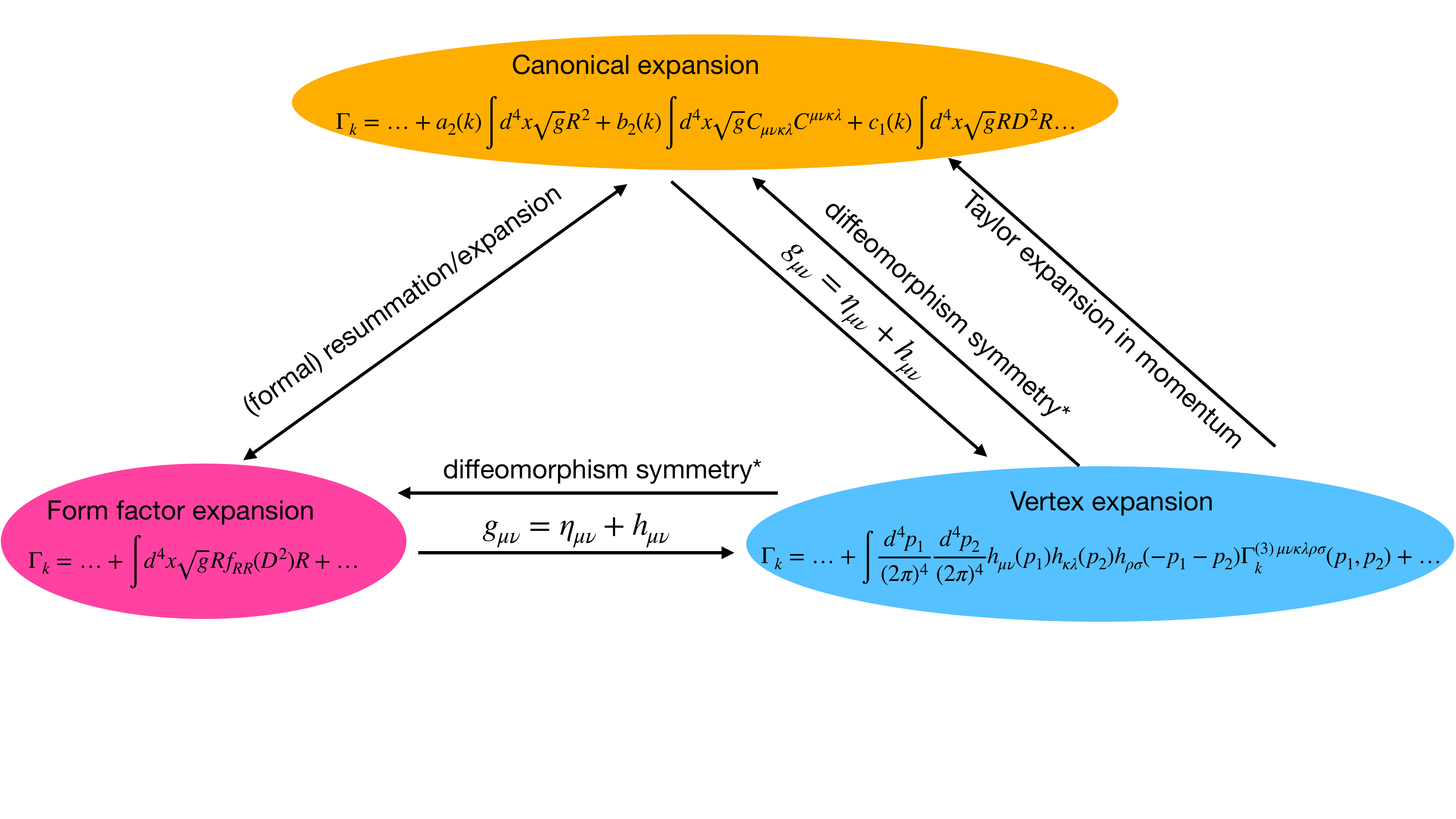}
\caption{\label{fig:expansions} The three expansions of the effective action are related to each other. From the canonical expansion, one obtains a vertex expansion by expansion about a (not necessarily flat) background. The inverse path is based on diffeomorphism symmetry, which dictates how terms in a Taylor expansion of $n$-graviton vertices in terms of the momenta correspond to terms in the canonical expansion. From the canonical expansion, one can also obtain the form-factor expansion, by a an identification of $a_2(k)$, $c_1(k)$ etc as the leading terms in a (formal) expansion of $f_RR$, and similarly for other form factors; the inverse path is based on performing this (formal) expansion, which may not necessarily converge. Subtleties in such formal expansions may arise due to non-local terms, such as, e.g., $\ln (-D^2)$ or similar.\\
Diffeomorphism symmetry is marked with an asterisk in this figure, because it is in practice broken by the regulator, implying that the vertex expansion contains strictly more terms than the canonical expansion or the form-factor expansion.}
\end{figure*}

To extract the dependence of $\Gamma$ on $k$ as well as on physical scales, two expansions are more suitable than the  expansion in canonical dimensions in Eq.~\eqref{eq:effaction_expansion_dimension}, namely the vertex expansion and the form-factor expansion.
In the vertex expansion, one considers fluctuations in the metric $h_{\mu\nu}$ about a given background, most often chosen to be flat. This results in quantities such as the three-graviton vertex $h_{\mu\nu}(p_1)h_{\kappa\lambda}(p_2)h_{\rho\sigma}(-p_1-p_2)\Gamma^{(3)\,\mu\nu\kappa\lambda\rho\sigma}(p_1,p_2)$ being the central objects of study. In addition, there is the expansion in form factors, including, e.g., $R f_{RR}(D^2)R$ etc. The expansions in canonical dimensions, vertex functions and form factors are related to each other, cf.~Fig.~\ref{fig:expansions}.

In the vertex expansion, one groups interactions not into diffeomorphism invariant terms, but instead organizes them by the power of the field $h_{\mu\nu}$, with the coefficients being functions of $k$ as well as the physical momenta,
\begin{eqnarray}
\Gamma_k&=& \int \!d^4x\, \left( h_{\mu\nu}\Gamma_k^{(2)\,\mu\nu\kappa\lambda}(-\partial^2)h_{\kappa\lambda} + \dots \right)\\
&=& \prod_{i=1}^n\int\!\! \frac{d^4p_i}{(2 \pi)^4} \Gamma_k^{(i)\,\mu_1\nu_1\dots \mu_n\nu_n}(p_1,\dots,p_{n-1})\cdot\nonumber\\
&{}& \quad \cdot h_{\mu_1\nu_1}(p_1)\dots h_{\mu_n\nu_n}(-p_1-\dots -p_{n-1}).\label{eq:Gamma_vertex_expansion}
\end{eqnarray}
The physical limit of these vertex functions, in which they retain a dependence on the physical (momentum) scales, is $k\rightarrow 0$.
In such an expansion, a given coupling in a diffeomorphism invariant term, e.g., the Newton coupling $\frac{1}{16\pi G_N} \int d^4x\sqrt{g}R$ gives rise to infinitely many ``avatars'' of a coupling. Examples of such avatars have first been studied in \cite{Manrique:2009uh,Donkin:2012ud} and the nomenclature was introduced in \cite{Dona:2015tnf}. 
From a given vertex function $\Gamma^{\mu_1\nu_1\dots \mu_n\nu_n}(p_1,\dots,p_{n-1})$, such an avatar can be extracted by identifying the tensor structure associated to the $n$th term in the expansion of the diffeomorphism invariant term into fluctuations $h_{\mu\nu}$. Whether the distinct tensor structures within a vertex function form an orthogonal basis, such that an isomorphism between the expansions \eqref{eq:Gamma_vertex_expansion} and \eqref{eq:effaction_expansion_dimension} exists, is, to the best knowledge of the author, unproven, but would be expected on the grounds of diffeomorphism symmetry. 

In practice, there is the added complication that gauge-fixing terms must be included in the effective action to use the Wetterich equation, and these break diffeomorphism symmetry, such that, in addition to avatars of couplings of diffeomorphism invariant terms, the vertex expansion contains additional tensor structures in all vertices.

To understand which approximations are typically associated to the vertex expansion, it is useful to contrast i) a Taylor expansion of a vertex function in momenta, ii) a vertex function that has a full momentum dependence and iii) a vertex function in which the momentum dependence is complete. Taylor expansions underlie the relation of vertex expansion to the canonical expansion, cf.~Fig.~\ref{fig:expansions}. Working with the vertex expansion, one often retains the full momentum dependence that the Wetterich equation generates on the right-hand-side \emph{within} a given truncation. This is strictly larger than what is captured in Taylor expansions, which are by no means guaranteed to converge and even if they do, to have an infinite radius of convergence. This \emph{full} momentum dependence is to be contrasted with the \emph{complete} momentum dependence of a vertex function, which is the momentum dependence that it has in principle, if calculated without any truncation. In the literature, one finds examples of full momentum dependence (see below) and should understand this terminology in contrast to the complete momentum dependence which is in practice very difficult to attain.

A systematic approach to the flow equation in terms of the vertex expansion starts from the flow equation for the two-point function, $\Gamma^{(2)}$, which depends on the three- and four-point functions. No higher $n$-point functions enter due to the one-loop structure of the Wetterich equation; and in general the flow of an $n$-point function depends all $m$-point functions with $m=2,\dots,n+2$. In addition to the choice of truncation in $n$, one can retain only some part of the physical momentum dependence. For instance, as a first step, only a $k$-dependent wave-function renormalization for the graviton as well as the Faddeev-Popov ghost fields arising in gauge fixing was retained \cite{Eichhorn:2010tb,Codello:2013fpa}, before the $p$-dependence of the graviton propagator was first studied in \cite{Christiansen:2014raa} and explored further in \cite{Knorr:2021niv}. As a next step, the graviton three-point function was studied in \cite{Christiansen:2015rva}, providing the first evidence for an asymptotically safe fixed point for the Newton coupling extracted from the vertex expansion.
In \cite{Denz:2016qks}, a dynamical four-point function was included, from which information on quadratic-curvature terms was extracted. The results in \cite{Denz:2016qks} confirm the results quoted in Tab.~\ref{tab:FP}, with three relevant directions. Importantly, the vertex expansion can also be used on a non-flat background, such that, e.g., the dependence on the Ricci scalar has been evaluated \cite{Christiansen:2017bsy,Burger:2019upn}, similarly to the $f(R)$ calculations discussed above.
Results in this approach are reviewed in \cite{Pawlowski:2020qer,Pawlowski:2023gym} and have more recently opened up the possibility to study asymptotic safety in Lorentzian signature \cite{Fehre:2021eob,Kher:2025rve,Pawlowski:2025etp}, see Sec.~\ref{sec:Lorentzian} and explore scattering processes, see Sec.~\ref{sec:particlescattering}, finding scattering cross-sections with asymptotically safe UV behavior \cite{Pastor-Gutierrez:2024sbt,Chiesa:2026tlz}.\newline

The second alternative expansion, well-suited to capture physical scale-dependence -- in this case through dependence on curvature and on covariant derivatives --  is in terms of form factors, reviewed in \cite{Knorr:2022dsx} and reads
\begin{eqnarray}
\Gamma&=&\frac{1}{16\pi G_N} \int d^4x\, \sqrt{g}\Biggl(2 \Lambda - R - \frac{1}{6}R\, f_{RR}\left(\Box \right)R \nonumber\\
&{}&+ \frac{1}{2}C_{\mu\nu\kappa\lambda}f_{CC}\left(\Box \right) C^{\mu\nu\kappa\lambda}\Biggr)+\dots\label{eq:Gammaformfactor}
\end{eqnarray}
The two form factors $f_{RR}$ and $f_{CC}$ enter the graviton propagator around a flat background, exemplifying the close relation of form factors to vertex functions. The form factor expansion has been developed in \cite{Knorr:2019atm}. In \cite{Bosma:2019aiu}, using non-local heat-kernel techniques \cite{Codello:2012kq},  $f_{CC}$ was calculated. From such a form factor, one can directly infer quantum corrections to the potential between two point sources and \cite{Bosma:2019aiu}, that the potential can be rendered finite in this way. This may be interpreted as an indication that asymptotic safety lifts spacetime singularities, see also Sec.~\ref{sec:BH}. A key point about form-factors is that they can be evaluated in the physical limit $k \rightarrow 0$, while retaining physical momentum dependence \cite{Knorr:2026vax},
see also \cite{Glaviano:2026lew} for a recent example using the proper-time flow equation.\newline

Taken together, the results obtained in the vertex expansion and with form factors not only complement and support results obtained in the expansion in canonical dimensions. In addition, they provide critical information on the physics of asymptotic safety, by giving access to the physical scale dependence of the effective action. At the moment of writing, these results, although not conclusively established beyond the first truncations, are consistent with well-behaved physical observables, e.g., cross-sections with asymptotically safe UV behavior and the absence of spacetime singularities.\newline

We close by mentioning supporting evidence for the existence of a fixed point from other methods. First, in a perturbative loop expansion based on the Einstein action, a fixed point for the Newton coupling is found \cite{Niedermaier:2009zz,Kluth:2024lar,Falls:2024noj}. 
The leading critical exponent from these studies is $\theta=2.8$ \cite{Kluth:2024lar}, $\theta=2.3 $ \cite{Falls:2024noj}, respectively. While robust error estimates from the functional RG as well as perturbative calculations are not straightforward to achieve, these values lie well within the typical range of critical exponents from functional RG studies.\\
Second, in Causal Dynamical Triangulations, several higher-order phase transitions are found in the phase diagram \cite{Ambjorn:2011cg,Ambjorn:2017tnl,Ambjorn:2018qbf}, and recently,  \cite{Ambjorn:2024qoe} provided support for the idea that an asymptotically safe fixed point may be connected to at least one of those transitions. 
\newline

Overall, the results from the functional RG, supported by perturbation theory and Dynamical Triangulations, constitute robust and compelling evidence that an asymptotically safe fixed point exists in Euclidean quantum gravity, with very likely 3 relevant directions\footnote{We are not making the distinction between essential and inessential couplings at this stage. Only the essential couplings, which cannot be removed by a field redefinition, need to exhibit a fixed point \cite{Baldazzi:2021orb}. This means that the number of relevant directions found above is an upper bound.} and accordingly very high predictive power. \\
Given that it is beyond reasonable doubt that an asymptotically safe fixed point exists in the Euclidean regime for pure gravity, it is critical to advance the next steps towards a more realistic theory and study a) the impact of matter, in particular Standard Model matter and b) the effect of a change of Lorentzian signature. We survey advancements in both directions next.

\subsection{Fixed point for gravitational couplings in gravity-matter systems}\label{sec:mattermatters}
\emph{...where we present the evidence for asymptotic safety in gravity-matter models which include the Standard Model. In this section, we focus on the simplest version of such a fixed point, at which matter is minimally interacting. This section is twinned with Sec.~\ref{sec:particlephysics}, which discusses the properties of the couplings in the matter sector in more detail.}\\

It is one of the most fundamental observations about our universe that it contains not just gravity, but also matter, where ``matter'' also encompasses the fields associated to the other fundamental forces. In a realistic theory of the quantum structure of gravity, applicable to our universe, matter must therefore be accounted for. This may take different forms, e.g., matter may be unified with gravity as in string theory, or it may emerge as a low-energy degree of freedom as in a Kaluza-Klein compactification, or it may be added as an independent degree of freedom. The latter is the approach taken to matter in asymptotic safety.  \\

As soon as we write kinetic terms for matter fields, there are interactions with gravity. This is a consequence of the fact that any form of energy gravitates, including the kinetic energy of matter fields. At a practical level for calculations, the kinetic terms for matter fields include explicit factors of the metric (and, in the case of fermions, also the connection).
\begin{eqnarray}
\Gamma_{{\rm kin}\, \phi}&=& \frac{1}{2}\int \!d^4x \sqrt{g}g^{\mu\nu}\partial_{\mu}\phi \partial_{\nu}\phi,\\
\Gamma_{{\rm kin}\, \psi} &=& i \int d^4x \!\sqrt{g}\bar{\psi}\gamma^a e^{\mu}_a(\partial_{\mu}+ \frac{[\gamma^{b},\gamma^c]}{8}\omega_{\mu\, bc})\psi,\\
\Gamma_{{\rm kin} A}&=& \frac{1}{4}\int \!d^4x \sqrt{g}g^{\mu\nu}g^{\kappa\lambda}F_{\mu\kappa}F_{\nu \lambda}\\
&=&\frac{1}{4}\int\! d^4x \sqrt{g}g^{\mu\nu}g^{\kappa\lambda}(\!\partial_{\mu}A_{\kappa}- \partial_{\kappa}A_{\mu}\!)(\!\partial_{\nu}A_{\lambda}- \partial_{\lambda}A_{\nu}\!)\nonumber.
\end{eqnarray}
where we have assumed a real scalar in the first line, a Dirac fermion\footnote{For Weyl fermions, gamma matrices have to be substituted with sigma matrices.} in the second line and a massless vector field, i.e., a gauge field, in the third line. In the second line, we have introduced the vielbein $e^a_{\mu}$ which translates between tangent space (with Latin indices) and spacetime (with Greek indices), with $g_{\mu\nu} = e_{\mu}^a e_{\nu}^b\eta_{ab}$ and the spin connection $\omega_{\mu}^{ab}= - (\partial_{\mu}e_{\nu}^a)e^{\nu b}+ \Gamma_{\nu\mu\sigma}e^{\nu a}e^{\sigma b}$.\footnote{The fluctuations of vielbeins can be translated into metric fluctuations in a suitable gauge for the local Lorentz transformations in tangent space \cite{vanNieuwenhuizen:1981uf,Woodard:1984sj,Eichhorn:2011pc}. These results are equivalent to using the spin-base invariant formalism \cite{Gies:2013noa,Gies:2015cka,Lippoldt:2015cea}.} For the gauge fields, we make clear that the connection drops out due to the antisymmetry of $F_{\mu\nu}$.

There are two effects that immediately follow: First, from these kinetic terms, there are contributions to the flow of the gravitational couplings. Second, from these kinetic terms, gravity necessarily generates matter self-interactions as well as non-minimal interactions, which cannot be set to zero at the gravitational fixed point \cite{Eichhorn:2011pc, Eichhorn:2012va}. These \emph{induced} interactions are all dimension-6 and higher interactions and do not include any of the dimensionless Standard Model couplings.
We focus on the resulting fixed point, called \emph{maximally symmetric asymptotically safe fixed point} \cite{Eichhorn:2017eht} in the present section, and discuss the addition of Standard Model interactions in Sec.~\ref{sec:particlephysics}.
\footnote{It is self-consistent to not include the Standard Model interactions when starting from the kinetic terms, i.e., the gauge and Yukawa couplings as well as Higgs potential can self-consistently be set to zero at the fixed point and are not generated from diagrams which are based on the vertices constructed from the kinetic terms, only. The question whether the non-vanishing IR values of these couplings, known from experiment, can be achieved, starting from such a fixed point, is addressed in  Sec.~\ref{sec:particlephysics}.} Below, we first discuss the gravitational properties of this fixed point, before commenting on the induced matter interactions.

The matter contributions to the $k$ dependence of the dimensionless Newton coupling $G=G_N k^2$ and dimensionless cosmological constant $\lambda= \Lambda k^{-2}$ have been found in \cite{Dona:2013qba}
\begin{eqnarray}
\beta_G\Big|_{\rm matter}&=& \frac{G^2}{6\pi} \left(N_S+2N_D-4N_V \right),\label{eq:betaGMatter}\\
\beta_{\lambda}\Big|_{\rm matter}&=& \frac{G}{4\pi}\left(N_S - 4N_D +2 N_V \right)\nonumber\\
&{}&+ \frac{G \cdot \lambda}{6 \pi}\left(N_S + 2N_D - 4N_V \right),
\end{eqnarray}
where $N_S$ is the number of scalars, $N_D$ the number of Dirac fermions and $N_V$ the number of massless vectors. 
When working with $N_W$ Weyl fermions, the correct result is obtained through the substitution $N_D \rightarrow 2 N_W$. Explicit mass terms do not play a role here, because they vanish at the maximally symmetric fixed point.
For the massless vector fields, gauge symmetry is assumed; therefore the correct result requires accounting for the Faddeev-Popov ghosts arising from gauge fixing. These Faddeev-Popov ghosts couple to gravity even in the Abelian case, where they do not couple to the gauge field. We rush to add that \cite{Dona:2013qba} is calculated in the simplest approximation, relying on the Einstein-Hilbert truncation and a background approximation. Studies going beyond these approximations include \cite{Meibohm:2015twa,Christiansen:2017cxa,Alkofer:2018fxj,Eichhorn:2018ydy,Pastor-Gutierrez:2022nki}.

Of particular interest is the fate of the gravitational fixed point with SM matter content, as well as the case with the SM and several additional fields, e.g., to describe dark matter. Studies which account for the Einstein-Hilbert part of the gravitational action, but in part also include higher-order gravitational interactions, conclude that a fixed point in the gravitational couplings persists in the presence of such matter content. This fixed point can be continuously connected to the gravitational fixed point without matter, when  $N_{S/D/V}$ are treated as continuous parameters \cite{Dona:2013qba, Biemans:2017zca, Christiansen:2017bsy, Alkofer:2018fxj, Wetterich:2019zdo, Sen:2021ffc, Pastor-Gutierrez:2022nki}.

The contributions in Eq.~\eqref{eq:betaGMatter} can be discussed in terms of screening and antiscreening: scalars and fermions screen the gravitational coupling \cite{Narain:2009fy, Dona:2012am,Dona:2013qba, Percacci:2015wwa, Labus:2015ska, Meibohm:2015twa, Meibohm:2016mkp,Dona:2015tnf, Biemans:2017zca, Alkofer:2018fxj, Eichhorn:2018ydy, Eichhorn:2018nda,Eichhorn:2018akn,Wetterich:2019zdo, Daas:2020dyo, Daas:2021abx,Laporte:2021kyp, Sen:2021ffc}, see also \cite{Kabat:1995eq, Larsen:1995ax,Christensen:1978sc} for results from perturbation theory. In the absence of gravitational degrees of freedom, this increases the fixed-point value; in contrast, vectors antiscreen the gravitational coupling \cite{Dona:2013qba, Biemans:2017zca, Christiansen:2017cxa, Alkofer:2018fxj, Wetterich:2019zdo, Sen:2021ffc}, driving the fixed-point value towards zero.
 One should, however, be careful when drawing conclusions from this about the large-$N$-behavior of the full gravity-matter system, because the fixed-point value of $G$ is also controlled by gravitational fluctuations and the additional gravitational couplings. Robust results on the large-$N$-behavior do not yet exist, see \cite{Christiansen:2017cxa,Eichhorn:2018akn,Eichhorn:2022gku} for discussions of this point.\\
The corresponding contributions have been evaluated previously in perturbation theory \cite{Larsen:1995ax,Christensen:1978sc} and a large-$N$-expansion \cite{Smolin:1981rm}, where $N$ is the number of matter fields. With the functional RG, a seeming ambiguity in the sign of fermionic contributions was observed \cite{Percacci:2002ie,Percacci:2003jz} and resolved in \cite{Dona:2012am}, establishing agreement with perturbative results. \\

Similarly to Eq.~\eqref{eq:Gamma_vertex_expansion}, the vertex expansion has been used in gravity-matter systems. It  provides supporting evidence for the existence of a fixed point for gravity coupled to SM matter \cite{Meibohm:2016mkp,Christiansen:2017bsy,Pastor-Gutierrez:2022nki}. In addition, the vertex expansion and form-factor approach are particularly useful in assessing the physical scale dependence for asymptotically safe gravity-matter systems, which is key to understanding the behavior of observables such as scattering cross sections \cite{Pastor-Gutierrez:2024sbt,Chiesa:2026tlz,Knorr:2026vax}, see also Sec.~\ref{sec:particlescattering}.\\

Let us now turn to the induced matter interactions. 
The maximally symmetric fixed point is a \emph{shifted Gaussian fixed point}. It arises from the fully Gaussian fixed point that each matter theory has in the absence of gravity. This fixed point exists because, once all interactions are set to zero, they stay zero. In the presence of gravity, even the kinetic terms of matter give rise to interactions with metric fluctuations and thus, a matter theory can never be truly non-interacting in the presence of gravity \cite{Eichhorn:2011pc, Eichhorn:2012va}. If gravity is asymptotically safe, the resulting interactions between matter and gravity induce matter self-interactions and non-minimal couplings with gravity. The induced interactions are characterized by the following properties \cite{Eichhorn:2017eht}, confirmed through explicit calculations for scalars, fermions and gauge fields in \cite{Eichhorn:2011pc, Eichhorn:2012va,Eichhorn:2013ug,Eichhorn:2016esv,Meibohm:2016mkp,Eichhorn:2017sok,Eichhorn:2017eht,Christiansen:2017gtg,Eichhorn:2018nda,Eichhorn:2019yzm,deBrito:2020dta,Laporte:2021kyp,deBrito:2021pyi,Eichhorn:2021qet,deBrito:2023kow,deBrito:2023myf,Knorr:2024yiu,Eichhorn:2024wba,deBrito:2025nog}:
\begin{itemize}
\item The induced interactions exhibit all the global symmetries that the kinetic terms for matter (at vanishing gauge couplings, if gauge fields are considered) exhibit. 
\item These induced interactions exhibit fixed points that are parametrically controlled by the fixed-point value of the Newton coupling. 
\item The induced interactions are all canonically irrelevant, i.e., these are higher-order interactions.
\item All induced interactions stay irrelevant, if gravity is sufficiently weakly coupled, i.e., the fixed point is near-perturbative enough, see Sec.~\ref{sec:nearpert}.
\end{itemize}
The most extensively studied interactions of this type are four-fermion interactions \cite{Eichhorn:2011pc,Meibohm:2016mkp,Eichhorn:2017eht,deBrito:2020dta,deBrito:2023kow}, schematically $(\bar{\psi}\gamma_{\mu}\gamma_5\psi)^2$, scalar derivative interactions $(\partial_{\mu}\phi \partial^{\mu}\phi)^2$ \cite{Eichhorn:2012va,Laporte:2021kyp,deBrito:2021pyi,deBrito:2023myf}, gauge field interactions\footnote{For a non-Abelian gauge field, the gauge coupling vanishes at this fixed point, i.e., the field strength effectively looks like $N_V$ copies of an Abelian gauge field.} $(F_{\mu\nu}F^{\mu\nu})^2$ and $(F_{\mu\nu}\tilde{F}^{\mu\nu})^2$ \cite{Christiansen:2017gtg,Eichhorn:2021qet,Knorr:2024yiu,Eichhorn:2024wba}; interactions between sectors, including non-minimal couplings with gravity, have also been studied \cite{Eichhorn:2016esv,Eichhorn:2017sok,Eichhorn:2018nda,Laporte:2021kyp,Knorr:2024yiu,deBrito:2025nog}.

At the maximally symmetric fixed point, the ``back-reaction" of induced interactions on the gravitational fixed point is expected to be small, unless gravity is very strongly coupled, see the discussion of Sec.~\ref{sec:nearpert}. 
The argument for this goes as follows: Gravitational loops induce these couplings, thus, for such an induced coupling $g_i$, the fixed-point value scales with $g_{i\, \ast}\sim G/(16\pi)$, where the factor $1/(16\pi)$ arises from a combination of a $1/(16\pi)^2$ for a loop effect and a $16 \pi G$, which is the prefactor for the gravitational coupling. In turn, the ``back-reaction'' of induced couplings comes with another $1/(16\pi)^2$ from loop effects; which is suppressed by this factor compared to any ``direct'' gravitational effect; see \cite{Eichhorn:2017eht}.\\

Finally, integrating towards $k\rightarrow 0$, the values of all these higher-order couplings are predicted, because they do not add new relevant directions.\footnote{This can in principle be circumvented if gravity fluctuations leave the near-perturbative regime and shift critical exponents in the direction of relevance by two (for fermions) or four (for scalars and vectors), see \cite{deBrito:2023myf}.}
Below the Planck-scale, the canonical scaling terms in their beta functions dominate. They lead to a Planck-scale suppression of these couplings at low energies. This generic expectation can only be circumvented by interacting fixed points for these couplings in the pure-matter regime, i.e., fixed points that can dominate the RG flow below the Planck scale \cite{Brenner:2024bps}. Thus, the effective action resulting from the maximally symmetric fixed point necessarily contains interactions, but these are generically Planck-scale suppressed. In addition, there may be marginal couplings, which run logarithmically, but none of them is necessarily induced. For a phenomenologically viable scenario, they must of course be present, see Sec.~\ref{sec:SM} for details. 

In summary, there is promising evidence that a maximally symmetric asymptotically safe fixed point exists for gravity together with SM matter fields (as well as some extensions of the SM). The evidence is less robust than in the case of pure gravity, because truncations of the functional RG have not been pushed to such high orders in the gravitational couplings. Nevertheless, many interactions that turn out to be irrelevant at the fixed point  \cite{Eichhorn:2011pc, Eichhorn:2012va,Eichhorn:2013ug,Eichhorn:2016esv,Eichhorn:2017sok,Eichhorn:2017eht,Christiansen:2017gtg,Eichhorn:2018nda,Eichhorn:2019yzm,deBrito:2020dta,Laporte:2021kyp,deBrito:2021pyi,Eichhorn:2021qet,deBrito:2023myf,Knorr:2024yiu,Eichhorn:2024wba,deBrito:2025nog} have been studied, reinforcing the stability of the first results.

In addition, there are first studies of matter fields in dynamical triangulations \cite{Ambjorn:2021uge}; with a quenched approximation providing information on the gravitational effect on matter \cite{Catterall:2018dns}. 
Initial results indicated a subleading effect of matter \cite{Ambjorn:2021uge}, but more recently, it has been observed that particular scalar fields may drive changes in the effective spacetime topology \cite{Ambjorn:2021fkp}.

\subsection{Gravitational fixed point in Lorentzian signature}\label{sec:Lorentzian}
\emph{...where we discuss the difficulties of setting up Renormalization Group flows in Lorentzian signature. We then
review first steps to find and characterize an asymptotically safe fixed point in Lorentzian signature. Existing results, so far limited to small truncations, are compatible with the assumption that the change in signature only has mild effects on the gravitational fixed point, but more work is needed to robustly establish a Lorentzian fixed point.}\\

Asymptotic safety, if it is to describe Nature, must be realized in Lorentzian, not in Euclidean signature.\footnote{We are assuming that there is no ``phase transition" associated to a change in signature in the early universe, which would reconcile a Euclidean transplanckian theory, with a Lorentzian, semi-classical theory.} This brings with it added challenges -- both conceptual as well as technical, as compared to the Euclidean theory.

A direct formulation of the Wetterich equation in Lorentzian signature faces the following challenges:

First, on a general metric background, expectation values depend on the choice of state. This is a well-known result of quantum field theory on curved backgrounds, underlying the physics of particle production in cosmological spacetimes as well as Hawking radiation. For the Wetterich equation it results in the necessity to choose a state \cite{Banerjee:2022xvi,DAngelo:2022vsh}. Thus, methods of algebraic quantum field theory, developed in the context of quantum field theory on curved backgrounds, underlie the development of a Lorentzian Wetterich equation in \cite{DAngelo:2022vsh,DAngelo:2023tis}.

Second, the choice of cutoff in the flow equation becomes more tricky. Even on a flat background, a momentum-cutoff cannot be imposed without breaking (local) Lorentz invariance, because one either restricts $p^2= p_0^2-\vec{p}^2< k^2$, but this does not restrict the energy/momentum at all in the case of 4-momenta that are null; or one restricts $\vec{p}^2<k^2$ and thereby breaks (local) Lorentz invariance. In \cite{Fehre:2021eob}, it was shown that this can be solved by using a Callan-Symanzik cutoff $\sim k^2$ for all modes, instead of choosing a cutoff function that depends on the momenta of the modes. In the Wetterich equation, the property that the regulator insertion $k\,\partial_k\, R_k$ acts as a UV regulator is lost in this case and additional UV regularization is needed. On a non-flat background, \cite{DAngelo:2022vsh} instead introduce a regulator function that satisfies spacetime locality. Both of these choices modify the interpretation of the flow equation; the simple Wilsonian picture of subsequent ``momentum shells'' of quantum fluctuations contributing to the change of $\Gamma_k$ is lost. 

\begin{figure}[!t]
\includegraphics[width=\linewidth]{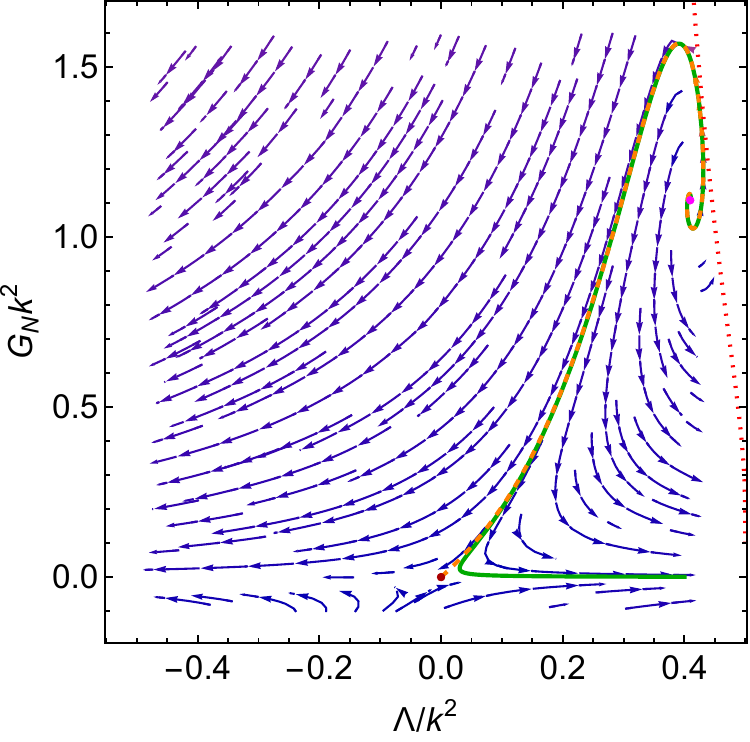}
\caption{\label{fig:Lorentzianflow} We show the RG flow in the Einstein-Hilbert truncation, calculated in Lorentzian signature in \cite{DAngelo:2023wje}. There is an interacting fixed point at $g \approx 1.1, \, \lambda \approx 0.4$ (magenta dot), connected to the free fixed point by a separatrix (orange dashed line). An RG trajectory with a long classical regime is indicated in green. In the deep IR, it terminates at a singularity of the RG flow (red dotted line). The qualitative similarity to Fig.~\ref{fig:EHplane} is clear.}
\end{figure}

These challenges have recently been overcome; \cite{Fehre:2021eob,Pawlowski:2025etp} as well as \cite{DAngelo:2023wje} (based on \cite{DAngelo:2022vsh,DAngelo:2023tis}) compute Renormalization Group flows in Lorentzian signature and identify an asymptotically safe fixed point. As a concrete example, we reproduce the results from \cite{DAngelo:2023wje} in Fig.~\ref{fig:Lorentzianflow}, which should be compared to Fig.~\ref{fig:EHplane}. While there are clearly some deformations of the flow, we highlight that similar deformations can be achieved in the Einstein-Hilbert truncation in Euclidean signature by changing the regulator. Similarly, the fixed point identified in \cite{Fehre:2021eob} is well compatible with Euclidean results. While these results require corroboration within larger truncations, the large degree of similarity to Euclidean results is promising.\\

There is a second pathway to explore the effect of Lorentzian signature on asymptotic safety. 
It is based on a restricted Euclidean configuration space in which each configuration can be analytically continued from Euclidean to Lorentzian signature. The resulting configuration space is not the full configuration space of Lorentzian metrics, because not every metric in Lorentzian signature is reachable by an analytical continuation from Euclidean signature \cite{Knorr:2022mvn}. An important example is a black-hole geometry, of which only the metric outside the horizon can be analytically continued. The horizon has no Euclidean counterpart, because the definition of an event horizon is as the boundary of the \emph{causal past} of future (timelike and null) infinity. Thus, an analytical continuation only gives rise to the part of the Schwarzschild geometry outside the horizon. Additionally, the continuation from Lorentzian to Euclidean signature is often ambiguous, even in maximally symmetric spacetimes such as de Sitter \cite{Baldazzi:2018mtl}.  

The restricted configuration space in which every configuration can be Wick-rotated (and is therefore foliatable) is used both in Causal Dynamical Triangulations  and more recently in the functional RG. As discussed in Sec.~\ref{sec:methods}, Causal Dynamical Triangulations provides evidence for a higher-order phase transition \cite{Ambjorn:2011cg,Ambjorn:2017tnl} supporting a continuum limit, with recent results showing how to recover the Gaussian fixed point and constituting grounds for optimism that an asymptotically safe fixed point may also be recovered
\cite{Ambjorn:2024qoe}.

Similarly, one may study the restricted Euclidean configuration space with functional RG methods, using a parameterization of metric fluctuations from the ADM formalism \cite{Manrique:2011jc} or other 3+1 splits of the metric \cite{Knorr:2018fdu}.  
Results indicate that an asymptotically safe fixed point persists in such a setting \cite{Manrique:2011jc,Rechenberger:2012dt,Biemans:2016rvp,Houthoff:2017oam,Biemans:2017zca,Saueressig:2023tfy,Korver:2024sam,Saueressig:2025ypi}.

Finally, when using the vertex expansion of the flow equation about a flat background, see Sec.~\ref{sec:methods}, one can reconstruct Lorentzian $n$-point functions from Euclidean ones by analytical continuation \cite{Bonanno:2021squ,Pastor-Gutierrez:2024sbt,Chiesa:2026tlz}, finding indications for an asymptotically safe fixed point even in some gravity-matter systems.

Taken together, these results constitute very promising evidence that a restricted Lorentzian configuration space in which each configuration can be foliated and analytical continuation from Euclidean to Lorentzian signature is possible, admits an asymptotically safe fixed point. \\

Confirming asymptotic safety in Lorentzian signature may also benefit from the use of other approaches to quantum gravity. First, it has been proposed in \cite{Eichhorn:2017bwe,Eichhorn:2019xav} to use causal sets, an intrinsically Lorentzian formulation. The  fundamental spacetime discreteness of causal set theory \cite{Bombelli:1987aa} is reinterpreted as a regularization. If a higher-order phase transition in the phase diagram of causal sets is found, a continuum limit can be taken in full analogy to Dynamical Triangulations, cf.~Sec.~\ref{sec:discretetechniques}, see \cite{Surya:2011du,Cunningham:2019rob}  for first studies with, however, no confirmed second-order transition yet.

More recently, it has been proposed \cite{Thiemann:2024vjx,Ferrero:2024rvi,Ferrero:2025idz,Ferrero:2025efd} to use methods from Loop Quantum Gravity, in which a canonical quantization of gravity in the Hamiltonian framework is performed \cite{Ashtekar:1987gu}. From the Hamiltonian formulation, a corresponding path integral can be derived. The Hamiltonian formulation relies on Lorentzian signature; thus giving access to asymptotic safety in Lorentzian signature.\\

In summary, understanding asymptotic safety in Lorentzian signature is a relatively recent frontier, with several complementary strategies to formulate Lorentzian flow equations, in addition to using analytical continuation and even ideas that borrow from other quantum-gravity approaches. Such a broad range of tools, together with the promising first results obtained in small truncations are grounds for optimism that it is possible to achieve a formulation of asymptotic safety in Lorentzian signature.

\subsection{Summing up: Establishing asymptotic safety -- the story so far}\label{sec:summing_up}
Since the proposal of asymptotic safety by Weinberg in \cite{Weinberg:1980gg}, there has been immense progress towards an increasingly realistic asymptotically safe theory of gravity. This progress has accelerated over the last decade or so, providing grounds for optimism that it will be possible to establish whether quantum gravity in our universe can be asymptotically safe.\\
Specifically, studies around two dimensions and at large $N$, reviewed in \cite{Martini:2022sll}, were superseded by studies of four-dimensional Euclidean gravity following the seminal paper by Martin Reuter \cite{Reuter:1996cp}. An asymptotically safe fixed point for pure gravity in four dimensions and Euclidean signature is now established beyond reasonable doubt and called Reuter fixed point. \\
Following the insight that ``matter matters'', individual studies included matter fields already early on, but an increased effort to establish the asymptotically safe Standard Model was made over the last decade, reviewed in \cite{Eichhorn:2022gku}. By now, there is promising evidence for the asymptotically safe Standard Model, with no counterindications discovered in the numerous studies that have been conducted of gravity-matter models. Going beyond the Standard Model to address open questions such as the nature of dark matter has also been done in an increasing number of studies, cf.~Sec.~\ref{sec:dm}.\\
Finally, one of the last big challenges, namely a formulation in Lorentzian signature, has been tackled over the last few years, with several complementary ideas being pursued simultaneously. First results  justify an optimistic outlook, namely that the asymptotically safe Standard Model (with suitable extensions for dark matter) could be formulated in Lorentzian signature, providing a theory of quantum gravity and matter that has all the ingredients required for a realistic description of our universe.\\

Overall, this development, and its acceleration over the last decade, with hard problems being tackled directly, speaks to a very healthy state of the field.

\section{Technicalities that matter}\label{sec:technicalities_matter}
\emph{...where we highlight selected technical aspects of studies on asymptotically safe gravity. These are only seemingly technicalities, because below the surface, they are intimately connected to the physics of asymptotic safety and the conceptual underpinnings of a Wilsonian approach to quantum gravity.}

\subsection{Is asymptotically safe gravity perturbative or non-perturbative?}\label{sec:nearpert}
\emph{...where we discuss the notion of near-perturbativity and the dynamical emergence of a background spacetime from the gravitational path integral as a prerequisite for such a near-perturbative regime of gravity. We then review the evidence for the dynamical emergence of a background and for the near-perturbative behavior of asymptotic safety and point out that the fixed point is ``as perturbative as it gets'' \cite{Falls:2013bv}.}\\

In a break with convention, the answer to the question in the title is probably ``neither". In other words, there is not a strict dividing line between perturbative and non-perturbative physics, but a transition between the two, and asymptotically safe gravity lies within this transitional regime.

Asymptotic safety is clearly not fully perturbative in the sense that the dynamics at (trans-)planckian scales is not well-approximated by the Einstein-Hilbert term, and higher-order terms are present.\footnote{It is, however, not \emph{necessary} to include these higher order terms to see asymptotic safety; the Reuter fixed point is already visible in the Einstein-Hilbert truncation.} 
This does, however, not imply that the theory is strongly coupled. Quadratic gravity, which contains quadratic curvature terms and can become asymptotically free precisely due to their presence, clearly exemplifies that the presence of higher-order terms does not necessarily result in a non-perturbative theory in the UV. There are also different notions of perturbativity. For example, one may require that quantities can be computed in perturbation theory. On the one hand, asymptotic safety appears to satisfy this -- after all, one can obtain a gravitational fixed point from perturbative calculations even in four dimensions \cite{Niedermaier:2009zz,Falls:2024noj,Kluth:2024lar}. On the other hand, asymptotically safe physical scaling of, e.g., scattering cross-sections in the UV (see Sec.~\ref{sec:particlescattering}) appears challenging to obtain from a purely perturbative calculation.

To avoid a discussion focused on semantics and instead be concrete, we specify a measure of perturbativity that defines what we mean by perturbative behavior and that is particularly useful in the asymptotically safe context. 
Values of couplings are not such a measure, because couplings can be rescaled arbitrarily without changing the physics -- a well-known example is the different choice of conventions for $n$ in for $\lambda/n \phi^4$ theory. Instead, a measure that is universal, and applicable to determine the degree of non-perturbativity of an RG fixed point, is based on the spectrum of critical exponents. For a free fixed point, i.e., a fully perturbative fixed point, the critical exponents agree exactly with the canonical dimensions of couplings. We can therefore define a measure of near-perturbativity \cite{Eichhorn:2020sbo}
\begin{equation}
\Delta_N = 
\sqrt{\frac{1}{N}\sum_{i=1}^N \left({\rm Re}(\theta_i) - d_{\bar{g}_i} \right)^2}.
\end{equation}
Herein, $N$ refers to the number of couplings that are being included, and it is understood that the labelling of critical exponents is by the size of their real part, i.e., starts from the most relevant interactions. There is no sharp boundary between perturbative and non-perturbative behavior in $\Delta_N$. Instead, growing $\Delta_N$ indicates behavior that is increasingly far away from perturbative. 

In fact, for $f(R)$ truncations, a small $\Delta_N$, connected to an approach to canonical scaling, was observed \cite{Falls:2013bv,Falls:2014tra,Falls:2017lst,Falls:2018ylp,Kluth:2020bdv,Becker:2024tuw} and enables a bootstrap for consistent truncations: under the assumption of near-perturbativity, one can know a priori whether an interaction is going to be relevant or irrelevant for all but a small handful of interactions (those with canonical dimension close to zero). 

Going beyond pure gravity, in gravity-matter systems, near-perturbativity has been supported by several distinct results: First, \cite{Eichhorn:2020sbo} explored the measure of near-perturbativity across the gravitational coupling space, arguing that $\Delta_N$ for matter couplings remains small for typical gravitational fixed-point values. Second, explicit studies of dimension-5 \cite{deBrito:2021akp,deBrito:2025ges,Assant:2025gto} and dimension-6 \cite{Eichhorn:2011pc}  (and higher \cite{Eichhorn:2016esv,Eichhorn:2017eht,Christiansen:2017gtg,Eichhorn:2021qet,deBrito:2021pyi,deBrito:2023myf}) matter interactions as well as dimension-6 non-minimal couplings \cite{Eichhorn:2017sok,Eichhorn:2018nda,Laporte:2021kyp} lend support to the scenario of a near-perturbative fixed point, because all these interactions contribute irrelevant directions at the asymptotically safe fixed point.\footnote{This does not mean that such an interaction may not also be part of a superposition of couplings that is relevant at the fixed point; it means that the interaction is not associated to a new relevant direction that is not already there due to a lower-order interaction.}\\

There is another measure of near-perturbativity, which is tied to the breaking of diffeomorphism symmetry by the regulator, see Sec.~\ref{sec:local_coarse_graining}. 
This breaking is, very loosely speaking, large, if quantum fluctuations dominate over the classical behavior of the theory.
Then, couplings that are related by diffeomorphism symmetry, and are called
``avatars'', introduced in \cite{Manrique:2010am,Donkin:2012ud}, with the terminology coined in \cite{Dona:2015tnf}, can actually differ at an asymptotically safe fixed point. The relation between such couplings is controlled by (modified) symmetry identities. These symmetry identities can (in principle) be used to restore diffeomorphism symmetry in the physical limit $k \rightarrow 0 $. How non-trivial these identities become in the UV reflects the non-perturbative character of the fixed point. At a perturbative fixed point, which is dominated by classical behavior rather than dominated by quantum fluctuations, these symmetry identities are more trivial than at a highly non-perturbative fixed point.
Thus, \cite{Eichhorn:2018akn} introduced the notion of ``effective universality'', whereby the avatars of a given coupling are close to each other at the asymptotically safe fixed point, further investigated in \cite{Eichhorn:2018ydy,Eichhorn:2018nda}. These results  support the scenario of a near-perturbative fixed point from a different type of diagnostic.\\

We now turn our attention to  near-perturbative behavior in the context of the gravitational path integral. We expect that near-perturbative behavior implies that the path-integral is dominated by a flat (or de-Sitter like, with very small cosmological constant) background with controlled fluctuations about this background. This is in contrast to a non-perturbative path integral, in which configurations contribute which are far from such a background, and for which the expectation value of fluctuations is large. In order for the gravitational path integral to actually describe nature, it must at least contain a regime in which it is effectively near-perturbative, because there is good observational evidence for a regime, in which spacetime is well-approximated by a de-Sitter-like background with very small cosmological constant and small fluctuations. However, it is by no means obvious that the full gravitational path integral actually gives rise to such a regime. This makes results in dynamical triangulations, which show the emergence of such a regime \cite{Ambjorn:2004qm,Ambjorn:2008wc,Bassler:2021pzt,Maas:2025rug}, highly non-trivial. They lend support to the idea that, while the full gravitational path integral contains very non-classical configurations, a regime with effectively classical behavior and small fluctuations, emerges dynamically, see  \cite{Maas:2022lxv} for a conceptual underpinning. With the functional RG in its continuum formulation, such emergence is difficult to study, because it requires that highly non-perturbative configurations are fully accounted for. It is technically simpler to work in a near-perturbative regime, where the existence of a dynamically emergent preferred background is taken for granted. The above results on the scaling exponents characterizing the gravitational fixed point suggest the self-consistency of the assumption of near-perturbativity; and the lattice results \cite{Ambjorn:2004qm,Ambjorn:2008wc,Bassler:2021pzt,Maas:2025rug} support that such a regime emerges from a full path integral.

Finally, we add another, very different piece of evidence for the near-perturbative nature of asymptotic safety, namely that the Standard Model of particle physics is perturbative at the Planck scale, with two-loop terms already being very small compared to one-loop terms. This implies that gravitational contributions, required to UV complete the Standard Model, can be small, resulting in near-canonical critical exponents \cite{Eichhorn:2017ylw,Eichhorn:2018whv,Alkofer:2020vtb,Eichhorn:2022vgp,Eichhorn:2025sux}. 
Conversely, if we were to assume that asymptotically safe quantum gravity is non-perturbative, there is no reason why it should result in a perturbative matter theory at the Planck scale. Rather, one would expect that a non-perturbative quantum-gravity regime beyond the Planck scale, with, e.g., highly non-canonical scaling exponents, gives rise to a non-perturbative matter theory below the Planck scale. This piece of evidence only relies on fully understood and experimentally supported physics (and a well-understood RG flow between the Planck scale and the electroweak scale). It may thus be interpreted as an (indirect) hint from experiment about the properties of quantum gravity at the Planck scale.

We close by highlighting that near-perturbativity is first of all a statement about the physical properties of asymptotic safety. Second, and crucial for the construction of consistent functional RG truncations, is brings the technical implication that truncations can be constructed following a bootstrap principle \cite{Falls:2013bv}: near-perturbativity is assumed, interactions are selected accordingly, and the resulting fixed point can be checked for consistency with the initial assumption. For the study of gravity-matter models, this has direct implications for the free parameters: We expect that some of the marginal interactions may introduce free parameters, but expect that at best few canonically irrelevant interactions do, and only those with mass dimension -1 or -2 for the coupling (i.e., 5 or 6 for the interaction terms).

\subsection{How to reconcile background independence with local coarse graining?}\label{sec:local_coarse_graining}
\emph{...where we discuss the apparent incompatibility between setting up a Renormalization Group transformation or coarse-graining scheme that is local, i.e., requires a background spacetime, but respecting background independence. We discuss both, conceptual  as well as technical aspects. The conceptual aspects are related to the discussion in Sec.~\ref{sec:nearpert}.}\newline

RG flows in flat-space QFT take into account quantum fluctuations in the path integral in a Fourier basis and sort them by their four-momentum $p^2$, distinguishing high-momentum modes, with $p^2>k^2$, from low-momentum modes, with $p^2<k^2$.\footnote{This is routinely done in Euclidean signature, see Sec.~\ref{sec:Lorentzian}.} In the presence of a nontrivial background geometry, this can be generalized to the d'Alembertian $-D^2$, such that loop diagrams can be evaluated either on backgrounds for which the spectrum of $-D^2$ is known exactly \cite{Lauscher:2001ya}, or through heat-kernel techniques \cite{Vassilevich:2003xt,Benedetti:2010nr,Codello:2012kq}.

However, there is a deeper conceptual challenge in quantum gravity: this is the idea that quantum gravity should \emph{explain}, where the spacetime geometry of our universe actually originates in the first place. In other words, why is there a preferred spacetime background for our universe? \\
In quantum gravity, there is a range of different perspectives one may take on the existence of a background. The two extreme points in this range are the following: On the one hand, one may assert that diffeomorphism symmetry prohibits the existence of a preferred background. All metrics must be treated on the same footing (e.g., in a gravitational path integral), and background independence is a key requirement of any sensible quantum theory of gravity. On the other hand, one may note that our universe is very well described by a preferred background on top of which quantum fluctuations exist, but are small. Thus, to describe quantum gravity effects in our universe, one may actually start from a given background, and investigate fluctuations around it.

Out of a path integral over all spacetime geometries (with fixed topology), a specific geometry can emerge in the following way:
The effective action $\Gamma$ is a functional for the \emph{expectation value} of the metric $\langle g_{\mu\nu}\rangle$, which includes the effect of all quantum fluctuations. Its extremization provide the quantum equations of motion for $\langle g_{\mu\nu}\rangle$,
\begin{equation}
\frac{\delta \Gamma[\langle g \rangle]}{\delta \langle g_{\mu\nu}\rangle}=0.
\end{equation}
The solution to these equations of motion, obtained with particular initial/boundary conditions, constitutes a \emph{dynamically emergent}, special background. 

The challenge in asymptotic safety lies in the fact that we aim to compute  $\Gamma[\langle g \rangle]$ while accounting for the relations between the couplings that arise as a consequence of an asymptotically safe fixed point. To do so, we calculate the RG flow of the couplings that arises from a \emph{local} coarse-graining procedure. By a local coarse-graining procedure, we mean a setup in which quantum fluctuations are sorted by a suitable notion of ``wavelength'', i.e., using the eigenvalues of a suitable differential operator. To define the latter, a background metric $\bar{g}_{\mu\nu}$ is needed.
Thus, we run into the apparent tension between \emph{background independence} and \emph{local coarse graining}.\footnote{Locality here is meant in the sense that a Wilsonian integration over subsequent momentum shells is local, i.e., constitutes coarse-graining over increasingly large wavelengths. Different notions of locality can also make sense in RG flows for quantum gravity \cite{DAngelo:2022vsh}. All notions have in common that they require a metric to define the differential operators the eigenvalues of which are compared to the RG scale to sort modes into UV and IR modes.}

This tension can be resolved by the background field method, where a background $\bar{g}_{\mu\nu}$ is introduced, but plays a purely auxiliary role \cite{Reuter:1996cp} and is never set to any specific metric. In other words, background independence is respected not by avoiding the introduction of a background, but by \emph{treating all backgrounds on an equal footing} \cite{Becker:2014qya}. In practice, this works by using background independent heat-kernel techniques \cite{Benedetti:2010nr,Knorr:2021slg}. These make it possible to evaluate the trace of quantum fluctuations in the Wetterich equation on a background $\bar{g}_{\mu\nu}$ that at no stage in the calculation needs to be specified exactly.
 There is therefore no actual incompatibility between background independence and local coarse-graining; the solution is, however, technically challenging.

In practice, many calculations rely on the technically simpler setup to specify a particular choice for the auxiliary background, with Euclidean de Sitter or Anti de Sitter as well as flat Euclidean spacetime constituting by far the most popular choices and, e.g., \cite{Benedetti:2009rx,Gies:2016con} using less symmetric backgrounds. 

The auxiliary background is typically introduced through a linear split of the metric
\begin{equation}
g_{\mu\nu} = \bar{g}_{\mu\nu} +h_{\mu\nu}.\label{eq:linearsplit}
\end{equation}
This can be implemented as a shift of variables in the path integral,
\begin{equation}
\int \mathcal{D}g_{\mu\nu}e^{-S[g_{\mu\nu}]} \rightarrow \int \mathcal{D}_C h_{\mu\nu}e^{-S[\bar{g}_{\mu\nu}+ h_{\mu\nu}]}.
\end{equation}
By the subscript $C$ in the measure over $h_{\mu\nu}$ we indicate that constraints have to be imposed to ensure that the configuration space corresponds to that of all metrics of fixed signature; because some choices of $h_{\mu\nu}$ in Eq.~\eqref{eq:linearsplit} can result in a signature-change in $g_{\mu\nu}$, see \cite{Nink:2014yya,Knorr:2022mvn}.
The path integral over $h_{\mu\nu}$ can be regularized through an IR cutoff quadratic in $h_{\mu\nu}$, but dependent on the background-covariant Laplacian $-\bar{D}^2$, with a cutoff term 
\begin{equation}
\Delta S_k = \frac{1}{2} \int d^4x \sqrt{\bar{g}}\,h_{\mu\nu}R^{\mu\nu\kappa\lambda}_k(-\bar{D}^2)h_{\kappa\lambda}.\label{eq:IRcutoff_h}
\end{equation}
By introducing a source term for $h_{\mu\nu}$, one can perform the Legendre transform and derive a flowing action that depends on two fields, $\bar{g}_{\mu\nu}$ and $\langle h_{\mu\nu}\rangle$, see, e.g., \cite{Becker:2014qya,Christiansen:2017bsy,Eichhorn:2018akn,Burger:2019upn} for works explicitly tracking the dependence on both fields and \cite{Pawlowski:2020qer,Pawlowski:2023gym} for reviews.
The flow equation for the resulting $\Gamma_k[\bar{g}_{\mu\nu}, \langle h_{\mu\nu}\rangle]$ reads
\begin{eqnarray}
&{}&k\partial_k\, \Gamma_k[\bar{g}_{\mu\nu}, \langle h_{\mu\nu}\rangle] \nonumber\\
&{}&\quad\quad = \frac{1}{2}{\rm STr} \left(\Gamma_k^{(0,2)}[\bar{g}_{\mu\nu}, \langle h_{\mu\nu}\rangle]+R_k \right)^{-1}k\partial_k\, R_k.\label{eq:fluctuationflow}
\end{eqnarray}
We highlight that, because $h_{\mu\nu}$ is the integration variable in the path integral, the second derivative of $\Gamma_k$ must be taken with respect to $\langle h_{\mu\nu}\rangle$ and not $\bar{g}_{\mu\nu}$. Thus the couplings showing up on the right-hand-side of the flow equation are those arising from vertices with $\langle h_{\mu\nu}\rangle$ on the internal legs. The two derivatives, $\Gamma_k^{(2,0)}$ and $\Gamma_k^{(0,2)}$ differ due to the gauge-fixing and regulator term.\footnote{There is an often used approximation, called the background-field approximation. In this approximation, one starts from an ansatz that differs in its dependence on $h$ and $\bar{g}$ through gauge-fixing and regulator terms. One then takes the second derivative with respect to $\langle h_{\mu\nu}\rangle$ and plugs this into the Wetterich equation. The approximation comes in the next step, where one identifies the couplings on the right-hand-side with the couplings of an action written purely in terms of $\bar{g}_{\mu\nu}$. In a treatment without this approximation, one must distinguish these couplings from corresponding ``avatars'' that mix $\bar{g}$ with $h$ and have $h$ on internal and $\bar{g}$ on external lines. More details can be found in \cite{Pawlowski:2020qer,Pawlowski:2023gym}.}

A main advantage of the background field method that is used here is the status of diffeomorphism symmetry. When using the flow equation, we must introduce a gauge-fixing term that breaks diffeomorphisms, because otherwise $\Gamma_k^{(2)}$ has zero modes and is not invertible to obtain the propagator. Under a background split,
diffeomorphism transformations come in two distinct incarnations: there is an auxiliary background diffeomorphism symmetry, under which $\bar{g}_{\mu\nu}$ transforms, but $h_{\mu\nu}$ does not, and a ``physical" diffeomorphism symmetry, under which $g_{\mu\nu}$ transforms and thus both $\bar{g}_{\mu\nu}$ as well as $h_{\mu\nu}$ transform \cite{Reuter:1996cp, Reuter:2019byg}. Once the full path integral has been evaluated, only $\bar{g}_{\mu\nu}$ remains, because one sets the expectation value of $h_{\mu\nu}$ to zero. Then, the auxiliary background diffeomorphism symmetry becomes physical.

The split of $g_{\mu\nu}$ into $\bar{g}_{\mu\nu}$ and $h_{\mu\nu}$ is controlled by the so-called split Ward-identity, which tracks the separate dependence of $\Gamma_k[\bar{g}_{\mu\nu}, h_{\mu\nu}]$ on its two arguments. Because of the regulator and the gauge-fixing term, both of which treat $\bar{g}_{\mu\nu}$ and $h_{\mu\nu}$ differently from one another, 
\begin{equation}
\left(\frac{\delta}{\delta \bar{g}_{\mu\nu}}- \frac{\delta}{\delta h_{\mu\nu}}\right) \Gamma_k[\bar{g}_{\mu\nu}, h_{\mu\nu}] \neq 0,
\end{equation}
as would be the case otherwise. Instead, the right-hand-side of this equation depends on the regulator and gauge-fixing, and only vanishes in the limit $R_k \rightarrow 0$ and $S_{\rm gf} \rightarrow 0$.

In practice, aside from a few attempts at reconciling the flow equation and the split Ward identity in truncations \cite{Morris:2016spn,Ohta:2017dsq,Nieto:2017ddk}, the split Ward identity is largely left aside. Whether the existing functional RG calculations should therefore truly be understood as a fully background-independent calculations, or whether they should be interpreted rather as relying, at least to some extent, on a pre-existing background, is therefore open for debate.

Instead, there are other methods which are better suited to search for the dynamical emergence of a background from a fully non-perturbative setting, including dynamical triangulations, see Sec.~\ref{sec:discretetechniques}. Indeed, in causal (and Euclidean) dynamical triangulations, it is well-established that a de-Sitter-like phase, characterized by a de-Sitter-like profile for the average scale factor  \cite{Ambjorn:2004qm}, with relatively small expectation value of fluctuations \cite{Ambjorn:2008wc}, emerges from the fully non-perturbative path integral, in which highly non-classical configurations are included.

We close by highlighting that there are also versions of the functional RG equations which are derived in a pre-geometric setting, namely in matrix and tensor models, and in which the notion of ``scale'' is tied to the number of degrees of freedom (aka the matrix/tensor size) \cite{Eichhorn:2013isa,Eichhorn:2017xhy,Eichhorn:2018phj,Castro:2026zch}, but has no interpretation as a local coarse graining scale. Such abstract notions of ``scale'' are the only possible notions if one insists to realize background independence by forbidding the existence of a background. This is in contrast to the (more subtle) realization of background independence discussed above, in which background independence does not forbid the existence of a background, but states that all backgrounds must be treated on an equal footing.

\subsection{How should one think about the dependence of beta functions on the parameterization of metric fluctuations?}\label{sec:parameterization}
\emph{...where we discuss, why there is a dependence of beta functions on the choice of metric parameterizations and to what extent parameterization dependence is physical.}\\

The choice of parameterization of metric fluctuations is related to the definition of the path-integral measure or the gravitational configuration space that one integrates over. We work in the background-field formalism to set up the RG flow for gravity, see Sec.~\ref{sec:local_coarse_graining}, such that the gravitational configuration space is defined with respect to a background metric $\bar{g}_{\mu\nu}$.
The general idea is that we start from a formal path integral expression \emph{over ``all metrics''} of a given spacetime topology and in four spacetime dimensions, $\int \mathcal{D}g_{\mu\nu}\, \exp(i S[g_{\mu\nu}])$
 and rewrite it into a path integral $\int \mathcal{D}h_{\mu\nu}\, \exp(iS[\bar{g}_{\mu\nu}, h_{\mu\nu}])$, such that we can use $\bar{g}_{\mu\nu}$ to define a ``momentum" scale for the RG flow and to gauge-fix, see Sec.~\ref{sec:local_coarse_graining} and \ref{sec:gaugedep}.
The relation between the full metric $g_{\mu\nu}$, the background metric $\bar{g}_{\mu\nu}$ and the fluctuation field $h_{\mu\nu}$ defines the parameterization. For instance, one can use the linear parameterization, first used for the flow equation in \cite{Reuter:1996cp}
\begin{equation}
g_{\mu\nu} = \bar{g}_{\mu\nu} + h_{\mu\nu},
\end{equation}
or exponential parameterization, first used for the flow equation in a unimodular context (see Sec.~\ref{sec:MG}) in \cite{Eichhorn:2013xr}
\begin{equation}
g_{\mu\nu} = \bar{g}_{\mu\kappa}{\rm exp}\left(h_{..} \right)^{\kappa}_{\,\,\,\nu} = \bar{g}_{\mu\nu}+ h_{\mu\nu} + \frac{1}{2}h_{\mu \kappa}h^{\kappa}_{\nu}+\dots\, .
\end{equation}
The difference between the two has first been investigated in \cite{Nink:2014yya}, and consists in the fact that
in the linear parameterization, choices of $h_{\mu\nu}$ exist, such that $g_{\mu\nu}$ can become degenerate or even change signature. This is not possible for any choice of $h_{\mu\nu}$ in the exponential parameterization. Such signature changes can be prevented by imposing additional constraints on $h_{\mu\nu}$ in the linear parameterization \cite{Knorr:2022mvn} which in practice has not yet been achieved.

Given this difference, one may wonder why any calculations using the flow equation are done in the linear parameterization. The reason lies in the one-loop structure of the Wetterich equation, which is crucial for the practical aspects of calculations in asymptotic safety.
This one-loop structure hinges on the regulator term being quadratic in the field that is the integration variable in the path integral. Thus, for the path integral over all metrics $\int \mathcal{D}g_{\mu\nu}\, \exp(i S[g_{\mu\nu}])$, the regulator is given by $\Delta S_k = \frac{1}{2}\int d^4x\, \sqrt{\bar{g}} g_{\mu\nu}R^{\mu\nu\kappa\lambda}_k(-\bar{D}^2) g_{\kappa\lambda}$, and gives rise to the familiar one-loop flow equation first derived for gravity in \cite{Reuter:1996cp}. After the regulator term has been introduced, one can perform a change in the integration variable in the path integral, $g_{\mu\nu} \rightarrow h_{\mu\nu}$, based on the linear split. This gives rise to the flow equation in Eq.~\eqref{eq:fluctuationflow}.\\
In contrast, for the exponential (or another non-linear split), the regulator term  for the metric, $\Delta S_k = \frac{1}{2}\int d^4x\, \sqrt{\bar{g}} g_{\mu\nu}R^{\mu\nu\kappa\lambda}_k(-\bar{D}^2) g_{\kappa\lambda}$, gives rise to a regulator term for $h_{\mu\nu}$ that contains higher-order terms in $h_{\mu\nu}$. This results in the flow equation no longer being of one-loop structure \cite{Pawlowski:2005xe}, making the flow equation much more challenging to use in practice. Thus, it is preferable to first perform the split of the metric into fluctuation and background and then regularize the fluctuation field. In this case, the derivation of a one-loop flow equation goes through. As a consequence, the starting point is, however, no longer the regularized path integral over all metrics.\\
In addition, a non-linear split results in a Jacobian factor in the path-integral measure related to $\frac{\delta g_{\mu\nu}}{\delta h_{\kappa\lambda}} \neq \mathds{1}$.

From this discussion, we conclude that different choices of parameterization may potentially not correspond to the same path integral, and that there appears to be a preference for the linear parameterization.\footnote{This holds except for a unimodular setting, see Sec.~\ref{sec:MG}, where the exponential parameterization is the only parameterization compatible with the unimodularity constraint.} However, there is a different point of view: the expression $ \int \mathcal{D}g_{\mu\nu}\, \exp(i S[g_{\mu\nu}])$ is of course ill-defined and (at best) the (UV and IR) regularized expressions are well-defined, i.e., the path integral over $h_{\mu\nu}$ should be understood as the starting point. In this point of view, calculations within different parameterizations differ by the Jacobian that relates the corresponding fluctuation fields.
Because one would not expect such a Jacobian to carry key physical information on an asymptotically safe fixed point, one can use parameterization dependence to test the robustness of the fixed point.
In this spirit, several studies have compared results in the linear and the exponential split (e.g., \cite{Dona:2015tnf}) and others have introduced general parameters that control the parameterization \cite{Gies:2015tca,Ohta:2016npm,Ohta:2016jvw,deBrito:2018hur,Falls:2024noj,Bonanno:2025tfj}. One may expect that at high orders in truncations, the results become sensitive to the (subtle) differences in the underlying path integrals, but this point has not yet been studied in depth.

\section{Frequently asked questions about asymptotic safety}\label{sec:FAQ}
\subsection{What are useful resources to start doing calculations within asymptotically safe quantum gravity?}\label{sec:codebase}
For practical calculations, the lecture notes in \cite{Reichert:2020mja,Basile:2024oms} as well as the textbooks \cite{Percacci:2017fkn,Reuter:2019byg} may provide useful introductions. There is, however, a gap between state-of-the-art calculations and those calculations reviewed in the lecture notes and textbooks. In particular, most of the results in recent years have been achieved on the basis of codes that evaluate the right-hand-side of the Wetterich equation, given a truncation as an input. A codebase can be found at \url{https://as-codebase.quantum-spacetime.net/home}, together with some tutorials and points of contact for the different codes.

\subsection{To run or not to run -- is that the question?}\label{sec:notionsofrunning}
\emph{\dots where we discuss how distinct notions of ``running" for couplings lead to conflicting statements about whether the Newton coupling runs or not. We also discuss how different notions of running are all relevant in asymptotic safety, but different notions pertain to distinct questions.}\\

Based on the functional RG, several hundreds of papers calculate, cross-check, and confirm, how the Newton coupling $G$ runs as a function of the RG scale $k$. Yet, there are papers which show, based on a sound analysis, that $G$ does not run \cite{Anber:2010uj,Anber:2011ut,Donoghue:2019clr}. Similarly, there is a collection of papers showing that quantum gravity contributes to the running of gauge couplings, e.g.,  \cite{Robinson:2005fj,Toms:2010vy} and a corresponding collection claiming that there is no gravitational contribution to this running, e.g., \cite{Ebert:2007gf,Ellis:2010rw}.
What is the resolution to these apparent contradictions? These apparent contradictions are actually entirely based on a broad use of the word ``running'', without specifying what mathematical operation has to be performed to determine whether or not a coupling runs. There are distinct notions of running, not all of them equivalent.

All notions of running relate to the Renormalization Group, but to distinct conceptual aspects and technical implementations. We have already touched on this distinction in Sec.~\ref{sec:WhatisAS}, but come back to it here.

Given any path integral that gives rise to an effective action, whether gravitational or not, there are two pertinent questions:

First, by how many free parameters is the effective action defined? This relates to the \emph{predictivity} of the theory, and is tied to the running in an auxiliary scale $k$. This auxiliary scale controls the actual integration in the path integral. One can imagine it as a (generalized) cutoff in the path-integral. To determine how predictive the path integral is, we need to determine the freedom in the ``initial condition'', i.e., the number of relevant directions of a fixed point which provides the UV completion. We achieve this by deriving beta functions with respect to the cutoff scale $k$ and using their zeros to provide us with initial conditions for the $k$-dependence of couplings. At large $k$, all couplings are equal to their fixed-point values plus small deviations along the relevant directions.\\
We stress that, because a partially completed path integral does not contain the full effect of quantum fluctuations, only the limit $k \rightarrow 0$, i.e., $\Gamma_{k \rightarrow 0}$ can be used to determine physical observables. Thus, $k$-running is never visible in physical observables. Its role is exclusively to determine the number of free parameters in the effective action.

Second, how do observables depend on physical scales?
This is tied to the running as a function of physical scales. This running is entirely encoded in $\Gamma_{k \rightarrow 0}$, which depends on physical scales (e.g., curvature scales, expectation values of scalar fields, physical momenta of particles...). Some of these scales can act similarly to infrared cutoffs, e.g., the momenta of external particles in a scattering process appear on (a subset of) internal lines in scattering amplitudes. However, in general this physical scale dependence is distinct from the cutoff dependence. \\

To determine whether a theory is asymptotically safe, we need to understand both notions of scale: We need the cutoff dependence to determine the number of free parameters in the effective action $\Gamma_{k \rightarrow 0}$ and we need the physical scale dependence to understand whether the observables (determined in terms of these free parameters) have a well-behaved UV limit.
We come back to these points in Sec.~\ref{sec:particlescattering}, when we discuss the status of scattering in asymptotically safe gravity.

There are special situations in which the cutoff dependence tracks the physical scale dependence. These situations arise in the case of logarithmic scale dependence, where both scale dependences tend to track each other. However, in general these scale dependences are simply distinct from each other and should not be mixed up \cite{Buccio:2023lzo,Buccio:2024hys}. 

One way to disentangle the dependence on $k$ and physical scale dependence is to work in a vertex expansion about a flat background and account for the physical momentum dependence, as done in \cite{Christiansen:2014raa,Christiansen:2015rva,Denz:2016qks,Eichhorn:2018akn,Eichhorn:2018nda,Knorr:2021niv,Bonanno:2021squ}, even in Lorentzian signature \cite{Fehre:2021eob,Kher:2025rve,Pawlowski:2025etp,Chiesa:2026tlz}, see \cite{Pawlowski:2020qer,Pawlowski:2023gym} for reviews. It is also possible to extend this setup to account for the dependence on another physical scale, namely the spacetime curvature \cite{Christiansen:2017bsy,Burger:2019upn}.

Against this background, Renormalization Group improvement can lead to misleading results. Under a Renormalization Group improvement, couplings in a classical action (or classical equations of motion or even solutions) are substituted with their $k$-dependent counterparts. Subsequently $k$ is identified with a physical scale of the problem and the resulting action (equations of motion or solution) are then examined in terms of physical consequences. This leads to phenomenological models which can be interesting in their own right, but have no guarantee to be related to the actual phenomenology of an asymptotically safe model. We come back to this point when discussing the status of black holes in asymptotic safety in Sec.~\ref{sec:BH}.


\subsection{Are truncations of the functional Renormalization Group for gravity inherently ill-controlled?}\label{sec:truncations}
\emph{...where we refute claims that truncations result in a lack of control and lack of reliability of results.
We review an example in which the functional Renormalization Group produces quantitatively precise results. We then explain that an ordering principle is needed for truncations to produce such impressive results. For gravity, the ordering principle that has emerged from many different works is near-perturbativity, see also Sec.~\ref{sec:nearpert}. It implies that there is a small parameter that provides control of truncations. This small parameter is the deviation of the critical exponents from the canonical scaling dimensions of couplings.}\\

The answer to the question in the section title is an emphatic ``no''. The most impressive -- but by no means the only --  demonstration of this fact is arguably the precise determination of the leading critical exponent for the Ising model, i.e., three dimensional $\phi^4$ theory in \cite{Balog:2019rrg}. From the conformal bootstrap, reviewed in \cite{Rychkov:2023wsd}, one obtains $\nu=0.629971(4)$ \cite{Kos:2014bka}, compared to $\nu=0.6300(2)$ from the derivative expansion of the functional RG. To interpret these results, let us note that i) the conformal bootstrap is widely regarded as the ``gold standard'' for the Ising model, ii) the conformal bootstrap is not applicable to gravity and in fact more limited in its scope than the functional RG. Thus, it is probably fair to consider the conformal bootstrap a more specialized method, applicable to a smaller range of theories, but capable of producing high-precision results in these theories. In contrast, the functional RG is a more versatile tool, applicable to a wide range of theories \cite{Dupuis:2020fhh}, for which it is often doubted whether high-precision results can be achieved. The fact that the functional RG result for $\nu$ agrees with the conformal bootstrap result within the  (sub-per-mille) error of the functional RG calculation, shows that this doubt is unwarranted.

Given this impressive demonstration of the capabilities of the functional RG, as well as many further examples, see \cite{Dupuis:2020fhh}, we may ask what it takes to achieve such results. The answer hinges on a well-controlled expansion scheme which in turn relies on a good understanding of the physics of the system. This provides an ordering principle for the infinitely many interactions in the flowing action. Without such an ordering principle, one does in fact risk that truncations can become ill-controlled.

For gravity, an ordering principle was not a priori clear in early works of the field, but is by now well established. It consists of the near-perturbative nature of the asymptotically safe fixed point, see Sec.~\ref{sec:nearpert}. This entails that the spectrum of critical exponents is close to Gaussian scaling, i.e., quantum corrections to scaling dimensions are small compared to the canonical scaling dimensions. This pattern becomes visible, once truncations extend far enough and account for the first few irrelevant interactions \cite{Falls:2013bv,Falls:2014tra,Falls:2017lst,Falls:2018ylp,Kluth:2020bdv}; whereas works stopping at quartic order in derivatives typically produce critical exponents with huge deviations from canonical scaling \cite{Lauscher:2002sq,Codello:2008vh,Knorr:2021slg}.

Therefore, there is a small parameter that governs the results and enables control of approximations. This small parameter is not a coupling, but the deviation of critical exponents from canonical scaling dimensions. 

Based on this ordering principle, one can formulate and then test expectations about interactions with canonical mass dimensions greater than four, i.e., couplings with negative mass dimensions. There is no dimension-five-interaction in gravity, but there are dimension-five interactions for (beyond-Standard-Model) matter sectors. For the corresponding couplings, one may expect that the associated critical exponents are negative, i.e., the couplings irrelevant. This expectation has indeed held up for the three different dimension-five-operators that have been studied to date \cite{deBrito:2021akp,deBrito:2025ges,Assant:2025gto}. Similarly, dimension-six-operators, both for gravity, see, e.g.,  \cite{Machado:2007ea,Codello:2008vh,Falls:2013bv,Falls:2014tra,Falls:2017lst,Falls:2018ylp,Kluth:2020bdv}, as well as for matter, see, e.g., \cite{Eichhorn:2011pc,Eichhorn:2017sok,Eichhorn:2018nda}, stay irrelevant. They can only become relevant, if the strength of metric fluctuations is increased (e.g., by increasing the fixed-point value of the Newton coupling), beyond values that are credible for the actual fixed-point values.\\

On top of providing an ordering principle, the functional RG provides several ways to test the robustness of results. Besides the obvious way to successively extend results and check for apparent convergence, as reviewed in Sec.~\ref{sec:FP}, one can use the response to variations in the regulator function. First, the choice of regulator shape function is not physical. Thus physical results, which include, besides more phenomenologically accessible observables also the critical exponents, do not depend on this choice within an untruncated setting. Truncations introduce dependence on the regulator even in physical observables. The amount of variability is informative regarding the reliability of results and provides an estimate of the systematic uncertainty. In addition, one can use the principle of minimum sensitivity to find regular choices for which the physical observables are as insensitive to variations in the regulator as possible within a truncation. This requires using a regulator with a continuously varying parameter. If physical observables exhibit an extremum as a function of this parameter, the corresponding value constitutes the best estimation of the observable. Impressive recent studies in asymptotic safety along those lines can be found in \cite{Baldazzi:2023pep,deBrito:2025nog}.

\subsection{Can asymptotic safety satisfy unitarity?}
\emph{\dots where we briefly discuss what unitarity could mean in a quantum gravitational context, review what the Ostrogradsky theorem does and does not say and finally review what is known about the poles of the propagator of metric fluctuations.}\\

The origin of this question is related to the higher-order derivative terms that appear in asymptotic safety. 
In classical mechanics, the Ostrogradsky theorem states that, under a non-degeneracy condition, the Hamiltonian of a system is unbounded from above and below, if higher-order time derivatives are present. In quantum mechanics, unboundedness can be avoided by a redefinition of creation and annihilation operators. This results in negative-norm states, and thus unitarity violation. Based on these results, a folk ``theorem'' has developed and the presence of higher-order derivatives is often taken to imply that the corresponding theory is necessarily a ``sick'' theory and not physically interesting.\\
More specifically, in a QFT on a flat background, higher-order derivatives can, under certain conditions, lead to additional poles in the propagator. If those poles have the opposite sign to the standard poles, they are classified as ghosts. If those ghosts are stable, they must be counted among the asymptotic states and make an appearance in the optical theorem, where they violate the unitarity of the $S$-matrix.\footnote{A quantum mechanical counterexample to this understanding highlights that there may be ways to evade this conclusion, e.g., in the presence of constants of motion \cite{Deffayet:2026cnu}. We also stress that the condition that the ghosts should be stable is relevant for the optical theorem, e.g., \cite{Donoghue:2019fcb}. This condition is sometimes underappreciated.}

There are several points worth stressing in the context of quantum gravity: \\
First, this notion of unitarity is tied to QFT on a flat background and notions like the $S$-matrix, which cannot be defined on a general background \cite{Bousso:2004tv}. Therefore, it is in principle unclear how much one should conclude from a violation of perturbative unitarity in quantum gravity. Here, we take the point of view that, in connection with the near-perturbativity of asymptotic safety, see Sec.~\ref{sec:nearpert}, such flat-space notions may remain meaningful. We therefore require that no stable ghost poles appear in the graviton propagator at finite values of the mass \cite{Becker:2017tcx}.\\
Second, even in QFT on a flat background, unitarity in a gauge theory only constrains the modes which are not gauge modes. Thus, for gravity it is the propagator related to the spin-2-mode, i.e., the transverse traceless part of metric fluctuations, as well as a scalar mode (although the latter is only an off-shell mode, at least in GR), which needs to be examined.\\
Third, as emphasized above, higher derivatives do not automatically lead to additional poles, but only under certain conditions. In particular, there is an important difference between a finite order of derivatives and an infinite order, already at the classical level \cite{Barnaby:2007ve}, because additional poles can be avoided at infinite order. An instructive example in this context, discussed in \cite{Platania:2020knd}, is the one-loop effective action in Quantum Electrodynamics (QED). We know two facts about QED: First, it is a unitary theory and second, its one-loop effective action contains higher-order terms.
\begin{equation}
\Gamma_{\rm QED,\,1-loop}= -\frac{1}{4}\int d^4x F_{\mu\nu}P(D^2)F^{\mu\nu},
\end{equation}
with 
\begin{equation}
P(z) = 1- \frac{\alpha}{3\pi}\ln \left(\frac{-z+m^2}{m^2} \right),\label{eq:QEDP}
\end{equation}
where $F_{\mu\nu}$ is the field-strength tensor, $\alpha$ is the fine-structure constant and $m$ is the mass of the degree of freedom that is integrated out (e.g., the electron mass). The expansion of $P(z)$ to $N$th order reads
\begin{equation}
P(z) = 1+ \frac{\alpha}{3\pi}\sum_{n=1}^N \frac{z^n}{n}.\label{eq:QEDP_expanded}
\end{equation}
From this example, we glean several lessons: First, Eq.~\eqref{eq:QEDP_expanded} clearly gives rise to higher-order derivatives, but these  \emph{cannot} render the theory non-unitary, because the unitarity of QED is established. Thus, any argument that makes a direct connection between higher-order time derivatives and unitarity-violations is clearly oversimplified \cite{Woodard:2015zca}.\footnote{Related oversimplifications already haunt the discussion of stable time evolution in classical theories. There, Ostrogradsky's theorem states that the Hamiltonian of a theory with finitely many higher-order time derivatives, if it is non-degenerate, is unbounded from above and below. This is often interpreted as synonymous with a ``catastrophic'' instability of the theory. Proven counterexamples to this exist \cite{Deffayet:2023wdg,Deffayet:2025lnj,Held:2025fii}, and even quadratic gravity, long believed to be a ``posterchild'' for a non-unitary theory, appears to admit stable time evolution even in the strong-gravity regime of a black-hole inspiral \cite{Held:2023aap,Held:2025ckb}. } 
This fits to the general expectation that in an EFT expansion of a unitary theory, there may be ``ghostly'' poles beyond the range of validity of the EFT, including in gravitational examples \cite{Burgess:2003jk,Burgess:2007pt,Creminelli:2010qf,deRham:2014fha}, without invalidating the unitarity of the theory.
Second, the expansion in Eq.~\eqref{eq:QEDP_expanded} contains additional poles (in the complex plane), even though the full expression  Eq.~\eqref{eq:QEDP} only contains a pole at $z=0$. Similarly, we expect that assessing unitarity in a truncation to finite order in derivatives may lead to misleading results, because it may show extra poles in the complex plane (some of them of ``ghostly'' nature), which are truncation artifacts. Finally, in this example, unitarity is manifestly restored, once the full expression for the propagator is considered beyond the expansion. The mechanism behind this is as follows: as one goes to higher order in the expansion, the residues of the additional poles tend to zero.\\

To discuss results on the graviton propagator in asymptotic safety, we first briefly review the analytic structure of a general propagator.
For a general propagator $\Delta(p^2)$, we can write
\begin{eqnarray}
\Delta(p^2)&=& i \Bigl( \sum_{n_p} \frac{R_{n_p}^{({\rm particle})}}{p^2-m_{n_p}^2} +  \sum_{n_g} \frac{R_{n_g}^{({\rm ghost})}}{p^2-m_{n_g}^2}\nonumber\\
&{}&+ \sum_{n_c} \left( \frac{R_{n_c}}{p^2 - m_{n_c}^2}+\frac{R_{n_c}^{\ast}}{p^2 - \left(m_{n_c}^2\right)^{\ast}} \right)\nonumber\\
&{}&+ \int_{m_{\rm th}^2}^{\infty} d \mu^2 \frac{\sigma(\mu^2)}{p^2-\mu^2}\Bigr).
\end{eqnarray}
In here, $n_p$ labels the single-particle states, for which the residue at the pole, $R_{n_p}^{\rm particle}$ must be positive. In contrast, for ``ghost'' states, labeled by $n_g$, it holds that $R_{n_g}<0$. Both particles and ghosts can in principle be tachyonic, for $m_{n_p}^2<0$ or $m_{n_g}^2<0$. 
Further, $n_c$ sums over complex poles, for which $m_{n_c}^2$ is complex and the residue is also in general complex. Finally, in the last line, we see the branch cut associated to multi-particle-states. For theories with a mass gap, the branch cut starts at $m_{\rm th}^2>0$, whereas it starts at $m_{\rm th}^2=0$ in the massless case.

The standard lore is that there should not be a) any (stable) ghosts, b) any tachyons, c) any complex poles to ensure unitarity and causality, see e.g., Tab.~[2] in \cite{Platania:2022gtt}. Then, the spectral density can be defined, 
\begin{equation}
\rho(p^2)= \sum_{n_p}R_{n_p}^{({\rm particle})}\delta(p^2-m_{n_p}^2) + \sigma(p^2)\theta(p^2-m_{\rm th}^2).
\end{equation}
In here, $\sigma(\mu^2)$ must be non-negative for physical asymptotic states. Finally, the spectral density $\rho(p^2)$ should be normalizable, although exceptions exist, see Sec.~IV B in \cite{Kher:2025rve}.

Against this background, discussions of unitarity in asymptotic safety can be found, e.g., in \cite{Bonanno:2020bil,Draper:2020bop,Platania:2020knd,Platania:2022gtt}. More recently, the graviton propagator has been studied explicitly in Lorentzian signature \cite{Bonanno:2021squ,Fehre:2021eob,Pawlowski:2025etp,Assant:2026dca}, see Sec.~\ref{sec:Lorentzian} and here we highlight the analytical results obtained in \cite{Knorr:2026jcg}. The input to these calculations is the Einstein-Hilbert action as a ``seed action'' from which the vertices and propagator entering the loop diagrams form. This does, however, not imply that the output is automatically given by a classical propagator. Already in perturbation theory, starting from the Einstein-Hilbert action generates higher-order curvature terms through loop contributions. Similarly, the calculations in \cite{Knorr:2026jcg} give rise to momentum-dependence of the propagator beyond Einstein-Hilbert. Accordingly,  the results in \cite{Bonanno:2021squ,Fehre:2021eob,Pawlowski:2025etp,Knorr:2026jcg,Assant:2026dca}  constitute a highly non-trivial test of unitarity.

\begin{figure}[!t]
\includegraphics[width=\linewidth]{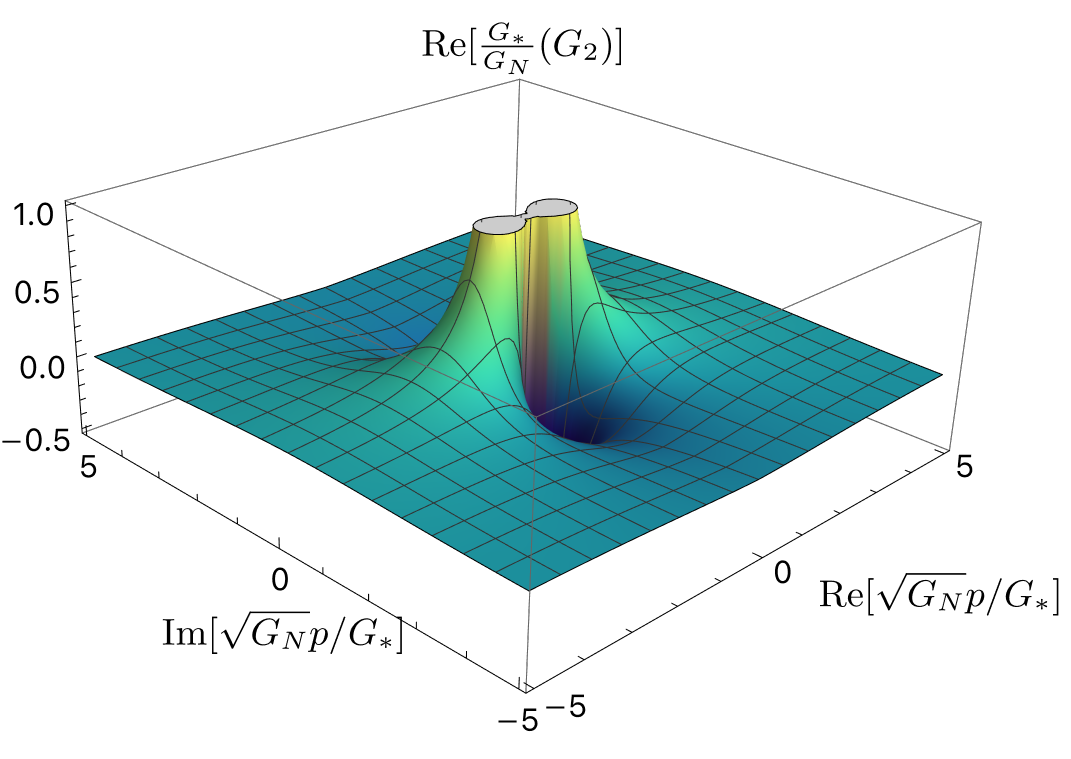}\\
\vspace{1cm}
\includegraphics[width=\linewidth]{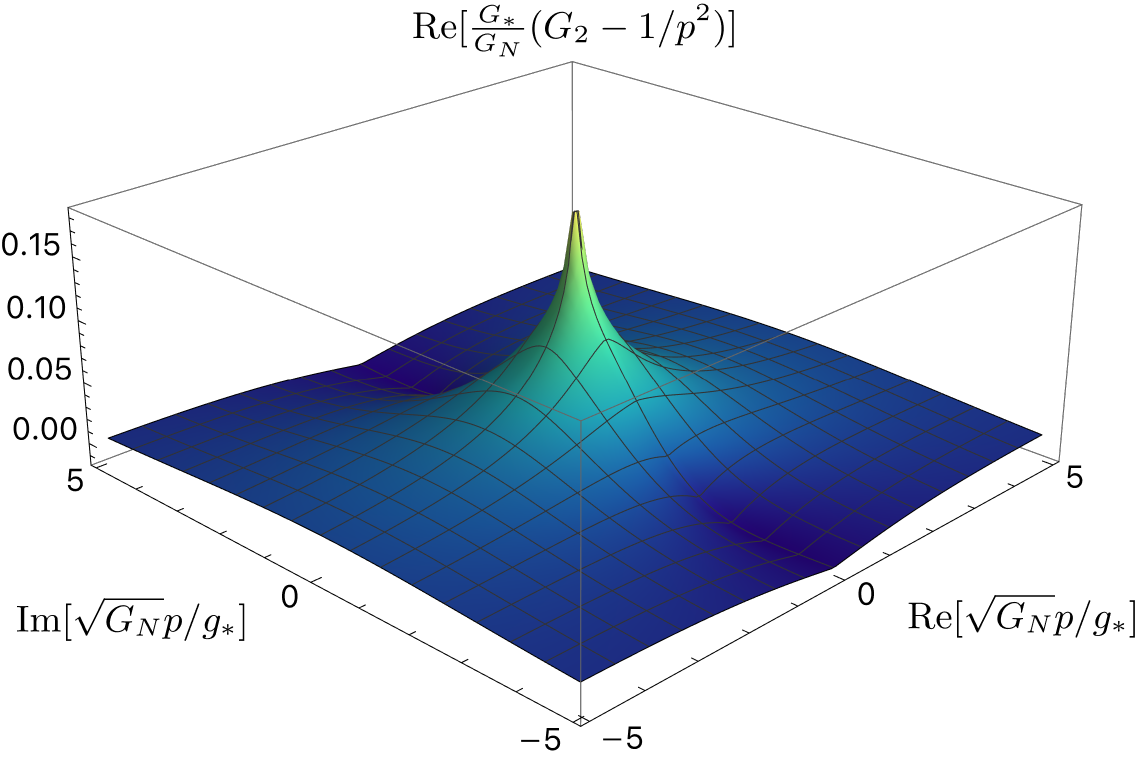}
\caption{\label{fig:Regravprop} We show the real part of $G_2(p^2)$ in the complex $p$ plane, based on the results in \cite{Knorr:2026jcg}. The upper panel shows the full result for the real part of $G_2(p^2)$, from which a singular structure around $p^2=0$ and no further singularities are visible. To better appreciate that there is really just a $\frac{1}{(Re(p)+ i\, Im(p))^2}$ pole, the lower panel shows the real part of $G_2(p^2)$ with this single pole at zero subtracted. The remaining results is clearly non-singular, i.e., there is no evidence for additional poles.}
\end{figure}

We now focus on the graviton propagator for the spin-2-mode, which consists of a transverse traceless project $\Pi_{\mu\nu\kappa\lambda}(p)$ times a scalar function that encodes the momentum dependence, $G_2(p^2)$. We are interested in testing whether the above conditions are all fulfilled. 

\begin{figure}[!t]
\includegraphics[width=\linewidth]{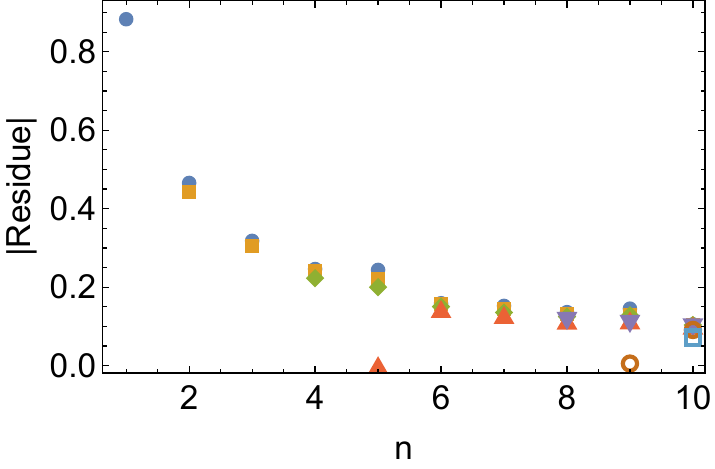}
\caption{\label{fig:residue} We show the residues of the poles that arise in the expansion of the full propagator from \cite{Knorr:2026jcg}. The full propagator does not have any poles beyond the massless graviton pole. Additional poles arise in the expansion. The absolute value of their residues tend towards zero, as the order $n$ of the expansion is increased.}
\end{figure}

This question has been studied using numerical results \cite{Bonanno:2021squ,Fehre:2021eob,Pawlowski:2025etp} as well as analytical results \cite{Knorr:2026jcg}. The results can be summarized by stating that the graviton propagator for the spin-2-mode only has a massless pole and no further poles, cf.~Fig.~\ref{fig:Regravprop}.\footnote{I thank B.~Knorr for sharing his results already prior to publication of his paper.} This is a highly nontrivial indication that unitarity may be preserved. Let us emphasize that the result also has a highly nontrivial origin, because the momentum-dependence of the graviton propagator is not captured by a simple $1/p^2$, instead, higher-order terms are present. Based on the discussion above, the absence of additional poles is therefore not guaranteed and was certainly not expected by many. 
The results in \cite{Knorr:2026jcg} are analytic and thus enable a more detailed study of the mechanism that prevents additional poles. In fact, one can perform a derivative expansion and obtains additional poles in the propagator at each finite order in the derivative expansion. As the order of the derivative expansion is increased, the residues of the poles tend to zero, cf.~Fig.~\ref{fig:residue}. Thus, the full propagator, beyond a finite expansion, no longer features any additional zeros. This is, in fact, the mechanism observed in \cite{Platania:2020knd} for the case of QED.

\begin{figure}[!t]
\includegraphics[width=\linewidth]{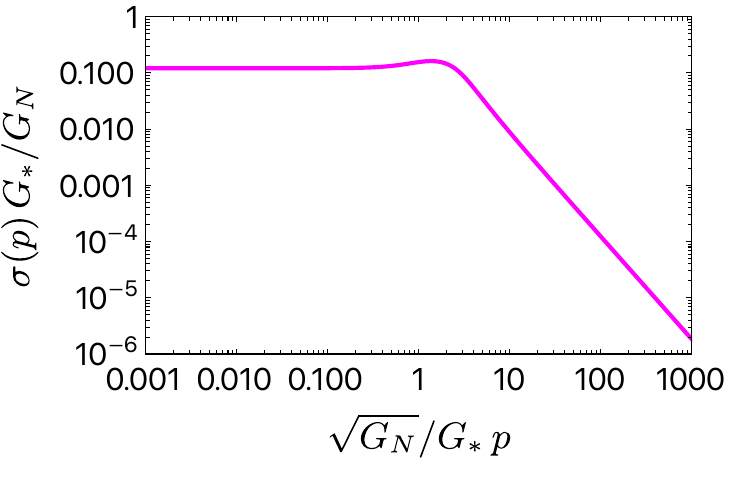}
\caption{\label{fig:sigma} We show the spectral function $\sigma(p)$ obtained in \cite{Knorr:2026jcg}, which is non-negative for the spin-2-mode. Beyond the Planck-scale, it exhibits a scaling regime. The non-negativity is a non-trivial and crucial result. Qualitatively, a similar result was achieved in \cite{Pawlowski:2025etp}, although the asymptotic scaling behavior differs at the quantitative level.}
\end{figure}

In addition, it is a crucial check whether the spectral function $\sigma(\mu^2)$ is non-negative. This is indeed observed numerically in \cite{Pawlowski:2025etp} and also in the analytical results in \cite{Knorr:2026jcg}, which we show in Fig.~\ref{fig:sigma}. In \cite{Assant:2026dca}, it is pointed out that this result is not automatic within asymptotic safety, but provides a non-trivial constraint on gravitational fixed-point values, that are in fact satisfied by the results in \cite{Assant:2026dca}.
In contrast to the behavior of the graviton, negative parts of the spectral function of scalar and photon propagators have been found in first approximations in \cite{Kher:2025rve}, clearly calling for further investigations. In addition, for gravity there is also a scalar mode which does not propagate in GR, but that can become dynamical beyond. It exhibits the wrong-sign kinetic term, connected to the conformal-factor problem in the Euclidean path integral.

A complementary, powerful piece of evidence for the unitarity of asymptotic safety comes from the lattice: The CDT formulation of gravity admits a reflection positive transfer matrix in the Euclidean \cite{Ambjorn:2001cv}. In a lattice theory, this is a sufficient condition to obtain a unitary time evolution in Lorentzian signature, which in CDTs is achievable through a Wick rotation. More generally, reflection positivity is the relevant property to check in Euclidean signature and may be violated by specific choices of higher-order derivatives \cite{Arici:2017whq}.\\

In summary, it is important to note that an understanding of unitarity in quantum gravity is challenging and by no means as simple as has sometimes been suggested.
Even if flat-space notions of unitarity actually apply to quantum gravity, there may be various underappreciated subtleties which invalidate often-made assumptions about the conditions under which a theory violates unitarity. Against this background, the results in \cite{Bonanno:2021squ,Fehre:2021eob,Pawlowski:2025etp,Knorr:2026jcg}  hint at the possibility of a straightforward form of unitarity in pure gravity; supported by the non-perturbative evidence from the lattice \cite{Ambjorn:2024qoe}. Nevertheless, as highlighted by the results in  \cite{Kher:2025rve} of negative parts in the spectral functions for certain matter propagators within a first approximation, unitarity is not settled in asymptotic safety. One may, however, note that the absence of additional poles in the spin-2-propagator, the mechanism \cite{Platania:2020knd} that dynamically removes such poles in \cite{Knorr:2026jcg} and the positivity of the spectral function for the spin-2-propagator are highly nontrivial results. These explicitly contradict often-voiced expectations that unitarity cannot be achieved in asymptotic safety. The technical advancements in the field over the last few years have made it possible to tackle this challenging question by explicit calculations, the results of which are very promising indeed.

\subsection{How should one think about gauge dependence of beta functions?}\label{sec:gaugedep}
\emph{...where we discuss how gauge dependence can be used to estimate systematic uncertainties of calculations.}\\

In a gauge-fixed path integral with the background-field method, there are two choices one has to make, the choice of gauge fixing and the choice of parameterization for metric fluctuations, see Sec.~\ref{sec:local_coarse_graining}. 

The gauge-fixing, for covariant gauges, generically depends on two parameters (although more elaborate gauge conditions can in principle also be written), $\alpha$ and $\beta$, and reads
\begin{equation}
\mathcal{F}_{\mu} = \bar{D}^{\nu}h_{\mu\nu} - \frac{1+\beta}{4}\bar{D}_{\mu}h,\quad S_{\rm gf} = \frac{1}{\alpha}\int d^4x\sqrt{\bar{g}}\bar{g}^{\mu\nu}\mathcal{F}_{\mu}\mathcal{F}_{\nu}.\label{eq:gaugefixing}
\end{equation}
To impose the gauge condition sharply, one should work in the Landau-gauge limit $\alpha \rightarrow 0$, which constitutes a fixed point of the RG flow \cite{Knorr:2017fus}.\footnote{For practical purposes, harmonic gauge, $\alpha=1=\beta$ is sometimes used, because the metric propagator has a particularly simple tensor structure in this gauge. Qualitatively, results in this choice of gauge are in agreement with results in the Landau-gauge limit, see, e.g., \cite{Reuter:2001ag, Dona:2013qba}.} The parameter $\beta$ determines, which of the two scalar modes, $h$ or $\sigma$, contributes to the RG flow. These two modes arise 
in a decomposition of $h_{\mu\nu}$ into irreducible representations of $SO(4)$, 
\begin{eqnarray}
h_{\mu\nu}(p)&=&h_{\mu\nu}^{TT}(p)+ i \frac{p_{\mu}}{\sqrt{p^2}}v_{\nu}+ i \frac{p_{\nu}}{\sqrt{p^2}}v_{\mu}\nonumber\\
&{}&+ \left(\frac{p_{\mu}p_{\nu}}{p^2} - \frac{1}{4} \delta_{\mu\nu} \right)\sigma(p)+\frac{1}{4} \delta_{\mu\nu}h(p).
\end{eqnarray}
For $\beta =0$, the trace mode $h(p)$ contributes, for $\beta \rightarrow -\infty$, the second scalar mode $\sigma(p)$ remains. The $\sigma$ mode, together with a transverse vector $v_{\mu}$, forms the four-vector component included in $h_{\mu\nu}$. Other choices of $\beta$ retain a scalar that is a mixture of these two. A priori, no choice of the gauge parameter $\beta$ is preferred.

Observable quantities must not depend on $\beta$ (or $\alpha$), if they are calculated without any approximations. Within approximations, a dependence on gauge parameters may arise, but should of course decrease with increasing quality of the approximation. 
One may therefore estimate systematic uncertainties from the gauge dependence, keeping in mind that the systematic uncertainty that arises from varying $\beta$ is only a lower bound on the full systematic uncertainty.\\
Within a functional RG approach, many (intermediate) quantities are calculated which are not observables; the most prominent example are the beta functions of couplings. Gauge dependence in off-shell quantities is not unexpected in a gauge theory; examples of off-shell quantities include beta functions and \cite{Benedetti:2011ct} discusses how the beta function of an essential coupling becomes gauge-independent when taken on shell.
Because they are intermediate quantities in calculations of observables, one may hope that an increasing quality of the approximation can also be observed in a decreasing dependence of off-shell beta functions and fixed-point values on the gauge parameter $\beta$.\\
In this spirit, $\beta$-dependence has been investigated for pure gravity, e.g., in \cite{Gies:2015tca,Ohta:2016npm,Ohta:2016jvw,Knorr:2021niv,Bonanno:2025tfj}, and for various gravity-matter systems, e.g., in \cite{Eichhorn:2016esv,Held:2019vmi,Ohta:2021bkc,Riabokon:2025ozw,Eichhorn:2024wba,deBrito:2025ges}.

\subsection{Are there other examples of asymptotically safe theories?}\label{sec:ASotherexamples}
\emph{...where we address the misconception that asymptotic safety is an ``exotic" scenario in quantum field theory, and point out examples of asymptotically safe theories in various dimensions and with different choices of field content.}\\

Because the Standard Model does not have any asymptotically safe sectors among its perturbatively renormalizable interactions, only asymptotically free ones, asymptotic safety is sometimes erroneously perceived as an ``exotic'' scenario, unlikely to be realized in quantum field theories. In reality, asymptotically safe quantum field theories exist across different dimensions, choices of field content as well as symmetries, see \cite{Eichhorn:2018yfc} for an overview of examples. It is worth to highlight that even Standard-Model like theories, i.e., theories with non-Abelian gauge symmetries and fermionic as well as bosonic matter with Yukawa couplings and quartic scalar couplings in four dimensions, can become asymptotically safe \cite{Litim:2014uca}.

Closely related are conformal field theories, of which numerous examples exist, although it is worth highlighting that conformal symmetry is strictly larger than scale symmetry. Scale symmetry implies conformal symmetry for low-dimensional settings, but under conditions that are too restrictive to directly apply to gravity. Thus, it is not settled whether asymptotically safe quantum gravity corresponds to a conformal field theory.\footnote{This is also relevant to appreciate in the context of arguments against asymptotic safety in gravity, which are based on the assumption of conformal symmetry \cite{Shomer:2007vq}.} We therefore focus only on scale symmetry in its incarnation as asymptotic safety, below.

There are several mechanisms from which asymptotic safety can arise, which all achieve a balance of screening and antiscreening contributions in beta functions, such that a nontrivial fixed point arises, see the review \cite{Eichhorn:2018yfc}. \\
For a coupling with non-vanishing mass dimension, if the loop contributions have the net opposite sign from the canonical scaling term, an interacting fixed point arises, e.g., if higher-loop orders can be neglected and the one-loop term is cubic:
\begin{equation}
\beta_{g_i} = -d_{\bar{g}_i}g_i + \beta^{({\rm 1-\ell})} g_i^3,\, \Rightarrow\, g_{i\,\ast}= \sqrt{\frac{d_{\bar{g}_i}}{\beta^{({\rm 1-\ell})}}}\,.\label{eq:beta_dimensional_FP}
\end{equation}
 For this scenario, for a canonically (ir)relevant coupling, the loop contributions must (anti)screen the coupling in question.\footnote{Interestingly, this applies to the marginally irrelevant couplings in the SM, with a quantum-gravity contribution taking the role of ``effective'' scaling dimension, see Sec.~\ref{sec:SM}.}\\
 For a coupling with vanishing mass-dimension, there must be different loop contributions that can balance against each other. Those can be, e.g., different loop orders or contributions from different fields (e.g., fermions vs.~bosons).\\
  Because the canonical dimension of a coupling changes with spacetime dimensionality, different mechanisms for asymptotic safety are available for a given coupling in different spacetime dimensionalities. For instance, a non-Abelian gauge coupling can become asymptotically safe in $d=5$ through a balance of canonical scaling term (which is positive) with the loop contributions (which are negative)\footnote{Lattice simulations have so far not succeeded in confirming the results from continuum quantum field theory, see \cite{Florio:2021uoz} for the most recent update.}, see \cite{Gies:2003ic,Morris:2004mg}, as in Eq.~\eqref{eq:beta_dimensional_FP}. In $d=4$, where the canonical term is absent, asymptotic safety can be achieved through a balance of one-loop and higher-loop contributions with appropriately chosen matter content \cite{Litim:2014uca}, so that the one-loop term changes sign, $\beta^{({\rm 1-\ell})}>0$, but the higher-loop term remains negative, $\beta^{({\rm 1-\ell})}<0$, i.e., 
\begin{equation}
\beta_{g_i} = \beta^{({\rm 1-\ell})} g^3+ \beta^{({\rm 2-\ell})}g^5 \, \Rightarrow \, g_{i\, \ast} = \sqrt{\frac{\beta^{({\rm 1-\ell})}}{-\beta^{({\rm 2-\ell})}}}.
\end{equation}

Asymptotic safety can be particularly robustly established in systems in which there is a small parameter that controls the fixed-point values as well as anomalous contributions to scaling dimensions. 
The $1/N$ expansion in the large-$N$-limit is one particular example for this idea, and provides evidence for fixed points, e.g., in fermionic theories such as the Gross-Neveu model in three dimensions, see, e.g., \cite{Rosenstein:1988pt,deCalan:1991km,Braun:2010tt,Cresswell-Hogg:2022lgg} and references therein. Similarly, the $\epsilon$ expansion around the critical dimension $d_c$ of a coupling is an example, as long as the fixed point in question deforms into the free fixed point for $\epsilon \rightarrow 0$. Many well-established fixed points, starting from the Wilson-Fisher fixed point \cite{Wilson:1971dc}, fall into this category.

In summary, asymptotic safety is by no means ``exotic". Instead, interacting fixed points are very common in quantum field theories and numerous examples with interacting UV fixed points exist, see, e.g., \cite{Pelissetto:2000ek,Bond:2017lnq,Eichhorn:2018yfc,Senthil:2023vqd} for collections of examples. That no subsector of the Standard Model is asymptotically safe in four dimensions may thus appear as somewhat of an ``unlucky accident'' -- or as a hint that we have missed a degree of freedom that is needed to achieve the balance between screening and antiscreening contributions to make the Standard Model asymptotically safe. In Sec.~\ref{sec:SM}, we will review the evidence for quantum gravity supplying the necessary contributions.

\subsection{Can one have asymptotic safety in gravity with gravitational degrees of freedom other than the metric?}\label{sec:MG}
\emph{...where we discuss the status of asymptotic safety in settings with non-metric gravitational degrees of freedom, some of which can be used to formulate theories which are classically equivalent to General Relativity, but may differ at the quantum level.}\\

Asymptotic safety is a general concept, that may be applicable in many quantum field theories, see Sec.~\ref{sec:ASotherexamples}. Thus, one may look for asymptotic safety in gravitational theories that are not based on metric gravity. This is motivated by phenomenological reasons, i.e., the hope that the resulting effective action describes some aspect of observational gravity better. It is also motivated by fundamental reasons, e.g., the hope that one can learn whether asymptotic safety is generic in gravitational theories and which degrees of freedom it hinges on.

Classically, one can write different gravitational theories that are equivalent to General Relativity on shell, see, e.g., \cite{Krasnov:2017epi,BeltranJimenez:2019esp,Krasnov:2020lku} for reviews, but do not contain the same field content, i.e., differ in their off-shell degrees of freedom.\footnote{We restrict ourselves to purely gravitational theories here; scalar-tensor (or vector-tensor) theories which cannot be transformed into a frame in which the dynamics can be written purely in terms of geometric quantities, are instead understood as theories with additional, non-gravitational degrees of freedom and discussed in Sec.~\ref{sec:cosmo}.} These constitute particularly interesting candidates to search for asymptotic safety. \\
Among such theories, there is a single example with a reduced field content, namely unimodular gravity, in which the determinant of the metric is fixed and not dynamical\footnote{There are also studies of asymptotic safety in ``conformally reduced gravity" \cite{Reuter:2008wj,Reuter:2008qx,Machado:2009ph,Bonanno:2012dg,Dietz:2015owa,Dietz:2016gzg,Nagy:2019jef,Gegeny:2020iyb,Knorr:2020ckv,Gegeny:2022fgu,Bonanno:2023ghc,Bonanno:2023fij,Giacometti:2024qva} in which only the conformal mode is treated as a quantum field. While this setting is computationally simpler to study than full metric gravity, and can shed some light on various questions in asymptotic safety, it is not a phenomenologically viable theory of gravity due to its lack of gravitational waves.} 
\begin{equation}
{\rm det} g_{\mu\nu} = \rm const.
\end{equation}
At the quantum level \cite{Henneaux:1989zc}, unimodular gravity is interesting for three reasons: \\
First, there is no term $\Lambda\ \int d^4x \sqrt{-g}$ in the action, because such a term is just a constant, but has no dynamical content. Therefore, there is no renormalization of the cosmological constant \cite{Weinberg:1988cp,Alvarez:2015sba}. The cosmological constant instead arises as a constant of integration at the level of the equations of motion derived from $S = \frac{1}{16\pi G}\int d^4x \sqrt{-g}R + S_{\rm matter}$,
\begin{eqnarray}
R_{\mu\nu} - \frac{1}{4}g_{\mu\nu}R = 9 \pi G \left(T_{\mu\nu} - \frac{1}{4}g_{\mu\nu}T^{\lambda}_{\lambda} \right),\label{eq:eomUG}
\end{eqnarray} 
which are the trace-free Einstein equations and $T_{\mu\nu}$ is the stress-energy tensor arising from $S_{\rm matter}$. These do not imply conservation of stress-energy, which must be imposed, $D^{\mu}T_{\mu\lambda}=0$. This implies, together with the Bianchi identities $D^{\mu}\left(R_{\mu\nu} - \frac{1}{2}g_{\mu\nu}R \right)=0$, that
\begin{equation}
D_{\nu}\left( -\frac{1}{4}R +2 \pi G T^{\lambda}_{\lambda}  \right)= - \frac{1}{2}D_{\nu}R.
\end{equation}
This equation can be integrated and gives rise to a constant of integration, $\Lambda/(16 \pi G)$, which we can use to rewrite Eq.~\eqref{eq:eomUG} into the standard Einstein equation with cosmological constant. This procedure generalizes beyond the Einstein action \cite{Eichhorn:2015bna} and implies that, while the phenomenology of GR, including a cosmological constant, remains intact, there is no fine-tuning or technical naturalness problem for $\Lambda$, because it is completely decoupled from any other scale in the theory, not impacted by quantum corrections and only chosen at the level of the equations of motion.\\
Second, there is a reduced configuration space compared to full General Relativity, together with a reduced local symmetry, namely transverse diffeomorphisms. In fact, transverse diffeomorphisms -- not the full diffeormorphism group -- come out of an analysis of the little group for massless spin 2 fields \cite{vanderBij:1981ym}.\footnote{The question of the precise difference between the path integrals for unimodular gravity and ``standard" gravity is debated outside of asymptotic safety \cite{deLeonArdon:2017qzg,deBrito:2021pmw,Carballo-Rubio:2022ofy}. Within asymptotic safety, an assessment of the (in)equivalence of the two theories is complicated by the fact that the full effective action is not yet known in either setting and (in)equivalence at the level of a truncation does not imply (in)equivalence of the full theory.} \\
Third, there are indications that unimodular gravity is asymptotically safe \cite{Eichhorn:2013xr,Eichhorn:2015bna,Benedetti:2015zsw,deBrito:2019gdd,deBrito:2020rwu,deBrito:2020xhy}. This also suggests that the conformal mode, which is absent in unimodular gravity and gives rise to the conformal-mode instability in the standard Euclidean setting, is not relevant to generate an asymptotically safe regime. \newline
 
Among theories that are classical equivalent to GR, there are several examples with an increased field content, typically based on choosing a more general connection than the Levi-Civita connection.  There is, at present, no indication that the standard field content used in asymptotic safety may be insufficient to generate a fixed point with viable phenomenology, such that an enlarged field content at present appears unnecessary. Nevertheless, an important motivation for searches for asymptotic safety in settings with an independent connection is the greater similarity to Yang-Mills theory that an independent connection provides. This may, for instance, enable the use of lattice gauge theory as a method to search for asymptotic safety \cite{Asaduzzaman:2019mtx}. \\
A general connection $\Gamma^{\kappa}_{\mu\nu}$ can be written as
\begin{equation}
\Gamma^{\kappa}_{\mu\nu}= \{\!\!\!\phantom{I}^{\,\,\kappa}_{\mu\nu} \} + K^{\kappa}_{\mu\nu}+ L^{\kappa}_{\mu\nu}.\label{eq:generalconnection}
\end{equation} 
The first term, $\{\!\!\!\phantom{I}^{\,\,\kappa}_{\mu\nu} \} $ is the Levi-Civita connection that can be written as the standard Christoffel symbol. The contortion part $K^{\kappa}_{\mu\nu}$ encodes the torsion, which vanishes for a symmetric connection.
Finally, $ L^{\kappa}_{\mu\nu}$ gives rise to the non-metricity of the connection, i.e., encodes the failure of the generalized covariant derivative (denoted $\nabla^{\mu}$ to distinguish it from the covariant derivative based on the Levi-Civita connection, $D^{\mu}$) of the metric to vanish
\begin{equation}
\nabla_{\mu}g_{\alpha\beta} = \partial_{\mu}g_{\alpha \beta} - \Gamma^{\kappa}_{\alpha\mu}g_{\kappa \beta} - \Gamma^{\kappa}_{\mu\beta}g_{\alpha\kappa} \neq 0.
\end{equation}
We thus have the possibility to search for asymptotic safety in theories with the pair $(g_{\mu\nu}, \Gamma^{\kappa}_{\mu\nu})$, where one may choose to set different parts of the general connection to zero. For an appropriate choice of action, namely the Hilbert-Palatini action, $S= \frac{1}{16\pi G} \int d^4x \sqrt{-g}R(\Gamma^{\kappa}_{\mu\nu})$, where $R(\Gamma^{\kappa}_{\mu\nu})$ is built from the generalized Riemann tensor based on Eq.~\eqref{eq:generalconnection}, the torsion and non-metricity vanish on-shell, such that the theory becomes classically equivalent to GR. This does not imply equivalence at the quantum level, because off-shell degrees of freedom contribute to the RG flow and because contributions to the action beyond the lowest order in general differ.
For instance, in \cite{Gies:2022ikv}, an on-shell reduction of an extension of the Hilbert-Palatini action to second order in curvature results in the triplet $(g_{\mu\nu}, \{\!\!\!\phantom{I}^{\,\,\kappa}_{\mu\nu} \}, A^{\mu})$, where $A^{\mu}$ is a vector field contributing to the trace of the torsion, $T^{\kappa}_{\mu\kappa}= 3 A_{\mu}$. Accounting for the quantum fluctuations of $A^{\mu}$, indications for an asymptotically safe fixed point are found \cite{Gies:2022ikv}.

Another motivation to go beyond the metric and Levi-Civita connection is fermionic matter. To consistently couple fermions to gravity, one works with the vielbein $e^{\mu}_a$ and the spin connection $\omega^{\mu}_{ab}$. This does not automatically entail different quantum fields; in fact, it is possible to reduce the quantum fluctuations in these fields to $h_{\mu\nu}$.\footnote{The tangent-space indices $a, b$ are subject to local Lorentz transformation, such that the group of gauge transformation is enlarged from diffeomorphisms on the manifold, ${\rm Diff}(\mathcal{M})$ to ${\rm Diff}(\mathcal{M})\times {\rm SO}(3,1)$, where the $ {\rm SO}(3,1)$ acts locally, not globally, and $\omega^{\mu}_{ab}$ is the corresponding connection, taking values in the Lie algebra of $ {\rm SO}(3,1)$. The additional gauge freedom can be gauge-fixed such that the fluctuations of the fields $(e^{\mu}_a, \omega^{\mu}_{ab})$ can be rewritten purely in terms of metric fluctuations \cite{vanNieuwenhuizen:1981uf,Woodard:1984sj}. At the technical level, this allows the treatment of fermionic fields in asymptotic safety based on the metric and Levi-Civita connection as degrees of freedom \cite{Eichhorn:2011pc}, see also \cite{Gies:2013noa,Gies:2015cka,Lippoldt:2015cea}.} 
 As long as the spin connection is not treated as an independent variable, the differences between the space of couplings based on $(g_{\mu\nu}, \{\!\!\!\phantom{I}^{\,\,\kappa}_{\mu\nu} \})$ and that based on $(e^{\mu}_a,\{\!\!\!\phantom{I}^{\,\,\kappa}_{\mu\nu} \}))$ are only in the gauge-fixing sector see \cite{Harst:2012ni}.  However, as in the case of the pair $(g_{\mu\nu}, \Gamma^{\kappa}_{\mu\nu})$ discussed above, one can also keep the connection more general while working with the vielbein, giving rise to a pair $(e^{\mu}_a, \Gamma^{\kappa}_{\mu\nu})$. 
The Hilbert-Palatini-action then generalizes to the Holst action, which contains an additional coupling, the Immirzi parameter $\gamma$, which vanishes if the torsion vanishes. The RG flow of the Newton coupling, cosmological constant and Immirzi parameter shows candidates for asymptotically safe fixed points \cite{Daum:2010qt,Daum:2013fu}. One can also take inspiration from Loop Quantum Gravity, where (anti)-self dual connections play an important role \cite{Ashtekar:1986yd,Ashtekar:1987gu} (see \cite{Sahlmann:2023eqt} for a recent overview) and impose a corresponding constraint on the generalized connection. This reduces the set of degrees of freedom within the generalized connection and may still allow for asymptotic safety \cite{Harst:2015eha}.

At the technical level, accounting for an independent connection is in general significantly more complex than the standard setup based on the pair $(g_{\mu\nu}, \{\!\!\!\phantom{I}^{\,\,\kappa}_{\mu\nu} \})$, see \cite{Harst:2014vca}, such that studies have not progressed to higher orders in truncations of the action.
 
To close this overview,  we note that there is even the possibility to remove the metric entirely and work purely with an affine connection \cite{Knorr:2020bjm}, although asymptotic safety has not yet been studied in this setting.
Finally, one can also go beyond the notion of geometry underlying General Relativity and generalize the standard metric that measures lengths to an area-metric \cite{Borissova:2023yxs}, for which RG flows have recently been studied \cite{Borissova:2025frj}.\newline

Summarizing the tentative results from studies that change the gravitational field content, it appears that asymptotic safety may arise more generally than just in metric gravity. However, enlarging the number of degrees of freedom as well as the level of technical complexity should be justified by a compelling physical reason. As long as standard metric gravity does not run into any consistency problems or inability to reproduce observed phenomenology, there appears at best to be a justification for unimodular gravity, given the reduction in field content and status of the cosmological constant.

\section{Physics of asymptotic safety, across all scales}\label{sec:pheno}
\emph{...where we explain the mechanism by which asymptotic safety can make predictions for particle physics (in and Beyond the Standard Model), cosmology (both in the early and the late universe) and black-hole physics. These predictions all hinge on the predictive power of asymptotic safety for the couplings in the effective action. We review the status of these predictions, critically discuss underlying assumptions and point out open questions for future research.}\\

The Planck scale $M_{\rm Planck}$ is much higher than scales accessible in current particle-physics experiments, and is significantly higher than curvature scales in cosmology, starting from inflation until today, as well as much larger than the curvature at the horizon of astrophysical black holes. 
Thus, quantum gravity is often viewed as disconnected from particle physics and cosmology; and its range of applicability to black-hole physics is often viewed as limited to theoretical problems such as the so-called information paradox or the fate of the curvature singularity, which are expected to be of little relevance for astrophysical black holes.

Yet, as a general principle in physics, microphysics determines macrophysics. Hence, the physics of cosmology, black holes and particle physics  at large distances (or low energies) is determined by the physics at short distances (or high energies), which is ultimately the physics of quantum gravity with matter. To make this connection useful in practice, we need to understand how exactly macrophysics is determined by microphysics. A useful analogy is given by hydrodynamics. It is an effective description of fluids at large scales and has free parameters, such as the viscosity of a given fluid. These free parameters are determined by the physics of the atoms and molecules that make up the fluid. Similarly, our models of particle physics, cosmology and black holes are effective theories, which come with free parameters. In many cases, these free parameters include not just the couplings in an effective Lagrangian, but also the choice of effective Lagrangian, e.g., which degrees of freedom, symmetries and interactions are included. A predictive quantum theory of gravity and matter selects among these effective Lagrangians and fixes (some of) the free parameters in a given Lagrangian. As we will see, this does not lead to a unique effective Lagrangian with fully determined couplings in asymptotic safety. However, as we will also see the predictive power is strong enough to place meaningful and strong constraints and even rule out various proposals for new physics.\footnote{This comes with the caveat of several technical assumptions that are typically being made at the time of writing of this review, such as the choice of Euclidean signature for virtually all of these studies. We will spell these caveats out clearly in what follows.}

Thus, the ``landscape'' of asymptotically safe theories is expected to be limited and significant efforts have been made to start mapping this landscape over the past decade, starting from numerous studies of concrete models and whether they lie in the landscape, summarized for particle physics in \cite{Eichhorn:2018yfc,Eichhorn:2022jqj,Eichhorn:2022gku}, as well as tools to map out this landscape \cite{Knorr:2024yiu,Saueressig:2024ojx,Eichhorn:2025xbb,DelPorro:2025wts}, to connections to other research efforts of delineating landscapes and separating them from swamplands \cite{deAlwis:2019aud,Basile:2021krr,Eichhorn:2024rkc,Basile:2025zjc}.

\subsection{Predictive power of asymptotic safety for particle physics}\label{sec:particlephysics}
\emph{...where we review what is known about the predictive power of asymptotic safety for particle physics. We review results for the Standard Model, which constitute an important consistency check: if the Standard Model (and all phenomenologically viable extensions) with the measured values of couplings cannot be accommodated in asymptotic safety, this is a problem for the viability of the theory. Conversely, if the Standard Model can be accommodated and some of the couplings can be ``post-dicted'' because they are associated with irrelevant directions, then this partially realizes an old dream about quantum gravity, namely that it may provide explanations for (parts of) the structure of the Standard Model. \\
Motivated by the need to explain the nature of dark matter, we then extend to settings beyond the Standard Model. There, the predictive power of asymptotic safety tentatively rules out several models of dark matter and portal interactions of dark with visible matter.\\
Finally, we discuss scattering in asymptotically safe gravity. This is of less direct relevance for phenomenology, but addresses key questions about the structure of asymptotic safety, including unitarity as well as the role of black holes in the theory.}

\subsubsection{Motivation: Why connect quantum gravity to Standard Model physics and beyond?}
The Standard Model (SM) is highly successful in making predictions for numerous scattering cross-sections, particle lifetimes, decay widths and other measurements in particle physics. These predictions rely on the 26 free parameters (3 gauge couplings, 6 quark Yukawa couplings, 4 quark mixing parameters, the Higgs quartic coupling and Higgs vacuum expectation value, the QCD theta parameter, 6 lepton Yukawa couplings and 4 neutrino mixing parameters).\footnote{We include right-handed neutrinos in the SM. While its original version contains only massless, left-handed neutrinos; neutrino oscillations are a well-established fact and require neutrino masses as well as mixing parameters.} One of these parameters is the subject of particularly intense scrutiny, namely the Higgs vacuum expectation value (or Higgs mass), which is technically ``unnatural'', having led to numerous attempts to solve the resulting hierarchy problem which typically rely on new physics close to the electroweak scale \cite{Giudice:2017pzm}. Another set of couplings, namely the Yukawa couplings, also have their own version of a hierarchy problem, termed the flavor problem.  The different Yukawa couplings are of rather different orders of magnitude, resulting in orders of magnitude of mass differences between the Standard Model fermions. For the neutrino masses, the number of orders of magnitude (at least 6 to the electron mass and 11 to the top quark mass), is often considered too large. This has triggered various proposals of seesaw mechanisms to explain this difference by introducing a high new-physics scale, see  \cite{King:2003jb,Mohapatra:2005wg} for reviews. \\
Observations to date indicate that no hierarchy problem (including that between the cosmological constant and the Planck scale) in nature appears to be resolved by new physics at the corresponding scale. This may suggest that we may have taken hierarchy problems too seriously. Instead, it is incumbent upon us to ask whether \emph{any} of the 26 free parameters of the SM, whether linked to a hierarchy problem or not, can be predicted from quantum gravity.\\
 One may object that quantum gravity is indifferent to the existence of distinct flavors or distinct gauge groups, because it does not couple to the corresponding internal indices. Thus, it is not clear that quantum gravity has enough distinguishing power to generate the structures of couplings in the SM. As we will see below, asymptotically safe quantum gravity, together with the Abelian hypercharge coupling, has enough distinguishing power to generate fixed points from which some part of the SM structure is fixed, and thus the number of free parameters is reduced -- albeit many of the free parameters of the SM remain. \\

Exploiting the same predictive power is of great interest for  beyond the Standard Model (BSM) physics, because the space of BMS models is huge.  BSM models are devised to answer questions left open by the SM on the nature of dark matter, dark energy and inflation, the origin of neutrino masses, the generation of matter-antimatter asymmetry. Each of these models has its own parameter space, spanned by the couplings in an effective Lagrangian. Typically, all but a handful of couplings are ignored, because perturbative renormalizability is demanded, or, for perturbatively non-renormalizable settings, one often focuses on the lowest-dimensional operators. The finitely many remaining couplings in these models are typically constrained by phenomenological considerations, e.g., couplings in dark-matter models are constrained in order to avoid an overabundance of dark matter and inflationary models are constrained by an upper bound on the tensor-to-scalar ratio of primordial fluctuations as well as a measurement of the spectral index of scalar fluctuations. In addition, there are direct and indirect searches for the corresponding fields and interactions. However, even the combination of phenomenological constraints and (in)direct searches is rarely enough to rule out the full parameter space of a given model. \\
This situation is different in asymptotic safety, where the added requirement is imposed that, together with gravity, the model should exhibit an interacting fixed point beyond the Planck scale. This requirement can suffice to fully rule out a model, or at least reduce its parameter space significantly. Thus, asymptotic safety may serve as a theoretical guiding principle to make the search for a viable model more effective. Conversely, asymptotic safety thereby generates testable predictions, e.g., for dark-matter searches. Thereby, we can confront asymptotic safety with observations, even though experiments do not reach the Planck scale directly.

In summary, in establishing connections of quantum gravity to particle physics in and beyond the SM, we do \emph{not} propose some exotic new ``smoking-gun'' effects of quantum gravity at particle-physics scales. Rather, we take the much more conservative point of view that the effective Lagrangians used in particle physics have free parameters which are determined by the underlying, microscopic quantum field theory which includes quantum gravity. This makes quantum gravity testable at particle-physics scales without the need to invoke exotic effects.

\subsubsection{Bridging the gap between scales: how does asymptotic safety impact physics at scales relevant for particle physics?}\label{sec:bridgingthegap}
\emph{...where we discuss a seeming tension between distinct claims. The first claim is that physics at different scales \underline{decouples}.  Thus, quantum gravitational degrees of freedom, being \underline{dynamically irrelevant} at sub-Planckian energy scales, do not show up in the effective field theories for particle physics. The second claim is that asymptotic safety makes predictions for particle physics. 
The seeming tension is resolved by recognizing that the couplings in the effective field theory are free parameters and not determined \underline{within} the effective field theory. They are determined by the UV completion -- in the case we are interested in, asymptotic safety predicts (some of) their values. The physics of decoupling, resulting in the suppression of higher-order interactions, tells us which of these predictions are in practice testable.}\\

The typical scale of quantum gravity is expected to be around the Planck scale, which is much higher than typical scales of quantum gravity. Thus, to understand how asymptotic safety makes predictions for particle physics, we discuss three questions below:
\begin{itemize}
\item Given that gravitons are massless, but their interaction strength is governed by the Planck scale, how do gravitons decouple from the rest of the SM?
\item Given decoupling, why is quantum gravity not simply negligible for the couplings in the SM?
\item Given predictions from asymptotic safety for the values of couplings at the Planck scale, for which category of couplings can we translate into predictions that are testable in practice at typical particle-physics scales?
\end{itemize} 

\noindent{\emph{Decoupling and quantum gravity}}\\
Effective field theory relies on a powerful mechanism, namely that of \emph{decoupling}. It implies that fields which are dynamically irrelevant, i.e., for which the energy in a system is insufficient to generate the corresponding on-shell degrees of freedom, can be neglected in an effective description of the system. For instance, knowledge of the Standard Model of particle physics is not required to build stable buildings, even though at a fundamental level, the elementary particles of the Standard Model do of course constitute the actual building blocks. 

Gravitational degrees of freedom are massless and one may thus suspect that they do not decouple -- similarly to photons. However, their interaction strength is set by the Planck mass.  Thus, decoupling follows, because their interaction is suppressed. 

Within the functional RG, decoupling is built in automatically, because the contribution of any field $\phi$ to a beta function generically comes in the schematic form
\begin{equation}
\beta\Big|_{\phi}\sim\left(\frac{\bar{g}_{\phi}}{k^{d_{\bar{g}_{\phi}}}}\right)^{n_1}\frac{1}{\left(1+\frac{\bar{m}_{\phi}^2}{k^2}\right)^{n_2}} \cdot f,
\end{equation} 
where $\bar{m}_{\phi}$ is the dimensionful mass of the field and $\bar{g}_{\phi}$ its dimensionful coupling. 
Both $n_1$ and $n_2$ are positive integer powers and $f$ stands for a factor coming from other fields in the setting; in the simplest case, $f=1$.
In general, both coupling as well as denominator can come with a positive integer power, but this is not important to understand how decoupling is built into the FRG. \\
Let us first assume that $d_{\bar{g}_{\phi}}=0$, i.e., the coupling is canonically marginal, as for many Standard Model fields. Then, for $k^2 > \bar{m}_{\phi}^2$, the denominator is effectively 1, and the field in question contributes to the beta function. Once the RG scale $k$ drops below the explicit mass scale of the field, $k^2 <\bar{m}_{\phi}^2$, however, the denominator suppresses the contribution. Then, the contribution scales as $\frac{\bar{g}_{\phi}}{k^{d_{\bar{g}_{\phi}}}}\frac{1}{1+\frac{\bar{m}_{\phi}^2}{k^2}} \sim \frac{k^2}{\bar{m}_{\phi}^2}$ and thus effectively becomes negligible over few orders of magnitude in $k$. \\
Let us now assume that $\bar{m}_{\phi}=0$, as is the case for gravity. Then, the denominator can never suppress the term. However, for gravity, $\bar{g}_{\phi}$ is the Newton coupling, $\bar{g}_{\phi}= G_N = \frac{1}{M_{\rm Pl}^2}$ and $d_{\bar{g}_{\phi}}=-2$. This results in two very distinct regimes. In the fixed-point regime above the Planck scale\footnote{In this regime, the Planck mass scales, $M_{\rm Pl}^2 \sim k^2$.}, $G_N\cdot k^2 = \rm const$. 
Thus, quantum-gravity effects can dominate in a beta function above the Planck scale.
In the regime below the Planck scale, $G_N = \rm const$, such that the gravitational contribution is suppressed, $G_N\cdot k^2 = \frac{k^2}{M_{\rm Pl}}$, very similarly to the suppression of a generic massive field, once $k^2$ has dropped below the mass-scale in question.\footnote{This argument for decoupling has subtleties when it comes to the gravitational contribution to higher-order couplings \cite{Knorr:2026vax}. For particle physics, these subtleties do not play a role on scales relevant for experiments.} 

Given the decoupling mechanism, and its automatic implementation in the functional RG, why should one expect that quantum gravity plays a role for physics below the Planck scale? Dynamically, of course it does not; e.g., the gravitational contribution to particle scattering at the LHC is suppressed by factors of $E_{\rm LHC}^2/M_{\rm Pl}^2$, such that it is negligible unless one would want to extract scattering cross-sections with fantastical precision. However, dynamical suppression of quantum gravity does not mean absence of imprints of quantum gravity. Specifically, the couplings in any effective field theory always carry a ``memory'' of the UV completion, even if the degrees of freedom of the UV completion are dynamically suppressed. This means for our case that the couplings in the Standard Model can carry the ``memory'' of physics in the transplanckian regime. More specifically, the values of some of the couplings in the SM are fixed by asymptotic safety not just in the transplanckian regime, but at all scales.

The main mechanism through which couplings are fixed, is the one presented in Sec.~\ref{sec:predpower}, i.e., they correspond to irrelevant directions at the fixed point. We need to distinguish two cases. The first case is higher-order couplings, for which these predictions would typically need extreme experimental precision to be tested.\footnote{For specific dimension-5-operators, such precision is actually close to being achievable in dedicated experiments.} The second case is dimensionless couplings, for which Planck-scale predictions are translated into testable predictions at particle-physics scales through logarithmic running.\\

\begin{figure*}[!t]
\includegraphics[width=\linewidth,clip=true,trim=0cm 4cm 0cm 0cm]{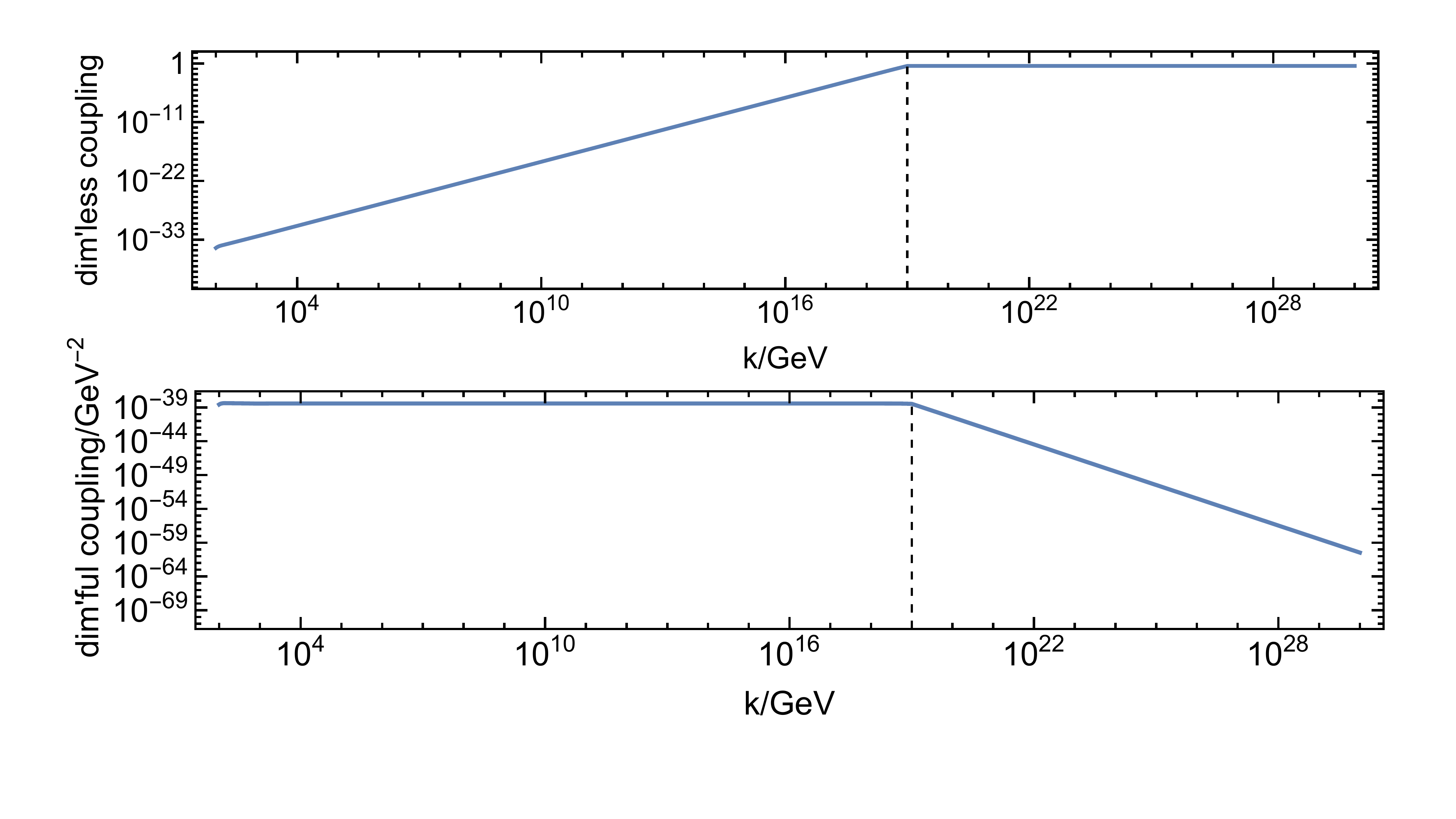}
\caption{\label{fig:predictionho} 
We show a coupling which has a non-zero fixed-point value and corresponds to an irrelevant direction, such that the Planck-scale value of its dimensionless counterpart is fixed. The coupling has a non-vanishing, negative mass-dimension (-2 in the example that is shown), such that below the Planck scale, its RG flow is determined by the corresponding scaling behavior. The dimensionless version of that coupling accordingly scales with the canonical mass dimension below the Planck scale. As a consequence, the dimensionful counterpart of the coupling is constant below the Planck scale and exhibits Planck-scale suppression.\\
This simple scaling analysis does not hold, if other terms in the beta function below the Planck scale matter, Eq.~\eqref{eq:betag_irrel}.}
\end{figure*}

\noindent{\emph{Predictions for higher-order couplings: Confronting Planck-scale suppression}}\\
First, couplings that are canonically irrelevant, i.e., correspond to higher-order interactions, generally stay irrelevant at the fixed point, assuming the fixed points retains its near-perturbative character, see Sec.~\ref{sec:nearpert}. Then the Planck-scale value of their dimensionless counterparts is fixed. Below the Planck scale, these couplings typically run according to their canonical dimension, because loop effects are suppressed compared to the canonical scaling term in a perturbative regime. 
\begin{equation}
\beta_{g}= -d_{\bar{g}}g+\dots, \quad \Rightarrow\,\, g(k) = g_{\ast} \left(\frac{k}{M_{\rm Pl}}\right)^{-d_{\bar{g}}},\label{eq:betag_irrel}
\end{equation}
with $d_{\bar{g}}<0$.
The dimensionful coupling is therefore generically Planck-scale suppressed in the IR\footnote{This suppression does not hold, if the scale dependence below the Planck scale is not simply determined by the canonical term.}, with a numerical coefficient that is predicted, cf.~Fig.~\ref{fig:predictionho},
\begin{equation}
\bar{g}(k) = g(k)\cdot k^{d_{\bar{g}}} = \frac{g_{\ast}}{M_{\rm Pl}^{-d_{\bar{g}}}}, \quad \mbox{with } d_{\bar{g}}<0.
\end{equation}
 Due to the generic Planck-scale suppression, this prediction is rather challenging to test. Notably, e.g., for some proposals for dark matter, highly suppressed values of couplings, e.g., of dimension-5-interactions, are actually testable. Similarly, proton decay may be mediated by dimension-six interactions which are already so tightly constrained \cite{ParticleDataGroup:2024cfk} that only few orders of magnitude are missing to testing predictions of the coupling from quantum gravity \cite{Eichhorn:2023jyr}.\\
 There is a caveat to the above reasoning: The Planck-scale suppression of dimensionful couplings holds, if the subplanckian RG flow of the coupling is dominated by the canonical scaling term. If higher-order, e.g., quadratic or cubic terms, partially balance the canonical scaling term, the subplanckian RG flow can result in a much lower effective suppression scale. A toy-model example of this is given by four-fermion couplings \cite{Brenner:2024bps}, which are dimension-six interactions, but can circumvent Planckian suppression through interacting fixed points in the purely matter-driven part of the RG flow below the Planck scale. 
 
 We discuss higher-order couplings and physical information about quantum gravity that one can extract from them in Sec.~\ref{sec:SMEFT}.\\

\begin{figure*}[!t]
\includegraphics[width=\linewidth,clip=true,trim=0cm 19.5cm 0cm 0cm]{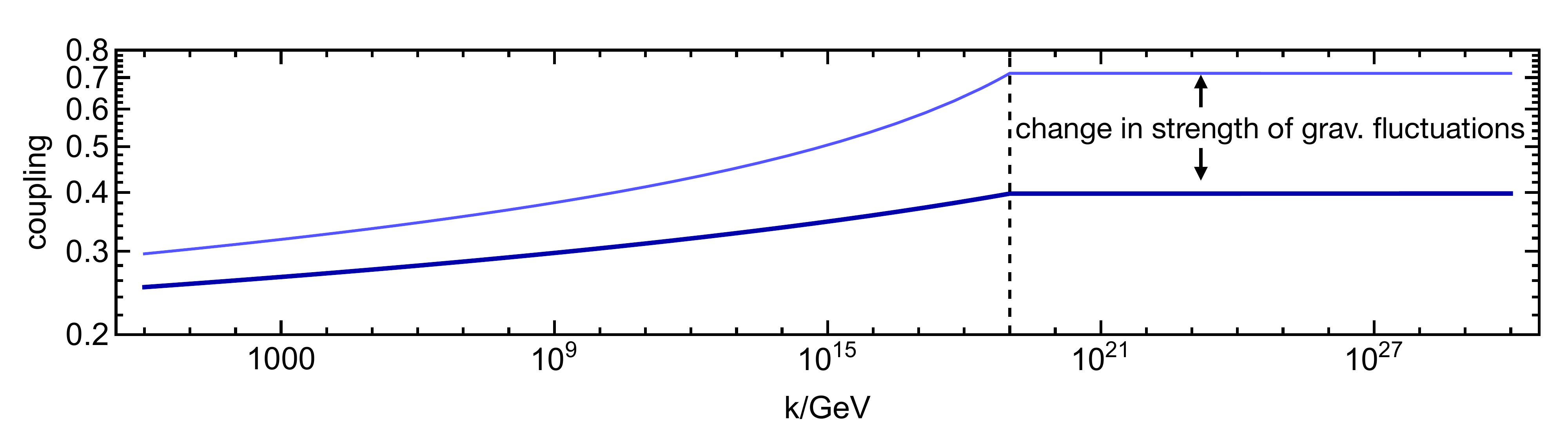}
\caption{\label{fig:predictions_illustration} We show a coupling which has a non-zero fixed-point value and corresponds to an irrelevant direction, such that its Planck-scale value is fixed. Below the Planck scale, it runs logarithmically, so that a change in fixed-point value (caused by a change in gravitational fixed-point values) results in a change of similar size at low scales, as one can see by comparing the thin and thick lines.}
\end{figure*}

\noindent{\emph{Predictions for dimensionless couplings: Testability saved by logs}}\\
Second, couplings that are canonically marginal can either become relevant or irrelevant. In fact, as we will see, both cases are realized within the SM. If a dimensionless coupling is irrelevant, its Planck-scale value is fixed. Below the Planck scale, the coupling depends on the scale logarithmically. As a consequence, its IR-value is fixed, cf.~Fig.~\ref{fig:predictions_illustration} and it differs from the Planck-scale value by an $\mathcal{O}(1)$ amount. 
In other words, logarithmic running circumvents Planck-scale suppression. Thus, experimental measurements of the IR-value are sensitive to the coupling's Planck-scale value: a change of $\mathcal{O}(1)$ at the Planck scale translates into a corresponding change of $\mathcal{O}(1)$ at IR scales, simply because on a logarithmic scale, the distance between, e.g., the electroweak scale and the Planck scale is not very large.\footnote{For the SM, it is not necessary to flow to $k\rightarrow 0$ because of one-loop universality between distinct notions of running for these logarithmically running couplings. Thus, one can compare the values of the couplings extracted from RG flows from the Planck scale down to $k=173\, \rm GeV$ to those extracted from experiment and reported at $\mu =173\, \rm GeV$, see, e.g., \cite{ParticleDataGroup:2024cfk}.} Thus, predictions from quantum gravity can meaningfully be confronted with experimental data for this case. The first examples for SM couplings for which there is evidence that they could fall into the category of calculable quantities in asymptotic safety include the Higgs quartic coupling \cite{Shaposhnikov:2009pv}, the Abelian gauge coupling \cite{Harst:2011zx,Eichhorn:2017lry} and the top Yukawa coupling \cite{Eichhorn:2017ylw}. \newline

Obviously, for all couplings, the predictions depend on thresholds that the RG flow passes. At a threshold scale, degrees of freedom with masses at that scale decouple. Thus, a threshold changes the IR prediction of a coupling corresponding to a given UV prediction.

For the SM, the simplest assumption is that of a desert, i.e., no degrees of freedom with masses between the electroweak scale and the Planck scale. This assumption is consistent with the current experimental situation at the LHC, and requires new physics (e.g., for dark matter) to be sufficiently weakly coupled to not generate sizable corrections to the SM RG flow. 

In the absence of a desert, e.g., under the assumption of additional, heavy degrees of freedom, the Planck-scale predictions can still be translated into predictions at low scales. One simply has to account for the extra heavy degrees of freedom at the fixed point. If a fixed point exists and the model is not ruled out by absence of a fixed point, the extra degrees of freedom and their thresholds have to be accounted for during the RG flow. 

One can even use this to constrain heavy new physics: in general, the Planck-scale predictions of couplings is at most compatible with the electroweak values of Standard Model couplings for relatively narrow ranges of the masses of new particles, see, e.g., \cite{Kowalska:2020gie} for a specific example. While the specific case of flavor-anomalies in \cite{Kowalska:2020gie} is generally no longer seen as compelling case for new physics, the idea that asymptotic safety constrains the masses of heavy new degrees of freedom  generalizes to other cases, see \cite{Eichhorn:2017ylw}.

\subsubsection{What does asymptotic safety imply for the couplings in the Standard Model of particle physics?}\label{sec:SM}
\emph{...where we review work towards establishing the asymptotically safe Standard Model, i.e., a candidate theory of all known degrees of freedom (SM plus gravity). The asymptotically safe Standard Model is UV complete. It may be even more predictive than the SM without gravity. We discuss the evidence for this candidate theory and point out open questions.}\\

The Standard Model (SM) is not a UV complete theory, because it is not asymptotically free (nor asymptotically safe) on its own. Instead, Landau poles in the Higgs quartic coupling $\lambda_H$ and the Abelian hypercharge coupling $g_Y$ signal the need for new physics.\footnote{Strictly speaking, these Landau poles signal the breakdown of perturbation theory. However, at the non-perturbative level, there is no indication for a UV completion and thus these Landau poles translate into triviality problems: to shift the scale of the Landau pole to infinitely high values and render the SM couplings finite at all scales, one has to accept vanishing Higgs quartic and Abelian hypercharge coupling in the IR, i.e., accept a trivial theory \cite{Cabibbo:1979ay,Gockeler:1997dn}.} Since the measurement of the Higgs boson mass at 125 GeV \cite{ATLAS:2012yve,CMS:2012qbp}, it is known that the scale at which new physics must at the latest exist, is transplanckian. This is a highly nontrivial result from the LHC, because a somewhat higher Higgs mass results in a Landau pole \emph{before} the Planck scale \cite{Hambye:1996wb}.
Accordingly, the SM is self-consistent and even perturbative up to the Planck scale. It only calls for new physics at transplanckian scales, at which we already know that quantum gravity must be incorporated into it.\footnote{Depending on the exact value of the top quark mass, that is not known with sufficient precision \cite{Bezrukov:2014ina}, the Higgs quartic coupling may be negative between approximately $10^{10}\, \rm GeV$ and the Planck scale. This is usually interpreted as signalling a metastability of the Higgs potential \cite{Bezrukov:2012sa,Buttazzo:2013uya}. The lifetime of the metastable state with a Higgs vacuum expectation value of 246 GeV exceeds the lifetime of the universe. Thus, even if the top quark mass lies at a value such that absolute stability of the electroweak vacuum is excluded, this is not unequivocal evidence for new physics, because a metastable state of large enough lifetime is phenomenologically acceptable.} As a consequence, it becomes crucial to understand the effect of quantum gravity on the gauge couplings, Yukawa couplings, Higgs quartic coupling as well as mixing matrices in the Standard Model. All couplings must be rendered UV safe under the impact of quantum gravity, i.e., they must start from a joint fixed point (at which some couplings may also vanish) in the UV. If this cannot be realized, additional new physics besides quantum gravity is required. \\

The gravitational contribution to the Standard Model can be described in a very economic fashion, because gravity is ``blind" to internal symmetries. Therefore we have
\begin{itemize}
\item A single gravitational contribution to all three gauge couplings, $g_Y$, $g_2$ and $g_3$, because gravity is insensitive to whether the gauge bosons carry color, weak charge, or no internal charge.
\item A single gravitational contribution to all 12 Yukawa couplings $y_{t,b,\dots}$ (for the six quarks, 3 charged leptons and 3 neutrinos), because gravity cannot distinguish whether the fermion carries charge under $SU(3)$ and/or $SU(2)$ and whether or not it carries hypercharge.
\item There is no gravitational contribution to the mixing matrices, because mixing refers to how the mass eigenstates mix in the weak interactions, but gravity cannot distinguish different flavors.
\item A single gravitational contribution to the Higgs quartic coupling (and one to the non-minimal coupling).
\end{itemize}

In addition, we can infer the form of these contributions from the corresponding loop diagrams before actually calculating the diagrams: The direct gravitational contribution to the beta function of a SM coupling must always be proportional to the dimensionless Newton coupling $G$ and is always proportional to the coupling in question\footnote{Proportionality to the coupling itself follows from symmetry considerations: Setting gauge and Yukawa couplings and scalar potentials to zero leads to an enhancement of the global symmetries of a model to the maximal global symmetry of the respective, minimally coupled kinetic terms. Because RG flows do not generate symmetry-breaking terms, the RG flow of any symmetry-breaking couplings has to be proportional to itself (or another coupling that breaks that same symmetry).}, i.e., 
\begin{eqnarray}
\beta_{g_{Y/2/3}}&=& - f_g\, g_{Y/2/3}+ \mbox{ SM-loops },\label{eq:betagauge_general}\\
\beta_{y_{t,b,\dots}}&=& -f_y\, y_{t,b,\dots}+ \mbox{ SM-loops },\label{eq:betayukawa_general}\\
\beta_{\lambda}&=& - f_{\lambda}\lambda+\mbox{ SM-loops }\label{eq:betaquartic_general}, 
\end{eqnarray}
where $f_g$, $f_y$ and $f_{\lambda}$ depend on the gravitational couplings. Beyond the direct contribution, there may also be a contribution from non-minimal couplings, which we neglect in this discussion for simplicity. 

We can compare  Eqs.~\eqref{eq:betagauge_general}-\eqref{eq:betaquartic_general} to the simple beta functions used in Fig.~\ref{fig:predictivepower}. We note that if a) the contribution from SM-loops in Eqs.~\eqref{eq:betagauge_general}, \eqref{eq:betayukawa_general} or \eqref{eq:betaquartic_general}, respectively, is screening, i.e., positive, and b) the gravitational contribution, encoded in $f_g$, $f_y$ and $f_{\lambda}$, respectively, is antiscreening, i.e., negative (i.e., $f>0$), the Planck-scale value of the corresponding coupling can be predicted. This is because it corresponds to an irrelevant direction.\footnote{At the same time, this fixed point acts as an upper bound on all coupling values that are reachable from an asymptotically free fixed point that exists due to the antiscreening gravitational contribution.} Below the Planck scale, the gravitational contribution decouples (see Sec.~\ref{sec:bridgingthegap}). Then, a non-vanishing RG flow sets in that maps the prediction at the Planck scale to a prediction at experimentally accessible scales. 

We also note that, because the contribution from SM loops is different for different couplings in a given sector, the predictions for different couplings can take different values despite the fact that gravity cannot tell these couplings apart. Therefore, the system  Eqs.~\eqref{eq:betagauge_general}-\eqref{eq:betaquartic_general} has all the ingredients that are needed to predict at least a subset of the couplings in the SM.

To investigate whether such predictions exist and what the corresponding values are, the economic form in Eqs.~\eqref{eq:betagauge_general}-\eqref{eq:betaquartic_general} suggests the following strategy: First, our task lies in calculating the form of the three $f$'s as precisely as possible.\footnote{The $f$'s can be understood as a parameterization that includes not just the direct gravitational contribution, but also, e.g., contributions from non-minimal gravity-matter couplings, see \cite{deBrito:2025nog}.} This is, of course, challenging, even though recent progress combining high-order calculations with error estimates holds promise \cite{deBrito:2025nog}.
Second, because they are just three contributions for the entire SM, which contains 24 couplings, we can \emph{parameterize} gravitational effects and treat the three parameters $f_g$, $f_y$ and $f_{\lambda}$ as parameters to be fixed by three pieces of experimental data. Once this is done, there is no more freedom for the gravitational contribution to the remaining 21 couplings. 
Below, we will discuss both, the result of explicit calculations of the $f$'s as well as the results of the parameterized approach.

In the following, we discuss first the gauge sector, then the Yukawa couplings and mixing matrices and finally the Higgs quartic coupling. This is because we will focus on one-loop results together with gravity. At one loop, the other couplings of the SM do not appear in the beta functions for the gauge couplings. This is different for the Yukawa couplings, which depend on the gauge couplings at one-loop order. The Higgs quartic coupling also depends on the gauge couplings, as well as the Yukawa couplings. Quantitatively, working with the one-loop contributions is robust, as long as the UV completion is as perturbative as the SM, for which two-loop contributions produce only subleading corrections.
\newline

Let us preface this discussion by highlighting that all results that will be discussed in this section are based on RG flows in Euclidean signature. At a near-perturbative fixed point, there may not be an obstacle to analytically continuing to Lorentzian signature in the standard way, such that these results may remain valid in Lorentzian signature. Testing this with explicit studies of Lorentzian gravity-matter systems is an important point for future research.\\

\noindent {\bf Gauge couplings:}\\
A consistent coupling to quantum gravity must UV complete the SM (or a suitable extension, see Secs.~\ref{sec:SMEFT} and \ref{sec:dm}). To UV complete the Abelian gauge coupling, gravitational fluctuations must antiscreen this coupling, as they in fact do \cite{Harst:2011zx}, i.e., $f_g>0$ holds in Eq.~\eqref{eq:betagauge_general}. This automatically entails the possibility to fulfill more than the minimal requirement of UV completion: the coupling to asymptotically safe gravity can turn the IR value of the Abelian gauge coupling from a free parameter of the SM into a calculable parameter of the asymptotically safe SM, resulting in an upper bound on the Abelian gauge coupling \cite{Eichhorn:2017lry}. 

In the following, we discuss this mechanism at a heuristic level, before going back to the underlying beta functions and discussing the robustness of the results.
It is well-established that quantum fluctuations of charged matter screen the Abelian hypercharge coupling $g_Y$. Thus, as one lowers the RG scale $k$, matter fluctuations drive the coupling towards zero. Quantum gravity fluctuations have the opposite effect.
\footnote{This was first observed in perturbation theory \cite{Robinson:2005fj}. However,  this result was dismissed in part of the literature, see, e.g., \cite{Toms:2007sk,Ebert:2007gf,Anber:2010uj,Ellis:2010rw}, because such a gravitational contribution does not affect a logarithmic dependence of the coupling on physical momentum scales in scattering events. However, in our context, the $k$-dependence (i.e., dependence on an auxiliary cutoff, not on a physical momentum scale) is the relevant notion of scale to determine how predictive the theory is.
In addition, there are also technical subtleties related to dimensional regularization, which, when applied naively, is not a true regularization scheme, but instead a projection onto logarithmic divergences. Power-law divergences in the $k$-dependence can be crucial to determine the predictive power of a fixed point. Thus, dimensional regularization has to be implemented with caution (accounting for poles not just for $d \rightarrow 4$), in order to correctly count the free parameters of a theory \cite{Baldazzi:2020vxk}. In fact, one can show that in regularization schemes in which the naive contribution of gravity to a running coupling vanishes, critical exponents may still be nontrivial, because the gravitational fixed-point value may actually diverge in such a setting, such that critical exponents remain finite \cite{deBrito:2022vbr}.\\
 Thus, the results in perturbation theory, which find that the gravitational contribution may not affect the $p$-dependence, because it is not logarithmic, are not in contradiction to the results in asymptotic safety. Most importantly, those results are not in contradiction with the statement that there is a gravitational contribution to the $k$-dependence of the gauge coupling.} 
Thus, gravitational fluctuations drive the coupling towards larger values, as one lowers $k$. These two effects compete, such that they cancel out at a critical ratio of the gravitational coupling strength to the gauge coupling strength. At this ratio, the net effect of all fluctuations on the coupling is neither screening nor antiscreening. The coupling instead remains constant, i.e., there is a scale invariant regime, attainable when the strengths of the competing effects balance. \\
This regime is stable against perturbations, i.e., a gauge coupling strength which is slightly above or below the critical strength evolves back to the critical value. The reason for this stability -- rather than instability -- is that the gravitational, antiscreening effect dominates at very weak gauge interaction strength, whereas the matter, screening effect dominates at stronger gauge interaction strength. This ordering follows from the fact that virtual charged matter particles couple to the photon through the gauge coupling; thus, quantum corrections to the gauge interaction strength $g$ are proportional to higher powers of $g$. In contrast, virtual gravitons couple proportionally to the gravitational interaction strength $G$. Thus, quantum corrections to the interactions of charged matter with a photon come with a single factor of $g$ (multiplied by positive powers of $G$) and thus dominate at low gauge interaction strength.\\

\begin{figure}[!t]
\includegraphics[width=\linewidth]{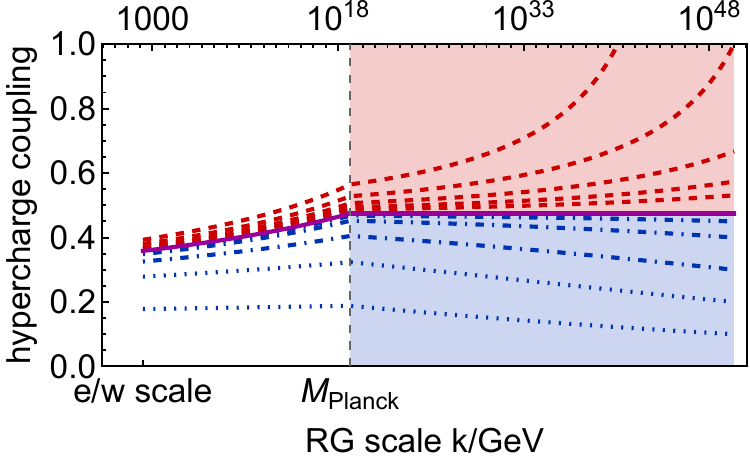}
\caption{\label{fig:upperbound} Transplanckian region: Screening effects dominate above the critical value of the coupling (red shaded region), $g_{\ast}$, resulting in trajectories being driven towards $g_{\ast}$ from above. Antiscreening effects dominate below $g_{\ast}$ (blue shaded region), resulting in trajectories being driven towards $g_{\ast}$ from below. Asymptotically free trajectories approximate $g_{\ast}$ well (dot-dashed blue lines), unless $g_Y(k\gg M_{\rm Planck})$ is chosen to be very close to zero, such that trajectories do not reach $g_{\ast}$ (dotted blue lines). The relative fraction of these trajectories, that do not complete the cross-over to the asymptotically safe fixed point is expected to be small compared to the remaining UV complete trajectories.\\
Trajectories are UV complete in the blue shaded region, but not the red shaded region. Antiscreening and screening effects balance out for the purple trajectory, which realizes scale-symmetry.\\
Subplanckian region: the antiscreening, gravitational effects are absent; screening matter effects dominate and drive the coupling towards lower values.\\ }
\end{figure}

As discussed above, gravitational contributions do not distinguish between couplings that are only distinguished by a change in internal symmetries or charge assignments. Thus, the gravitational contribution to gauge couplings is the same for Abelian and non-Abelian gauge couplings and does not depend on the non-Abelian symmetry group. Thus, for the three SM gauge couplings $g_{Y,2,3}$, where $g_2$ and $g_3$ are the SU(2) and SU(3) gauge couplings, we have Eq.~\eqref{eq:betagauge_general}, which can be written as
\begin{equation}
\beta_{g_{Y,2,3}}=-f_g \, g_{Y,2,3} + \beta_{g_{Y,2,3}}^{(1)}\, g_{Y,2,3}^3+\dots.\label{eq:beta_g}
\end{equation}
We restrict the discussion to the one-loop order for the matter contributions $\beta_{g_{Y,2,3}}^{(1)}$, because, as one can check, already the two-loop contributions are very small for the SM, and thus not relevant for our discussion below.
The one-loop coefficient $\beta_{g_{Y,2,3}}^{(1)}$ depends on the symmetry group as well as the number and representation of matter fields charged under the symmetry group. In contrast, $f_g$ does not depend on the symmetry group at all, because gravity is ``blind'' to internal symmetries. 
Because $f_g \sim G$ (and further gravitational couplings), it holds that, to a very good approximation 
\begin{equation}
f_g = \begin{cases} {\rm const},\, \quad& \mbox{for } k \geq M_{\rm Planck},\\
0,\, \quad &\mbox{for } k< M_{\rm Planck}.
\end{cases}
\end{equation}
This reflects the standard expectation that gravitational fluctuations are negligible for scales below the Planck scale. \footnote{Here, we are not concerned with the very deep IR, far below all SM mass scales, where massless fluctuations are the only remaining ones and IR logarithms may play a role \cite{Knorr:2026vax}.} At transplanckian scales, in the fixed-point regime, all gravitational couplings are constant, and so is $f_g$. In terms of the gravitational couplings, 
$f_g$ has been studied in \cite{Daum:2009dn,Daum:2010bc,Harst:2011zx,Folkerts:2011jz,Christiansen:2017gtg,Christiansen:2017cxa,Eichhorn:2017lry,deBrito:2019gdd,Eichhorn:2021qet,deBrito:2022vbr} and consistently found to be larger than zero.\footnote{Ref.~\cite{Folkerts:2011jz} argued that $f_g$ could also be zero for an appropriate choice of regulator. Such a regulator has, however, not been constructed explicitly. In addition, gravitational fluctuations indirectly contribute to $f_g$ through induced, higher-order field-strength terms \cite{Christiansen:2017gtg,Eichhorn:2021qet}.},\footnote{By using a Ward identity, the gravitational contribution to the gauge coupling can be read off from the scale dependence of the wave-function renormalization of the gauge field \cite{Daum:2009dn,Daum:2010bc,Harst:2011zx,Folkerts:2011jz,Christiansen:2017gtg,Christiansen:2017cxa,deBrito:2019gdd,Eichhorn:2021qet,deBrito:2022vbr} or from the scale dependence of vertices that couple the gauge field to charged matter \cite{Eichhorn:2017lry}.} 
Accordingly, asymptotic safety contains a mechanism to render all gauge couplings, including Abelian ones, asymptotically free, encoded in
\begin{equation}
\theta\Big|_{g_{Y,2,3,\,\ast}=0}= f_g.
\end{equation}
In addition, for an Abelian gauge coupling $g_Y$, Eq.~\eqref{eq:beta_g} admits an interacting fixed point at
\begin{equation}
g_{Y,\,\ast} = \sqrt{\frac{f_g}{\beta_{g_Y}^{(1)}}}, \quad \mbox{with }\theta\Big|_{g_{Y,2,3,\,\ast}}=-2 f_g.
\end{equation}
Here we assumed that higher-loop contributions are negligible, which is valid as long as $f_g \ll 1$. The interacting fixed point is IR attractive in $g_Y$, such that $g_{Y,\,\ast}$ constitutes an upper bound for $g_{Y}(k=M_{\rm Planck})$, cf.~Fig.~\ref{fig:upperbound}. The RG flow below the Planck scale, without gravitational fluctuations, translates this into an upper bound for $g_Y$ at observationally relevant scales.\footnote{In this regime, universality between different schemes holds, thus the dependence on $k$ and on the renormalization scale $\mu$ track each other. Therefore, the physically relevant scale is at $k=\mu=m_{\rm pole}$, with $m_{\rm pole}$ the pole mass of, e.g., the $Z$-boson or the top quark, which are used to report values of couplings extracted from experimental data \cite{ParticleDataGroup:2024cfk}.} IR values below the upper bound are compatible with asymptotic safety, but require the gauge coupling to stay very close to the free fixed point (which is IR repulsive) for a large range of (transplanckian) scales. The more ``natural'' expectation therefore is to find an IR value at or close to (but below) the upper bound.\\

In \cite{Eichhorn:2017lry}, the upper bound was estimated to lie about 30\% above the value of the hypercharge coupling inferred from measurements. This estimate should be considered in the light of large -- and difficult to quantify \cite{deBrito:2022vbr,Riabokon:2025ozw} -- systematic uncertainties arising from truncations which make the fixed-point values of the gravitational couplings, and thus the transplanckian value of $f_g$, uncertain. \\

Let us now discuss more concretely the results for $f_g$ and their implications. The analytic expression for $f_g$ from these calculations, within the Einstein-Hilbert truncation\footnote{The expression also depends on the choice of regulator function in Eq.~\eqref{eq:IRcutoff} and \eqref{eq:IRcutoff_h}; we quote results obtained with the Litim regulator \cite{Litim:2000ci,Litim:2001up}; results for an exponential regulator are provided in \cite{deBrito:2022vbr}.} is found to be 
\begin{equation}
f_g=\frac{5G}{18 \pi} \left(\frac{1-4 \lambda}{(1-2 \lambda)^2} \right) + \mathcal{O}(\beta^2),\label{eq:fgleading}
\end{equation}
where $\beta$ is the gauge parameter in Eq.~\eqref{eq:gaugefixing} and $\lambda = \Lambda k^{-2}$ is the dimensionless cosmological constant. $f_g$ depends on $k$ through the $k$-dependence of $G$ and $\lambda$ and is constant when these couplings assume their fixed-point values.
The gauge-dependent part can be set to zero in the gauge $\beta=0$; it cannot change the sign of $f_g$ compared to the gauge-independent part. Gauge dependence, as discussed in Sec.~\ref{sec:gaugedep}, is not a priori problematic, because $f_g$ evaluated away from gravitational fixed-point values is not a physical quantity. However, changes in $f_g$ impact the IR value of $g_Y$, which is compared to a value extracted from experimental results \cite{ParticleDataGroup:2024cfk}. This comparison only makes sense if, in large enough truncations, the gauge dependence of the function $f_g$ is fully compensated by a corresponding gauge dependence of gravitational fixed-point values. Thus, the gauge dependence of the upper bound on $g_Y$ at the Planck scale can be used to estimate the quality of the truncation, see \cite{Riabokon:2025ozw}.\\
The change of sign to $f_g<0$ at $\lambda=1/4$ was found in \cite{Christiansen:2017cxa} to be an artifact of the derivative expansion and not present otherwise. Fixed-point values for the gravitational coupling typically lie at lower values of $\lambda$ anyways.\\
The result $f_g>0$ is also in agreement with perturbative studies \cite{Robinson:2005fj,Toms:2010vy} and a recent calculation using proper-time-flows \cite{Giacometti:2026zrs}.\\
The leading-order result in Eq.~\eqref{eq:fgleading} was extended in \cite{deBrito:2019gdd} to a truncation with quadratic-gravity terms, $\Gamma_{\rm gravity} = -1/(16\pi G_N)\int d^4x\sqrt{g}\left( -R+ \bar{a}R^2+ \bar{b}R_{\mu\nu}R^{\mu\nu}\right)$, where the dimensionless coupling $b= \bar{b}/k^2$ enters the gauge-independent part:
\begin{equation}
f_g=\frac{G}{36 \pi} \left(\frac{10+7b-40 \lambda}{(1+b-2 \lambda)^2} \right) + \mathcal{O}(\beta).\label{eq:fgcurvsquared}
\end{equation}
From this expression, we note that a change of sign in $f_g$ is possible for large enough negative $b$. However, this occurs only after the denominator in Eq.~\eqref{eq:fgleading} has crossed a zero, i.e., the expression for $f_g$ has crossed a pole. We do not expect that this is a physically viable region. The other possibility for $f_g$ to change its sign, which happens at $\lambda=1/4$ for $b=0$ is shifted towards larger $\lambda$, but, just like for $b=0$, we may expect that this change in sign is an artifact of the derivative expansion  \cite{Christiansen:2017cxa}.\\
There are further higher-order contributions to $f_g$, coming only indirectly from gravitational fluctuations. They arise, because gravitational fluctuations shift the fixed-point values of higher-order gauge interactions to non-zero values \cite{Christiansen:2017gtg,Eichhorn:2019yzm,Eichhorn:2021qet,Knorr:2024yiu,Eichhorn:2024wba}, see Sec.~\ref{sec:SMEFT} for more details. Concretely, 
\begin{eqnarray}
\Gamma_{\rm higher\, order} &=&\int d^4x\sqrt{g}\Bigl(\frac{w_2}{8\,k^4}(F_{\mu\nu}F^{\mu\nu})^2\nonumber\\
&{}&+ \frac{v_2}{8\,k^4} (F_{\mu\nu}\tilde{F}^{\mu\nu})^2\Bigr)+\dots\,\,. 
\end{eqnarray}
It holds that $w_{2\, \ast}\sim G^2$ and $v_{2\, \ast}\sim G^2$, and 
\begin{equation}
f_g=\frac{G}{36 \pi} \left(\frac{10+7b-40 \lambda}{(1+b-2 \lambda)^2} \right) + \mathcal{O}(\beta)- \frac{w_2}{6\pi^2} - \frac{v_2}{24\pi^2}.\label{eq:fg_w2v2}
\end{equation}
Because $w_{2\,\ast}<0$ and $v_{2,\ast}\simeq - w_{2\,\ast}$ (at least for $\lambda$ sufficiently close to zero), these contributions also strengthen the antiscreening character of $f_g$. They are also numerically suppressed compared to the direct gravitational contribution. \\
Overall, there is no evidence for $f_g\lesssim0$; numerous robustness studies support  $f_g>0$. Nevertheless, further studies are of course needed, in particular to provide a quantitatively precise evaluation of $f_g$ and the resulting upper bound on $g_Y$, to robustly determine whether or not asymptotic safety passes a critical observational consistency test.\footnote{One may object that an Abelian gauge group is not needed in nature, and the SM may be embedded inside a grand unified theory, which in turn becomes asymptotically safe. While this is not logically excluded, it should be noted that first, it is not ensured that such a system features a fixed point at all, see \cite{Dona:2013qba} for a first study of this question, and second, that scalar potentials are highly constrained in such a setting \cite{Eichhorn:2019dhg,Held:2022hnw}. Accordingly, phenomenologically viable symmetry breaking chains may actually not be compatible with asymptotic safety.}\newline

\noindent{\bf Yukawa couplings and mixing matrices:}\\
Next, we turn our attention to the Yukawa sector of the SM. The gravitational contribution to the scale dependence of Yukawa couplings is encoded in $f_y$. In the Yukawa sector, we encounter not just the Yukawa couplings, but also the mixing matrices, which describe how the mass eigenstates mix in the weak interactions. The beta functions of the mixing matrices do not receive gravitational contributions \cite{Alkofer:2020vtb}, because mixing concerns \emph{internal} symmetry structures, that gravity is entirely insensitive to.

In the SM, we must distinguish between four types of Yukawa couplings, because of the different hypercharges of up-type quarks (top, charm, up), down-type quarks (bottom, strange, down), charged leptons (tau, muon, electron) and finally, neutrinos, which we discuss further below.\footnote{We consider Yukawa couplings for neutrinos as part of the SM, even though they are sometimes viewed as BSM physics, given that the first version of the SM was developed before neutrino oscillations were known and it was therefore consistent to assume that neutrinos are massless.}

For up-type-quarks ($i = u,c,t$), we have
\begin{eqnarray}\label{eq:betaYukawa1}
    \beta_{y_i} \!&=&\!-f_y\, y_i+\frac{y_i}{16\pi^2} \bigg[ \frac{3}{2}y_i^2 + \left(3 \sum_{j}y_j^2 + 3 \sum_{\kappa} y_{\kappa}^2 + \sum_{l}y_l^2 \right) \nonumber\\
    &{}&-\left(\frac{17}{12} g_Y^2 +\frac{9}{4}g_2^2 +8 g_3^2\right) - \frac{3}{2} \sum_{\kappa} y_\kappa^2 |V_{i\kappa}|^2\bigg]\,\,.
    \end{eqnarray}
For down-type quarks, ($\kappa = d,s,b$) one has:
\begin{eqnarray}\label{eq:betaYukawa2}
    \beta_{y_\kappa}\! &=&\!-f_y\, y_{\kappa} +\frac{y_\kappa}{16\pi^2} \bigg[ \frac{3}{2}y_\kappa^2 + \left(3 \sum_{j}y_j^2 + 3 \sum_{\kappa} y_{\kappa}^2 + \sum_{l}y_l^2  \right)\nonumber\\
    &{}& -\left(\frac{5}{12} g_Y^2 +\frac{9}{4}g_2^2 +8 g_3^2\right) - \frac{3}{2} \sum_{i} y_i^2 |V_{i\kappa}|^2\bigg]\,\,,
    \end{eqnarray}
    Herein, $V_{i\kappa}$ are the CKM mixing matrix elements.
    Finally, for the charged leptons, ($l=e, \mu, \tau$), we have
\begin{eqnarray}
 \beta_{y_l}&= &-f_y\, y_l+\frac{y_l}{16\pi^2} \bigg[3 \sum_{j}y_j^2 + 3 \sum_{\kappa} y_{\kappa}^2 + \sum_{l}y_l^2  \; \nonumber\\
 &{}& -\left(\frac{5}{4} g_Y^2 +\frac{9}{4}g_2^2 \right) + \frac{3}{2}y_l^2  -\frac{3}{2} \sum_{l} y_{l}^2 |U_{\nu_{l} l}|^2 \bigg],
\end{eqnarray}
where $U_{\nu_l l}$ are the PMNS mixing matrix entries.
 We note that, as anticipated, gravity is blind to internal symmetries (e.g., the distinction between flavors, the presence of SU(3) charge etc), with $f_y$ being the same for all fermions. Up-type and down-type quarks are distinguished from each other by the different hypercharge contribution, $-17/12 g_Y^2$ versus $-5/12 g_Y^2$. This provides the possibility to have a predictive fixed point for the top Yukawa coupling and thus top quark mass, which \cite{Eichhorn:2017ylw} found to agree with the measured top pole mass within the estimated systematic uncertainties. Moreover, \cite{Eichhorn:2018whv} even found the possibility to predict both top and bottom Yukawa coupling within 10 \% of their measured values.
 Quarks are distinguished from leptons by the presence of an SU(3) gauge coupling, which is absent for the leptons. From this, we can form the expectation that there is not enough structure for a fixed point which encodes the full flavor structure of the SM, i.e., requires the nine different Yukawa couplings to be different from each other and explains their values. Nevertheless, the ingredients for a mechanism that distinguishes Yukawa couplings partially is present \cite{Eichhorn:2018whv,Eichhorn:2019tcj}: At $g_{Y\, \ast}= (f_g/\beta_{g_Y}^{(1)})^{1/2}$, the contribution from the hypercharge coupling and gravity can be summarized into an anomalous scaling dimension that is different for up-type quarks and down-type quarks, such that (neglecting all other Yukawa couplings),
 \begin{eqnarray}
 y_{i\, \ast} &=& \frac{\sqrt{2}}{3}\sqrt{16 \pi^2 f_y +\frac{17}{12}g_{Y\,\ast}^2},\label{eq:ytFP}\\
 y_{\kappa\, \ast} &=&  \frac{\sqrt{2}}{3}\sqrt{16 \pi^2 f_y +\frac{5}{12}g_{Y\,\ast}^2}.\label{eq:ybFP}
 \end{eqnarray}
These fixed points are real-valued, unless $f_y$ becomes too negative. When they are real, they are IR attractive and the free fixed point is IR repulsive, realizing the analogue of Fig.~\ref{fig:upperbound} and constituting another concrete example for the predictive power of asymptotic safety. We note that, in the absence of the hypercharge coupling, i.e.,, when the hypercharge coupling is asymptotically free, $f_y>0$ appears to be required for a UV completion, otherwise there is only a free fixed point which is IR attractive. \\ 
Because the free fixed point is IR repulsive, a large enough $f_y$ allows to render all Yukawa couplings asymptotically free, while their IR values are free parameters. There is, however, the possibility of predictive power for a subset of the Yukawa couplings. The most recent study of the full SM, including the effect of mixing matrices, finds \cite{Eichhorn:2025sux}
\begin{equation}
f_y =-3.27 \cdot 10^{-4},
\end{equation}
as a prerequisite for maximal predictive power of asymptotic safety for the Standard Model \cite{Eichhorn:2025sux}.\footnote{We note that the sign of $f_y$ is actually positive, when similar studies are done in simplified settings, e.g., when not accounting for the CKM mixing matrix and working with a single generation of fermions only \cite{Eichhorn:2018whv}.} In that setting, the values of top and bottom quark are controlled by the IR attractive fixed points. The values of all other SM Yukawa couplings constitute free parameters, with the values inferred from experiment accommodated, but not explained. A large mass gap between lepton Yukawa couplings and neutrino Yukawa couplings is, however, generically present, see below.\\

There is also the possibility of a UV completion of the SM which does not lead to any emergent structures. Instead, the IR values of the Yukawa couplings are encoded in the choice of relevant parameters, with each Yukawa coupling introducing a relevant direction. This is realized if $f_y\gtrsim 3.7\cdot 10^{-3}$, because then the upper bound on the top quark Yukawa coupling lies above the measured value and all Yukawa couplings are asymptotically free. In this case, while a UV completion is achieved, no testable prediction in the Yukawa sector ensues and the flavor problem, the question of why neutrino masses are so small and the question why the quark mixing matrix is near-diagonal, while the lepton mixing matrix is not, remain unexplained. In contrast, \cite{Eichhorn:2025sux} has pointed out that these questions find partial resolutions in a more predictive UV completion, in which some of the Yukawa couplings are irrelevant, and IR fixed points can be approached and generate structures.\\
Thus, the quantitative determination of $f_y$ is not just crucial to determine whether or not the SM can be UV completed by gravity, but in particular to determine the predictive power of this setting.\\

The study of gravitational effects on the Yukawa sector was initiated in \cite{Zanusso:2009bs,Vacca:2010mj,Oda:2015sma,Eichhorn:2016esv}; the correct form of $f_y$ in the Einstein-Hilbert truncation was derived in \cite{Oda:2015sma,Eichhorn:2016esv} and extended to quadratic-gravity couplings   in \cite{Hamada:2017rvn,Eichhorn:2017eht,deBrito:2019gdd}. The robustness under the inclusion of various additional couplings was explored in \cite{deBrito:2025nog}. 
The result for $f_y$ does not have a fixed sign when gravitational fixed-point values are varied; instead, $f_y>0$ for negative enough $\lambda$ (where the threshold depends on the gauge parameter $\beta$) and a positive sign for larger $\lambda$. In the Einstein-Hilbert truncation,
\begin{equation}
f_y=-\frac{15 G}{16 \pi (1-2 \lambda)^2}+ \frac{7}{24\pi(3-4\lambda)}- \frac{7}{16\pi(3- 4 \lambda)^2} + \mathcal{O}(\beta),
\end{equation}
such that $f_y<0$ for $\lambda>-3.3$. Consistently with this result, in perturbative calculations, where the cosmological constant is absent, $f_y<0$ is found \cite{Rodigast:2009zj}. Because gravitational fixed-point values have uncertainties due to truncations, the sign of $f_y$ is not unequivocally established in asymptotic safety; in addition, it may change when gravitational fixed-point values change, e.g., under the impact of (dark) matter fields, see \cite{Eichhorn:2017ylw}.\footnote{There is also the possibility that $f_y<0$, such that $y_{\ast}=0$, but a Yukawa coupling is part of a superposition of couplings that is relevant \cite{deBrito:2025nog}. This can occur, if the stability matrix characterizing the fixed point has large off-diagonal contributions, as one would expect in the transition from a near-perturbative to a non-perturbative fixed point. This possibility was found recently in \cite{deBrito:2025nog}, for a simple Yukawa system, i.e., a single Dirac fermion coupled to a real scalar without gauge interactions or mixing effects.} While Ref.~\cite{Eichhorn:2017ylw} performed an assessment of systematic uncertainties by tracking typical variations of gravitational fixed-point values, Ref.~\cite{deBrito:2025nog} provided a more thorough  estimation of systematic uncertainties arising from truncations. The results indicate that systematic uncertainties are still large, at least in the regime studied in \cite{deBrito:2025nog}, where the critical exponents for matter couplings can be $\mathcal{O}(1)$.\newline

In summary, determining the gravitational fixed-point values with sufficient precision to know the fate of Yukawa couplings in asymptotic safety is one of the key challenges in asymptotic safety, but recent progress as in \cite{deBrito:2025nog} shows that this challenge can in principle be met. In the meantime, the various possibilities of how an asymptotically safe phenomenology looks like, can be mapped by treating $f_y$ as a free parameter, see Sec.~\ref{sec:principledparameterized}.\\

\noindent{\bf Neutrino Yukawa couplings and Dirac masses:}\\
Neutrino masses are not known experimentally, but are only constrained from above \cite{KATRIN:2024cdt,Jiang:2024viw}. Their differences are known from neutrino oscillations \cite{ParticleDataGroup:2024cfk}. Nevertheless, it is already established that the gap between the heaviest SM neutrino and the next-lightest SM fermion is at least five orders of magnitude. Accordingly, if neutrinos only have Dirac masses, the corresponding Yukawa couplings are $\mathcal{O}(10^{-11})$ or lower, while the top quark Yukawa coupling is $\mathcal{O}(1)$. Thus, heavy new degrees of freedom are typically invoked to render neutrinos light through the seesaw mechanisms, where neutrinos have both a Dirac and a Majorana mass term. The properties of Majorana mass terms in asymptotic safety have first been studied in \cite{DeBrito:2019rrh}.  

\begin{figure*}[!t]
\includegraphics[width=\linewidth]{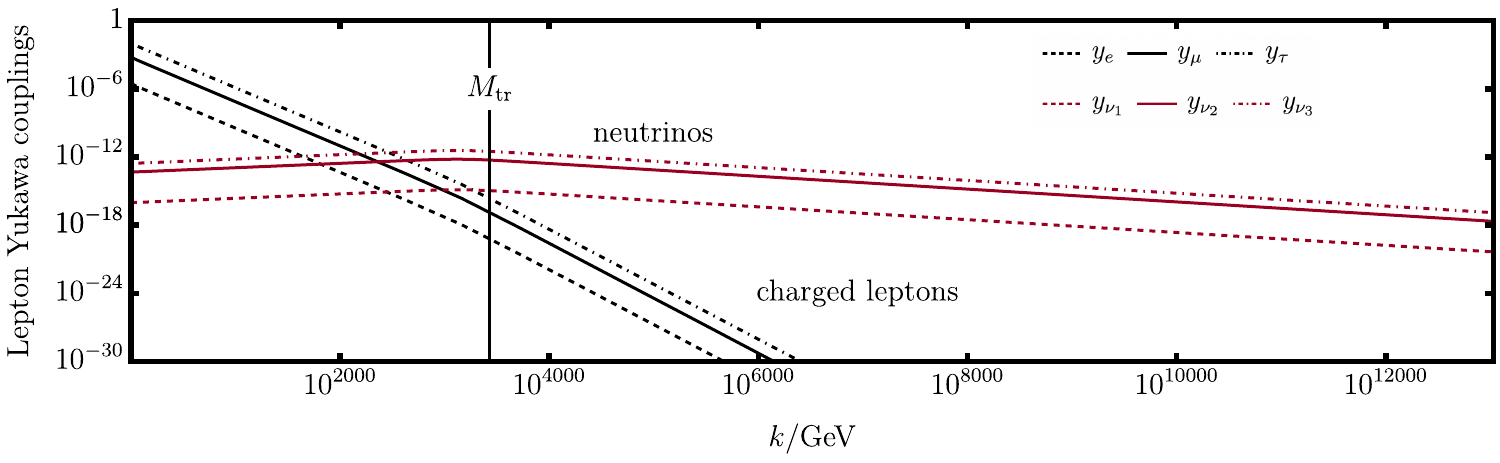}
\caption{\label{fig:numassgap} Due to the tiny critical exponents in Eq.~\eqref{eq:neutrinotheta}, neutrino Yukawa couplings change very little, even over huge ranges of scales. We show the results from \cite{Eichhorn:2025sux}, where all lepton Yukawa couplings are asymptotically free in the deep UV, in a regime dominated by a constant bottom Yukawa coupling; at $M_{\rm tr}$, the top-Yukawa coupling starts to dominate and the critical exponents for the neutrinos change sign. At either sign, because of the tiny absolute value, there is very little change. Thus, even though $y_{\nu}\gg y_{\ell}$ in the deep UV, there is nevertheless a mass gap between neutrinos and charged leptons in the IR.}
\end{figure*}

In the asymptotically safe SM, there may, however, be a new mechanism to keep neutrinos light dynamically \cite{Held:2019vmi,Kowalska:2022ypk,Eichhorn:2022vgp}. To see this mechanism, we consider the beta function for neutrino Yukawa couplings $y_{\nu_l}$, after having added three right-handed neutrinos to the SM (uncharged under any of the gauge groups):
    \begin{eqnarray} 
    \beta_{y_{\nu_l}} &=&-f_y\,y_{\nu_l}+\frac{y_{\nu_l}}{16\pi^2} \bigg[\frac{3}{2}y_{\nu_l}^2 +3 \sum_{j}y_j^2 + 3 \sum_{\kappa} y_{\kappa}^2 + \sum_{l}y_l^2 \nonumber\\
    &{}&+ \sum_{k} y_{\nu_k}^2 -\left(\frac{3}{4} g_Y^2 +\frac{9}{4}g_2^2     \right) -\frac{3}{2} \sum_{l} y_{l}^2 |U_{\nu_{l} l}|^2\bigg]\,\,,
\end{eqnarray}
where $U$ denotes the PMNS mixing matrix. Inspired by the above discussion on the remainder of the Yukawa sector, we assume that all Yukawa couplings, with the exception of the top and bottom Yukawa coupling, are asymptotically free, as is $g_2$. Then, the critical exponent for the neutrino Yukawa couplings at $y_{\nu_l\, \ast}=0$ becomes
\begin{equation}
\theta_{y_{\nu_l}}= f_y -\frac{1}{16\pi^2}\left(3 y_{t\,\ast}^2+3 y_{b\,\ast}^2 - \frac{3}{4}g_{Y\,\ast}^2 \right),\label{eq:neutrinoyukawaFPregime}
\end{equation}
which does not appear to have any remarkable features. However, for a fixed point at which $g_{Y\,\ast}>0$ and $y_{t\, \ast}>0$ or $y_{b\, \ast}>0$, $f_y$ is linked to these fixed-point values.
We plug in $y_{t\,\ast} (y_{b\,\ast})$ according to Eq.~\eqref{eq:ytFP} to obtain \cite{Eichhorn:2022vgp}
\begin{eqnarray}
\theta_{y_{\nu_l}}&=& f_y -\frac{1}{16\pi^2}\left(\frac{2}{3}\left(16 \pi^2 f_y+ \frac{17 (5)}{12}g_{Y\,\ast}^2 \right) - \frac{3}{4}g_{Y\, \ast}^2\right)\nonumber\\
&=& -\frac{1}{3}f_y- \frac{7(17)}{576\pi^2}g_{Y\, \ast}^2 \approx (-4)(+5) \cdot 10^{-4},\label{eq:neutrinotheta}
\end{eqnarray}
where the negative sign holds for $y_{t\,\ast}\neq 0$ and the positive sign for $y_{b\,\ast}\neq 0$.
Such a tiny critical exponent means that the rate of change of neutrino Yukawa couplings is entirely negligible, compared to the rate of change of, e.g., charged lepton Yukawa couplings, for which Eq.~\eqref{eq:neutrinoyukawaFPregime} contains a factor $-5/4 g_Y^2$, instead of $-3/4 g_Y^2$, which results in $\theta_l \approx 5 \cdot 10^{-3}$ \cite{Eichhorn:2025sux}. Thus, over a range of scales over which the lepton Yukawa coupling changes by a factor of $10^2$, the neutrino Yukawa only changes by a factor $1.25$; and while the neutrino Yukawa couplings changes by a factor of 10, the lepton Yukawa coupling has changed by a factor of $10^{10}$. These are clearly the ingredients necessary to generate a mass gap, cf.~Fig.~\ref{fig:numassgap}.\\

\noindent{\bf Mixing matrices:}\\
The beta functions for the mixing matrices receive no gravitational contributions, because mixing is a phenomenon entirely connected to the internal symmetries, which gravity is blind to. The beta functions for the CKM mixing matrix in the quark sector and the PMNS mixing matrix in the lepton sector are identical under an exchange of quark with lepton Yukawa couplings. Both admit a fully attractive IR fixed point in a no-mixing configuration, which is the diagonal configuration \cite{Pendleton:1980as,Alkofer:2020vtb,Dolan:2026gvw}. Thus, generic initial conditions for the RG flow in the UV result in no or very little mixing, if the RG flow is non-vanishing over a large enough range of scales. The speed of the flow is set by the Yukawa couplings and turns out to be very low for the SM values of the couplings, such that no appreciable running occurs between the Planck scale and the electroweak scale.\\
In \cite{Eichhorn:2025sux}, it was discovered that appreciable running occurs in the CKM matrix, if the transplanckian regime features a constant Yukawa coupling. Specifically, \cite{Eichhorn:2025sux} discovered a fixed-point cascade, starting from a bottom-Yukawa-dominated regime in the very deep UV, followed by a top-Yukawa-dominated regime and then the subplanckian regime without gravity. The fixed-point cascade starts from a UV fixed point for the full SM, achieved by treating $f_g$, $f_y$ and $f_{\lambda}$ as free parameters and fixing them such that $g_Y$, $y_t$ and $\lambda_H$ are predicted in accordance with experimental values. At the UV fixed point, all 8 mixing matrix parameters (four for the CKM and four for the PMNS, due to unitarity of these matrices) introduce relevant directions. Towards the IR, the CKM is driven towards a near-diagonal configuration in agreement with experimental values. The fixed-point cascade leaves an additional imprint, in that the mixing between first and third generation quarks is required to be significantly smaller than the mixing between second and third generation quarks \cite{Eichhorn:2025sux}. This also matches observations \cite{ParticleDataGroup:2024cfk}.\\
Along the same fixed-point cascade, the PMNS matrix can avoid a similar fate to the CKM matrix, if neutrino masses are far enough from degeneracy. Otherwise, as previously realized in a context without gravity, see \cite{Lindner:2005as} and references therein, the PMNS matrix elements start to run appreciably, once the differences between the neutrino Yukawa couplings become too small. In \cite{Eichhorn:2025sux}, the quantitative bound for this was identified in the $\mathcal{O}(1-10)\rm eV$ region, close to the current upper bound on the electron neutrino mass from KATRIN \cite{KATRIN:2024cdt}. As discussed above, asymptotic safety with a constant top- (or bottom) Yukawa coupling contains a mechanism to drive the neutrino masses to small values, automatically avoiding the conditions that would result in a no-mixing, diagonal PMNS configuration.\\

There is also the possibility that, if $f_g$ and $f_y$ are both sufficiently large, all gauge and Yukawa couplings are asymptotically free. Then, no appreciable running of the CKM matrix occurs, because the range of scales over which Yukawa couplings are appreciably different from zero is too small to generate much running in the CKM sector. Such a  completely asymptotically free fixed point is therefore a consistent possibility for a UV completion \cite{Pastor-Gutierrez:2022nki}, albeit, just like for the Yukawa sector, it does not provide explanations for the observed structures of couplings in the SM.\\

\noindent{\bf Higgs quartic coupling:}\\
The beta function for the Higgs quartic coupling reads, up to one-loop order,
\begin{eqnarray}
 \beta_{\lambda_H} &=&-f_{\lambda}\lambda_H+\frac{1}{16 \pi^2} \Big[24 \lambda_H^2 - 3 \lambda_H \left(3 g_2^2+g_Y^2\right) \;\;\\ 
    &+& \frac{3}{8}\left(2 g_2^2 g_Y^2+3 g_2^4+g_Y^4\right)+ 4 \lambda_H \left(3\!\! \sum_{\rm quarks} \!\!y_i^2 + \sum_{\rm leptons}\!\! y_l^2 \right) \nonumber \\ 
    & -& 2 \left(3 \sum_{\rm quarks} y_i^4 + \sum_{\rm leptons} y_l^4 \right)\Big], 
\end{eqnarray}
where the sum over leptons includes charged leptons and neutrinos (even though the neutrino contribution is in practice negligible).
We first set all SM couplings except $\lambda_H$ to zero and observe that for $f_{\lambda}<0$, for which there is compelling evidence \cite{Narain:2009fy,Oda:2015sma,Eichhorn:2017als,Pawlowski:2018ixd,deBrito:2019gdd,Wetterich:2019rsn,Wetterich:2019zdo,Eichhorn:2020sbo,Ohta:2021bkc,Eichhorn:2021tsx}, there is only an IR attractive fixed point at
\begin{equation}
\lambda_{H\, \ast}=0,\,\quad \theta_{\lambda_H}= f_{\lambda}<0.
\end{equation}
This fixes the quartic coupling for the Higgs to zero at the Planck scale.\footnote{In practice, there is a small deviation from zero, because, as the gauge and Yukawa couplings increase from asymptotically free fixed points towards their Planck-scale values, they pull along the Higgs quartic coupling. However, his effect is $\sim \frac{1}{16\pi^2}g_i^4$ and $\sim \frac{1}{16\pi^2}y_i^4$ and thus extremely small.} Such a scenario was proposed in \cite{Shaposhnikov:2009pv}, assuming that the remaining SM couplings become asymptotically free under the impact of quantum gravity. $\lambda_H(M_{\rm Pl})=0$ translates into a Higgs mass of around $126\, \rm GeV$, where the precise value is highly sensitive to the top quark mass \cite{Bezrukov:2014ina}: within a 3 $\sigma$ region around the central experimental value, $\lambda_{H}(M_{Pl})$ can be in agreement with the experimental value of $125.5\, \rm GeV$ \cite{ATLAS:2015yey,ATLAS:2023oaq}, but masses several GeV above the experimental value result from the current central value for the top quark mass. Thus, a more accurate determination of the top quark mass is needed, before it can be decided whether or not the prediction in \cite{Shaposhnikov:2009pv} -- predating the discovery of the Higgs boson \cite{CMS:2012qbp,ATLAS:2012yve} by three years -- matches experimental data or not.

It is important to stress that this prediction is very robust in the sense that it does not depend on the magnitude of $f_{\lambda}$, just on its sign, which is rather robustly established. This is an important distinction between the quantum-gravity prediction for the Higgs quartic coupling and all the other SM couplings, for which predictions only match observations for very specific values of $f_g$ and $f_y$.\\

As we have reviewed above, there is also the possibility that some of the gauge and Yukawa couplings of the SM are not asymptotically free, but safe. In that case, the fixed point of the Higgs quartic coupling gets shifted to a non-zero value, which depends on the non-zero couplings and $f_{\lambda}$. In \cite{Eichhorn:2025sux}, it has been shown that this can also result in a Higgs mass compatible with experiment.\\

Let us turn to the second parameter in the Higgs sector, namely the mass parameter. The mass parameter sets the electroweak scale, because it determines the vacuum expectation value of the Higgs. There is an associated hierarchy problem, because the mass parameter has canonical dimension two. This hierarchy problem is expected to persist in asymptotic safety, because the Higgs mass parameter remains relevant. However, because quantum-gravity fluctuations shift all terms in the Higgs scalar potential uniformly towards irrelevance, by the same amount $f_{\lambda}$, a sufficiently non-perturbative asymptotically safe fixed point can even render the Higgs mass parameter irrelevant \cite{Wetterich:2016uxm}. 
In that case, the electroweak scale would be completely fixed and depend on $f_{\lambda}$ as well as (through the fixed-point values for $g_Y$ and/or $y_t$) on $f_g$ and $f_y$. A detailed quantitative study of the resulting scale has not yet been performed, because this scenario requires gravitational fluctuations to be strong, and likely not compatible with the near-perturbative nature of the fixed point.\\

\noindent{\bf Summing up: Asymptotically safe Standard Model}\\
In summary, there is good evidence that asymptotically safe quantum gravity can render the Standard Model UV complete and possibly even increase its predictive power \cite{Shaposhnikov:2009pv,Harst:2011zx,Eichhorn:2017ylw,Eichhorn:2017lry,Eichhorn:2018whv,Eichhorn:2025sux}. 

To achieve this, gravity needs to antiscreen gauge couplings. The required sign of gravitational contributions to gauge couplings is well established \cite{Daum:2009dn,Daum:2010bc,Harst:2011zx,Folkerts:2011jz,Christiansen:2017gtg,Christiansen:2017cxa,Eichhorn:2017lry,deBrito:2019gdd,Eichhorn:2021qet,deBrito:2022vbr,Riabokon:2025ozw}. For an appropriate magnitude of the gravitational contribution, a prediction of the hypercharge gauge coupling results. Otherwise, there is an upper bound on the gauge coupling \cite{Eichhorn:2017lry}, required the magnitude of the gravitational contribution to exceed a minimal strength, such that the upper bound on the gauge coupling does not lie below the experimental value of the hypercharge gauge coupling.\\

In addition, there is robust evidence that gravity screens the Higgs quartic coupling \cite{Narain:2009fy,Oda:2015sma,Eichhorn:2017als,Pawlowski:2018ixd,deBrito:2019gdd,Wetterich:2019rsn,Wetterich:2019zdo,Eichhorn:2020sbo,Ohta:2021bkc,Eichhorn:2021tsx}, leading to a prediction of the Higgs quartic coupling which sets the ratio of Higgs mass to electroweak scale \cite{Shaposhnikov:2009pv}. This prediction is robust in the sense that it is largely independent of the magnitude of the gravitational contribution and only depends on its sign, which is rather robustly established.\\

For the Yukawa couplings, there are several possible scenarios \cite{Zanusso:2009bs,Vacca:2010mj,Oda:2015sma,Eichhorn:2016esv,Hamada:2017rvn,Eichhorn:2017eht,deBrito:2019gdd,Pastor-Gutierrez:2022nki,deBrito:2025nog}: \\
First, if all gauge couplings are asymptotically free (i.e., the upper bound on the hypercharge coupling exceeds the experimental value), then the gravitational contribution should be antiscreening to render the Yukawa sector free of poles. Similarly to the hypercharge coupling, an upper bound on the top Yukawa coupling ensues, which bounds the gravitational contribution from below. For an appropriate magnitude of the gravitational contribution, a prediction of the top quark mass ensues \cite{Eichhorn:2017ylw}.\\
Second, if the gravitational contributions to the hypercharge coupling and the Yukawa couplings are of the appropriate magnitude, a prediction for the hypercharge coupling and top quark mass ensues, together with several emergent structures: first, neutrino Yukawa couplings generically develop a gap towards the other Yukawa couplings, generically resulting in Dirac neutrino masses several orders of magnitude below the lepton masses, without any need for finetuning \cite{Held:2019vmi,Kowalska:2022ypk,Eichhorn:2022vgp}. Second, the CKM mixing matrix is generically driven towards a diagonal (no-mixing) configuration, with mixing between the first and third generation being significantly smaller than mixing between the second and third generation \cite{Eichhorn:2025sux}. These structures match observations, making this scenario the most predictive one. The required values of $f_y$ and $f_g$ can be interpreted as a prediction for the expected values of gravitational contributions.\\
We highlight that this predictive power carries over to the case of effective asymptotic safety, see Sec.~\ref{sec:eff_AS}, such that the predictions discussed above remain robust and do not rely on asymptotic safety being realized up to arbitrarily high scales.\footnote{Those predictions associated to larger values of critical exponents hold with greater accuracy in this case, see \cite{Held:2020kze}.}

\subsubsection{What does asymptotic safety imply for the Standard Model effective field theory (SMEFT)?}\label{sec:SMEFT}
\emph{...where we explain what can be learned from connecting asymptotically safe gravity to the Standard Model Effective Field Theory (SMEFT). First, it constitutes a framework in which one can organize all gravitationally induced non-zero, higher-order matter interactions. Second, consistency bounds on effective-field-theories, such as causality bounds, are formulated for the Wilson coefficients in the SMEFT. Investigating whether these bounds are respected in asymptotic safety gives us insights into the physical viability of asymptotic safety. Third, the SMEFT is a general framework to parameterize effects of new physics. Experimental results, e.g., from the LHC, are used to place constraints on SMEFT coefficients. Investigating whether asymptotic safety obeys the expected Planck-scale suppression of all gravitationally induced SMEFT couplings provides us with a prediction for  measurements of these coefficients.}\\

The Standard Model effective field theory (SMEFT) \cite{Buchmuller:1985jz,Grzadkowski:2010es}, reviewed in \cite{Ellis:2021kzk,Falkowski:2023hsg}, is based on an effective Lagrangian, containing the perturbatively renormalizable Standard Model Lagrangian as the leading term, followed by a series of higher-order operators, compatible with the gauge symmetries of the SM
\begin{equation}
\mathcal{L}_{\rm SMEFT} = \mathcal{L}_{\rm SM} + \sum_{D>4}^{\infty}\sum_i\frac{c_i}{\Lambda_{\rm UV}^{D-4}}\mathcal{O}_{i,D}.
\end{equation}
Here, $D$ is the dimension of the interaction $\mathcal{O}_{i,D}$, which is an interaction term built from the SM fields. The sum over $i$ extends over all interactions at a given mass-dimension; $c_i$ are the dimensionless Wilson coefficients associated to an interaction and $\Lambda_{\rm UV}$ is the scale of new physics. 

In practice, the SMEFT is used as a framework to constrain new physics beyond the SM at the LHC by placing bounds on the $c_i$ in a (largely) model-agnostic fashion \cite{Isidori:2023pyp}.\footnote{It is typically assumed that the coefficients obey naturalness such that, e.g., coefficients of $D=6$ and $D=8$ operators are constrained from LHC data, but higher-order operators are neglected.}

One particular type of new physics which generates nonzero $c_i$ is quantum gravity.\footnote{There is a conjecture that quantum gravity generates both those $c_i$ compatible with the global symmetries of the SM as well as those associated to the breaking of global symmetries \cite{Kallosh:1995hi}. Within asymptotic safety, there are no indications for the breaking of global symmetries by quantum gravity fluctuations, see the discussions in \cite{Eichhorn:2017eht,Eichhorn:2020sbo,Laporte:2021kyp,Eichhorn:2022gku}. Therefore, a subset of the $c_i$ in the SMEFT, namely those that break global symmetries (such as $B-L$) are not present in asymptotic safety. For practical purposes, the two possibilities may not be easy to distinguish, because in string-theoretic settings, in which the no-global-symmetries conjecture appears to be realized, the $c_i$ are typically suppressed exponentially, $c_i \sim \exp(-M_{\rm Pl}^2/\Lambda_{\rm UV}^2)$ if the cutoff of the EFT, $\Lambda_{\rm UV}$, lies below the Planck scale $M_{\rm Pl}$, \cite{Fichet:2019ugl,Daus:2020vtf}; and even in the absence of such an exponential suppression of the dimensionless Wilson coefficient, any quantum-gravity effect comes with the dimensional suppression ${\Lambda_{\rm UV}^{D-4}}$ of the higher-order operator in question.} 
The reason is that even a seemingly non-interacting theory of matter fields, with only kinetic terms present, contains interactions with gravity. Integrating out gravity results in those terms in the SMEFT that can be understood as $(\sum_{\rm fields}{\rm kinetic\, term})^2$. This in particular happens in asymptotically safe gravity, where the corresponding couplings are necessarily non-zero at a fixed point at which $G_{\ast}\neq 0$ \cite{Eichhorn:2012va,Christiansen:2017gtg,Laporte:2021kyp,Eichhorn:2021qet,deBrito:2021pyi,deBrito:2023myf,Eichhorn:2024wba,Knorr:2024yiu,deBrito:2025nog}. For fermions, which are the only matter field that requires a gravitational connection\footnote{For the gauge field, the Christoffel connection drops out of the field strength due to antisymmetry.}, the spin connection, in its kinetic term, the resulting terms include four-fermion operators, specifically $(\bar{\psi}\gamma_{\mu}\gamma_5\psi)^2$ \cite{Eichhorn:2011pc, Meibohm:2016mkp,Eichhorn:2017eht,deBrito:2020dta,deBrito:2023kow}. For the other fields, the resulting terms are all derivative interactions, e.g., $(\partial_{\mu}\phi\partial^{\mu}\phi)^2$ at lowest order in the SMEFT for scalars \cite{Eichhorn:2012va}.
These interactions are all irrelevant; thus, their IR values are predicted (and in general non-zero, because the fixed-point values are non-zero).
The associated scale is clearly $\Lambda_{\rm UV} \approx M_{\rm Planck}$, such that experimental tests of these predictions  are not imminent, unless one of the interactions is relevant below the Planck scale \cite{Brenner:2024bps}. \\

In writing the SMEFT, it is assumed that CPT symmetry and Lorentz symmetry remain intact. This, in fact, constrains (trans)Planckian physics, because there are several interactions of matter fields of dimension lower than four, which break one or both of these symmetries. These interactions can all be consistently set to zero at an asymptotically safe fixed point, but some do, in fact, correspond to relevant directions \cite{Eichhorn:2019ybe,Eichhorn:2025ilu}. Thus, they do not constitute emergent symmetries. If violated even by a very small amount by quantum gravity, they would result in the Standard-Model-extension \cite{Colladay:1996iz,Colladay:1998fq}, which includes such interactions, as the correct framework for IR physics. For many of the couplings in the Standard-Model-extension, experimental bounds are very strong \cite{Kostelecky:2008ts} and only compatible with quantum-gravity violations of Lorentz and/or CPT symmetry under the assumption of extreme degrees of fine-tuning \cite{Eichhorn:2025ilu}. For asymptotic safety, this simply means that the flow should be constrained to a hypersurface in which CPT and Lorentz invariance hold. This can be achieved self-consistently, because any symmetry that is imposed on the RG flow at any given scale is respected at all other scales; the RG flow does not generate explicit breaking of symmetries.\\

There are even interactions within the SMEFT which are already constrained to scales not far below the Planck scale, e.g., the four-fermions interactions mediating proton decay. For such interactions, the results from asymptotic safety \cite{Eichhorn:2023jyr} are closer to reaching phenomenological relevance, but indicate that, due to the associated violation of the global $B-L$ symmetry, proton-decay mediating interactions vanish at an asymptotically safe fixed point and are even more irrelevant than expected canonically. Therefore, if asymptotic safety is realized in the UV, bounds on the proton lifetime from future measurements \cite{JUNO:2015zny,DUNE:2016hlj,Hyper-Kamiokande:2018ofw} can be expected to continue increasing.

\begin{figure}[!t]
\includegraphics[width=\linewidth]{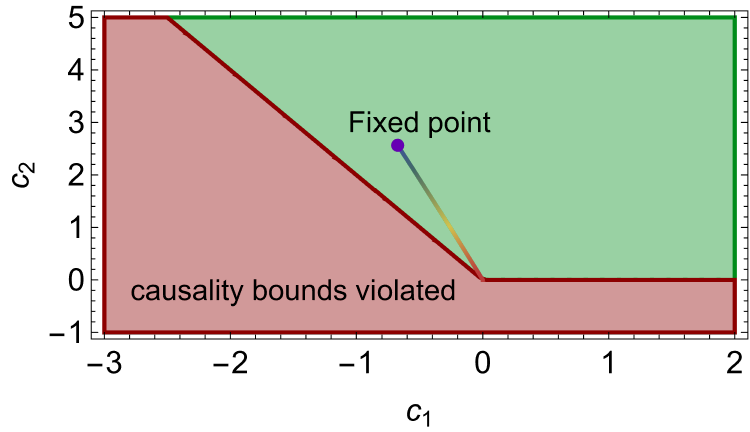}
\caption{\label{fig:causality_bounds} We show the $c_1, c_2$ plane with the asymptotically safe fixed point lying at non-zero values due to quantum-gravity fluctuations. Results are calculated including up to dimension-twelve-interactions in \cite{Eichhorn:2024wba}. The resulting RG flow lies entirely within the wedge in which causality bounds are respected (green region).}
\end{figure}

With few exceptions such as proton decay, the calculation of the coefficients $c_i$ is therefore not of interest for a direction connection to experimental efforts, but rather for a connection to other theoretical work, in which Wilson coefficients are constrained through various consistency conditions, such as positivity bounds \cite{Adams:2006sv} as well as causality bounds \cite{CarrilloGonzalez:2023cbf}, reviewed in \cite{deRham:2022hpx}.
For concreteness, we review here an example from the interplay of gravity with photons. It has been known for some time that photon self-interactions cannot be set to zero in the UV regime of asymptotic safety \cite{Christiansen:2017gtg,Eichhorn:2021qet,Eichhorn:2024wba,Knorr:2024yiu}. 
Specifically, in 
	\begin{align}
		\label{eq:gammakone}
		\mathcal{L}^{\mathrm{photons}} =& \frac{1}{4} \int d^4x\sqrt{g} 	\mathcal{F}_2\,   \nonumber\\
		& + \int d^4x\sqrt{g}  \left( \frac{c_1}{k^4}\, \left(	\mathcal{F}_2\right)^2 +\frac{c_2}{k^4}	\mathcal{F}_4\right) \nonumber\\
		& + \int d^4x\sqrt{g}  \left( \frac{d_1}{k^8}\, \left(	\mathcal{F}_2\right)^3 +\frac{d_2}{k^4} \mathcal{F}_2\,\mathcal{F}_4\right)\,, 
	\end{align}
	with
	\bea
		\mathcal{F}_2&=&g^{\mu\nu}g^{\kappa\lambda}F_{\mu\kappa}F_{\nu\lambda}\,,\\
				\mathcal{F}_4&=&g^{\mu\nu}g^{\kappa\lambda}g^{\rho\sigma}g^{\zeta \tau}F_{\mu\kappa}F_{\lambda \rho}F_{\sigma \zeta}F_{\tau \nu},
	\eea
we have the beta functions (written here for $\Lambda =0$, see \cite{Eichhorn:2024wba} for the full result)
\bea
\label{eq:betaschem}
\beta_{c_1}&=& 4c_1 -\frac{53}{18 \pi}G\, c_1 - \frac{23}{36 \pi} G \,c_2 + \frac{5}{2}G^2 +\mathcal{O}(c_{1,2}^2),\\
\beta_{c_2}&=& 4c_2 + \frac{19}{18 \pi}G\, c_2 + \frac{40}{9 \pi}  G\,c_1- 10 G^2+\mathcal{O}(c_{1,2}^2),
\eea
and similar results, with terms $\sim G^2$, for the higher-order couplings.
The requirement that photon propagation at low energies is subliminal\footnote{The same bounds arise as positivity bounds; however, positivity bounds do not apply in the presence of gravity as a mediator \cite{Alberte:2020jsk}, unless one makes additional assumptions that cancel a problematic, gravity-induced pole in the scattering amplitude \cite{Tokuda:2020mlf}.}, restricts
\be
4c_1> -3c_2,\quad \quad \frac{4c_1+3c_2}{|4c_1+c_2|}>1.\label{eq:poscons}
\ee
Starting from the UV fixed point, one can integrate towards the IR. Causality bounds must be respected at $k \rightarrow 0$, but to be maximally conservative, one would require them to hold at all values of $k$, assuming that there is a qualitative connection between the behavior of the theory at finite $k$ to a corresponding finite value of physical scales. Fig.~\ref{fig:causality_bounds} shows the result, highlighting two points: first, connecting asymptotic safety to the SMEFT is highly relevant, even if the predicted values for the higher-order couplings are Planck-scale suppressed. Because causality bounds (and other consistency requirements on the effective field theory) typically only constrain the dimensionless Wilson coefficients, and not the absolute scale of the coupling, they provide non-trivial checks of the physical consistency of asymptotically safe gravity. Second, the causality bounds from \cite{CarrilloGonzalez:2023cbf} are not violated in \cite{Eichhorn:2024wba}.\footnote{The truncation used in \cite{Eichhorn:2024wba}, including interactions up to dimension twelve, is not large enough to entirely prevent a gauge dependence of this result, so that violations may be found in some gauges \cite{Knorr:2024yiu}.
Recently, important progress was made in \cite{Knorr:2026vax} on understanding the logarithmic scaling that dimensionless combinations of higher-order couplings can have below the Planck scale, and which was treated at a more heuristic level in \cite{Knorr:2024yiu,Eichhorn:2024wba} to compare to causality bounds.}

\subsubsection{What does asymptotic safety imply for dark matter?}\label{sec:dm}
\emph{...where we explain that asymptotic safety has predictive power for dark-matter models: it requires new relations between couplings, can force some couplings to vanish and may thereby even produce new stabilization mechanisms that generate dark-matter candidates in models which do not have a dark-matter candidate without asymptotic safety.}\\

Asymptotic safety has two important implications for dark matter: first, the predictive power can constrain, or even rule out, dark-matter models, in the sense that a given dark-matter model cannot become asymptotically safe. This has been studied in \cite{Eichhorn:2017als,Reichert:2019car,Eichhorn:2020kca,Kowalska:2020zve,Hamada:2020vnf,Eichhorn:2021tsx,deBrito:2021akp,Boos:2022pyq,deBrito:2023ydd,Assant:2025gto}. Second, asymptotic safety can stabilize particles by ruling out their decay channels and thus generate a viable dark-matter candidate in a model which does not have a stable dark-matter candidate without asymptotic safety. This has first been proposed in \cite{Chikkaballi:2025pnw} in the context of a grand unified theory, but is in fact a mechanism which generalizes to other settings.

As an example of how the predictive power of asymptotic safety eliminates dark-matter candidates or reduces their parameter space, we consider the arguably simplest model of dark matter, namely a  particular weakly interaction massive particle (WIMP): we consider a massive scalar field $\chi$, coupled to the Higgs field $H$ of the SM through a portal with a dimensionless coupling $\lambda_H$ \cite{Silveira:1985rk,McDonald:1993ex,Burgess:2000yq} , such that the Euclidean version of the Lagrangian reads
\begin{eqnarray}
\mathcal{L}&=& \int d^4x\sqrt{g}\Bigl(\frac{1}{2}g^{\mu\nu}\partial_{\mu}\chi\partial_{\nu}\chi + \frac{\bar{m}_{\chi}^2}{2}\chi^2 + \lambda_\chi \chi^4+ \xi_{\chi} R \chi^2\nonumber\\
&{}& \quad + \lambda_H\, H^{\dagger}H\, \chi^2 \Bigr)+ \mathcal{L}_{\rm SM}.
\end{eqnarray}
 Through the portal, $\chi$ can be produced thermally in the early universe and can be searched for at the LHC  as well as dedicated dark-matter experiments, see  \cite{Athron:2018ipf} for a summary of constraints and \cite{Arcadi:2019lka} for a review. In this way, most of the model's parameter space has been excluded \cite{Athron:2018ipf}. This result is in fact compatible with the expectation from asymptotic safety, where, under the impact of gravity, 
\begin{equation}
\lambda_{H\, \ast}=0,\quad \theta_{\lambda_H}<0,
\end{equation}
i.e., $\lambda_H$ vanishes at the fixed point and corresponds to an irrelevant direction \cite{Eichhorn:2017als}. 
In fact, because gravity is ``blind'' to internal symmetries, the gravitational contribution to the portal coupling is the same as to a Higgs quartic coupling, which has first been observed to be irrelevant in \cite{Narain:2009fy}, i.e.,
\begin{equation}
\beta_{\lambda_{H}}= -f_{\lambda}\lambda_H+ \mathcal{O}(G^2 m_{\chi}^2 m_H^2)+\mathcal{O}(G^2 \xi_{\chi}^2 \xi_H^2)+ \beta^{\rm (matter)}.
\end{equation}
At the fixed point, $\xi_{\chi}= 0=m_{\chi}$, such that the term proportional to $f_{\lambda}$ is the main one.
In the IR, $\lambda_H \approx 0$, such that $\chi$ cannot become a thermal relic.\footnote{$\lambda_H$ does not vanish exactly in the IR, because of the terms $\sim - G^2\, m_{\chi}^2\, m_{H}^2$ in $\beta_{\lambda_H}$. Accordingly, as the mass of the Higgs as well as the dark scalar increase from their fixed-point values at $m_{\chi\, \ast}=0$, $m_{H\, \ast}=0$, at which they correspond to relevant directions, $\lambda_H$ grows. Quantitatively, $\lambda_H$ stays far below the $\mathcal{O}(0.1)$ values required for $\chi$ to become a thermal relic. Instead, $\chi$ might be a feebly-interacting dark matter candidate \cite{Pauly:2021ozy}.}

This no-go-result can be circumvented by adding additional fields to the dark sector, which either i) give rise to a richer fixed-point structure with fixed points at which $\lambda_{H\, \ast}\neq 0$ \cite{Eichhorn:2020kca,Eichhorn:2020sbo,Eichhorn:2021tsx} or ii) contain new relevant interactions such that $\lambda_H$ can be regenerated below the Planck scale, even if it vanishes above \cite{Reichert:2019car,Hamada:2020vnf}. However, such models of the dark sector are less compelling, precisely because they contain extra degrees of freedom and free parameters which only serve the purpose of circumventing the result in the more minimal model.\\

In a similar spirit, further dark-matter models based on dimension-4-interactions have been studied in asymptotic safety  and have been either ruled out \cite{deBrito:2023ydd} or found to be strongly constrained by the requirement of asymptotic safety \cite{Kowalska:2020zve,Boos:2022pyq}, because couplings of dimension-4-operators may be irrelevant at the fixed point.\footnote{In a similar spirit, asymptotic safety without gravity (which one can think of as an additional requirement on a theory that remains an effective description of nature, valid only below the Planck scale) has been used to constrain dark-matter phenomenology, see, e.g., \cite{Wang:2015sxe}.} Each such irrelevant coupling reduces the dimensionality of the parameter space by one compared to an effective-field-theory treatment of the same model.

More recently, dark-matter models based on dimension-5-operators as portals have risen in popularity, chief among them axion-like particles \cite{OHare:2024nmr}, which are described by a pseudoscalar field $a$ that couples to the electromagnetic field strength $F_{\mu\nu}$ through a coupling 
\begin{equation}
\mathcal{L}_{\rm ALP}= \bar{g}_a\, a\ F_{\mu\nu}F_{\kappa\lambda}\epsilon^{\mu\nu\kappa \lambda}.\label{eq:ALP}
\end{equation}
 The beta function for the corresponding dimensionless coupling $g_a= \bar{g}_a\, k$ starts with a dimensional term that reflects the canonical irrelevance of the coupling, 
\begin{equation}
\beta_{g_a} = g_a + f_{g_a}\, g_a+\mathcal{O}(g_a^2),
\end{equation}
where $f_{g_a}\sim G$ is the gravitational contribution. There is a fixed point, $g_{a\, \ast}=0$, at which the critical exponent is $\theta_{g_a}= -1+f_{g_a}$. Accordingly, $f_{g_a}>1$ must be realized in order for $g_a$ to be nonzero in the IR. In \cite{deBrito:2021akp}, it was found that while the sign of $f_{g_a}$ is indeed positive, rather large values of $G$ are required to achieve $f_{g_a}>1$, such that there does not appear to be a viable way to accommodate the dimension-5-operator in Eq.~\eqref{eq:ALP} in asymptotic safety as a fundamental interaction. This does, of course, not preclude that Eq.~\eqref{eq:ALP} only arises below the Planck scale, when some non-gravitational degrees of freedom are integrated out, as is, indeed the case in models with a QCD axion.

A similar result was found for the dimension-five-interaction $\phi\, F_{\mu\nu}F_{\kappa\lambda}g^{\mu\kappa}g^{\nu\lambda}$ of the field strength with a real scalar $\phi$  \cite{Assant:2025gto}. Thus, from an asymptotically safe perspective, one may expected that searches for ALPs as well as scalars with dimension-5-interactions will only make a discovery, if the dark sector is non-minimal, and contains extra fields that provide ways to circumvent the no-go-results from the minimal models.\\

In summary, while huge parts of the dark-matter parameter space are as of yet unexplored in asymptotic safety, the outlook is promising for the program to strongly reduce this parameter space by imposing asymptotic safety. 
This program resonates strongly with the string-inspired swampland program, where one similarly attempts to constrain the properties of dark sectors \cite{Montero:2022jrc,Montero:2022prj,Lehnert:2025izp}.
Similarly, other BSM scenarios, either motivated by experimental results, or motivated by fundamental open questions, can be constrained by asymptotic safety \cite{Kowalska:2020gie,Boos:2022pyq,Chikkaballi:2022urc,Chikkaballi:2023cce,Eichhorn:2023gat,Kowalska:2024lnm,Carone:2025wjw}.

\subsection{Imprints of asymptotic safety in particle-physics dynamics: particle scattering}\label{sec:particlescattering}
\emph{...where we discuss particle scattering in asymptotic safety. This is motivated by fundamental questions, including unitarity, the imprints of asymptotic safety in scale-dependent observables, as well as the role of black holes and the question whether black-hole production ultimately shields us from probing the asymptotically safe regime through scattering.}\\

We have highlighted in Sec.~\ref{sec:problems_of_perturbation_theory} that one of the two major problems of perturbatively quantized gravity is the increase of scattering cross-sections with energy, exemplified by the total tree-level cross-section for $e^{+}e^{-} \rightarrow \mu^{+}\mu^{-}$,
\begin{equation}
\sigma(s)= \frac{4\pi \alpha^2}{3s} + \frac{\pi}{20}G_N^2\, s.\label{eq:sigma_epem}
\end{equation}
Herein, $\alpha^2$ is the QED fine-structure constant and the Mandelstamm variable $s=(p_{e^+}+p_{e^-})^2$ is given in terms of the momenta of the positron, $p_{e^+}$ and the electron $p_{e^-}$. The problem with this scattering cross-section is its increase with $s$ at asymptotically large $s$, which ultimately leads to a unitarity violation in the perturbatively quantized theory.

This problem may be resolved in a somewhat trivial way, if scattering cross-sections  simply do not constitute meaningful observables in the Planckian and trans-Planckian regime.
Certainly making sense of a scattering cross-section requires fluctuations in the fields to be small enough to be meaningfully associated with particles and such a regime may simply not exist at trans-Planckian scales.  In addition, the $S$-matrix is not defined on general backgrounds, see \cite{Bousso:2004tv,Marolf:2012kh,David:2019mos,Melville:2023kgd} for discussions of the issues and proposed definitions of an $S$-matrix on de Sitter spacetime. Thus, an (asymptotically) flat spacetime should be a dynamically emergent, preferred background, otherwise scattering cross-sections may be computable, but physically uninformative.
 
Motivated by the near-perturbative properties of the gravitational fixed point, we will here assume that scattering cross-sections stay meaningful observables.\footnote{More recently, it has been proposed that functional renormalization could be implemented directly on the $S$-matrix, giving more direct access to scattering \cite{Freidel:2025uws}.} Then it becomes a pressing question to understand the large $s$-asymptotic behavior of Eq.~\eqref{eq:sigma_epem} and the behavior of other cross-sections at large values of the Mandelstam variables for $2-2$ scattering, and more generally large values of momentum exchange.\\

Early work on scattering cross-sections in asymptotic safety \cite{Hewett:2007st,Litim:2007iu,Gerwick:2011jw,Dobrich:2012nv} was based on the idea of Renormalization Group improvement, in which the $k$ dependence of $G_N$ (and other couplings) is equated with a physical dependence. In Eq.~\eqref{eq:sigma_epem}, this appears to be straightforward, because $\sigma$ only depends a single physical scale, suggesting the replacement 
\begin{eqnarray}
G_N \rightarrow G_N(k^2)=\frac{G_N}{1+ \frac{k^2}{G_{\ast}\cdot M_{\rm Pl}^2}}
 \rightarrow G_N(s) = \frac{G_N}{1+\frac{s}{G_{\ast}\cdot M_{\rm Pl}^2}}.
\end{eqnarray}
Even then, ambiguities remain in how precisely $k$ should be associated with $s$. In general, however, gravitational cross-sections depend on all three Mandelstamm variables and RG improvement obviously fails \cite{Anber:2010uj,Anber:2011ut,Knorr:2026vax}.\\

Thus, more recent research lines in asymptotic safety are devoted to calculating objects called \emph{form factors} \cite{Draper:2020knh,Knorr:2022dsx} or \emph{vertex functions} \cite{Pawlowski:2020qer,Pawlowski:2023gym}, introduced in Sec.~\ref{sec:EuclideanFP}, see Fig.~\ref{fig:expansions}. These are associated to correlation functions, but account for the \emph{physical scale dependence} of the correlation functions. One can think of these as a generalization of couplings to functions of the physical scales. To understand these objects, let us first neglect all spacetime indices, such that we denote the graviton as $h$. To simplify things further, we consider two scalar fields, $\phi$ and $\chi$, instead of a spinor field for the electron and another one for the muon. Now we use a crucial property of the effective action, namely that it already includes all loop effects. Thus, once the effective action has been calculated, all observables follow from it directly, without any further loop corrections to be computed. This means that we only need to write down the tree-level diagrams, as long as all vertices and propagators are evaluated from the effective action. Then, the vertex functions needed to calculate graviton-mediated $\phi\,\phi \rightarrow \chi\ \chi$-scattering, are the coupling of one $h$ to two $\phi$'s as well as the coupling of one $h$ to two $\chi$'s, in addition to the propagator for $h$. The crucial property is that this three-field vertex function is not just a coupling, but its momentum-dependent generalization, i.e., it is a function of two momenta (not three, due to momentum conservation). We may write, in Fourier space,
\begin{eqnarray}
\Gamma_{k \rightarrow 0}\Big|_{h\phi\phi}&=&\int \frac{d^4 p_1}{(2\pi)^4}\frac{d^4 p_2}{(2\pi)^4} f_{h\phi\phi}(p_1,p_2)\cdot \nonumber\\
&{}&\quad \cdot h(-p_1-p_2)\phi(p_1)\phi(p_2),\label{eq:exampleformfactor}
\end{eqnarray}
and analogously for $f_{h\chi\chi}$. Then it holds that the scattering cross-section for $\phi\,\phi \rightarrow \chi\ \chi$-scattering is determined by the two $f$'s, also called form factors:
\begin{equation}
\sigma_{\phi\,\phi \rightarrow \chi\ \chi} \sim f_{h\phi\phi}(p_1,p_2)\cdot f_{h\chi\chi}(p_1,p_2),
\end{equation}
with additional factors from the propagator and kinematics.
In a simple classical theory, $f_{h\phi\phi}(p_1,p_2)$ would reduce to a single coupling, multiplied by powers of the momenta if we were considering a derivative interaction. Due to quantum effects, $f_{h\phi\phi}(p_1,p_2)$  becomes a nontrivial function of the two momenta. Its high-momentum behavior is clearly crucial to determine the asymptotic dependence of the scattering cross-section on the Mandelstamm variables and cannot in general be inferred from its small-momentum behavior \cite{Knorr:2026vax}.
Therefore, the calculation of form factors has become a key endeavor in research on asymptotic safety in the last few years, e.g., \cite{Draper:2020knh,Knorr:2026vax,Chiesa:2026tlz}, see also the discussion around Fig.~\ref{fig:expansions}. In short, on proceeds as follows: First, one calculates the $k$-dependence as well as $p_i$-dependence of a form factor from the Wetterich equation. Second, one searches for a fixed-point solution, which is no longer given by a particular value for a coupling, but is given by a function of the physical momenta $p_i$. Third, starting from this fixed-point solution, one integrates to $k\rightarrow 0$ along a relevant perturbation of the fixed point. This provides the physical form factor that no longer depends on $k$, but retains a full physical momentum dependence, i.e., dependence on $p_i$.

We emphasize that the computation of scattering cross-sections is ultimately only meaningful in Lorentzian signature, which very recent works have tackled \cite{Pastor-Gutierrez:2024sbt,Chiesa:2026tlz}. \\

\begin{figure}[!t]
\includegraphics[width=\linewidth]{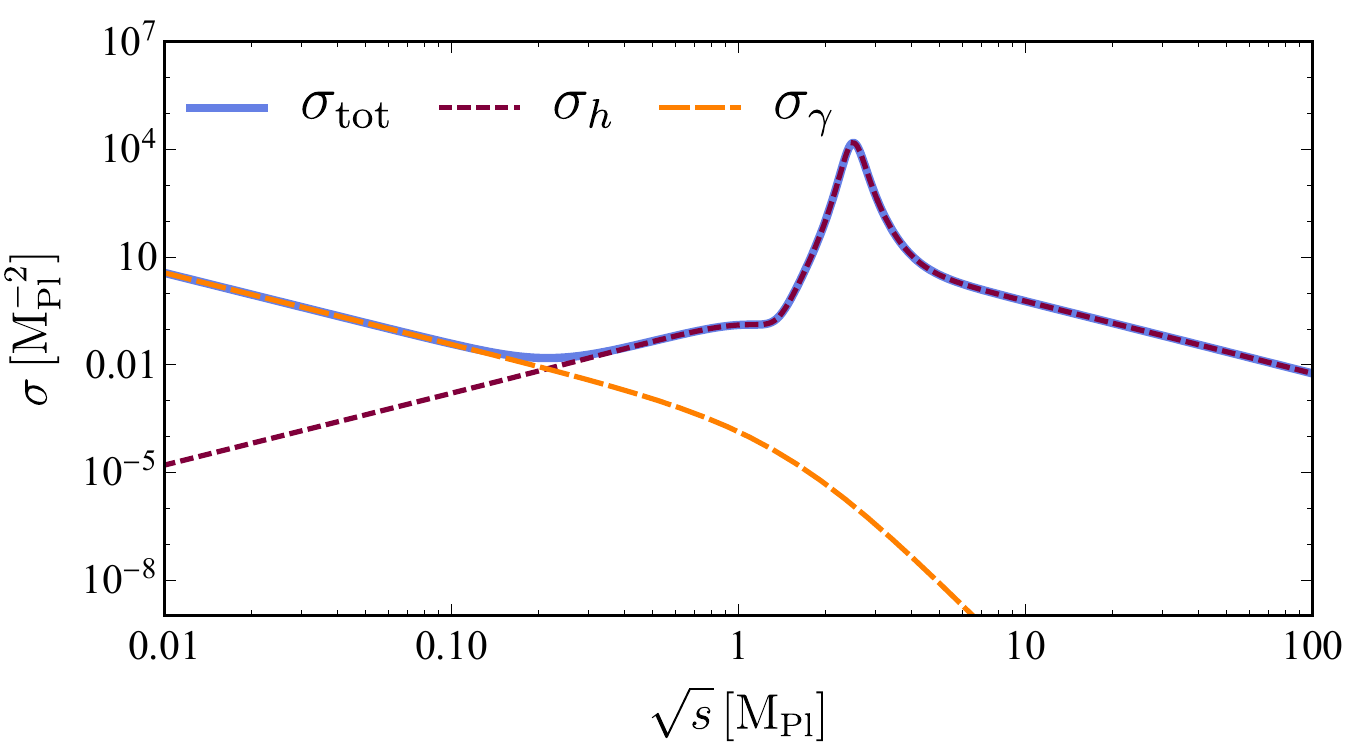}
\caption{\label{fig:crosssectionepem} This figure is reproduced from \cite{Pastor-Gutierrez:2024sbt} with the kind permission of the authors. The blue line shows the total cross-section, the orange dashed line the contribution from photons and the purple dashed line the gravitational contribution.}
\end{figure}

Let us now come back to the example in Eq.~\eqref{eq:sigma_epem}, which has been addressed in  \cite{Pastor-Gutierrez:2024sbt}. There, this cross-section has been computed, in a truncation that retains dependence on the momentum\footnote{Momentum dependence can be retained in various approximations: even if one starts from a single interaction term in the derivative expansion, the right-hand-side of the Wetterich equation generically constitutes a function of the momentum. This function is not necessarily well-approximated by its small-momentum expansion. Instead, one should retain the \emph{full} momentum dependence that comes out of the Wetterich equation. 
One may further improve the accuracy of the momentum-dependence by already using a form factor as an ansatz, in which case additional momentum-dependence is fed into the Wetterich equation. In this way, one can in principle approach the \emph{complete} momentum dependence of a vertex.}, see \cite{Pastor-Gutierrez:2024sbt} for further details. The result in shown in Fig.~\ref{fig:crosssectionepem}, reproduced from \cite{Pastor-Gutierrez:2024sbt} with the kind permission of the authors. We observe that in the UV, the total cross-section, which for $\sqrt{s}/M_{\rm Pl}>1$ is dominated by the gravitational contribution, shows a fall-off $\sim s^{-1}$, which obeys the Froissart bound. This calculation -- keeping in mind that it is within a truncation of the full dynamics -- therefore constitutes highly non-trivial evidence that asymptotic safety might be unitary.\\

In addition, $\sigma \sim s^{-1}$ is precisely the scaling that one expects in a scale-invariant regime based on dimensional analysis. Therefore, the result achieved in \cite{Pastor-Gutierrez:2024sbt} shows how asymptotic safety in the sense of a fixed point in $k$ leads to results for which the \emph{physical} scale-dependence exhibits scale-symmetry, i.e., dimensionful observables scale according to their canonical dimension.\\

 There is another important observation one can make from Fig.~\ref{fig:crosssectionepem}, namely the peak at $\sqrt{s} \simeq M_{\rm Pl}$. This may be interpreted as a hint that a gravitationally bound object is produced. This in facts fits with the ``folklore'' that scattering events at Planckian energies result in black-hole production. This folklore is based on the hoop conjecture, formulated by Thorne \cite{Thorne_hoop}. It states that an object of mass $M$ forms a black hole, if it is sufficiently compact, i.e., its diameter in any direction is smaller or equal to its Schwarzschild radius (neglecting the effect of spin). Extending the conjecture further, one may expect black-hole formation in ultrarelativistic scattering events at an impact parameter $b$ smaller than the Schwarzschild radius $R_S$ corresponding to the center-of-mass energy $E_{cm}$, \cite{tHooft:1987vrq,Banks:1999gd,Kohlprath:2002yh},
 \begin{equation}
 b< R_S = 2 G\, E_{\rm cm},
 \end{equation} 
 in units with $c=1$.
 Let us now discuss the caveats of such a conjectural extension of the hoop conjecture.\\
 First, even within the classical theory, ultrarelativistic scattering is not fully understood, see \cite{East:2012mb,Pretorius:2018lfb} and references therein. In fact, recent numerical simulations of ultra-relativistic black-hole encounters, with relativistic gamma factor of about 5, show  that up to 65 \% of the ADM energy is actually radiated away \cite{Zhu:2026mhn}. If similar results hold in ultrarelativistic scattering of particles, the scattering event may actually produce a large number of gravitons, into which a majority of the ADM energy goes, casting doubt on the formation of a black hole. \\
Second, the generalization from the scattering of classical particles or waves to quantum fields is also non-trivial \cite{Giddings:2009gj}, even assuming a regime in which quantum effects of spacetime itself are still negligible and in which semi-classical gravity may be applicable.  \\
Finally, the extension to quantum gravity is doubtful at best. At Planckian scales, the curvature at the horizon of a Schwarzschild black hole is Planckian,
\begin{eqnarray}
R_{\mu\nu\kappa\lambda}R^{\mu\nu\kappa\lambda} \Big|_{\rm Schw}= \frac{48 \, G_N^2 M^2}{r^6}\,\overset{\substack{r=R_S,\\M= M_{\rm Pl}}}{\longrightarrow}\,\frac{3}{4}M_{\rm Pl}^4.
\end{eqnarray}
Thus, higher-order curvature terms in the action are not negligible and, because some of these are Riemann curvature invariants, it is guaranteed that the Schwarzschild metric is no longer a solution to the corresponding equations of motion. Instead, one has to evaluate the horizon radius for the new solution; and there are general arguments why, if quantum gravity successfully solves the singularity problem, the new horizon is more compact than the Schwarzschild radius \cite{Eichhorn:2022oma}. \\
Moreover, even if black-hole formation were realized in the classical strong-gravity regime, this is insufficient to conclude that an $S$-matrix for quantum gravity is dominated by the same process \cite{Giddings:2009gj}. The efficiency of black-hole formation is subject to the same higher-order curvature corrections that change the expected horizon radius.
In summary, it is simply an unjustified assumption that the correct action to describe Planckian scattering in quantum gravity is that of GR. After all, the reason why the effective-field-theory description of quantum gravity breaks down at the Planck scale is that the Planck-scale suppression of higher-order curvature operators is absent and the GR action is no longer an adequate approximation. \\
Thus, no statement about black-hole formation in quantum gravity can be made in general. Rather, the presence or absence of black-hole formation  depends on details of the dynamics, e.g., whether the effective action at Planckian curvature scales even has black-hole solutions and what the radius of their event horizon is, see \cite{DelPorro:2025wts} for first steps to chart the phase diagram of compact objects in asymptotic safety.\\
In summary, it is unclear whether one should in general expect black-hole formation in Planckian scattering.\footnote{At highly transplanckian energies and impact parameters that are far larger than the Planck length, one is of course back in the classical gravity regime and will observe black-hole formation. However, decreasing the impact parameter towards the Planck length, one moves from the classical into the quantum gravity regime, where higher-curvature corrections to GR are not negligible.} The interpretation of the peak in \cite{Pastor-Gutierrez:2024sbt} is therefore a wide open, but highly relevant question, in order to understand the role of black-hole production in asymptotic safety.

\subsection{Predictive power of asymptotic safety in cosmology}\label{sec:cosmo}
\emph{...where we briefly review what is known about cosmology in asymptotic safety, both in the early and late universe. Asymptotic safety has constraining power for both. We also point out how that constraining power could be leveraged further, e.g., to constrain models of inflation as well as dynamical dark energy. We also discuss the compatibility of asymptotic safety with a de Sitter phase. The discussion in this section is in large parts aspirational: there exists proof-of-principle for the predictive power of asymptotic safety in cosmology, but many question are as of yet unanswered. These questions can be tackled based on the current state-of-the-art in the field; thus we point to several concrete open questions and ways forward.}\\

We provide a brief overview of some results here and a perspective for the future, but point to the reviews \cite{Platania:2020lqb,Wetterich:2022ncl,Bonanno:2024xne} for a more in-depth discussion and other perspectives.\\

The most obvious application of quantum gravity to cosmology lies in the Big-Bang-singularity, which quantum gravity is generally expected to resolve.
In asymptotic safety, a robust description of this regime hinges on the higher-order curvature terms in the effective action, which become dynamically important as one follows the evolution of the universe backwards towards the Big-Bang singularity. In asymptotic safety, the number of relevant parameters not yet fixed by observations is very small, and there is likely only a single one, related to a superposition of $R^2$ and $R_{\mu\nu}R^{\mu\nu}$. All other higher-curvature terms are predicted in asymptotic safety, i.e., the asymptotically safe \emph{landscape} in a purely gravitational settings is three-dimensional and, once Newton coupling and cosmological constant have been fixed using observations, only a single unknown direction remains. We stress that this does not mean that further, higher-order operators are zero, it simply means that their coefficient are determined in terms of the three relevant parameters. Once better calculational control of the high-curvature regime in Lorentzian signature has been achieved, it will accordingly be possible to determine whether asymptotic safety resolves the singularity and what it is substituted by. \\
It has been pointed out by Ashtekar, that the effective Friedmann equation in Loop Quantum Cosmology \cite{Ashtekar:2011ni,Agullo:2023rqq}, which gives rise to a big bounce rather than a singularity, 
\begin{equation}
\left( \frac{\dot{a}}{a}\right)^2 = \frac{8\pi G\, \rho}{3}\left(1-\frac{\rho}{\rho_{\rm crit}} \right),
\end{equation}
could be obtained by starting from a classical Friedmann equation, and promoting $G$ to a scale-dependent coupling that becomes asymptotically safe, with the appropriate scale given by the matter density $\rho$, see \cite{Bonanno:2017gji} for similar considerations. This heuristic argument motivates a more in-depth study of whether asymptotic safety can produce a big bounce.

At a heuristic level, one may also argue that a theory that has a scale invariant dynamics in the UV is likely to produce near-scale invariant spectra of fluctuations in the early universe, where curvature scales were high and the near-scale-invariant regime of the dynamics was probed. This may provide a mechanism to generate the observed spectrum of near-scale-invariant scalar fluctuations \cite{Planck:2018vyg} (and its tensor counterpart) even without an inflationary period. However, supporting this argument with detailed calculations is an outstanding challenge. \\
It has been found, however, that the quadratic-curvature terms present in asymptotic safety suffice to suppress inhomogeneities and anisotropies, when a finite-action principle is imposed \cite{Lehners:2019ibe}.\\

While the origin of the universe is a wide-open question in asymptotic safety, some constraints on its subsequent evolution are already understood. Chiefly, these constraints pertain to any components of the universe that can be modelled by additional fields and apply to, e.g., inflation, dark matter (see Sec.~\ref{sec:dm} and (dynamical) dark energy. These constraints are useful, because models for inflation, dark matter and dynamical dark energy have proliferated in number, and while some are constrained or even ruled out by observations \cite{Planck:2018jri}, a vast number of possibilities remains open. Additional theoretical principles to constrain or rule out models are thus desirable. Conversely, theoretical constraints from asymptotic safety amount to new predictions for observations and experiments from asymptotic safety, bringing the theory closer to being testable.
 
These constraints are all derived by following these steps:\footnote{Phenomenological model-building has also relied on RG-improvement, in which the $k$-dependence of couplings is identified with a dependence on a physical scale, e.g., the Hubble scale in an inflationary setting, see, e.g., \cite{Bonanno:2010bt,Contillo:2011ag,Copeland:2013vva,Bonanno:2015fga}. In Sec.~\ref{sec:notionsofrunning}, we have highlighted that different notions of running may not be universal and thus the results of RG-improvements procedures need to be tested a posteriori by comparing to results obtained along the lines outlined below.}
\begin{itemize}
\item[1)] Consider a model of inflation, dark matter or dark energy. To be phenomenologically viable, the couplings in this model have to satisfy certain constraints.
\item[2)] Calculate the additional constraints that asymptotic safety imposes on the couplings through the irrelevant directions. This step is affected by systematical uncertainties due to the choice of truncation and because it is typically done in Euclidean signature.
\item[3)] Impose the constraints from 1) and 2) on the couplings and confirm whether there is a phenomenologically viable part of the parameter space that is also asymptotically safe. If not, asymptotic safety may rule the model out, unless the constraints from 2) may be circumvented, e.g., by extending the model, e.g., through extra fields that give rise to a larger parameter space.
\end{itemize}
The success of these steps hinges on the high predictive power of asymptotic safety. The number of relevant directions is typically found to be \emph{smaller than the number of relevant and marginal couplings} in a given model.\\

Applied to inflation, one can, e.g., find, that Higgs inflation \cite{Bezrukov:2007ep} is not compatible with asymptotic safety,  because the phenomenologically required values of the non-minimal coupling of the Higgs field to the Ricci scalar cannot be achieved simultaneously with other phenomenological constraints on the Higgs sector \cite{Eichhorn:2020sbo}; see also \cite{Wetterich:2019rsn}.
More generally, scalar potentials are highly constrained in asymptotic safety. In terms of a Taylor expansion about the origin
\begin{equation}
V(\phi) = \sum_i \lambda^{(i)} k^{4-i} \phi^i,
\end{equation}
for a real scalar $\phi$, we have that
\begin{equation}
\beta_{\lambda^{(i)}}= \left(-(4-i)-f_{\lambda}\right)\lambda^{(i)}+\dots\, .\label{eq:ASpotential_noZ2}
\end{equation}
The fixed-point potential is simply flat, $\lambda^{(i)}_{\ast}
=0$ and thus, the critical exponents are $\theta^{(i)}= f_{\lambda}+(4-i)$. Because $f_{\lambda}<0$, quartic interactions always correspond to irrelevant directions, e.g., \cite{Narain:2009fy,Percacci:2015wwa,Eichhorn:2017als,Pawlowski:2018ixd,Eichhorn:2020sbo}, whereas cubic interactions  may be relevant or irrelevant, depending on the value of $f_{\lambda}$.
This removes much of the freedom of model-building with scalar fields and thus constrains inflationary models.

On the other hand, inflation -- if realized at all -- need not arise from the ad-hoc addition of a scalar field. Instead, the scalar degree of freedom that becomes dynamical once $R^2$ is added to the Einstein-Hilbert action, and that drives inflation in Starobinsky's model \cite{Starobinsky:1980te}, may do so in asymptotic safety, see \cite{Copeland:2013vva,Bonanno:2015fga,Gubitosi:2018gsl}. For this to work it is crucial that $R^2$ in fact is part of a relevant direction \cite{Lauscher:2002sq,Benedetti:2009rx,Falls:2020qhj}, and may even reach the large values required for Starobinsky inflation \cite{Gubitosi:2018gsl}. There are, however, two outstanding questions on Starobinsky inflation in asymptotic safety. First, the $R^2$ coupling is not aligned with the relevant direction, instead, the relevant direction contains a superposition of the $R^2$ and $R_{\mu\nu}^2$ couplings. It has not yet been studied whether the $R^2$ coupling can be made large without the $R_{\mu\nu}^2$ coupling becoming large as well and affecting the predictions of Starobinsky inflation. Second, recent observational constraints \cite{AtacamaCosmologyTelescope:2025blo,AtacamaCosmologyTelescope:2025nti} provide a hint that besides $R^2$, higher powers of $R$ may be needed \cite{Addazi:2025qra}. In case this hint is substantiated by future observations, it becomes paramount to study whether higher-order curvature terms can be accommodated at the right coupling values in asymptotic safety.\newline

The late universe is described by a de-Sitter-like phase, during which a cosmological constant, with the value inferred from observations, dominates the evolution. It is a crucial question whether this can be accommodated in asymptotic safety. Because the cosmological constant adds a relevant direction to the fixed point, both positive as well as negative values can be accommodated \cite{Reuter:2001ag}. This appears to remain the case also in more recent Lorentzian-signature studies \cite{DAngelo:2023wje,DAngelo:2025yoy}. However, a positive cosmological constant of the right size to match observations is not sufficient to ensure a de-Sitter-like phase. In addition, it must be ensured that this remains a solution of the full dynamics in the presence of higher-order curvature terms, see, e.g., \cite{Falls:2016wsa}. There is also corroborating evidence on a de-Sitter like phase from dynamical triangulations \cite{Ambjorn:2008wc}, albeit with overall spacetime volumes only a few orders of magnitude about a Planck volume, due to numerical limitations in simulating very large volumes.

Results of recent observational surveys targeting baryon acoustic oscillations may be interpreted as a tentative hint of an evolving dark energy, rather than a cosmological constant \cite{DESI:2024mwx,DESI:2025zgx}. If future observations substantiate these results, asymptotically safe settings with dynamical dark energy, which can, e.g., be modelled by an additional scalar field, become of strong interest. In this case, the scalar potential is strongly constrained, see Eq.~\eqref{eq:ASpotential_noZ2}. Therefore phenomenological models of dynamical dark energy, proposed to explain the DESI data, may not all be compatible with asymptotic safety. Additionally, such scenarios are also subject to so-called fifth-force constraints, which correspond to deviations from GR in the weak-field regime, well constrained, e.g., through solar-system constraints \cite{Will:2014kxa}. Such constraints have to be avoided \cite{Bonanno:2026aiw}, which may also be possible if screening mechanisms, reviewed, e.g., in  \cite{Burrage:2017qrf}, can be accommodated in asymptotic safety.

There is a scenario in which asymptotic safety imposes particularly strong constraints. In this scenario, all scales in the universe are set by an evolving scalar field, the dilaton. There is evidence that an asymptotically safe fixed point exists in dilaton quantum gravity \cite{Henz:2013oxa,Henz:2016aoh,Maitiniyazi:2025pou}.
In this case, it is possible that the fixed-point solution itself determines the physics, because the RG flow is started exactly on the fixed point, so that the effective action never departs from the scaling solution along any of the relevant directions. Physical scale dependence arises in this setting, as the dilaton evolves over cosmological timescales.
In such a setting, inflation and dynamical dark energy may be sourced by a single scalar, see \cite{Wetterich:2022ncl} for a review.\newline\\

The final question to address is how to connect from asymptotic safety to phenomenological models in cosmology not on a model-by-model basis, but rather at the level of the phenomenological framework. \\
This starts with the question whether there is a preferred set of fields beyond the metric, or whether modified gravity can be compatible with asymptotic safety in a scalar-tensor, as well as vector-tensor or other frameworks. So far, most studies have focussed on scalar-tensor theories, but recent works  \cite{Heisenberg:2026rdk,Pastor-Marcos:2026nyb} indicate that vector-tensor theories may also be compatible with asymptotic safety. \\
As a second question, the distinction between different settings for the effective field theory becomes relevant. So far, general EFT-frameworks for scalar-tensor as well as vector-tensor theories are typically built to avoid higher-order equations of motion. For instance, Horndeski gravity \cite{Horndeski:1974wa} provides the most general effective Lagrangian for scalar-tensor theories for a scalar $\phi$ with manifestly second-order equations of motion, 
\begin{eqnarray}
\mathcal{L}_H &=& \int d^4x\sqrt{-g}\Bigl[G_2(\phi,X) - G_3(\phi,X)\Box \phi + G_4(\phi,X)R\nonumber\\
&{}&+ G_{4,X}\left((\Box\phi)^2- (D_{\mu}\partial_{\nu}\phi)(D^{\mu}\partial^{\nu}\phi) \right)\nonumber\\
&{}&+ G_5(\phi,X)\left(R_{\mu\nu}- \frac{1}{2}g_{\mu\nu}R \right)(D^{\mu}\partial^{\nu}\phi)\nonumber\\
&{}&-\frac{G_{5,X}}{6}\Bigl((\Box \phi)^3- (3 \Box \phi)(D_{\mu}\partial_{\nu}\phi)(D^{\mu}\partial^{\nu}\phi)\nonumber\\
&{}& + 2(D^{\mu}\partial_{\alpha}\phi)(D^{\alpha}\partial_{\beta}\phi)(D^{\beta}\partial_{\mu}\phi) \Bigr)\Bigr],
\end{eqnarray} 
where $X = \partial_{\mu}\phi\partial^{\mu}\phi$ and $G_{i,X}= \partial_X G_i$. This includes, e.g., an inflaton as well as quintessence field, minimally (or non-minimally) coupled, with a potential, but also many other scalar-tensor theories discussed in cosmology, see \cite{Kobayashi:2019hrl} for a review.
It can be generalized to beyond-Horndeski and DHOST \cite{Langlois:2015cwa}, see \cite{Langlois:2018dxi} for a review, in which the equations of motion can be higher order, but a degeneracy condition in the Hamiltonian prevents the propagation of further modes beyond the metric and a scalar. If the quasi-local terms in the effective action from asymptotic safety can be shown to lie within one of those frameworks, two points are achieved: First, existing results on phenomenological constraints on, e.g., given Horndeski models, can be confronted  with predictions from asymptotic safety, see, e.g., \cite{Eichhorn:2022ngh}. Second, this result would establish directly that asymptotic safety does not propagate additional degrees of freedom, most importantly ghostly ones.\footnote{We highlight that the converse statement may not be true: a theory outside the DHOST theory space, which one may naively expect to propagate ghostly modes, may still not suffer from instabilities in its time-evolution, see \cite{Deffayet:2021nnt,Held:2023aap,Deffayet:2023wdg,Deffayet:2025lnj,Held:2025fii,Held:2025ckb,Deffayet:2026cnu}.}

Overall, the connection of asymptotic safety to cosmology has the potential to be very fruitful: On the one hand, asymptotic safety may provide a new theoretical principle to select among the numerous proposals to modify GR, e.g., to explain cosmic acceleration, see \cite{Silvestri:2009hh,Clifton:2011jh,Joyce:2014kja} for reviews. On the other hand, the phenomenological consequences of many of these proposals are already very well understood. Thus, testing whether these proposals are asymptotically safe provides a direct way to connect asymptotic safety to predictions for cosmological surveys.

\subsection{What does asymptotically safe gravity imply for the spacetime structure of black holes?}\label{sec:BH}
\emph{...where we review asymptotic-safety inspired model building for black-hole spacetimes, which gives rise to regular black holes. We highlight the difference between such model building and a proper derivation of a black-hole solution from the theory and outline first steps that have been taken to achieve such a derivation.}\\

Here, as in other parts of this review, we make the assumption that asymptotic safety is realized in Lorentzian, not just Euclidean, quantum gravity, supported by the tentative hints reviewed in Sec.~\ref{sec:Lorentzian} that a fixed point exists in the Lorentzian and is similar to its Euclidean counterpart. Otherwise the study of black-hole spacetimes would be largely meaningless, because only the spacetime region outside the event horizon of a stationary black-hole spacetime can be continued analytically.\\

Black holes are an important testing ground for quantum gravity, because they signal the breakdown of GR. This is signaled through several physical inconsistencies: first, some curvature invariants are singular in the center of a black hole; second, geodesics are incomplete; third, at nonzero spin, there is a Cauchy horizon in the interior of the black hole at which the initial-value problem breaks down. Note that these statements all apply to stationary vacuum solutions of GR. The situation might be different in a time-dependent gravitational collapse, in particular, when Hawking radiation is also accounted for \cite{Gralla:2025gzl,DiFilippo:2025kzh}.\\
It is generally expected that quantum-gravity effects render black holes physically consistent. Within asymptotic safety, we expect that a black-hole interior can be described in terms of a spacetime manifold equipped with a metric. This metric is the solution to the quantum equations of motion, $\delta \Gamma[\langle g_{\mu\nu}\rangle]/\delta \langle g_{\mu\nu}\rangle =0$, with appropriate boundary conditions; its interpretation is as the expectation value of the metric for a black-hole spacetime with a given set of Killing symmetries.\footnote{The fact that one can derive such a solution to the quantum equations of motion along the same lines as a classical solution does not mean that the interior of the black hole necessarily is classical. From the effective action, one can also evaluate the higher-order correlation functions and may in general expect that the expected amplitude of fluctuations about the expectation value is also large in the interior.}, \footnote{\label{footnotecirc}For instance, for a spherically symmetric, stationary spacetime, the corresponding equations of motion determine the two functions $f(r)$ and $h(r)$ in the general ansatz
\begin{equation}
ds^2 = f(r)dt^2 - h(r)^{-1}dr^2 -r^2 d\Omega_2^2,
\end{equation}
which is formulated in terms of the natural coordinates of an asymptotic observer. In turn, the functions $f(r)$ and $h(r)$ determine the black-hole structure, including its curvature invariants, but also special surfaces such as the horizon, determined by $h(r_H)=0$, or photon sphere, which also depends on the form of the geodesic equation that is modified, e.g., by non-minimal couplings.\\
For an axisymmetric, stationary vacuum spacetime, GR has an automatic special symmetry property, called circularity that guarantees that the event horizon is a Killing horizon. This is no longer guaranteed beyond GR and thus the most general ansatz, formulated in horizon-penetrating coordinates $v, r, \theta, \phi$ is
\begin{eqnarray}
ds^2 &=& g_{vv}(r,\theta)dv^2 + g_{rr}dr^2 + g_{\theta\theta}d\theta^2 + g_{\phi\phi}d\phi^2\nonumber\\
&{}& + 2 g_{v\phi}dv \,d\phi+ 2 g_{vr}dv \,dr + 2 g_{r\phi}dr\,d\phi,\label{eq:noncirc}
\end{eqnarray}
see \cite{Delaporte:2022acp,Babichev:2025szb}. We note in passing that a majority of work on phenomenological models of regular black holes is based on a simplified ansatz with $g_{vr}=0$ and $g_{r\phi}=0$. This imposes circularity, but is not even respected within an asymptotic-safety inspired construction of a regular black hole \cite{Eichhorn:2021iwq,Eichhorn:2022bgu}.}
A consistent black-hole interior in asymptotic safety must satisfy the following: in physically relevant black-hole spacetimes, i.e., black holes that arise from gravitational collapse (and generically carry spin), i) there must not be any singularities in curvature invariants, ii) any inextendible causal curves, including timelike and null geodesics, must cover an infinite range in proper time/affine parameter, iii) Cauchy horizons, at which the initial-value problem breaks down, must be absent.\footnote{There is of course also the more exotic possibility that the quantum equations of motion do not have black-hole solutions, and instead, there are extremely compact objects that serve as black-hole mimickers and cannot be distinguished from black holes by current observations \cite{Bambi:2025wjx}. } 

These conditions indirectly constrain the effective action in asymptotic safety and in particular its high-curvature properties. It is, however, not in general understood what conditions must hold for the effective action in order for i) to iii) be realized. Research efforts aimed at understanding which beyond-GR-actions give rise to regular black-hole solutions, such as, e.g., \cite{Bueno:2025zaj}, can be interpreted as a step towards delineating the space of possible effective actions for asymptotic safety. Intriguingly, the actions found in  \cite{Bueno:2025zaj} include terms at arbitrarily high order in curvature, as one would expect in an effective action from asymptotically safe quantum gravity. As anticipated in \cite{Knorr:2022kqp}, they cannot be written as an expansion of local terms, instead, non-local terms, e.g., $C_{\mu\nu\kappa\lambda}C^{\kappa\lambda\rho \sigma}C_{\rho \sigma}^{\,\,\,\,\,\,\mu\nu}R_{\alpha \beta}R^{\beta \gamma}R^{\alpha}_{\gamma} C^2/(C_{\eta \tau \xi \zeta}R^{\eta\xi}R^{\tau\zeta} C^2 - 2 C^3 R^2)$ are necessarily part of it. Similarly, based on destructive interference of singular black-hole configurations in the path integral, one can argue for higher-order terms in the action \cite{Borissova:2020knn} and even for the action to be of infinitely high order in curvature \cite{Borissova:2023kzq}.\\

From the above, we see that what is needed to fully understand black-hole (like) spacetimes, is the full effective action. At the same time, the discussion in Sec.~\ref{sec:FP} highlights that the full effective action in asymptotic safety is as of yet unknown.\footnote{We stress that while the existence of a fixed point is regarded as established, the step towards computing the full effective action, in particular at high curvature, is still highly non-trivial, even if the fixed point guarantees that such an action exists and has a finite number of free parameters.} Thus, the structure of black-hole spacetimes has so far only been studied in relatively small truncations of the effective action \cite{Pawlowski:2023dda,DelPorro:2025wts}. At the same time, studies of black holes in quadratic gravity \cite{Holdom:2002xy,Lu:2015cqa,Held:2022abx,Held:2023aap} and their phenomenological signatures \cite{Daas:2022iid,Held:2025ckb} can be interpreted as studies of leading-order effects from asymptotic safety on black holes, because quadratic curvature corrections are present in the effective action for asymptotic safety -- albeit with a specific relation that constrains one of the curvature-squared couplings in terms of the other \cite{Benedetti:2009rx,Falls:2020qhj}, which is usually not included in studies of asymptotic safety.\footnote{In pure gravity, quadratic curvature terms can be removed by field redefinitions. However, such redefinitions change the couplings of gravity to matter. All physical predictions for black holes, cosmology and particle physics are invariant under field redefinitions in the effective action, but only, if they are done consistently across all sectors; performing distinct field redefinitions in distinct sectors (e.g., to remove quadratic gravity terms, but without accounting for the corresponding changes in the gravity-matter part of the effective action) is not consistent.}\\

In the absence of having black-hole solutions of the theory, we may discuss what might be expected for black-hole spacetimes, if physical asymptotic safety is realized, in the sense that a scaling regime sets in, once the effective action $\Gamma_{k\rightarrow 0}$ is considered at high values of physical scales, e.g., curvature scales. In that case, once physical scales exceed the Planck, scale invariance should be realized, i.e., the Planck scale is expected to serve as the maximum physical scale that can be distinguished before one enters a scale-invariant regime. Thus, we may expect that there is a maximum to the value of curvature scales that can be realized in solutions to the full equations of motion. In turn, limiting curvature scales may be sufficient to result in regular black-hole interiors, although it may not in general be sufficient to guarantee geodesic completeness. In addition, static regular black holes, unless they are extremal, contain an interior Cauchy horizon \cite{Carballo-Rubio:2019nel}, likely rendering them inconsistent; see \cite{Carballo-Rubio:2025fnc} for a review of regular black holes and corresponding open issues. Therefore, it appears that asymptotic safety may not automatically solve all problems of classical black-hole spacetimes. This makes the task of finding black-hole spacetimes as solutions of (truncations of) the full effective action particularly relevant.\\

To date, black-hole spacetimes have been studied, starting from \cite{Bonanno:2000ep}, that are not derived, but \emph{inspired} by asymptotic safety. The vast literature on the topic is reviewed in \cite{Eichhorn:2022bgu} and \cite{Platania:2023srt}.

Concretely, asymptotic-safety-inspired black-hole spacetimes rely on the idea of RG improvement. This procedure can successfully be used in simple enough settings in QFT, and consists in two steps: first, the couplings in the classical action/equations of motion/solutions are substituted by their RG-scale-dependent counterparts; second, the RG scale is identified with a physical scale of a system.

Asymptotic-safety inspired black-hole spacetimes actually realize the expectation described above, because they are based on an upgrade of a classical spacetime,
\begin{equation}
ds^2 = \left(1-\frac{2G_N\, M}{r}\right)dt^2 -  \left(1-\frac{2G_N\, M}{r}\right)^{-1}dr^2 -r^2d\Omega_2^2,
\end{equation}
 through the substitution 
\begin{equation}
G_N \rightarrow G_N(k) = \frac{G_N}{1+\frac{G_N}{g_{\ast}}k^2} \rightarrow G_N(k_{\rm phys}) = \frac{G_N}{1+\frac{G_N}{g_{\ast}}k_{\rm phys}^2}. 
\end{equation}
Herein, $k_{\rm phys}$ is a physical scale of the system. In spherically symmetric, static spacetimes, all scales are related, therefore $k_{\rm phys}^4 = \zeta \left(R_{\mu\nu\kappa\lambda}R^{\mu\nu\kappa\lambda} \right)$, with a constant $\zeta$, results in the same black-hole interior as $k_{\rm phys} = d_{\rm r}^{-1}$, with $d_{\rm r}$ the radial geodesic distance.\footnote{There is, in general, a coordinate dependence within RG improvement \cite{Held:2021vwd}, and, in spacetimes with fewer Killing vectors, such as axisymmetric spacetimes, different choices of scale do not give rise to the same RG improved spacetime, as reviewed in \cite{Eichhorn:2022bgu}.} The scale is always evaluated within the original, classical spacetime.\footnote{\cite{Platania:2019kyx} implemented an iteration procedure, such that the physical scale is evaluated self-consistently within the RG-improved spacetime.} It turns out that the behavior $G_N(k) \sim k^{-2}$ at high scales implies a sufficient weakening of gravity to generically lift curvature singularities and limit the values of curvature scales to (appropriate powers of) the Planck scale.

Despite this success in agreeing with the intuition as to how black-hole spacetimes in an asymptotically safe theory should look like, it is important to remember that RG improvement is not a reliable procedure in the sense of yielding results that are equivalent to solving the quantum equation of motion. In fact, RG improvement works successfully only in special settings; asymptotically safe gravity is not expected to be one of them \cite{Held:2021vwd,Pastor-Gutierrez:2022nki,Knorr:2026vax}. Concretely, in systems in which quantum corrections are organized in the form $\ln(k^2/p^2)$, where $k$ is the RG scale and $p^2$ the physical momentum squared, the RG improvement procedure correctly reproduces the leading-order quantum correction with the correct physical scale dependence. Examples include, e.g., the Uehling potential in QED. However, in general $k$ is simply an auxiliary scale, keeping track of the extent to which the path integral has been performed. The physical limit is $k\rightarrow 0$. In this limit, physical scales, e.g., curvature scales, can  stay finite.  Therefore, RG improvement is not in general expected to provide the correct result for the leading order quantum corrections, unless there is a \emph{decoupling} mechanism, such that quantum fluctuations with momenta below $k_{\rm decoupling}$ no longer contribute to the RG flow. This may happen in systems with a distinct physical scale which acts as an IR cutoff in the path integral. In that case, the effective action at $\Gamma_{k = k_{\rm decoupling}>0}$ already encodes the full physics of the system. Decoupling has been argued to be relevant for black-hole spacetimes in \cite{Borissova:2022mgd}. However, we stress that even if there is decoupling, the physical suppression of low-momentum quantum fluctuations through a \emph{physical} cutoff is in general not expected to be accurately encoded in the suppression through a regulator function $R_k$.\\
In summary, the vast literature on RG improved black holes, reviewed  in \cite{Eichhorn:2022bgu} and \cite{Platania:2023srt}, should  be interpreted as phenomenological model-building, with inspiration from asymptotic safety. While such model-building efforts can surely be important to build intuition and even suggest observational signatures\footnote{In fact, RG improvement for a spinning black hole can result in a non-circular spacetime \cite{Eichhorn:2021iwq}, see Eq.~\eqref{eq:noncirc} in footnote~\ref{footnotecirc}. Circularity is a symmetry property of the Ricci tensor, and usually assumed, e.g., in parameterized studies of black holes beyond GR \cite{Johannsen:2011dh,Konoplya:2016jvv}. One might interpret the non-circular spacetime in \cite{Eichhorn:2021iwq} as a hint, that non-circularity should be included in parameterizations \cite{Delaporte:2022acp,Babichev:2025szb}; this is also supported by non-circular solutions to modified gravity theories \cite{Anson:2020trg,Fernandes:2026rjs} and semi-classical gravity \cite{Fernandes:2023vux}. In turn, non-circularity may, e.g., result in specific features in black-hole shadows \cite{Eichhorn:2021etc}.}, RG improved black holes do not constitute robust predictions of asymptotic safety. Thus, confronting them with observations is useful to test alternatives to Kerr black holes, but can neither rule out nor confirm asymptotic safety. Instead, more efforts along the lines of \cite{Pawlowski:2023dda,DelPorro:2025wts} are needed to derive the structure of black holes in asymptotic safety, complemented by studies in higher-derivative gravity as in \cite{Held:2022abx,Held:2023aap,Bonanno:2025ksz}, which can be interpreted as accounting for the leading-order terms in the effective action for asymptotic safety, in particular in studies as in \cite{DelPorro:2025wts}, where the relations between couplings coming from asymptotic safety are explicitly accounted for.

\subsection{Is there a phenomenological approach to determine the potential consequences of asymptotically safe gravity for beyond-Standard Model physics for particle physics and cosmology?}\label{sec:principledparameterized}
\emph{...where we review an approach for model-building within asymptotic safety and discuss the limitations and regime of validity of this approach.}\\

All existing methods to search for asymptotic safety, including the functional RG as well as lattice simulations, require a significant up-front investment to master them, in practice making it somewhat challenging for anyone interested in exploring physical consequences of asymptotic safety for particle physics or cosmology without a background in these methods. See, however, Sec.~\ref{sec:codebase} for a list of lecture notes, textbooks and a link to a code base.\\
In addition, even the most advanced calculations within both frameworks still come with sizable systematic uncertainties, hampering the derivation of physical predictions. \\
It is therefore useful to follow an approach which provides a ``shortcut'' to develop an approximate understanding of the phenomenology of asymptotic safety and which at the same time accounts for systematic uncertainties of the above approaches. Such an approach exists, but makes assumptions about the nature of asymptotic safety, most importantly, that it is near-perturbative, see Sec.~\ref{sec:nearpert}. This assumption is supported by concrete results, but is not unequivocally established. The application of this approach should therefore proceed in parallel with additional studies aimed at checking whether the underlying assumption holds or whether the approach requires revisions.\\

In our discussion in Sec.~\ref{sec:SM}, we have already used a parameterization of metric fluctuations through three parameters
$f_g$, $f_y$ and $f_{\lambda}$, to account for gravitational contributions to gauge couplings, Yukawa couplings and quartic scalar coupling. More generally, the beta function for any matter coupling receives contributions proportional to itself and to the gravitational couplings, i.e., gravitational fluctuations generate anomalous scaling dimensions for couplings. Because gravity is ``blind'' to any internal symmetry, it is only the \emph{spacetime structure} of the interaction that determines the gravitational contribution.\footnote{There may be higher-order contributions, which are proportional to higher-order matter couplings which are themselves generated by quantum gravity, such as the couplings of higher-order field strength interactions that contribute to $f_g$ in Eq.~\eqref{eq:fg_w2v2}. These are generally expected to be subleading \cite{Eichhorn:2017eht}.} As an example, we have used in our discussion of the SM that $f_y$ is the same for all Yukawa couplings in the SM, irrespective of the charge assignments of the corresponding fermions and scalars.

For all couplings of dimension-four-interactions of scalar, fermions and gauge fields (in four spacetime dimensions), these anomalous scaling dimensions are the only gravitational contribution that is necessarily non-zero.\footnote{Just as in Eq.~\eqref{eq:fg_w2v2}, contributions from induced matter interactions can be subsumed in the $f_i$'s in a near-perturbative regime.} 
This is rooted in a symmetry argument: within all calculations in asymptotically safe gravity-matter systems, global symmetries of matter fields are respected in the beta functions\footnote{Whether or not asymptotic safety thereby constitutes a counterexample to the most prominent and long-standing swampland conjecture, namely the no-global-symmetries conjecture, is not fully resolved, see \cite{Eichhorn:2017eht,Eichhorn:2020sbo,Eichhorn:2022gku,Eichhorn:2024rkc,Borissova:2024hkc}.}, see \cite{Eichhorn:2017eht} as well as \cite{Eichhorn:2022gku} and references therein.\footnote{This does not imply that these symmetries must be respected by the effective action; couplings which break global symmetries may be relevant and can thus take nonzero values in the IR. CPT-symmetry-violating interactions constitute an example: they vanish at the fixed point, but can remain relevant \cite{Eichhorn:2025ilu}. This implies that CPT symmetry must be imposed strictly on the theory, unless one is willing to incur CPT violations at low scales.}. We make the assumption that the near-perturbative nature of the fixed point implies that the mixing of different interactions in the eigendirections of the fixed point is a subleading effect; thus direct gravitational contributions to each coupling of interest are the terms we focus on.

Therefore, a phenomenological approach to asymptotic safety within a given model with a set of matter fields (by matter here we also mean gauge fields) and dimension-four-interactions proceeds as follows. Assuming that the interactions include gauge couplings $g_i$, Yukawa couplings $y_i$ and quartic couplings\footnote{These may also be quartic couplings that mix several different scalars, e.g., for a two-scalar model with scalar $\phi_1$ and $\phi_2$, there are three quartic couplings, associated to $\phi_1^4$, $\phi_2^4$ and $\phi_1^2\, \phi_2^2$.} $\lambda_i$, we have that
\begin{eqnarray}
\beta_{g_i}&=& -f_g\, \Theta(k-M_{\rm Pl}) \, g_i + \beta_{g_i}^{\rm(matter)},\\
 \beta_{y_i}&=& -f_y\, \Theta(k-M_{\rm Pl}) \, y_i + \beta_{y_i}^{\rm(matter)},\\
 \beta_{\lambda_i}&=& -f_{\lambda}\, \Theta(k-M_{\rm Pl}) \, \lambda_i + \beta_{\lambda_i}^{\rm(matter)}.
\end{eqnarray} 
In here, we have included $\Theta(k-M_{\rm Pl})$ to model how gravitational effects decouple at the Planck scale. In practice, because the gravitational contributions are all proportional to the Newton coupling, these contributions scale with $k^2$, once the RG flow has left the fixed-point regime. This leads to a very fast suppression of the gravitational effects below the Planck scale, which is well-approximated by a theta-function, see, e.g., \cite{Eichhorn:2017ylw}.\\
The above parameterized beta functions should be supplemented with 
\begin{equation}
f_g>0,\quad f_{\lambda}>0,
\end{equation}
both of which hold in all studies to date, see Sec.~\ref{sec:SM}. In contrast, because the sign of $f_{y}$ changes, depending on the gravitational fixed-point values, one can in principle use either sign for phenomenological studies. One can, however, as done, e.g., in \cite{Kowalska:2020gie,Kowalska:2020zve} also fix $f_y$ (as well as $f_g$) to obtain a highly predictive UV completion of the SM, as reviewed in Sec.~\ref{sec:SM} and then investigate consequences for the beyond SM part of the matter model in question. Phenomenologically, it is also viable to choose $f_g$ and $f_y$ beyond the lower bounds that are needed in order to prevent Landau poles in the SM; in this case, the SM couplings become asymptotically free and one may either also render all other couplings in the model in question asymptotically free or obtain predictions for some of them.

Thus, given a matter Lagrangian with dimension-four-interactions, we can use just the three parameters $f_g$, $f_y$ and $f_{\lambda}$, to map out the potential constraints coming from asymptotic safety. It is worth emphasizing that this is a very ``economical'' parameterization, with just three free parameters, even in settings with many gauge, Yukawa and scalar quartic couplings. The robustness of predictions resulting from this parameterization has been explored in \cite{Kotlarski:2023mmr}. 

Given the parameters $f_g$, $f_y$ and $f_{\lambda}$ (and the constraints $f_g>0$, $f_{\lambda}>0$, which are believed to be robust, given numerous explicit studies of them), their values can be varied within an interval such that $|f_i|<1$, in accordance with the assumption of near-perturbativity. This delineates the \emph{maximally possible landscape} of near-perturbatively asymptotically safe theories. The actual landscape is strictly smaller, because the values of the $f_i$'s cannot be varied arbitrarily.\footnote{Even by varying the number of matter fields in a dark sector with no dimension-four-interactions with the visible sector, while gravitational fixed-point values change, the resulting set of values of the $f_i$'s is not expected to be dense within the real numbers.}

Examples of work along these lines can be found in \cite{Eichhorn:2018whv,Reichert:2019car,Kowalska:2020gie,Kowalska:2020zve,Alkofer:2020vtb,Boos:2022pyq,Eichhorn:2022vgp,Kowalska:2022ypk,Chikkaballi:2022urc,Chikkaballi:2023cce,Kotlarski:2023mmr,deBrito:2023ydd,Kowalska:2024lnm,Carone:2025wjw,Chikkaballi:2025pnw,Chen:2025muq} for beyond SM settings, as well as \cite{Eichhorn:2025sux} in the SM. These works highlight that it is possible to exclude some phenomenologically interesting models by requiring consistency with asymptotic safety, while other models are severely constrained. \\

Note that we have not included a non-minimal coupling of the scalar fields to gravity here, which is another dimension-4-interaction. To include non-minimal couplings $\xi_i \phi_i^2 R$ between scalar fields $\phi_i$ and the Ricci scalar, we can similarly parameterize
\begin{equation}
\beta_{\xi_i} =- f_{\xi}\,\Theta(k-M_{\rm Pl}) \,  \xi_i + \beta_{\xi_i}^{\rm(matter)} +\mathcal{O}(\xi_i^2),
\end{equation}
and account for a $\xi_i^2$-term in $\beta_{\lambda_i}$, see, e.g., \cite{Eichhorn:2020sbo}. The sign of $f_{\xi}$ depends on the fixed-point values of the gravitational couplings. Except for a relatively small region of this parameter space, the sign is positive, rendering $\xi$ an irrelevant coupling.\newline

For higher-order couplings, for which the global symmetries of the kinetic terms (i.e., with covariant derivatives which only contain the gravitational connection, but with the couplings to all gauge connections set to zero) are respected, there is an additional gravitational contribution besides an anomalous scaling dimension. This additional contribution prevents the existence of a Gaussian fixed point for these couplings, see Sec.~\ref{sec:SMEFT}. At a heuristic level, this contribution follows the intuition that gravity mediates new interactions. That all these interactions are necessarily dimension-six or higher interactions follows from an inspection of the kinetic terms of scalar, fermions and vectors. The symmetries realized by these terms do not admit any dimension-4 (or 5) interactions. However, in a near-perturbative regime, the resulting shift of the Gaussian fixed point to an interacting, "shifted Gaussian'' fixed point, does not matter much: below the Planck scale the canonical scaling of such couplings drives them to zero, i.e., the dimensionful coupling (or corresponding Wilson coefficient) is generically Planck-scale suppressed. See Sec.~\ref{sec:SMEFT} for a further discussion of such couplings in phenomenological contexts.\\

Finally, let us highlight that in some settings, e.g., in cosmology, it is of interest to go beyond the quartic order in the potential for a scalar field. In fact, it holds that the beta function for each coupling $\lambda^{(i)}$ in a Taylor expansion of a scalar potential
\begin{equation}
V(\phi) = \sum_i\, \lambda^{(i)}\, \phi^i,
\end{equation}
 contains a term 
\begin{equation}
\beta_{\lambda^{(i)}}= - f_{\lambda}\, \Theta(k-M_{\rm Pl})\,\lambda^{(i)}+\dots\,\,.
\end{equation}
Therefore, each term is shifted towards irrelevance by the same amount. Depending on the size of $ f_{\lambda}$, the linear, quadratic and cubic term can remain relevant, but, because $f_{\lambda}>0$, the quartic one never does. Consequently, a given phenomenologically desirable form of the scalar potential, e.g., for the inflaton, may not be achievable in asymptotic safety.\footnote{We highlight that the effective potential, which is the potential obtained from the effective action upon setting the field to a constant, is strictly convex. In physical situations in which the potential should not be convex, e.g., in cases with spontaneous symmetry breaking, one obtains the convex hull.} While this has not been systematically explored and the landscape of allowed scalar potentials is not comprehensively mapped out, it is already clear that this leads to significant restrictions, e.g., for dark matter \cite{Eichhorn:2017als}, grand unified theories \cite{Eichhorn:2019dhg} as well as inflationary potentials.\\

To close, let us highlight that such a parameterized approach relies on several assumptions. First, it applies only in a near-perturbative regime, where, e.g., the effects of operator mixing at the fixed point are subleading compared to the direct quantum-gravity effects that are included. Second, its application to phenomenological questions assumes that results obtained in Euclidean signature on the signs of various $f$'s actually hold in Lorentzian signature as well. Keeping these caveats in mind, the parameterized approach can be powerful in charting the landscape of asymptotic safety approximately, e.g., in order to identify which phenomenological models of BSM physics or cosmological settings are particularly promising and deserve a more in-depth study based on a fully-fledged functional RG study.

\section{Relation to other approaches to quantum gravity}\label{sec:relations}
%
\emph{...where we motivate why a fruitful view of distinct quantum-gravity approaches is based on \underline{connection and collaboration}, rather than \underline{division and competition}. First, a concrete motivation from this comes from ``effective asymptotic safety'', which is a proposal for a universal, nearly scale-invariant regime within the effective-field-theory description of gravity, connecting several distinct UV theories to known physics in the IR. Second, it is motivated by the observation that the challenges faced by individual quantum-gravity approaches are typically different ones, and therefore a transfer of concepts and techniques between approaches could trigger progress in resolving these challenges}.\\

We start this section by discussing why it is even a relevant question to ask what the connection of asymptotic safety is to other approaches to quantum gravity. After all, distinct approaches to quantum gravity have traditionally been viewed as  being in competition with each other. This outlook on quantum gravity is suggested by the -- seemingly insurmountable -- differences between the key conceptual ideas and technical formulations of different approaches. One can easily find a set of seeming contrasts -- discrete versus continuous, local versus non-local, formulated in four spacetime dimensions or a higher number of spacetime dimensions, unified with matter or making do without matter, and others -- between popular contenders for a quantum theory of gravity, as reflected, e.g., in \cite{Buoninfante:2024yth}. However, concluding that, based on such contrasts, distinct approaches are actually distinguishable at the level of physical predictions, may be premature \cite{deBoer:2022zka}. After all, there are numerous examples of different formulations of a physical theory (e.g., canonical versus path-integral quantization of QFT, Heisenberg's matrix formulation of Quantum Mechanics versus Schrödinger's formulation based on differential operators, Hamiltonian mechanics versus Lagrangian mechanics and many more). In each of these examples, different concepts and even mathematical frameworks are used, but it is of course also understood in each of these cases how to derive one formulation of the physical results from the other. Similarly, in quantum gravity, the seeming distinctions between quantum-gravity theories may be there at the level of the mathematical formulation, but this need not imply that differences exist at the level of physical observables. To give a simplistic example, the distinction of discrete versus continuous, even if built into the foundations of the theory, may not be present at an observable level. For instance, the concept of a ``perfect action" is well-known in lattice theories, and refers to an action that yields perfect continuum physics even at finite lattice spacing \cite{Hasenfratz:1993sp,Bahr:2011uj}. In addition, even if differences in different approaches to quantum gravity exist in observables at the Planck scale, there may still be universality in the effective description of black holes, the early universe, as well as the constraints on matter theories etc. In other words, the ``landscapes'' of these approaches, spanned by those effective field theories that can arise from a consistent UV completion with gravity, may be universal, at least at the level of observables that do not probe the Planck scale.

Based on the above discussion, the possibility of universal physical predictions coming out of distinct approaches to quantum gravity should not be dismissed out of hand, before we actually have a complete enough understanding of these predictions. This motivates the study of relations between distinct approaches \cite{deBoer:2022zka}. \\
A useful quantity to base such a comparison on is the effective action, which can be defined in all physically viable approaches at sufficiently large scales. In the effective action, the full physics of a quantum system is encoded, because all quantum fluctuations are accounted for. Therefore, various details of how quantum fluctuations are parameterized -- which may well be very different in different quantum-gravity approaches -- are integrated out and universal aspects of physics are more directly accessible. In particular, effective actions have been computed (in various approximations) from many different approaches. Besides the functional RG approach to asymptotic safety, which is based on the effective action, an effective action has been computed from Causal Dynamical Triangulations \cite{Ambjorn:2014mra,Brunekreef:2022pns}, see also \cite{Knorr:2018kog}. In Loop Quantum Gravity and spin foams, effective actions have been computed in \cite{Borissova:2022clg,Ferrero:2025est}.\footnote{The effective action computed in \cite{Borissova:2022clg} is based on an area metric. To understand whether it can be connected to an effective action for a standard length metric, it is crucial to determine whether the additional degrees of freedom are massive and can be neglected at sufficiently low energies \cite{Borissova:2023yxs,Borissova:2025frj}.}. In string theory, higher-order curvature terms in the effective action have been computed, e.g.,  in \cite{Fradkin:1985ys,Gross:1986iv,Gross:1986mw,Metsaev:1986yb,Kikuchi:1986rk,Jack:1989vp,Tseytlin:1995bi,Green:1999pv,Green:2010wi}. At the moment, a comparison between the different effective actions is an aspirational projects, because different terms have been computed in different settings; however, a concerted effort aimed at a comparison may be feasible in the future.

A concrete proposal for how asymptotic safety may be related to other approaches, and how it may generate universality in quantum gravity is through the concept of \emph{effective asymptotic safety} \cite{deAlwis:2019aud}, discussed in Sec.~\ref{sec:eff_AS}. In short, effective asymptotic safety is the idea that scale symmetry is approximately realized in a quantum field theory that has a fundamental UV cutoff $\Lambda_{\rm UV}$. Beyond  $\Lambda_{\rm UV}$, there are other degrees of freedom and below $\Lambda_{\rm UV}$, the theory achieves approximate scale symmetry. 

There are two prerequisites for effective asymptotic safety to be realized in a quantum gravity approach: First, $\Lambda_{\rm UV}> k_{\rm trans}$, where $k_{\rm trans}$ is the transition scale between fixed-point scaling and the matter-dominated regime in which gravitational fluctuations are negligible. Often, $k_{\rm trans}$ is essentially equal to the Planck scale, although it may deviate, e.g., in settings with many matter fields \cite{Dona:2013qba}. Second, the values of couplings at $\Lambda_{\rm UV}$ must be such that they are very close to, but not exactly in the critical hypersurface of the fixed point and such that an RG trajectory towards the IR first leads towards the fixed point, stays in its vicinity for a range of scales, and then departs the fixed-point regime along a relevant direction. In practice, this second requirement is difficult to check, because it requires the evaluation of couplings at $\Lambda_{\rm UV}$ within the more fundamental theory. An example taking a step in this direction can be found, e.g., in \cite{Borissova:2025frj}, where additional, area-metric degrees of freedom from spin foams are included in the gravitational RG flow. Further, \cite{Basile:2021euh,Basile:2021krk} conducted a conceptually similar study, but for string theory, calculating RG flows in actions from string theory. This followed up on  \cite{deAlwis:2019aud}, which spelled out several conditions required for effective asymptotic safety to be realizable in string theory.

\begin{figure}[!t]
\includegraphics[width=\linewidth,clip=true, trim=3cm 9cm 29cm 3cm]{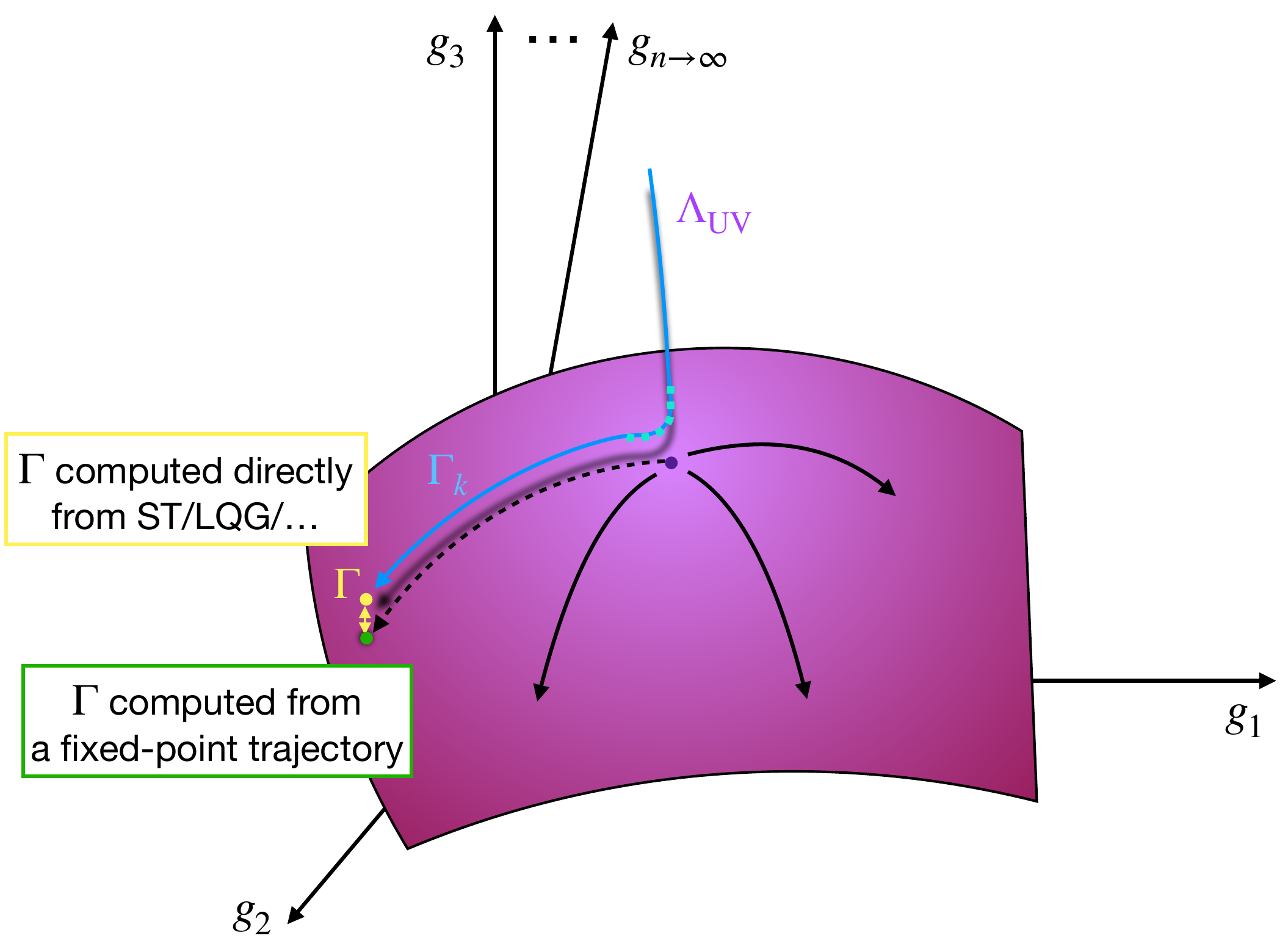}
\caption{\label{fig:convergence_approaches_Gamma} We show the critical hypersurface (purple surface) of an asymptotically safe fixed point (violet dot). ``True'' fixed-point trajectories are indicated in black and end with effective actions within the hypersurface, parameterized by the relevant directions. We also show an effective action $\Gamma$, in yellow, computed directly from another approach, such as string theory, Loop Quantum Gravity etc. This effective action lies close to the critical surface. Therefore, it can be reached by an RG trajectory (in blue) that, when traced ``backwards'' into the UV, passes close to the fixed point and features a nearly scale-invariant regime (indicated by the light blue dashes). Traced further into the UV, some coupling diverges at $\Lambda_{\rm UV}$, which indicates the fundamental scale associated to the more fundamental theory.}
\end{figure}

The effective action is a useful tool to assess whether or not effective asymptotic safety is present: Given an effective action $\Gamma$ that has been computed directly from a given theory (yellow dot in Fig.~\ref{fig:convergence_approaches_Gamma}), one can compare the values of the couplings in this effective action to values that can be achieved in a theory that is fundamentally asymptotically safe, i.e., calculated from a fixed-point trajectory (dotted black line). If these are close to each other, one expects to find an RG trajectory that is slightly ``detuned" from the fixed-point trajectory (blue line). This trajectory does not emanate from the fixed point.  Rather, it has a closest approach to the fixed point for which the distance along an \emph{irrelevant} direction is non-zero. Accordingly, tracing this trajectory backwards into the UV from the vicinity of the fixed point results in a divergence at a scale $\Lambda_{\rm UV}$, which can be associated to the fundamental scale of the theory from which the effective action $\Gamma$ (yellow dot) has originally been calculated.\\
If this construction can be achieved, the RG trajectory (blue) that results in the effective action has an effectively asymptotically safe regime, where the couplings are approximately constant, indicated by the light blue dashes in Fig.~\ref{fig:convergence_approaches_Gamma}. Most of the predictions obtained from this theory will be very close to those obtained from asymptotic safety; there will be a high degree of universality between those predictions.\\

Finally, even if approaches differ at the physical level, it is of course highly plausible, that methods or concepts can be transferred between approaches. An example of this is the spectral dimension, first evaluated in Causal Dynamical Triangulations \cite{Ambjorn:2005db} and asymptotic safety \cite{Lauscher:2005qz}, that has subsequently been studied in virtually all approaches and resulted in an improved understanding of the properties of quantum spacetime in these various approaches. A universal result, namely a dynamical dimensional reduction (of the spectral dimension) to two dimensions, has been found in many Euclidean settings and argued for on physical grounds \cite{Carlip:2017eud}, with \cite{Eichhorn:2013ova} pointing out the possibility of dimensional enhancement in Lorentzian settings.\\

In summary, the physical predictions of quantum gravity approaches -- in other words, their ``landscapes'' -- are not sufficiently well-understood to rule out the possibility of universal predictions, or, conversely, the existence of an ``absolute swampland'' \cite{Eichhorn:2024rkc}. Therefore, a much closer collaboration between quantum-gravity approaches is desirable. Against this background, we discuss in the next subsections, what is known about relations of asymptotically safe gravity to other approaches. The spirit of these sections is largely aspirational, because work in this direction is in its infancy. Thus, we survey which starting points for concrete connections have been established and provide a basis for future studies.

\subsection{Towards relations to string theory and holography}
\emph{...where we highlight that the technical basis for a first step to study the compatibility of string theory with asymptotic safety, namely the study of asymptotic safety in supersymmetric, higher-dimensional settings, exists. In addition, the holographic RG constitutes an explicit point of contact, albeit one which is so far underexplored.}\\

To establish a relation between string theory and asymptotic safety within the effective-asymptotic-safety-paradigm, as proposed in \cite{deAlwis:2019aud} and further studied in \cite{Basile:2021krr,Basile:2025zjc}, it is useful to know whether asymptotic safety persists in a supersymmetric setting and in settings with higher dimensions. If this is not the case, then effective asymptotic safety can only be realized at distance scales larger than the compactification scale and in settings in which supersymmetry is broken at these scales, e.g., by the compactification.
Using regulators in the functional RG that preserve supersymmetry is not straightforward, but has been solved in non-gravitational theories \cite{Synatschke:2008pv,Heilmann:2014iga,Feldmann:2017ooy}, in principle paving the ground for such studies, see, e.g., \cite{DallAgata:2022abm}. Another ingredient of a supersymmetric gravitational setting has been considered in \cite{Dona:2014pla}, where the contributions of $N_{\rm RS}$ spin-3/2 fields, understood as gravitinos (but with a regulator that breaks supersymmetry) have been added to the beta function for the Newton coupling and cosmological constant, resulting in
\begin{equation}
\beta_G = 2 G+ \frac{G^2}{6\pi}\left( -N_{\rm RS} - 46\right),
\end{equation}
where the 46 arises from graviton fluctuations and one can see that gravitinos in this treatment support a gravitational fixed point; see also \cite{Montero:2024sln} for another take on the compatibility of asymptotic safety and supergravity.\\
Regarding higher spacetime dimensions, one should note that the expected number of relevant couplings increases with spacetime dimensionality, at least if the fixed point retains its near-perturbative nature. Thus, less can be concluded from studies in small truncations. In principle, the spacetime dimensionality can be treated as an additional parameter in the functional RG, and settings with more than four dimensions have been studied in gravity in \cite{Fischer:2006fz,Falls:2015qga,Falls:2015cta,Dona:2013qba}. It has also been shown in \cite{Eichhorn:2019yzm} that asymptotic safety may not be achievable in $d>4$, if an Abelian gauge coupling is present, which must also be UV safe (or free).\\
Overall, there does not appear to be an unsolved technical challenge that would prevent a study of asymptotic safety in $d=10$ or $d=11$ and in the presence of supersymmetry. Conversely, under the hypothesis that effective asymptotic safety may be realized in compactifications that break supersymmetry, the question rather becomes whether any of the established properties of four-dimensional asymptotic safety prevents its compatibility with string theory. This question can also be addressed in the context of the swampland program, see below.

Holographic ideas in principle provide several points of contact to asymptotic safety, however, none of them has as of yet been investigated in depth or substantiated, see also the discussion in \cite{Bonanno:2020bil}. First, holographic dualities relate (quantum) gravitational theories to conformal field theories. Whether the full conformal group can be realized at an asymptotically safe fixed point is not clear, because scale symmetry plus Poincare symmetry is strictly less than full conformal symmetry, although the former implies the latter in low-dimensional settings. Accordingly, it is an intriguing and relevant question whether asymptotic safety fits into a holographic duality -- either as part of a boundary theory, which would exceptionally also include gravity, or as a special bulk theory which would already exhibit part of the conformal symmetry of the boundary theory, namely scale symmetry. At the time of writing, this question is largely unexplored, see \cite{Ferrero:2022dpk} for a proposal within a non-standard holographic setting.\\
In addition, in holographic setups the holographic RG \cite{Bianchi:2001kw} has been formulated to integrate along the radial direction in the bulk theory and connect to the boundary theory. This is another example of RG concepts and tools becoming relevant in quantum gravity and a specific point of contact to asymptotic safety \cite{Litim:2011qf} which deserves further study, see \cite{Heemskerk:2010hk,Radicevic:2011py} for ideas how the holographic RG may be connected to the Wilsonian RG.

In addition to the idea of effective asymptotic safety, there may also be a benefit in a transfer of methods between the approaches, e.g., it has been pointed out in \cite{Basile:2021euh} that functional RG techniques can also be useful in a string-theoretic context, specifically for the calculation of $\alpha'$ corrections.
%

\subsubsection{What is the relation of asymptotically safe gravity to the swampland program?}\label{sec:AS_swamp}
\emph{...where we highlight three current avenues of research into the relation of asymptotically safe models to the string-inspired swampland. We emphasize that the idea of the swampland resonates with attempts to quantify and leverage the predictive power of asymptotic safety. Mapping out the asymptotically safe landscape and understanding its relation to the string landscape therefore go hand in hand.}\\

The swampland program has its origin in string theoretic considerations \cite{Vafa:2005ui,Ooguri:2006in}. Its main assumption is that not all seemingly consistent effective field theories (i.e., effective field theories which satisfy standard quantum field theoretic consistency conditions, e.g., unitarity, stability, causality) are actually viable effective descriptions of a fundamental theory which includes quantum gravity. This idea clearly resonates with the assertion that asymptotic safety restricts the field content and imposes relations on the couplings of an effective field theory -- in the most extreme case, these relations imply that phenomenologically crucial couplings must vanish, see, e.g., \cite{deBrito:2021akp,Assant:2025gto} for examples. 

Against this background, three questions are under current investigation: \\
First, given specific phenomenologically interesting effective field theories (e.g., for dark matter or dark energy), can they find a UV completion within asymptotic safety as well as in string theory? While direct comparisons between the two in a single paper are still very rare, distinct papers from asymptotic safety and from string theory sometimes address the same models, for instance, attempting to reproduce or even explain the structure of Yukawa couplings in the Standard Model from quantum gravity \cite{Eichhorn:2017ylw,Eichhorn:2018whv,Constantin:2024yxh,Constantin:2025vyt,Eichhorn:2025sux}. \\
Second, do the string-inspired, general swampland conjectures, which, e.g., prohibit the existence of global symmetries \cite{Banks:1988yz,Kallosh:1995hi} or limit the degree of flatness of scalar potentials \cite{Ooguri:2018wrx,Garg:2018reu}, apply to asymptotic safety as well? Concrete studies on these questions can be found in \cite{deAlwis:2019aud,Basile:2021krr,Basile:2025zjc}. A sharpening of terminology that distinguishes the relative swamplands of different quantum-gravity theories from an absolute swampland and questions the universality of the swampland, is presented in \cite{Eichhorn:2024rkc}.\\
Third, how can one most efficiently map out the ``asymptotically safe landscape"? The landscape is spanned by several qualitatively distinct quantities. First, there are discrete choices (spacetime dimensionality, field content, local and global symmetry groups and representations), second, there are continuous ones (values of relevant couplings).  It may be possible to constrain the values of some of the discrete choices, e.g., spacetime dimensions larger than five appear to be incompatible with having a U(1) gauge theory at the fundamental level \cite{Eichhorn:2019yzm}, but the understanding of the resulting discrete landscape is far from complete. Mapping out the possible Wilson coefficients within a given effective field theory from the values of relevant couplings in the theory is done, e.g., in \cite{Basile:2021krr,Knorr:2024yiu,Saueressig:2024ojx}. In addition, many papers derive the relations between couplings resulting from asymptotic safety without phrasing these results in terms of ``mapping out the asymptotically safe landscape". An early example within pure gravity is \cite{Benedetti:2009rx}, an example within a dark-matter model can be found in \cite{Eichhorn:2020kca}. Such results give a hint at the relative smallness of the asymptotically safe landscape, e.g., in \cite{Eichhorn:2020kca}, within an eight-dimensional space of couplings, only two combinations of couplings constitute relevant directions.\\

Given the wealth of distinct swampland conjectures, see \cite{Palti:2019pca,vanBeest:2021lhn,Grana:2021zvf,Agmon:2022thq,Lehnert:2025izp} for reviews, there are many concrete questions to be investigated in asymptotic safety to determine whether or not there is overlap between the swamplands of string theory and asymptotic safety, as started in \cite{deAlwis:2019aud,Basile:2021krr,Knorr:2024yiu,Eichhorn:2024rkc,Basile:2025zjc}.

\subsection{Towards relations to discrete approaches through a universal continuum limit}
\emph{...where we discuss the relation of asymptotic safety to several distinct approaches that invoke spacetime discreteness at some level. Besides the idea of effective asymptotic safety that may be realized in such approaches at distance scales larger than the discreteness scale, there is a second paradigm for this relation. This is the paradigm of a universal continuum limit, in which the continuum physics is independent of the details of the discretization. In this paradigm, the discretization is interpreted as a regularization}.

\subsubsection{Towards a relation to Loop Quantum Gravity}\label{sec:LQG}
Loop Quantum Gravity \cite{Ashtekar:1986yd,Ashtekar:1987gu}, both in its canonical as well as its covariant (spin-foam) formulation, reviewed, e.g., in \cite{Ashtekar:2021kfp} shares various points of contact with asymptotic safety at the technical level: first, they share the assumption that quantum gravity can be described by a quantum field theory which must be treated beyond perturbation theory. Second, they share the idea that matter fields, needed to make contact with the ``real world", do not emerge from an underlying, fundamental structure that unifies gravity and matter, but rather, that they must be added to the theory by hand. There are also physical points of contact, e.g., the idea that the theory should give rise to regular black-hole solutions \cite{Bonanno:2000ep,Eichhorn:2022bgu,Platania:2023srt,Ashtekar:2023cod}. Commonalities have also been pointed out in the recent review \cite{Ferrero:2025efd}.

Nevertheless, at the conceptual level, there appears to be a tension between the assertion that geometric operators, such as the area operator, have a discrete spectrum and a smallest eigenvalue in Loop Quantum Gravity \cite{Rovelli:1994ge} and the assertion that asymptotic safety should give rise to a regime in which spacetime geometry is scale-invariant, i.e., free of distinct scales. However, RG techniques and the search for universality, generated by an RG fixed point, actually play an important role in Loop Quantum Gravity, and specifically its spin-foam incarnation: there, a given spin foam, loosely speaking a spacetime lattice ``decorated" with variables that encode geometric properties of spacetime, is interpreted as a regularization. Put differently, transition amplitudes between distinct spatial geometries depend on the spin network that is used to calculate the amplitude.
Therefore, the physical limit, in which diffeomorphism symmetry is restored \cite{Bahr:2011uj}, is the continuum limit, which should be universal and independent of the choice of regulator. This requires the continuum limit to correspond to an RG fixed point. To search for such a fixed point while respecting the key feature of spin foams, namely background independence, background-independent real-space RG techniques for spin-foam models have been developed \cite{Dittrich:2013xwa,Dittrich:2014ala}, see \cite{Steinhaus:2020lgb,Asante:2022dnj} for reviews. 
Numerical simulations within simplified configuration spaces provide evidence for a phase transition \cite{Bahr:2016hwc}. Most recently, \cite{Han:2026hyq} identified a UV fixed point.
Note that this continuum limit (or refinement limit) is expected to keep the property of a discrete area spectrum intact \cite{Rovelli:2011fk}. This exemplifies the difference between a formal continuum limit, which the mathematical structures used in constructing the theory satisfy, and physical discreteness, which the physical values of physical operators can exhibit in such a limit. \\
It should be stressed that in Loop Quantum Gravity, the RG flow does not have a straightforward interpretation in terms of momentum shells, because that would require a preferred, flat background, at odds with the background-independent construction of Loop Quantum Gravity. Thus, for instance in group field theory, an approach closely related to spin foams, functional RG techniques, pioneered in \cite{Benedetti:2014qsa}, generalize those developed for matrix models \cite{Eichhorn:2013isa} and rely on an RG flow that is ``pregeometric'' in that it does not refer to scales associated to a spacetime (or its tangent space). Studies of the RG flow in group field theories have been performed, e.g., in \cite{BenGeloun:2015xrk,BenGeloun:2016rqa,Carrozza:2016tih,Lahoche:2018oeo,Lahoche:2018hou,BenGeloun:2018ekd,Lahoche:2019vzy,Lahoche:2019orv,Pithis:2020kio,Geloun:2023ray,Yerima:2026jzx} to investigate phases and phase transitions in these theories. In particular, in some settings, evidence for an asymptotically safe fixed point has been uncovered \cite{BenGeloun:2016tmc,Juliano:2024rgu}, although in general not with an appropriate number of relevant directions to be a counterpart of the Reuter fixed point. 
\\
There is also a complementary approach, based on Hamiltonian renormalization in the canonical Loop Quantum Gravity framework \cite{Lang:2017beo,Thiemann:2020cuq,Zarate:2025erv}. \\
It is of course not guaranteed that a fixed point identified in spin foams or canonical Loop Quantum Gravity agrees with the Reuter fixed point. After all, asymptotic safety is simply a general concept for QFTs, but may be realized in distinct gravitational quantum theories, i.e., give rise to distinct universality classes. Therefore, a comparison of the critical exponents characterizing the distinct fixed-point candidates is needed in order to decide whether the distinct approaches all rely on asymptotic safety, but on different universality classes, or whether they recover the same universality class. In the second case, the theories would agree at the physical level.\\

Along a different line of approach, a concrete connection between Loop Quantum Gravity and asymptotic safety was recently established following the general derivation of a path integral from a propagator in a canonically quantized setting. In practice, starting from the Loop Quantum Gravity framework on a reduced phase space, a nonzero Hamiltonian can be derived, from which in turn a path integral can be deduced that can be studied with functional RG techniques \cite{Thiemann:2024vjx,Ferrero:2024rvi,Ferrero:2025idz}. This setting reproduces an asymptotically safe fixed point in the Einstein-Hilbert truncation that is compatible with results achieved in the standard functional RG approach.\\

As a last point of contact between asymptotic safety and Loop Quantum Gravity, functional RG techniques have been used in contexts explicitly motivated by Loop Quantum Gravity. This is a priori independent of physical agreement between the two settings, and merely implies that the same set of methods is useful in both settings. For instance, the RG flow of the Immirzi parameter (which governs, e.g., the spectrum of the area operator) has been computed \cite{Harst:2012ni,Daum:2013fu}, also in a setting with area metric degrees of freedom \cite{Borissova:2025frj}. Nevertheless, the development of the functional RG for these settings paves the ground for future studies that investigate whether effective asymptotic safety may be realized in these settings.\\

Overall, there is a remarkable convergence of ideas between Loop Quantum Gravity and asymptotically safe gravity, with Renormalization Group techniques and concepts growing in importance in both settings. It is, therefore, a distinct possibility that not only can there be a fruitful transfer of methods, but that the two approaches may share universal features at the physical level.

\subsubsection{Towards a relation to tensor models and (causal) dynamical triangulations}
See Sec.~\ref{sec:methods}. In these frameworks, the discretization is explicitly introduced as a regularization and so the theory only has a physical limit if a universal regime, i.e., asymptotic safety, exists. This does not guarantee that the universality class has to be that of the Reuter fixed point, because there may be distinct gravitational universality class, but it certainly appears to be the most conservative assumption. A first explicit comparison of RG flows and a critical exponent has been done in \cite{Ambjorn:2024qoe} and the effective action corresponding to the CDT de-Sitter-like phase reconstructed in \cite{Knorr:2018kog}.

\subsubsection{Towards a relation to causal sets}
Causal sets are traditionally viewed as a fundamentally discrete approach \cite{Bombelli:1987aa}. The Lorentzian path integral over all geometries becomes a Lorentzian path integral over all causal sets, which are discrete causal orders, see \cite{Surya:2019ndm} for a review. Nevertheless, there is no conceptual issue with re-interpreting the discreteness as a regularization. Most interestingly, it then constitutes a regularization of Lorentzian, not Euclidean geometries. This makes the search for a universal continuum limit, i.e., a higher-order phase transition, in this setting highly relevant, because it would constitute evidence for Lorentzian quantum gravity, as proposed in \cite{Eichhorn:2017bwe,Eichhorn:2019xav}. So far, the phase diagram has been mapped in restricted, low-dimensional configuration spaces \cite{Surya:2011du,Cunningham:2019rob}, but accounting for higher-curvature interactions may well be needed to recover a higher-order phase transition that agrees with the Reuter universality class \cite{deBrito:2023axj}.

\section{Current and future frontiers}\label{sec:frontiers}
Asymptotically safe quantum gravity has undergone a rapid development over the last decade. Several critical questions that have been open not many years ago \cite{Bonanno:2020bil}, are now being answered.  Challenging issues like the fate of the Reuter fixed point in Lorentzian signature, the unitarity of the theory, as well as its implications for (beyond) Standard Model physics are now being addressed. It is particularly encouraging that in this development, to date, no unexpected roadblocks have been uncovered. Rather, points that were long regarded as likely stumbling blocks, such as the question of unitarity, may turn out to not actually be problematic.\\
 Therefore, summarizing the developments reviewed above, the formulation of an asymptotically safe quantum gravity theory is progressing with an increasingly positive outlook.  In this context, it is worthwhile to stress that this does not mean that other approaches to quantum gravity must consequently be in decline. Rather, it is the perspective advocated in this review that a search for universal features of distinct approaches, both at the level of methods, but in particular at the level of physical consequences, is highly worthwhile to undertake. Asymptotic safety, in its incarnation of effective asymptotic safety, may provide a universal regime that is shared by a number of distinct approaches.\\

This review has focused on physical implications of asymptotic safety, advocating that the connection to observations -- difficult as it may be to establish in quantum gravity -- is critical to determine whether ideas pursued in asymptotically safe quantum gravity and other approaches are worthwhile to develop further. It is particularly encouraging that features of the Standard Model of particle physics may not just be recovered, but potentially even explained in an asymptotically safe setting. \\
This includes the setting of effective asymptotic safety, in which an approximately scale-invariant regime may constitute a universal part of several quantum-gravity approaches, thereby connecting them to testable predictions in particle physics. \\
Over the last decade, asymptotic safety has been remarkably successful in making contact with particle physics in and beyond the SM, and several works extend the predictive power to questions relevant to cosmology, such as the nature of dark matter or dark energy. Building on this success, it appears viable to hope that the landscape of asymptotically safe effective field theories of beyond SM matter and cosmological models is rather limited in size and can be mapped in significant parts. \\
There is scope for actual predictions in particle physics; for instance, while several popular dark-matter models are tentatively ruled out in asymptotic safety\footnote{This holds under several assumptions, including that Euclidean-signature results carry over to Lorentzian signature. These assumptions must be tested in future work.}, there is as of yet no compelling candidate for the true nature of dark matter in asymptotic safety. In particular, being guided by the generic structures that asymptotically safe models exhibit (e.g., the presence of higher-order interactions with a given set of global symmetries; the absence of many free parameters in scalar potential; the irrelevance of dimension-5-interactions), one may hope that a proposal for the nature of dark matter can be developed that is grounded within asymptotic safety. Such a construction, that goes beyond testing the theoretical viability of an existing phenomenological model of dark matter in asymptotic safety, has not yet been made.\\

Besides the existing research lines within asymptotic safety, there are many further questions to be explored. Besides the obvious -- and highly important goals -- of establishing whether the Reuter fixed point, together with SM matter, exists in Lorentzian signature and gives rise to a unitarity theory, a crucial goal is to establish the form of the full effective action. This can be understood as the goal to map out the landscape of asymptotic safety. Most importantly, having a robust understanding of the form of the effective action at high curvature opens the door to a study of the origin of the universe and black-hole spacetimes.
In fact, a number of related questions revolve around the role of black holes in the theory. \\
First, it is critical to discover whether the regular black holes, constructed as asymptotic-safety-inspired, phenomenological models through the procedure of RG improvement, are actually realized as solutions to the full gravitational dynamics. \\
Second,  understanding the role of black holes in scattering events is virtually unexplored in asymptotic safety and crucially connected to the question whether an asymptotically safe regime can in principle be accessed through high-energy scattering, or whether black-hole formation shields us from this regime. Moreover, black-hole thermodynamics is not understood in asymptotic safety.\footnote{Recent development from rather distinct directions highlight that in a non-stationary spacetime, an old black hole of a given mass may have a rather distinct interior from a young black hole \cite{Rovelli:2024sjl,Alexandre:2024nuo}; understanding how a black hole forms and evolves in asymptotic safety and what its thermodynamic properties are therefore appears critical \cite{Basile:2025zjc}.} This also links to the fact that swampland conjectures such as the no-global-symmetry-conjecture are not established in asymptotic safety and the fact that asymptotic safety may constitute an exception to this conjecture underlies many developments in asymptotically safe particle physics. 
Third, the role of black holes is more broadly connected to questions of UV/IR-mixing, which is often suggested to be at odds with asymptotic safety. We stress that UV/IR-mixing may refer to some properties of the theory, or specific sets of observables, but does not in general invalidate the idea of an asymptotically safe regime in the UV.  One particular aspect of UV/IR mixing is related to holographic ideas, which are not well-explored in asymptotic safety. On the one hand, it appears that starting from a gravity theory that already has scale symmetry, constructing a boundary theory that is a CFT should be easier than if the bulk theory does not have scale symmetry. However, except for several works that indicate that the Reuter fixed point is compatible with a hyperbolic background spacetime \cite{Falls:2016msz,Burger:2019upn}, this question has not been systematically explored. It also remains to be understood whether the Reuter fixed point itself defines an actual CFT or merely a scale-symmetric theory.\\
Fourth, astrophysical black holes constitute the most strong-gravity regime we have observational access to. A quick back-of-the-envelope estimate, based on the curvature at the horizon, would indicate that quantum-gravity effects are safely negligible. This conclusion may be premature. First, near-extremal black holes may be very sensitive even to Planck-scale corrections \cite{Eichhorn:2022bbn,Horowitz:2023xyl}, although it is unclear whether near-extremality is achievable astrophysically, see the discussion in \cite{Eichhorn:2022bbn}. Second, black-hole uniqueness is a special property of GR and does not generically hold beyond GR, where additional branches of black-hole solutions may exist. If such branches are selected based on the initial conditions during, e.g., an astrophysical gravitational collapse, then the horizon curvature is not a useful diagnostic to determine whether or not higher-order curvature terms in the effective action modify the solution. Gravitational-wave calculations in higher-order theories, such as \cite{Witek:2018dmd,Corman:2022xqg,Held:2025ckb,Capuano:2026lhs}, complemented by studies in asymptotic safety of admissible coupling values \cite{Falls:2020qhj,Marino:2026}, provide an explicit pathway to mapping out asymptotically safe imprints on gravitational-wave signatures. Finally, gravitational waves are not only generated by binary coalescences, but may also originate from particle physics beyond the SM. For instance, cosmic strings or first-order phase transitions result in stochastic gravitational wave backgrounds and are constrained by observations by pulsar timing arrays \cite{NANOGrav:2023hvm}. In turn, such new physics may be constrained by asymptotic safety \cite{Eichhorn:2023gat}, deserving further study.

Overall, many of the questions listed here can be tackled on the basis of the existing developments in the field. Thus, it appears not unrealistic to hope that over the next decade(s), it can be established robustly, whether asymptotic safety is an intrinsically consistent, theoretically viable approach to quantum gravity. Moreover, there is reason to believe that, due to its high predictive power, the approach can make genuinely testable predictions and can thereby not only be established theoretically, but either ruled out or supported by a confrontation with experimental or observational data.\\
 To close, it is worth highlighting that asymptotic safety need not be realized as a ``fundamental'' theory to have predictive power. Rather, its predictive power carries over to the scenario of ``effective asymptotic safety'' in which a scale-symmetric regime is approximately realized over a finite range of scales. It is a particularly attractive possibility that such a regime is universally shared between distinct approaches to quantum gravity, resulting in universal physical predictions.\\

{\bf Acknowledgements:}
I would like to express my gratitude for discussions and collaborations on asymptotically safe quantum gravity with numerous colleagues; these discussions and collaborations have shaped the views expressed in this review. These include in particular Diego Buccio, Raúl Carballo-Rubio, Alicia Castro, Gustavo de Brito, Pedro Fernandes, Moritz Gessner, Holger Gies, Zois Gyftopoulos, Aaron Held, Lidia Marino, Benjamin Knorr, Jan Pawlowski, Roberto Percacci, Antonio Pereira, Alessia Platania, Shouryya Ray, Manuel Reichert, Martin Reuter, Frank Saueressig, Marc Schiffer, Fabian Wagner, Christof Wetterich, Fabian Willaschek, Masatoshi Yamada as well as  all  former and current members of the quantum gravity group at Heidelberg University not named individually.

I would also like to thank Aaron Held, Benjamin Knorr, Antonio D.~Pereira, Alessia Platania and Marc Schiffer for feedback on a draft of this review.

Finally, I am also grateful to the organizers and participants of the two recent conferences, ``SCALE 2026: strings \& cosmology'' and ``Loops 26'' for constructive and inspiring discussions about the relations between different approaches to quantum gravity, which have influenced the writing of the corresponding section of the review.

 I acknowledge the European Research Council's (ERC) support under the European Union’s Horizon 2020 research and innovation program Grant agreement No.~101170215 (ProbeQG). I also acknowledge support by the Deutsche Forschungsgemeinschaft (DFG, German Research Foundation) under Germany’s Excellence Strategy EXC 2181/1 - 390900948 (the Heidelberg STRUCTURES Excellence Cluster).

\bibliographystyle{apsrev4-1}
\bibliography{references.bib}

\end{document}